\documentclass[nofootinbib, aps, prb, twocolumn, floatfix, longbibliography]{revtex4-1}
\usepackage{amsfonts,bm}
\usepackage{amsmath}
\usepackage{amssymb}
\usepackage{amsthm}
\usepackage{graphicx}
\usepackage{graphics}
\usepackage{subfigure}
\usepackage{color}
\usepackage{bm}
\usepackage[colorlinks, breaklinks=true,linkcolor=red, citecolor=blue, linktocpage=true]{hyperref}
\usepackage{rotating}
\usepackage{comment}

\renewcommand{\>}{\rangle}
\newcommand{\be}{\begin{equation} }
\newcommand{\ee}{\end{equation} }
\newcommand{\ba}{\begin{eqnarray} }
\newcommand{\ea}{\end{eqnarray} }

\newcommand{\al}{\alpha}
\newcommand{\ga}{\gamma}
\newcommand{\ka}{\kappa}
\newcommand{\bpm}{\begin{pmatrix}}
\newcommand{\epm}{\end{pmatrix}}
\newcommand{\bmm}{\begin{matrix}}
\newcommand{\emm}{\end{matrix}}

\begin{document}

\title{Multi-flavor string-net models}
\author{Chien-Hung Lin}
\affiliation{Department of Physics, University of Alberta, Edmonton, Alberta T6G 2E1, Canada}

\begin{abstract}
	We generalize the string-net construction to multiple flavors of strings, each of which is labeled by the elements of an abelian group $G_i$. The same flavor of strings can branch while different flavors of strings can cross one another and thus they form intersecting string-nets. We systematically construct the exactly soluble lattice Hamiltonians and the ground state wave functions for the intersecting string-net condensed phases. We analyze the braiding statistics of the low energy quasiparticle excitations and find that our model can realize all the topological phases as the string-net model with group $G=\prod_i G_i$. In this respect, our construction suggests several ways of building lattice models which realize topological order $G$. They correspond to intersecting string-net models with various choices of flavors of strings associated with different decomposition of $G$.
In fact, our construction concretely demonstrates the K\text{\"u}nneth formula by constructing various lattice models with the same topological order.
As an example, we construct the $G=\mathbb{Z}_2\times \mathbb{Z}_2 \times \mathbb{Z}_2 $ string-net model which realizes a non-abelian topological phase by properly intersecting three copies of toric codes. 

\end{abstract}

\maketitle

\affiliation{} 

\section{Introduction}
Topological phases are gapped quantum phases of matter which support quasiparticle excitations with fractional statistics \cite{WenBook}. The classical examples of topological phases include fractional quantum Hall states and spin liquids. These phases can not be understood by the Landau symmetry breaking theory. Thus it requires new approaches to study them. One useful approach is the construction of exactly soluble lattice models that realize these topological phases.

The toric code model of Ref. \onlinecite{KitaevToric} is one of the simplest examples of exactly soluble lattice model which realizes $\mathbb{Z}_2$ topological order. The model is a spin-1/2 system with spins living on the links of the square lattice. The Hamiltonian is a sum of commuting projectors and thus is exactly soluble. One interesting aspect of toric code model is that its ground state can be thought of as a closed loop condensate. Levin and Wen\cite{LevinWenstrnet} generalized this picture and constructed the ``string-net'' models whose low energy effective degrees of freedom are extended objects called string-nets---a network of strings.

Like the toric code, the string-net models\cite{LevinWenstrnet} are also exactly soluble lattice spin models which can realize a large class of topological phases such as phases whose low energy effective theories are finite lattice gauge theories and doubled Chern-Simons theories. 
Recently, the string-net construction was generalized\cite{LevinWenstrnet,LanWen13,HuWanWu12,MesarosRan13,KitaevKong,Kong12} to realize all topological phases which support a gapped edge. 

The string-net models provide a nice physical picture for realizing topological phases--condensation of string-nets.
In this paper, we extend the picture to multiple flavors of string-nets. 
One way to think of different flavors of string-nets is to imagine a multi-layer system where various string-nets sit on different layers. We can then obtain a two-dimensional system with multi-flavor string-nets by letting the layer spacing $d\rightarrow 0$. 
In this way, string-nets of the same flavor can branch and string-nets of different flavors cross one another. Thus they form intersecting string-nets.
We ask the question: what topological phases can be obtained from the intersecting string-nets? 

We answer this question for a simpler case where each flavor of string-nets is associated with an abelian group $G_i$. Depending on the interactions between different flavors of string-nets, we may obtain various topological phases. In the work, we restrict our attention to the subset of interactions which do not change the string types, namely the interactions are diagonal in the string-net state basis. 

Our analysis is based on an explicit construction: we systematically construct all intersecting string-net models with interactions between different flavors of string-nets. For each model, we analyze the quasiparticle braiding statistics. From this analysis, we find that the multi-flavor string-net model can realize all the topological phases as the string-net model with $G=\prod_i G_i$.
In this regard, multi-flavor string-net models associated with $\{G_i\}$ can be viewed as an alternative construction of the original string-net models with $G$. Moreover, our construction also provides several ways to build lattice models with topological order $G$ corresponding to various decomposition of $G$ into $G=\prod_i G_i$ and thus different flavors of string-nets.

Specifically, our construction starts with a set of string-net models with $\{G_i\}$ associated with each flavor of string-nets. We then intersect/stack the set of string-net models in a proper way so that the resulting model is exactly soluble. We find that the model realizes all topological phases with topological order $G=\prod_i G_i$. One can also start with the other set of string-net models with $\{G_i'\}$ such that $G=\prod_i G_i'$. Our construction then gives a different exactly soluble model which also realizes topological order $G$.

Intuitively, one can imagine that we decompose the string labeled by the elements of $G$ into multiple strings each of which is labeled by the elements of $G_i$ such that $G=\prod_i G_i$. We then put each component string into an individual string-net model. However, these component strings are not independent but satisfy certain constraints. These constraints dictate how different component strings intersect with one another. 
As a result, the model built from the intersecting string-net models with $\{G_i\}$ associated with the component strings gives a ``parton'' construction of the string-net model with $G=\prod_i G_i$. Like the usual parton construction of particles, we have various ways of decomposing strings while they all describe the same topological phases.

In contrast to the original string-net models whose input is a set of complex functions $F$ which satisfy 3-cocycle conditions, we encode the information of $F$ into simpler objects, called $F^{(2)},F^{(3)}$, which satisfy 2-cocycle and 1-cocycle conditions respectively. The objects $F^{(2)}$ and $F^{(3)}$ are associated with intersections between two and three flavors of string-nets. It turns out that the underlying mathematical structure of our construction is the K\text{\"u}nneth formula.    

The advantage of using $F^{(2)}$ and $F^{(3)}$ objects as input is that it provides a simple way to construct models which realize a ``twisted'' topological phase.
For example, one can start with 3 copies of toric codes and then intersect them with properly chosen $F^{(3)}$. The resulting model can realize a twisted $\mathbb{Z}_2\times \mathbb{Z}_2 \times \mathbb{Z}_2$ gauge theory. Interestingly, this model supports non-abelian quasiparticle excitations which will be discussed in detail later.


The paper is organized as follows. In Sec. \ref{section:review}, we review
some basics of string-net models and abelian string-net models. In Sec. \ref%
{section:1wavefunction}, we warm up by constructing ground state wave functions 
with one flavor of abelian string-nets. 
In Sec. \ref{section:mwavefunction},  \ref{section:H}, we generalize to construct ground state wave functions and lattice Hamiltonians for multi-flavor string-net models.
We analyze the low energy quasiparticle excitations of these
models in Sec. \ref{section:particle}. In Sec. \ref{section:statistics}, we
explicitly compute the quasiparticle braiding statistics for general abelian
string-net models. We discuss the relation between different constructions associated with different choices of flavors of string-nets in Sec. \ref%
{section:connection}. Finally we illustrate our new construction with
concrete examples in Sec. \ref{section:examples}. The mathematical details
can be found in appendices.

\section{Review of string-net models \label{section:review}}

In this section, we briefly review the basic structure of string-nets and
string-net models. We mainly focus on a special class of string-net
models--abelian string-net models. The materials in this section are adapted
from Ref. \onlinecite{LinLevinstrnet}.

A string-net is a network of strings. The strings can come in different
types and carry orientations. In this paper, we focus on trivalent networks
in two-dimensional space, namely, exactly 3 strings meet at each branch
point or node in the network. Thus, we can think of string-nets as trivalent
graphs with labeled and oriented edges in the plane (in the continuum or on
a lattice).

A string-net model is a quantum mechanical model which describe the
dynamics of the string-nets. To specify a string-net model, one has to
provide several pieces of data. First, one has to specify a finite set of
string types $\left\{ a,b,c,...\right\} .$ Second, one has to specify the
dual string type $a^{\ast}$ of each string type $a.$ A string $a$ with a
given orientation corresponds to the same physical state as a string $%
a^{\ast}$ with the opposite orientation. Finally, one has to specify the
branching rules. The branching rules are the set of all triplets of string
types $\left\{ \left( a,b,c\right) ,...\right\} $ which are allowed to meet
at a point.

It is also convenient to include the null string type into the formalism.
The null string type, denoted by $0,$ is equivalent to no string at all.
This string type is self-dual: $0^{\ast}=0$ and thus we will neglect the
orientation of the null string. The associated branching rule is that $%
\left( 0,a,b\right) $ is allowed if $a=b^{\ast}.$

The abelian string-net models are a special class of string-net models
associated with abelian groups. To construct an abelian string-net model
associated with a finite abelian group $G,$ we first label the string types
by the elements of the group $a\in G$ with null string being the identity
element $0.$ Second, we define the dual string $a^{\ast}$ as the group
inverse: $a^{\ast}=-a.$ Finally, we define branching rules by 
\begin{equation*}
\left( a,b,c\right) \text{ is allowed if }a+b+c=0.
\end{equation*}
(Here we use additive notation for the group operation.)

So far we focused on the Hilbert space of the string-net model. We also need
to specify the Hamiltonian. A typical string-net Hamiltonian is a sum of a
kinetic energy term and a string tension term. The kinetic energy term gives
an amplitude for the string-net states to move while the string tension term
gives an energy cost to large string-nets. Depending on the relative size of
the two terms, we have two quantum phases. When the string tension term
dominates, the ground state will contain only a few small strings. When the
kinetic energy term dominates, the ground state will be a superposition of
many large string-net configuration--a string-net condensed phase.

The string-net condensed phases are known to support excitations with
fractional statistics. The wave functions and the
corresponding exactly soluble Hamiltonians for these topological phases are
constructed systematically in Ref. \onlinecite{LevinWenstrnet}. 
In this paper, we will generalize their construction to the Hilbert space which consists of multiple flavors of string-nets, each of which is associated with a group $G_i$.

\section{Single-flavor string-net wave functions \label%
{section:1wavefunction}}

In this section, we review the wave functions and Hamiltonian for abelian
string-net condensed phases. The materials are adapted from Ref. %
\onlinecite{LinLevinstrnet}.

We start with the ground state wave functions for string-net condensed
phases. The wave functions are defined implicitly using local constraint
equations. More specifically, the local constraint equations take the
following graphical form: 
\begin{eqnarray}
\Phi \left( \raisebox{-0.16in}{\includegraphics[height=0.4in]{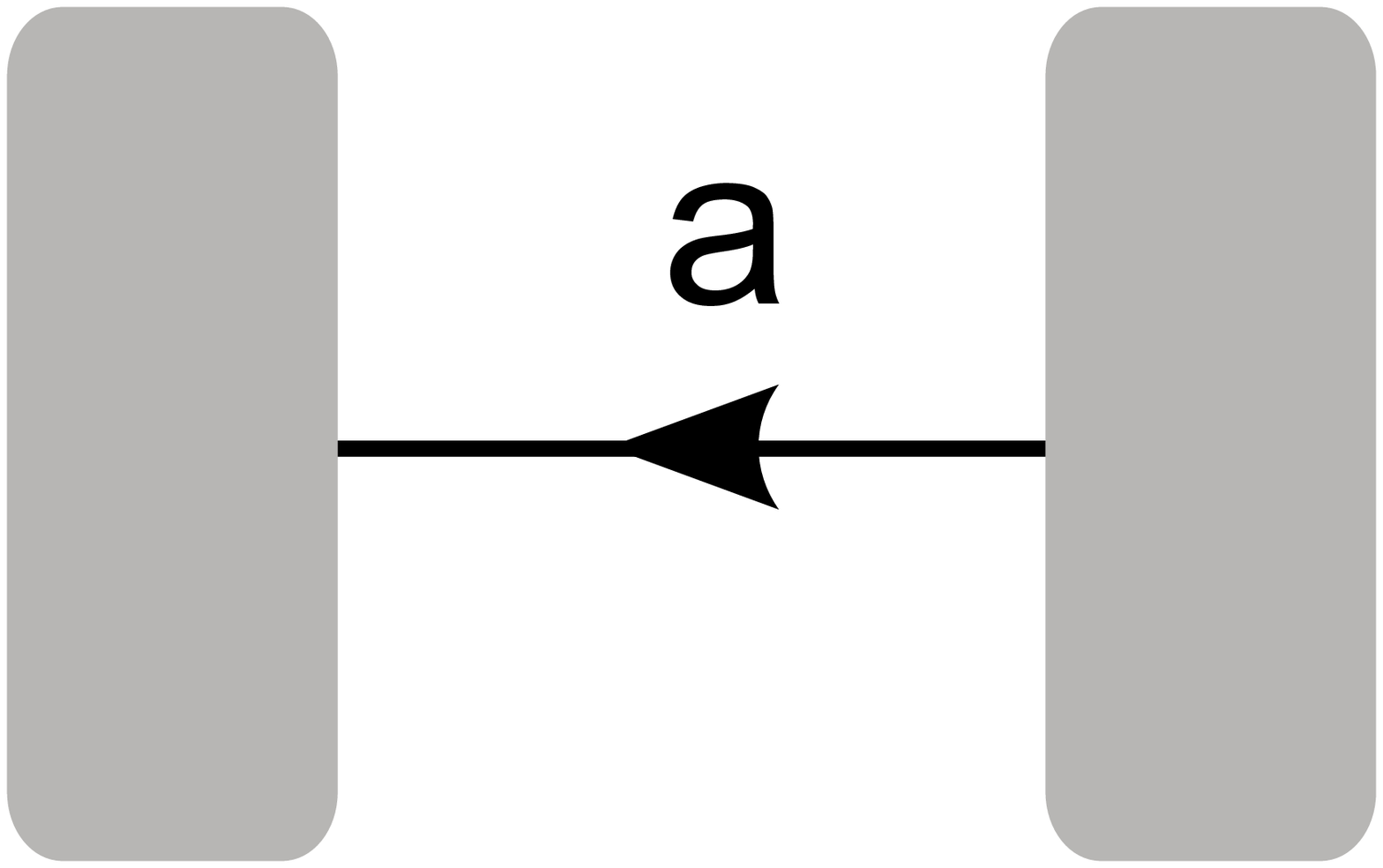}}%
\right) &=& \Phi \left( \raisebox{-0.16in}{%
\includegraphics[height=0.4in]{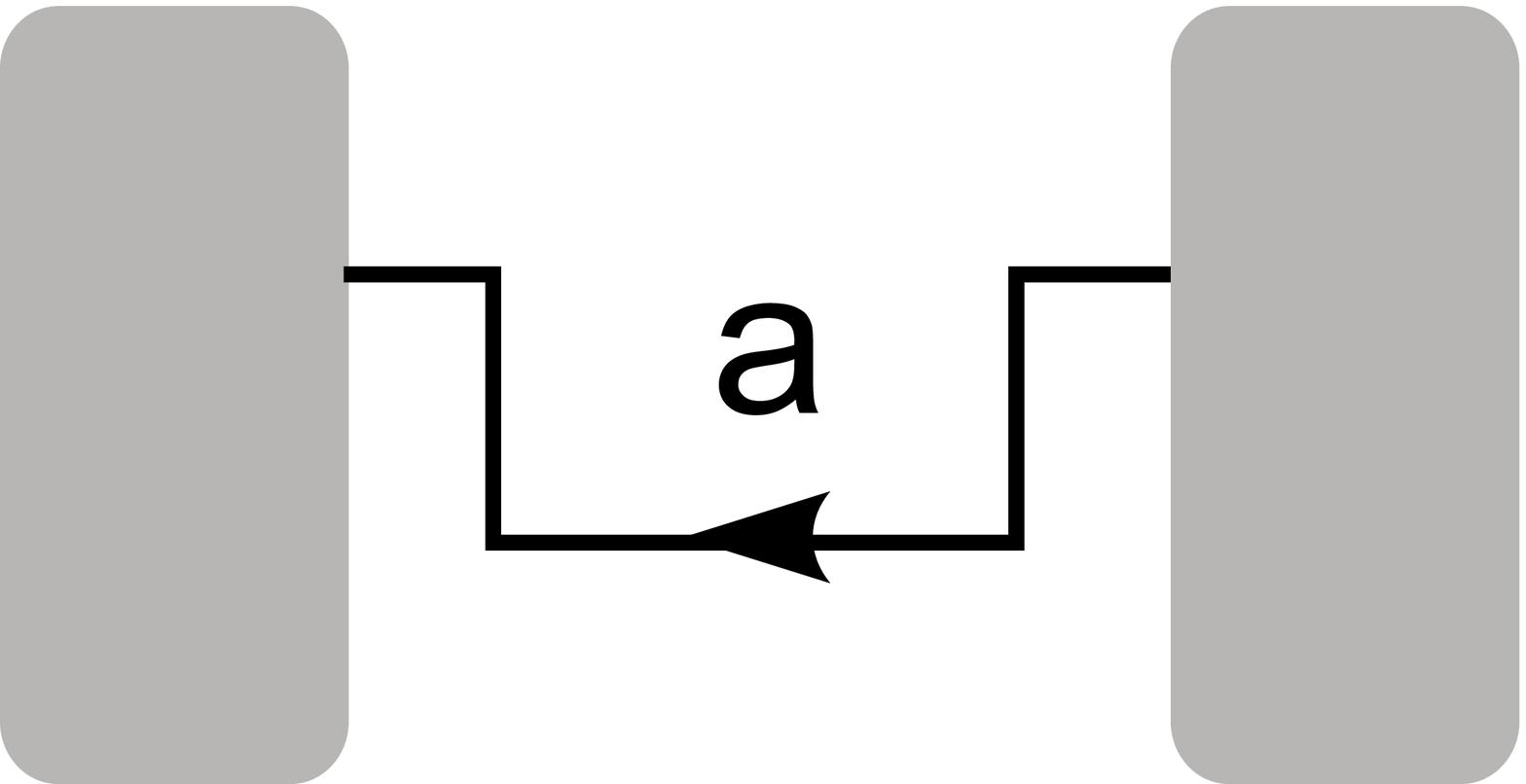}}\right),  \label{rule1a} \\
\Phi \left( \raisebox{-0.16in}{\includegraphics[height=0.4in]{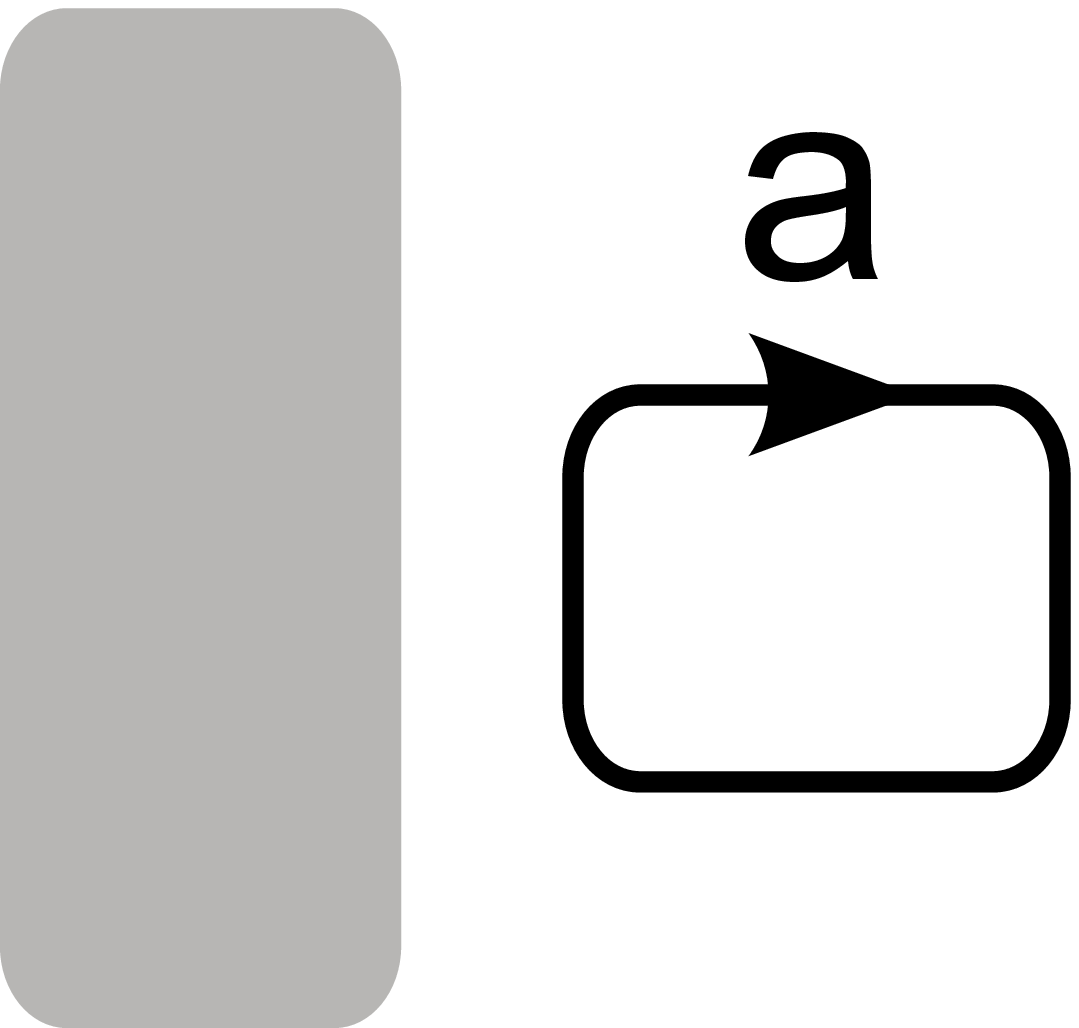}}%
\right) &=& d_{a}\Phi \left( \raisebox{-0.16in}{%
\includegraphics[height=0.4in]{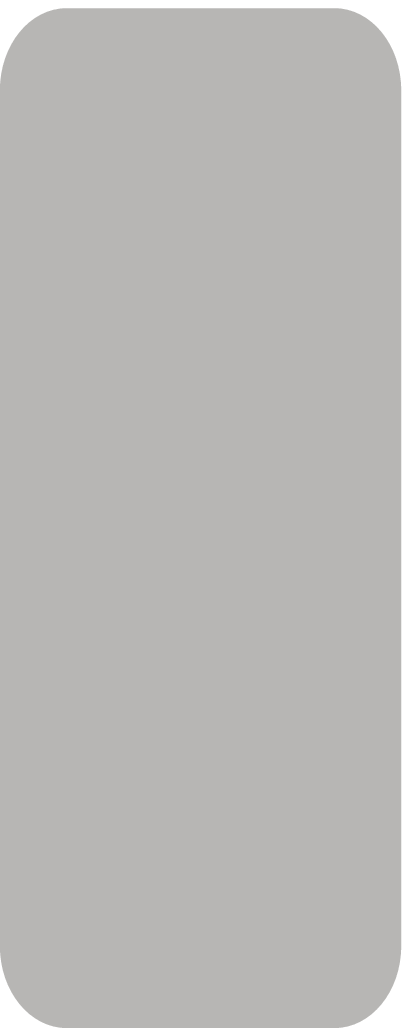}}\right),  \label{rule1b} \\
\Phi \left( \raisebox{-0.16in}{\includegraphics[height=0.4in]{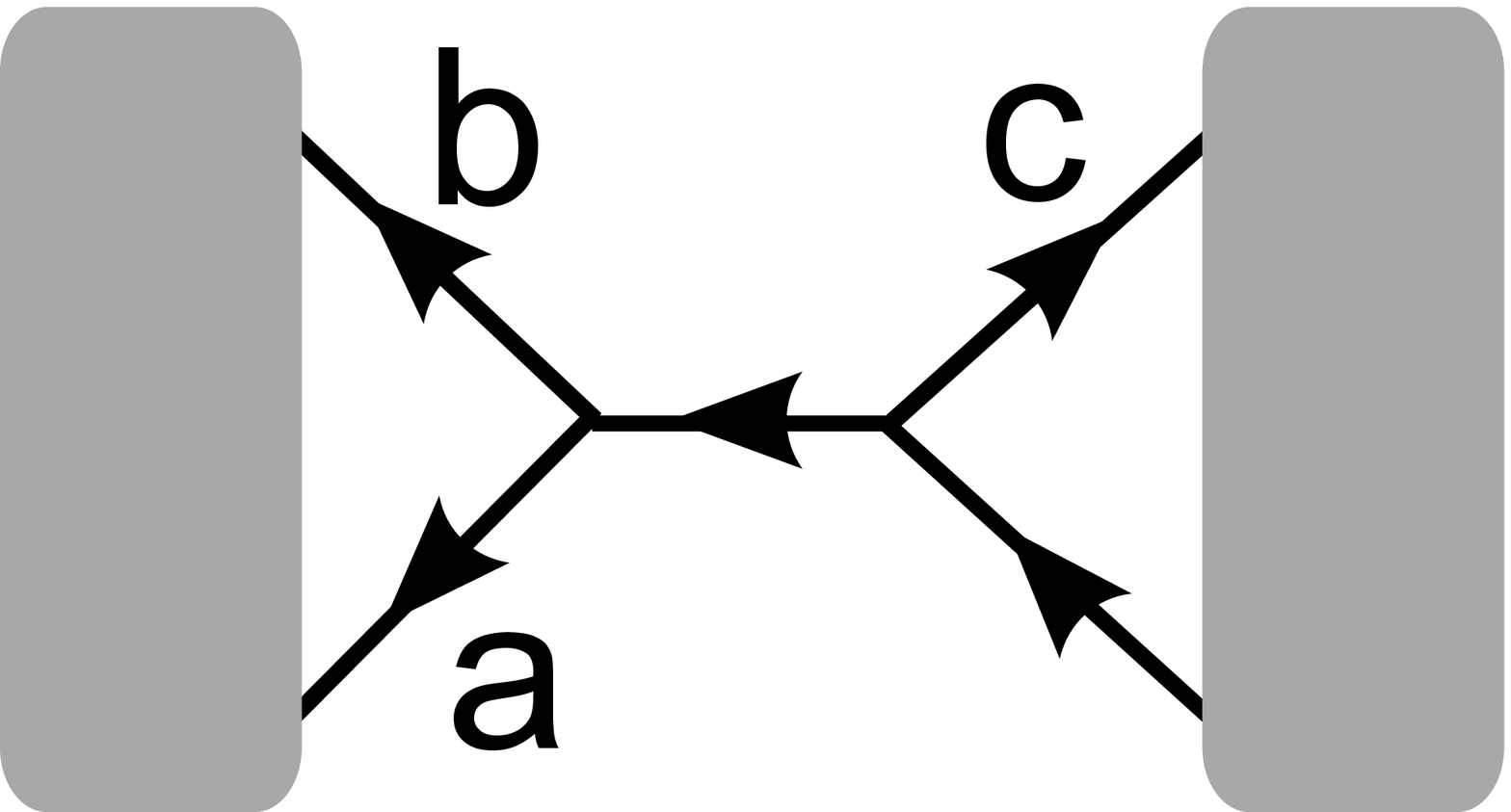}}%
\right) &=& F(a,b,c)\Phi \left( \raisebox{-0.16in}{%
\includegraphics[height=0.4in]{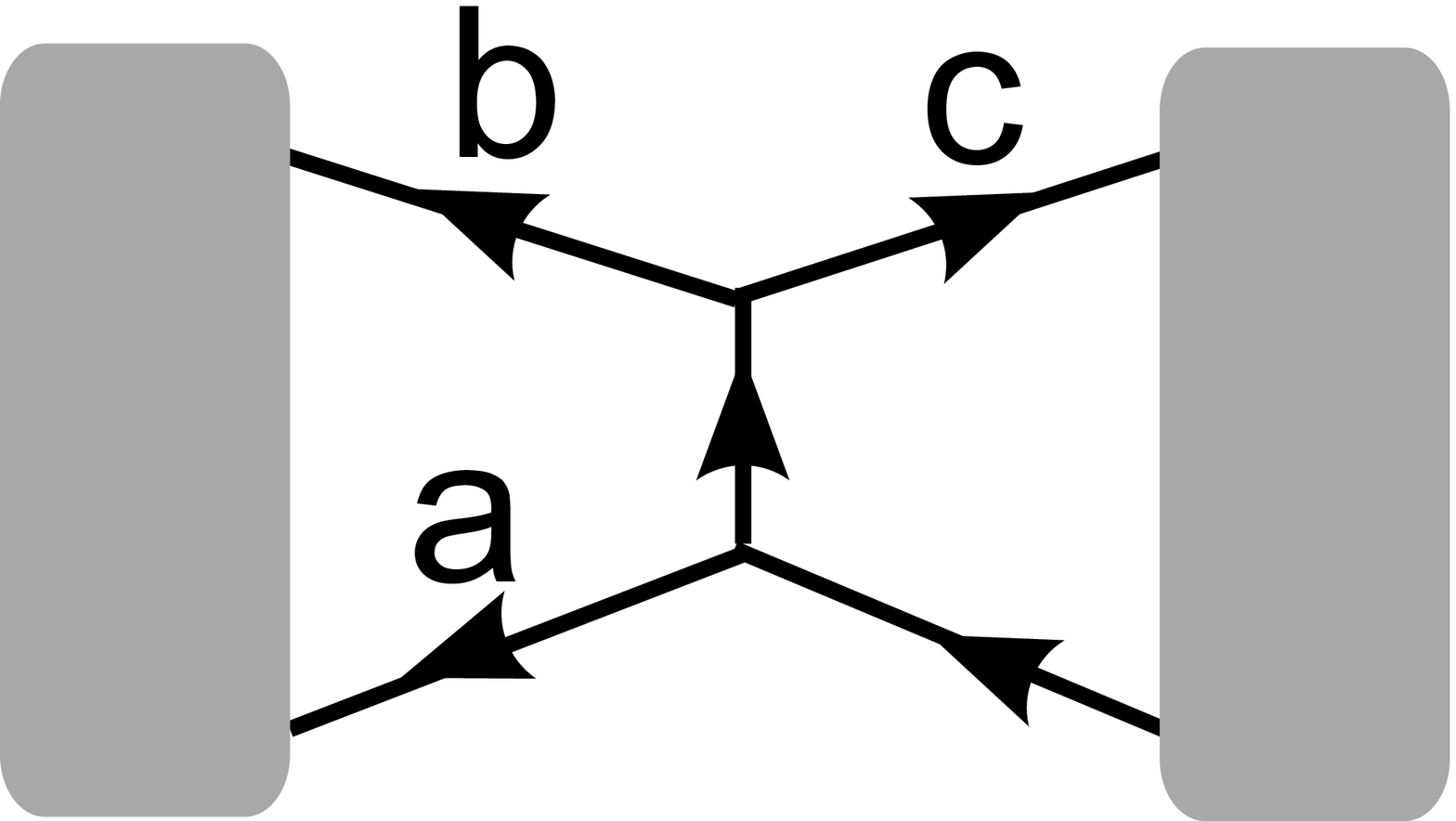}}\right).  \label{rule1c}
\end{eqnarray}
Here $a,b,c\,$are arbitrary string types and the shaded regions represent
arbitrary string-net configurations which is unchanged. The $d_{a}$ and $%
F\left( a,b,c\right) $ are complex numbers which satisfy certain algebraic
equations we will specify below.

The idea of Eqs. (\ref{rule1a}--\ref{rule1c}) is that we can relate the
amplitude of any string-net configuration to the amplitude of the vacuum
(no-string) configuration by applying the local rules multiple times. We use
the convention that%
\begin{equation*}
\Phi\left( \text{vacuum}\right) =1,
\end{equation*}
and then the amplitude of any configuration is fully determined. Thus, the
string-net wave function is fully determined once the parameters $%
d_{a},\gamma_a,\alpha(a,b),F(a,b,c) $ are given.

To construct the most general abelian string-net models, we need two
additional ingredients $\gamma ,\alpha $. First, the $\gamma $ is a $\mathbb{%
Z}_{2}$ phase factor associated with vertices with one null string and two
opposite oriented strings:%
\begin{align}
\left\langle \raisebox{-0.16in}{\includegraphics[height=0.4in]{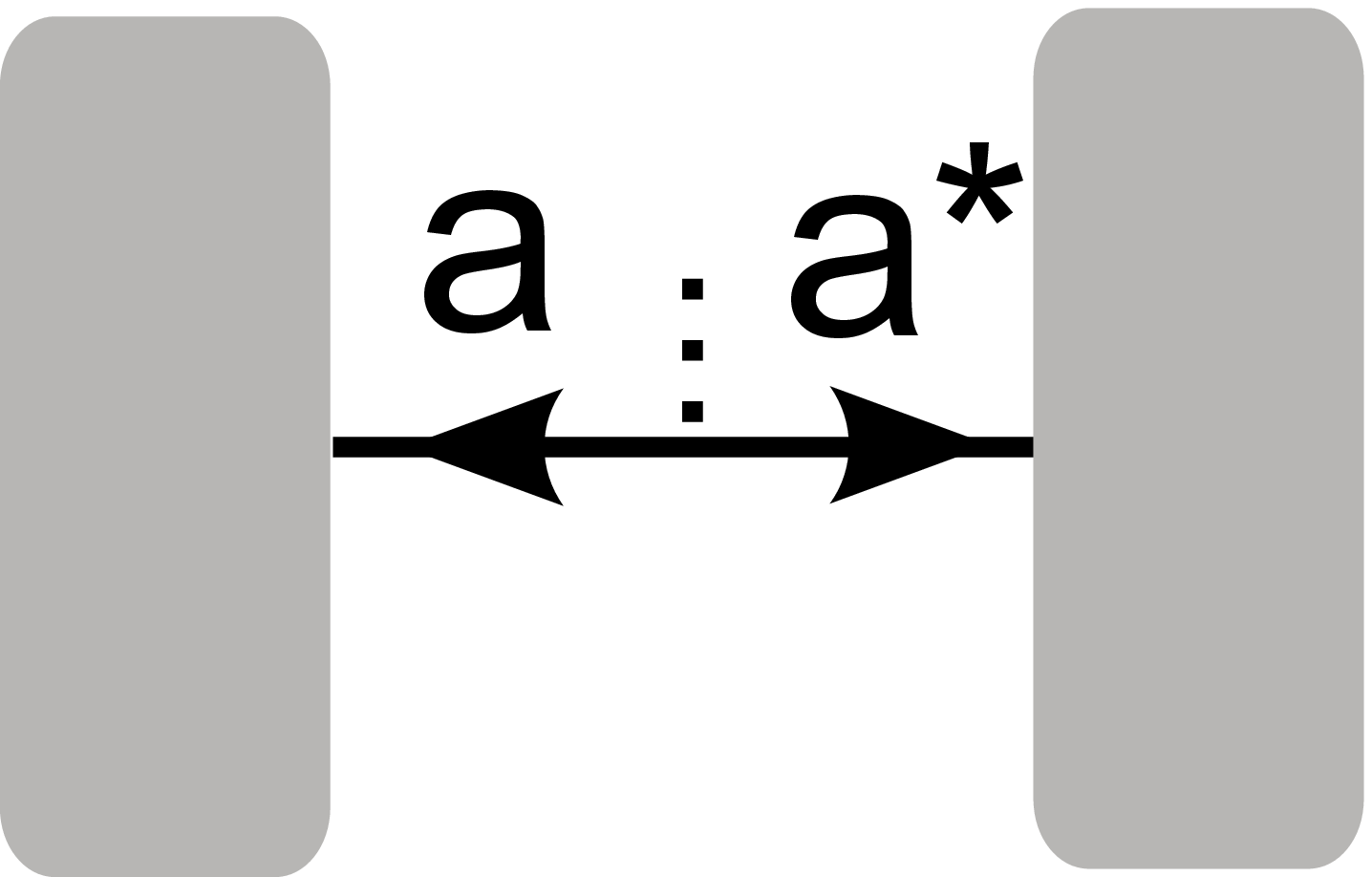}}%
\right\vert & =\gamma _{a}\left\langle \raisebox{-0.16in}{%
\includegraphics[height=0.4in]{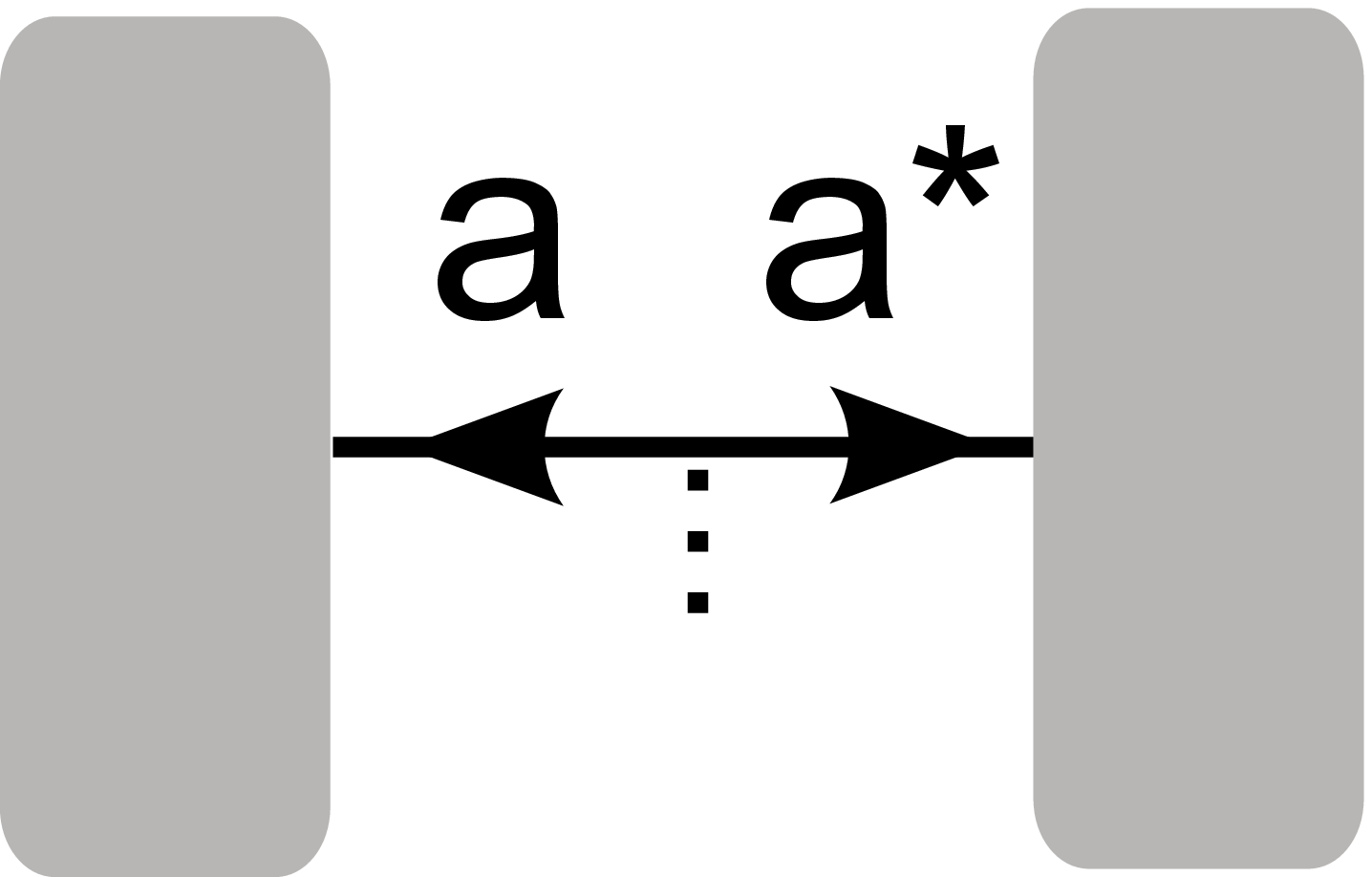}}\right\vert ,  \label{gamma1} \\
\left\langle \raisebox{-0.16in}{\includegraphics[height=0.4in]{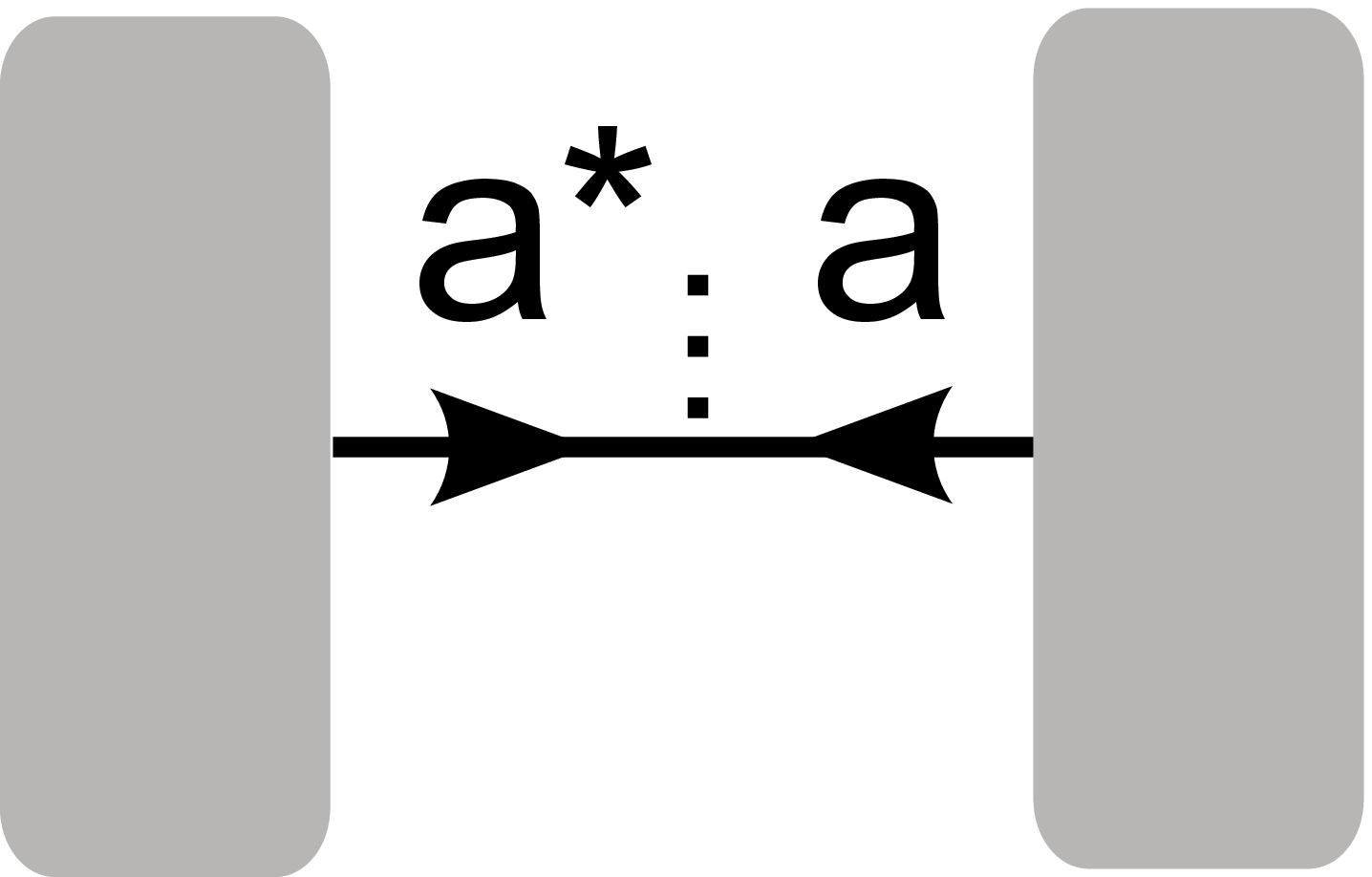}}%
\right\vert & =\gamma _{a}\left\langle \raisebox{-0.16in}{%
\includegraphics[height=0.4in]{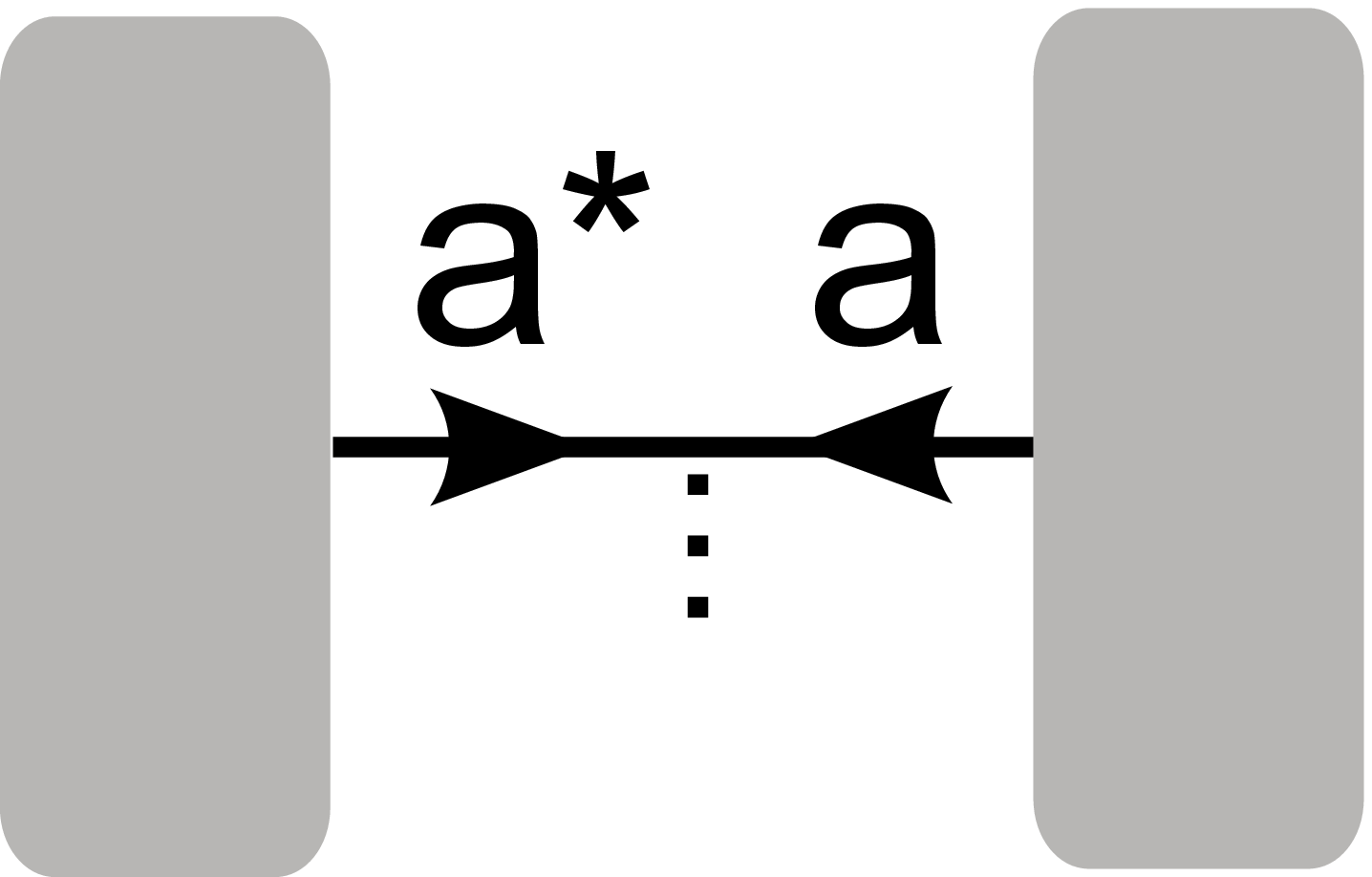}}\right\vert .  \notag
\end{align}%
Here $\gamma _{a}$ can be chosen to be $\pm 1$ without loss of generality. A
pair of null strings with opposite orientations can be erased in pairs
according to:%
\begin{equation}
\left\langle \raisebox{-0.16in}{\includegraphics[height=0.4in]{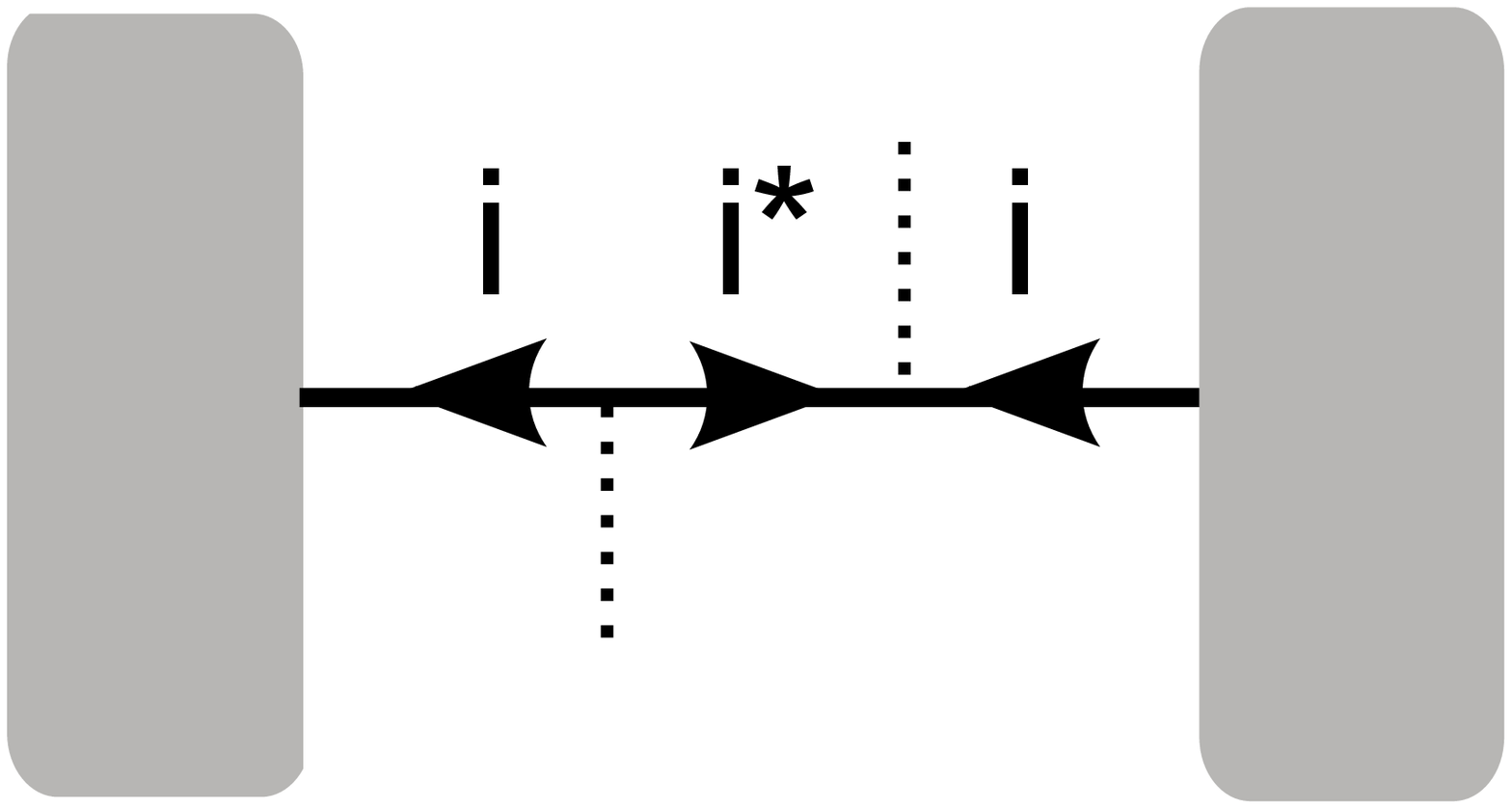}}%
\right\vert =\left\langle \raisebox{-0.16in}{%
\includegraphics[height=0.4in]{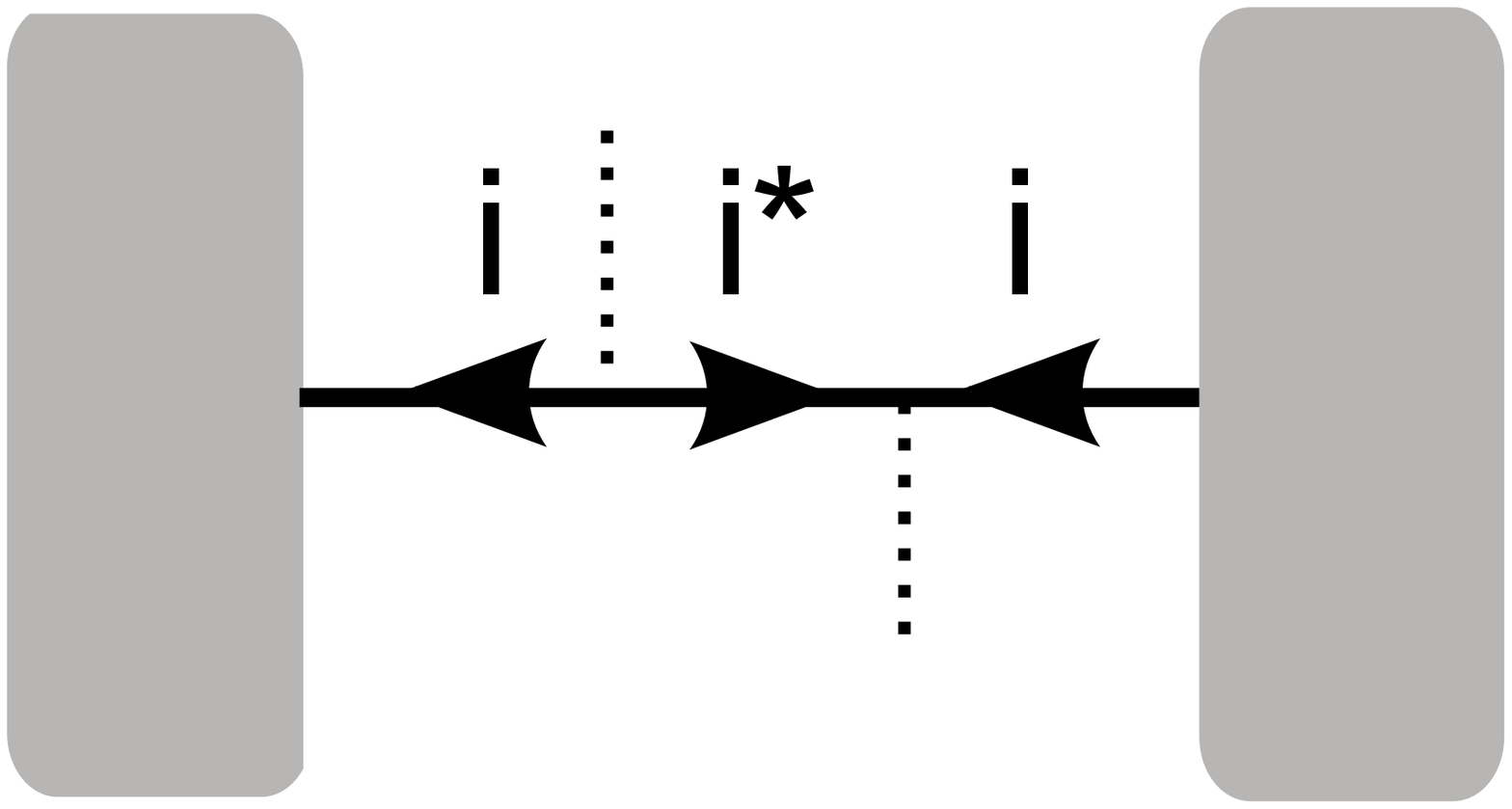}}\right\vert =\left\langle %
\raisebox{-0.16in}{\includegraphics[height=0.4in]{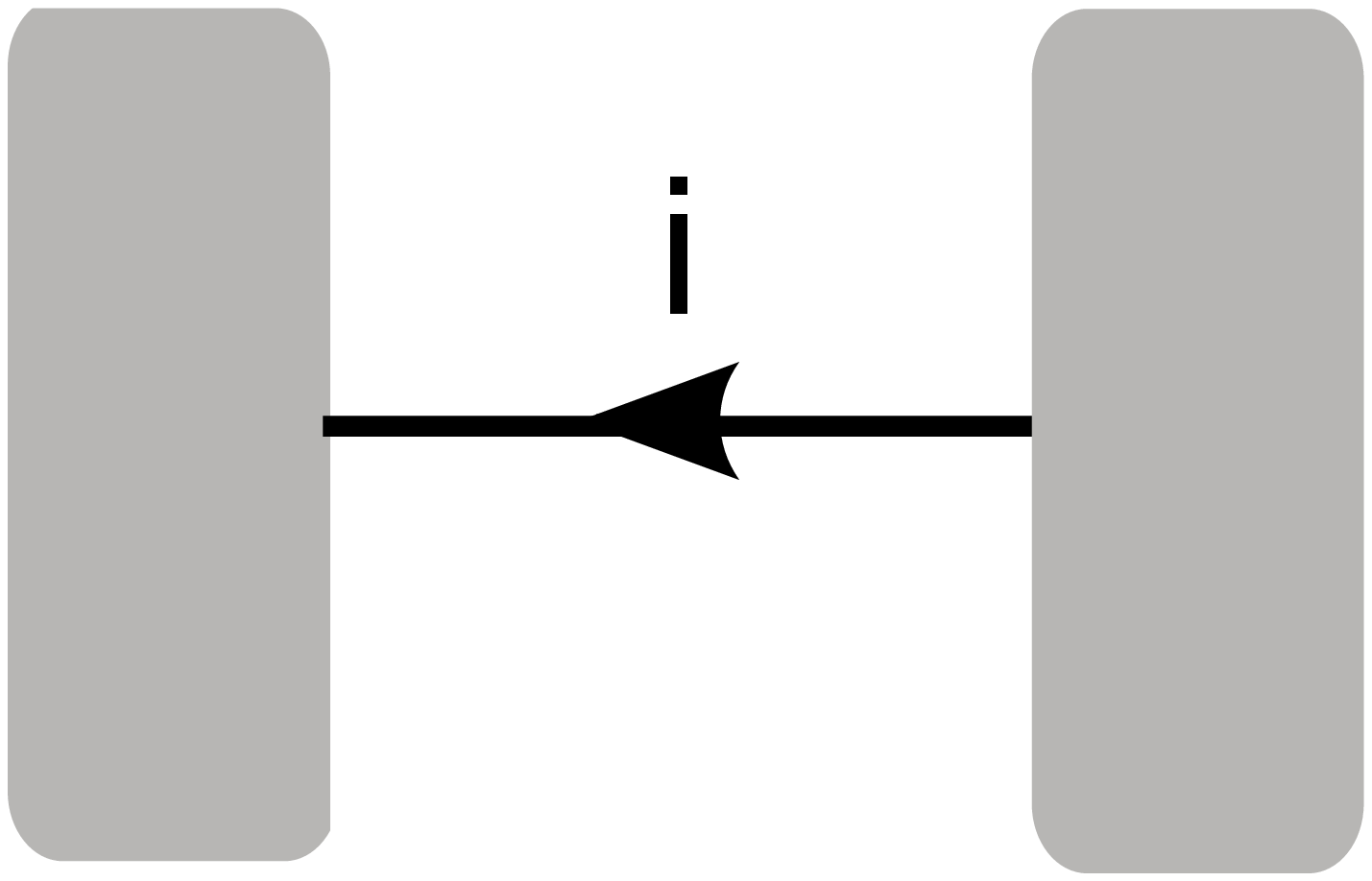}}\right\vert .
\label{gamma2}
\end{equation}

Second, to define $\alpha,$ we absorbed the end of the null strings into
vertices by defining:

\begin{align*}
\left\langle \raisebox{-0.16in}{\includegraphics[height=0.4in]{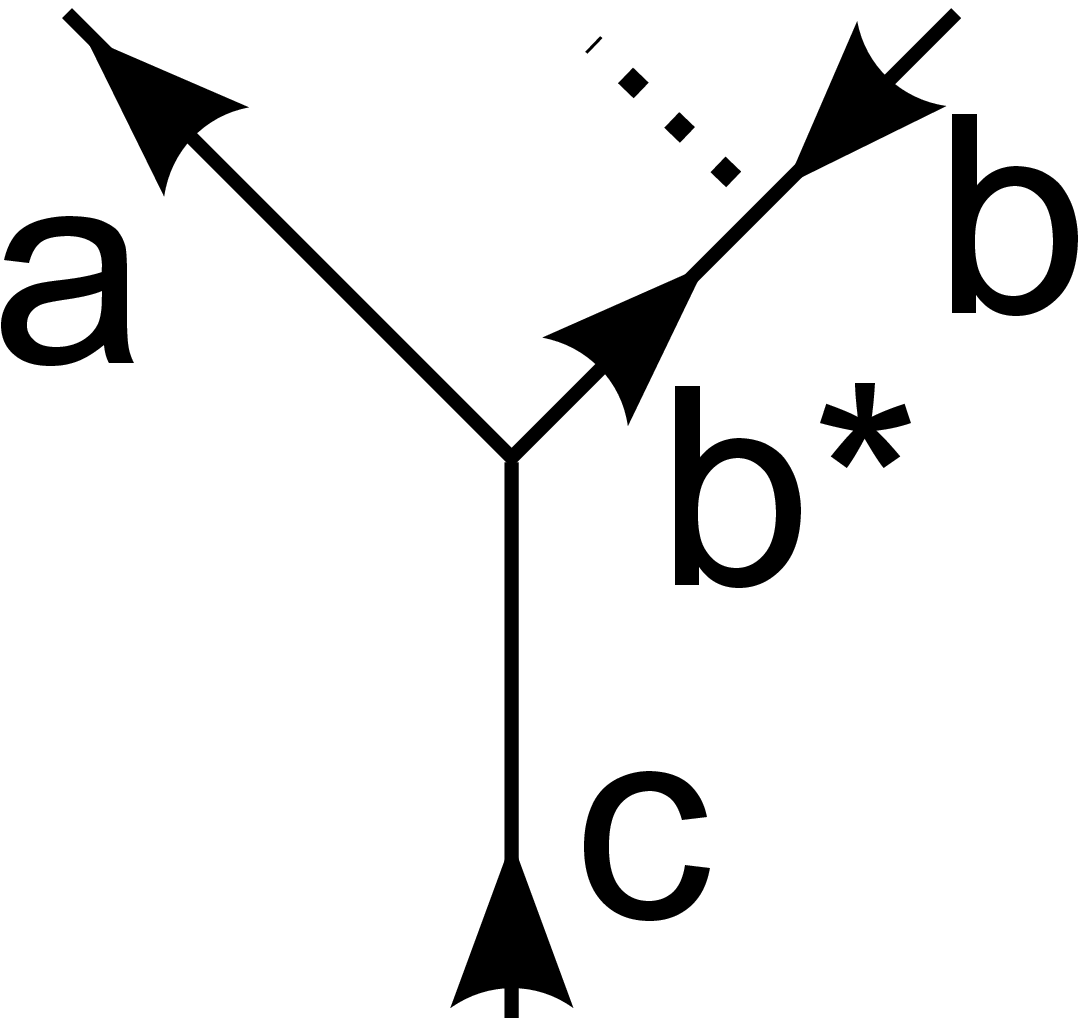}}%
\right\vert & =\left\langle \raisebox{-0.16in}{%
\includegraphics[height=0.4in]{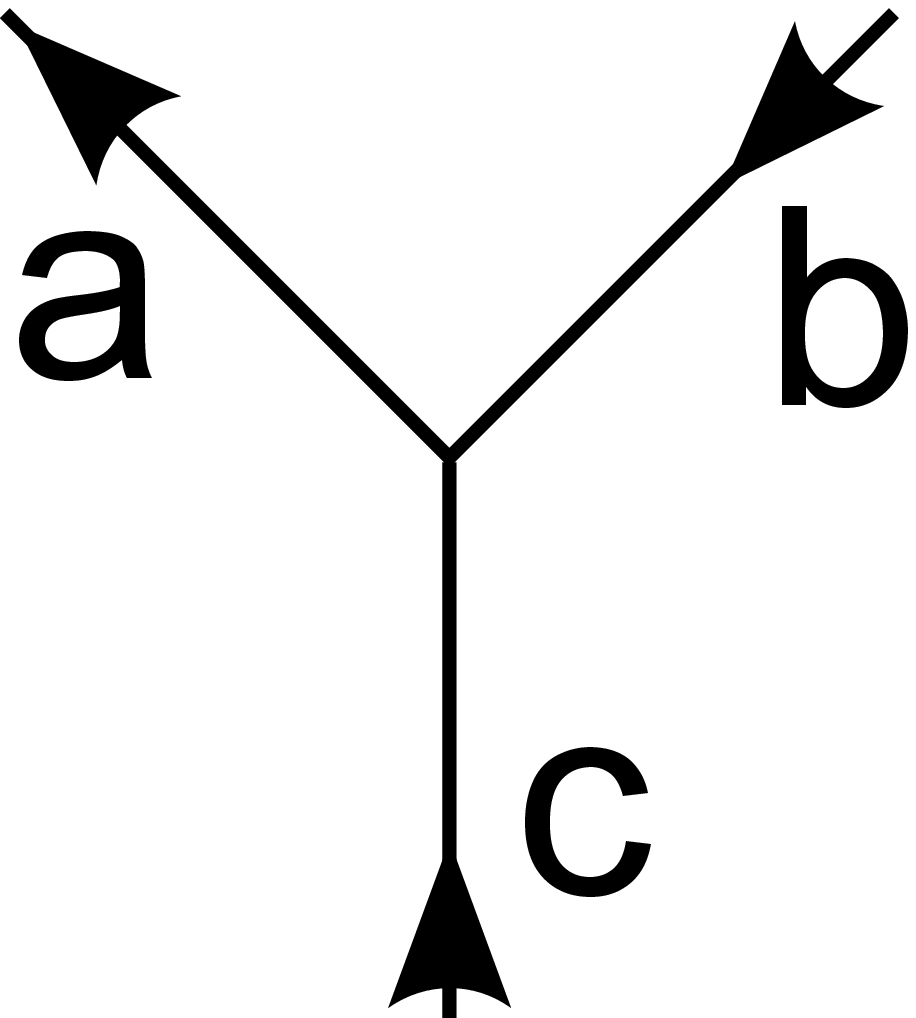}}\right\vert , \\
\left\langle \raisebox{-0.16in}{\includegraphics[height=0.4in]{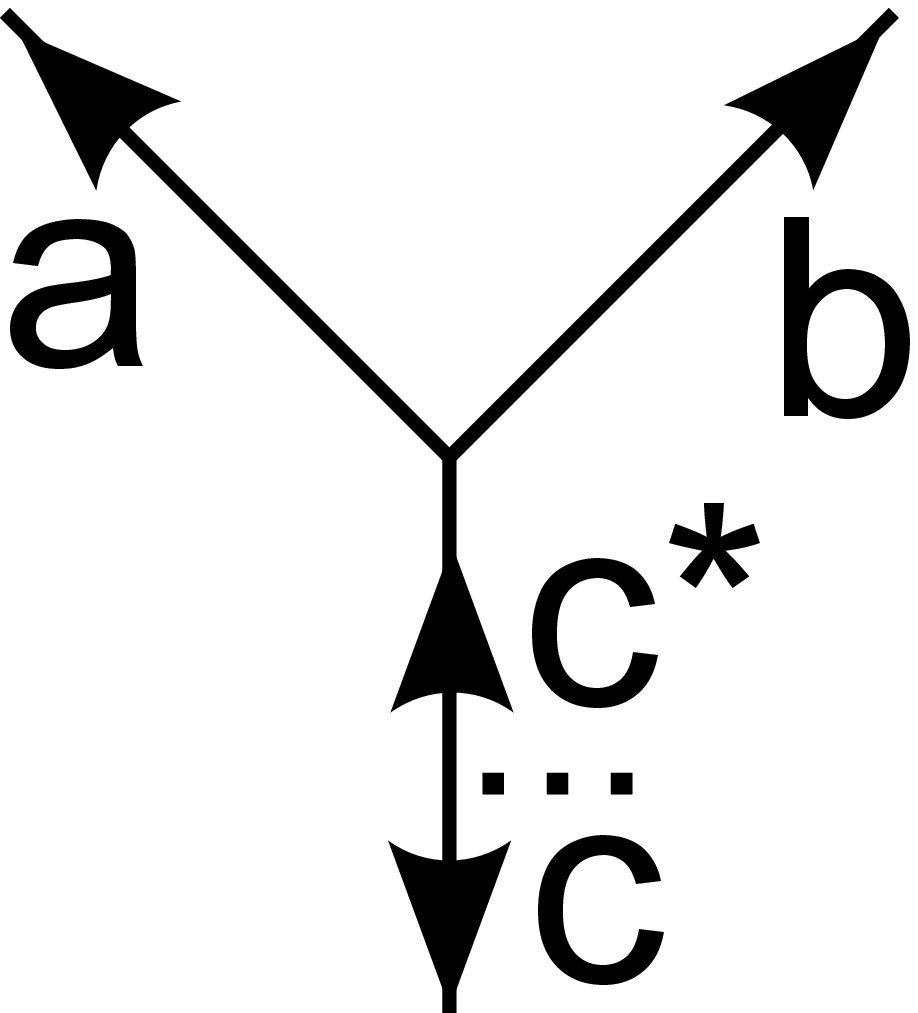}}%
\right\vert & =\left\langle \raisebox{-0.16in}{%
\includegraphics[height=0.4in]{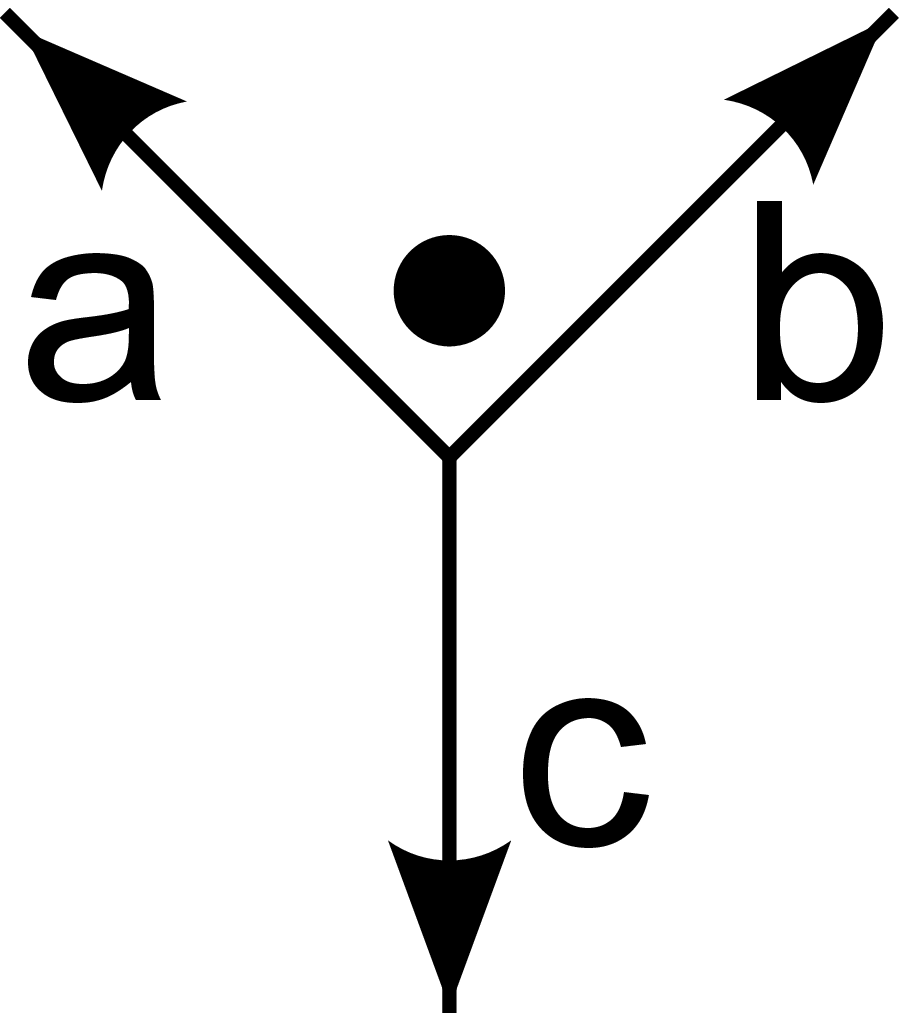}}\right\vert , \\
\left\langle \raisebox{-0.16in}{\includegraphics[height=0.4in]{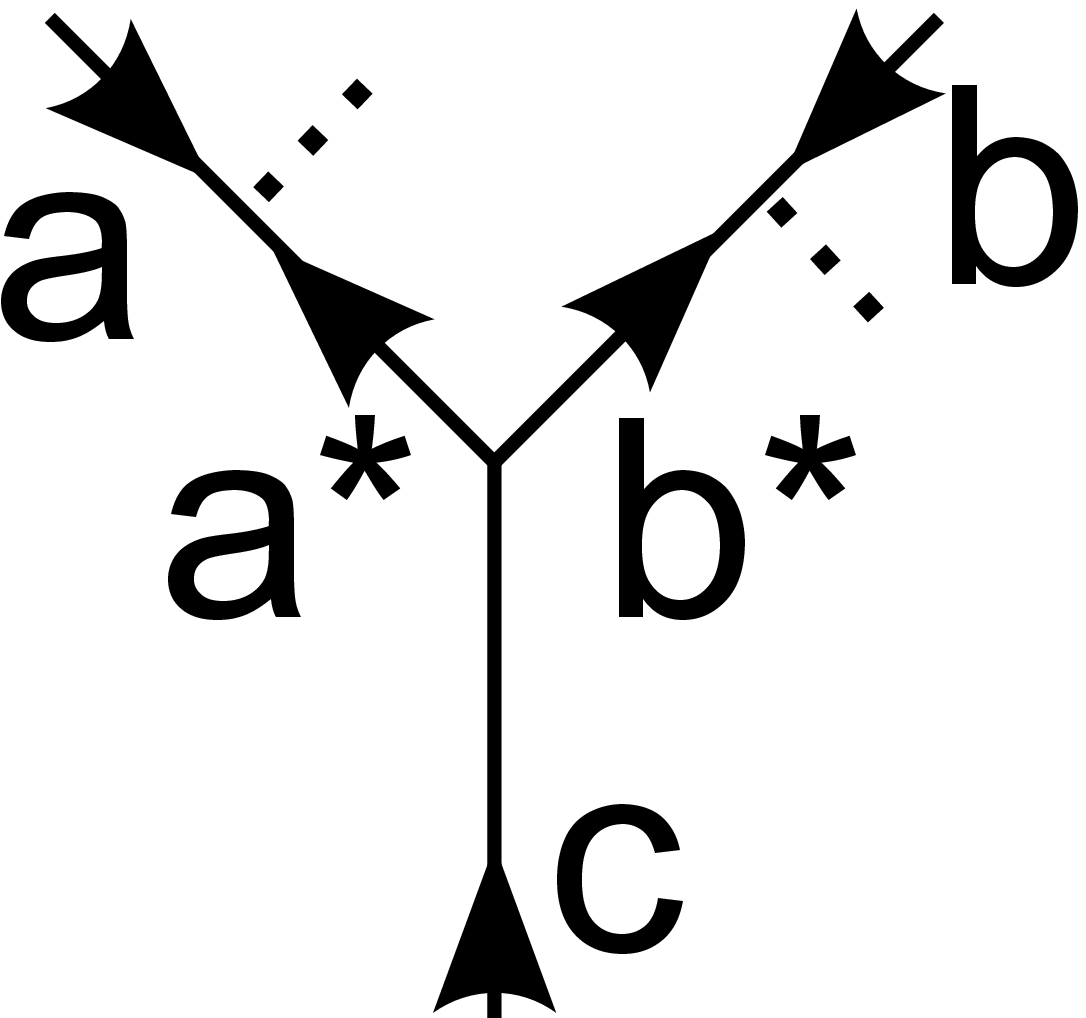}}%
\right\vert & =\left\langle \raisebox{-0.16in}{%
\includegraphics[height=0.4in]{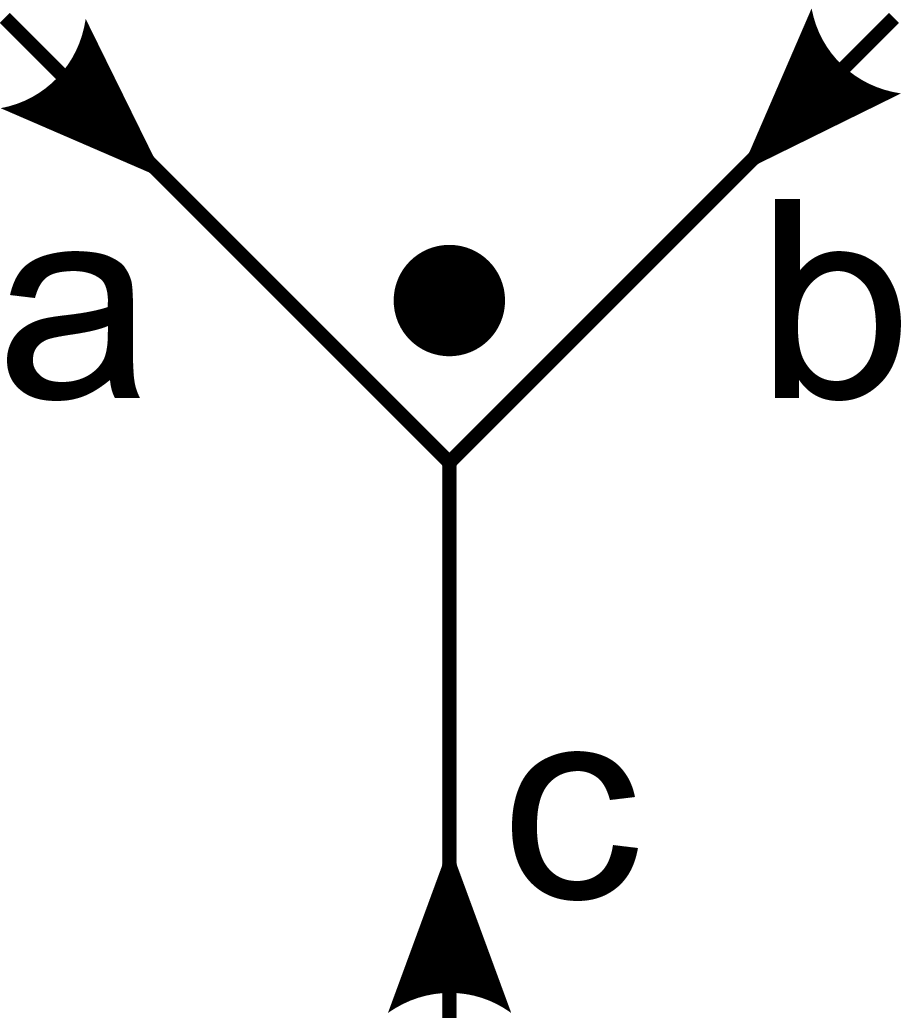}}\right\vert .
\end{align*}%
Here we decorate the vertices that have three incoming or three outgoing
legs with dots. The dots can be placed in any of the three positions near
the vertex. Then, the $\mathbb{Z}_{3}$ phase factor $\alpha $ is defined by%
\begin{align*}
\left\langle \raisebox{-0.16in}{\includegraphics[height=0.4in]{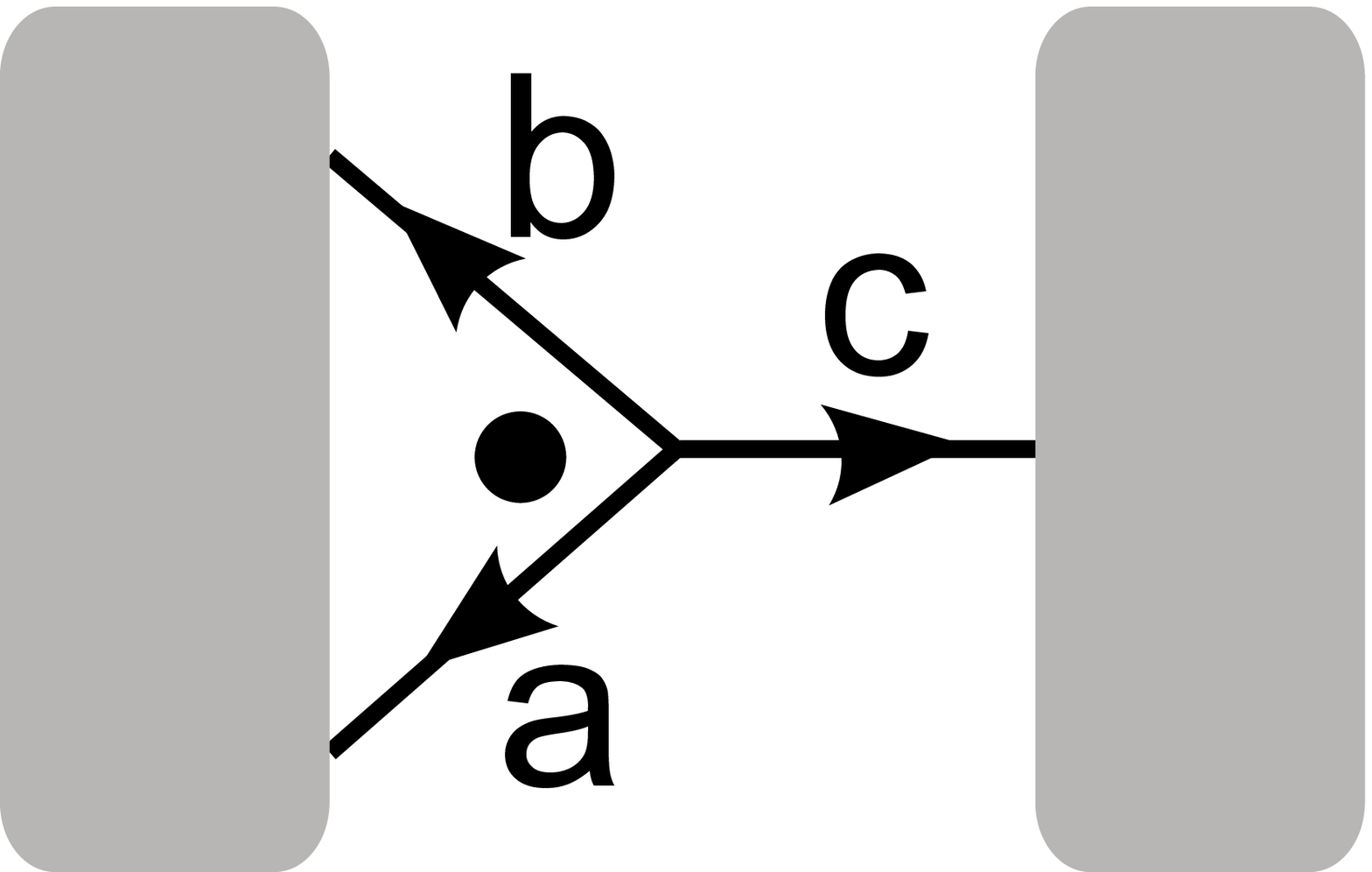}}%
\right\vert & =\alpha \left( a,b\right) \left\langle \raisebox{-0.16in}{%
\includegraphics[height=0.4in]{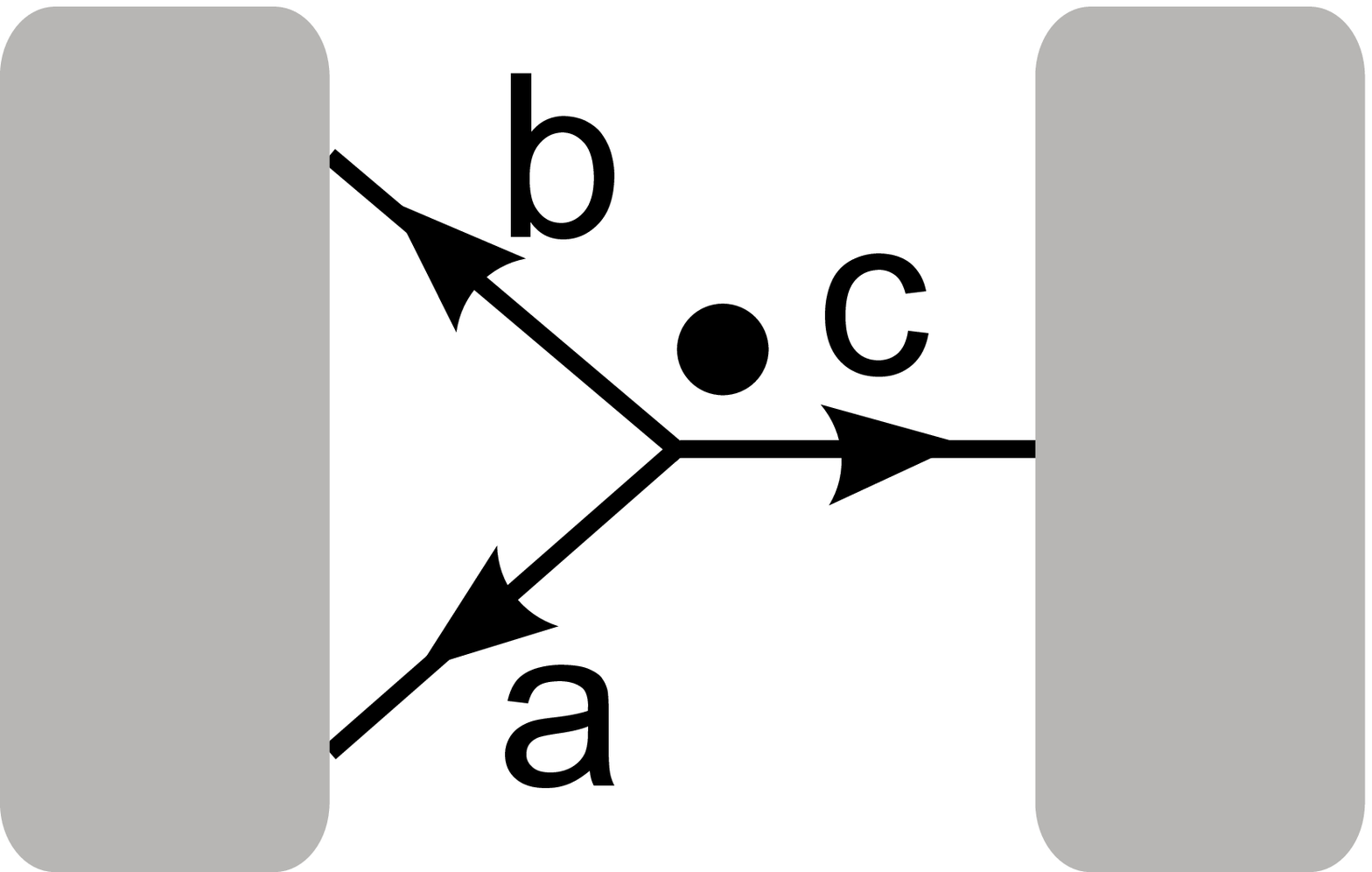}}\right\vert \\
\left\langle \raisebox{-0.16in}{\includegraphics[height=0.4in]{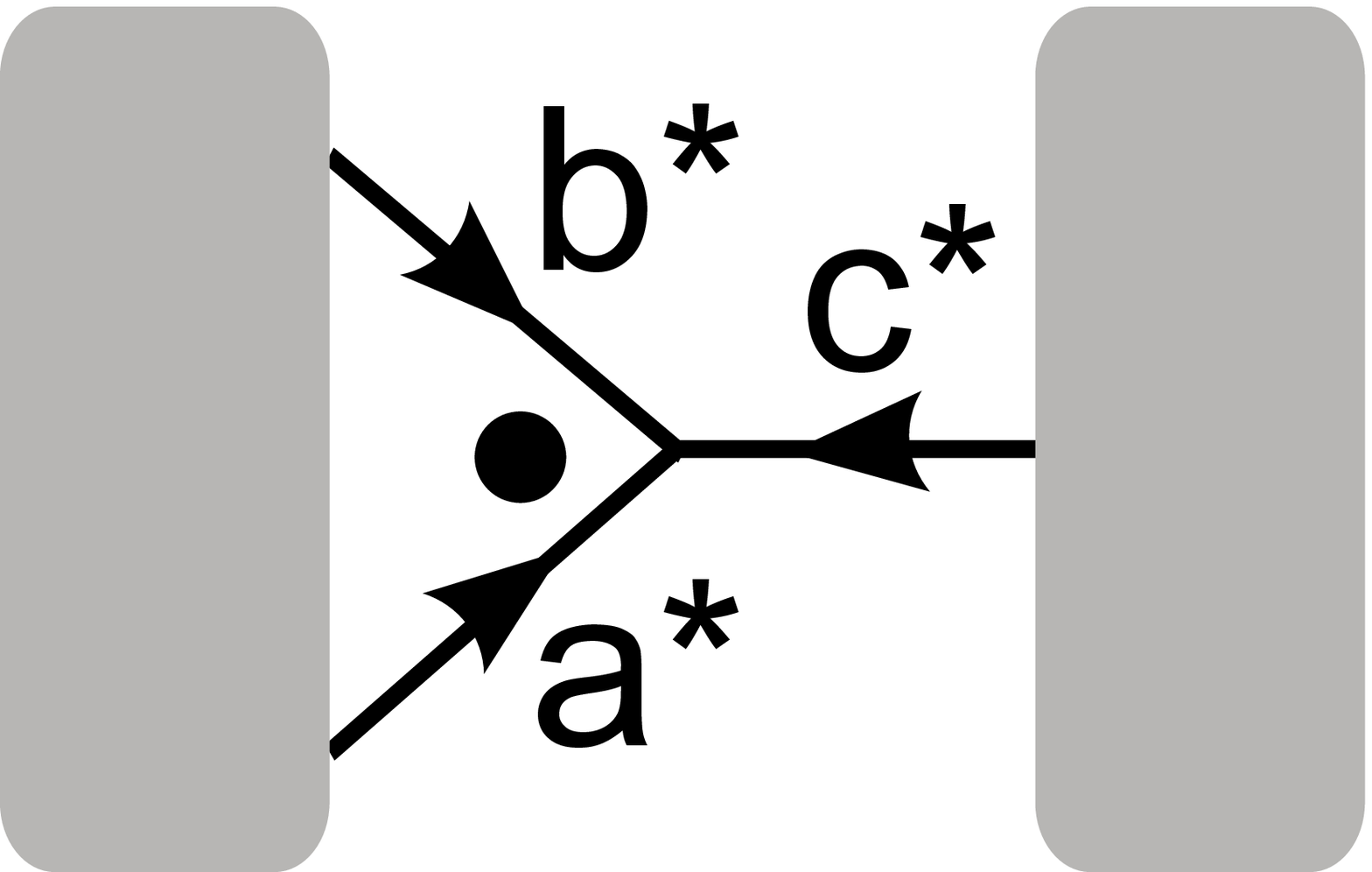}}%
\right\vert & =\alpha \left( a,b\right) \left\langle \raisebox{-0.16in}{%
\includegraphics[height=0.4in]{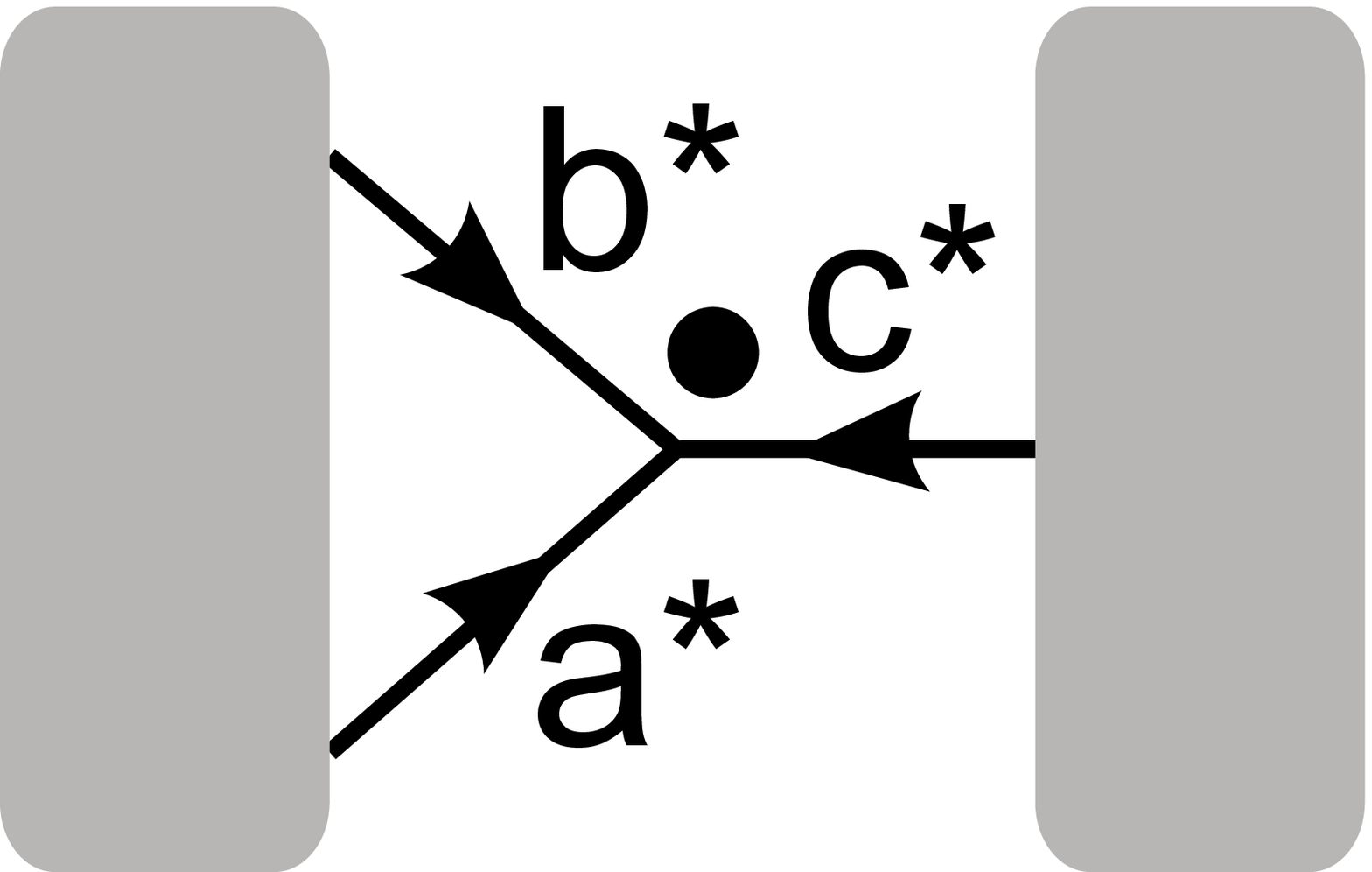}}\right\vert .
\end{align*}%
The $\alpha \left( a,b\right) $ can be chosen to be a third root of unity
without loss of generality. They are designed to keep track of the
orientation of vertices. Note that flipping the null string or moving the
position of the dot does not change the physical state, but it can introduce
a phase factor $\gamma _{a}$ or $\alpha \left( a,b\right) .$ These phases $%
\gamma _{a},\alpha \left( a,b\right) $ are closely related to so-called $%
\mathbb{Z}_{2}$ and $\mathbb{Z}_{3}$ Frobenius-Schur indicators\cite{KitaevHoneycomb}\cite{BondersonThesis}.

So far the parameters $\left\{ d_{a},F\left( a,b,c\right) ,\gamma
,\alpha\right\} $ are arbitrary. However, these parameters have to satisfy a
set of algebraic equations so that they lead to self-consistent local rules
and a well-defined wave function $\Phi:$%
\begin{subequations}
\begin{align}
F\left( a+b,c,d\right) &F\left( a,b,c+d\right)  = \\
&F\left( a,b,c\right) F\left( a,b+c,d\right) F\left( b,c,d\right) ,  \notag\\
F\left( a,b,c\right) & =1\text{ if }a\text{ or }b\text{ or }c=0,   \\
d_{a}d_{b} & =d_{a+b},  \\
\gamma_{a} & =F\left( a^{\ast},a,a^{\ast}\right) d_{a},  \\
\alpha\left( a,b\right) & =F\left( a,b,\left( a+b\right) ^{\ast }\right)
\gamma_{a+b}.  
\end{align}
\label{sfeqa}
\end{subequations}
In addition, to construct a consistent string-net model (Hamiltonian), we
need one more constraint%
\begin{equation*}
\left\vert F\left( a,b,c\right) \right\vert =1.
\end{equation*}
This constraint ensures the corresponding exactly soluble Hamiltonians to be
Hermitian.

For each solution to the above constraints, we can construct a well-defined
wave function and an exactly soluble Hamiltonian. However, if two sets of
solutions $\left\{ F,d,\alpha,\gamma\right\} ,\left\{ \tilde{F},\tilde {d},%
\tilde{\alpha},\tilde{\gamma}\right\} $ are related by gauge
transformations (see Ref. \onlinecite{LinLevinstrnet} for details), the two solutions are equivalent and the corresponding wave
functions $\Phi,\tilde{\Phi}$ and Hamiltonians can be transformed into one
another by a local unitary transformation.\ Thus it implies that $\Phi ,%
\tilde{\Phi}$ describe the same quantum phase. 
Therefore, one only need to consider one solution within each gauge
equivalence class to construct distinct topological phases.

\section{Multi-flavor string-net wave functions \label{section:mwavefunction}%
}
In this section, we study the wave functions for multiple flavors of
string-nets. For each flavor-$i$ of strings, we label the string types by the elements of the group $G_i:\{a_i,b_i,c_i,\dots\}$. The strings of the same flavor-$i$ can branch according to the branching rules $\{(a_i,b_i,c_i),\dots\}$ and form the string-nets of flavor-$i$.
On the other hand, different flavors of strings can intersect/cross one another and thus form the intersecting string-nets of multiple flavors. 
To describe the wave functions for the
intersecting string-nets, we need additional rules to describe how different
flavors of string-nets intersect with one another. In two dimensions, it is
sufficient to consider two kinds of intersections: intersections between two
flavors of string-nets and intersections among three flavors of string-nets.
In this section, we introduce new local rules for the multi-flavor
string-net wave functions. In section \ref{section:H}, we will show that
these wave functions are the ground states of the exactly soluble
multi-flavor string-net Hamiltonians, defined on a lattice.

\subsection{Local rules ansatz}

Let us consider $L$ flavors of string-nets. We label the string types of the 
$i$-th flavor by $\left\{ a_{i},b_{i},...\right\} $ with the flavor index $%
i=1,...,L$. Like single-flavor string-nets, each flavor of string-nets 
satisfy the original local rules (\ref{rule1a}--\ref{rule1c}) individually.
In addition, since different flavors of string-nets can intersect, 
we require new local rules to describe the wave
functions $\Phi $ for the intersecting string-nets. In particular, we
consider intersections between two flavors of string-nets with flavor
indices $i\neq $ $j$ and intersections among three flavors of string-nets
with $i\neq j\neq k$. Once we specify these two kinds of intersections, the
whole string-net configuration is uniquely determined.

Specifically, these new local rules for constructing intersecting string-net
wave function $\Phi $ can be put in the following graphical forms%
\begin{align}
\Phi \left( \raisebox{-0.16in}{\includegraphics[height=0.4in]{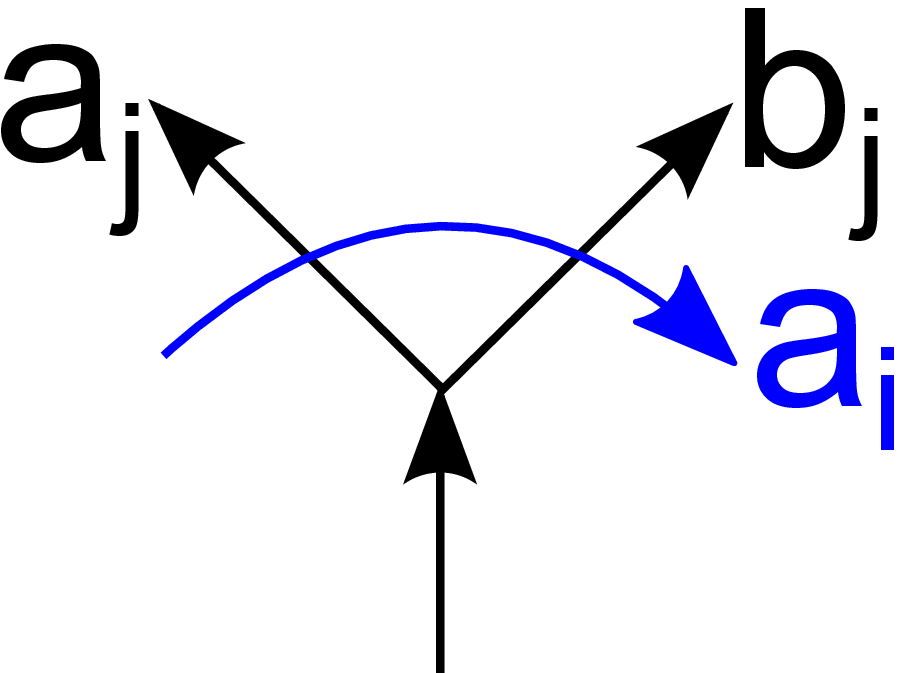}}%
\right) & =F_{a_{i}}^{\left( 2\right) }\left( a_{j},b_{j}\right) \Phi \left( %
\raisebox{-0.16in}{\includegraphics[height=0.4in]{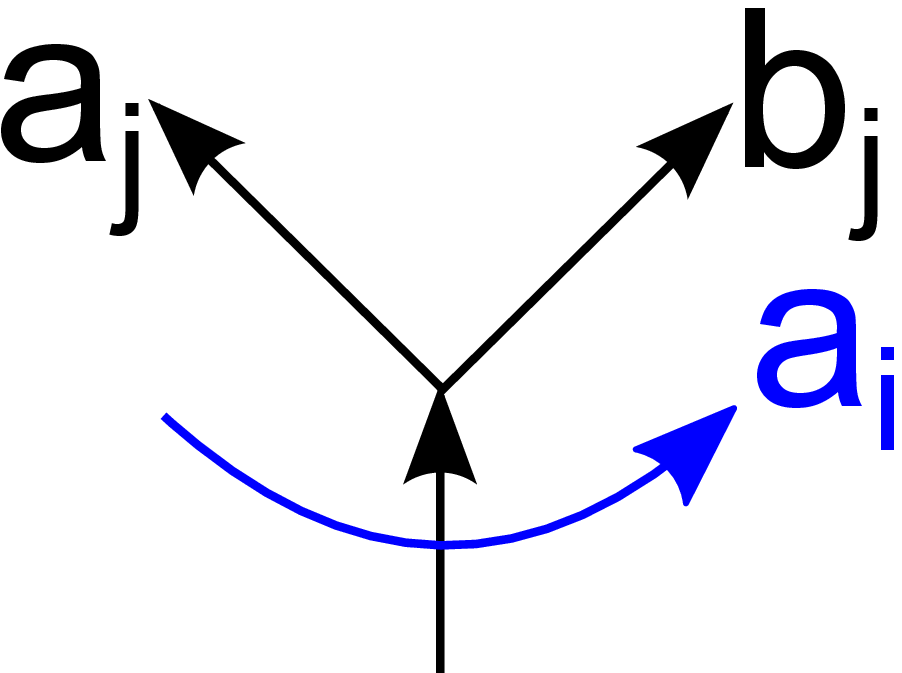}}\right) ,
\label{rule2a} \\
\Phi \left( \raisebox{-0.16in}{\includegraphics[height=0.4in]{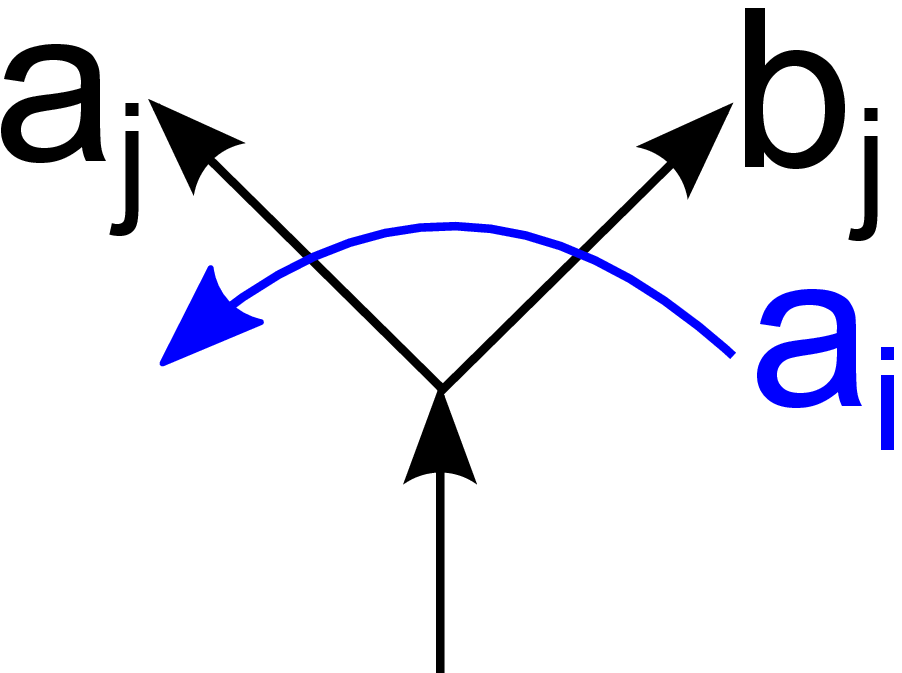}}%
\right) & =\bar{F}_{a_{i}}^{\left( 2\right) }\left( a_{j},b_{j}\right) \Phi
\left( \raisebox{-0.16in}{\includegraphics[height=0.4in]{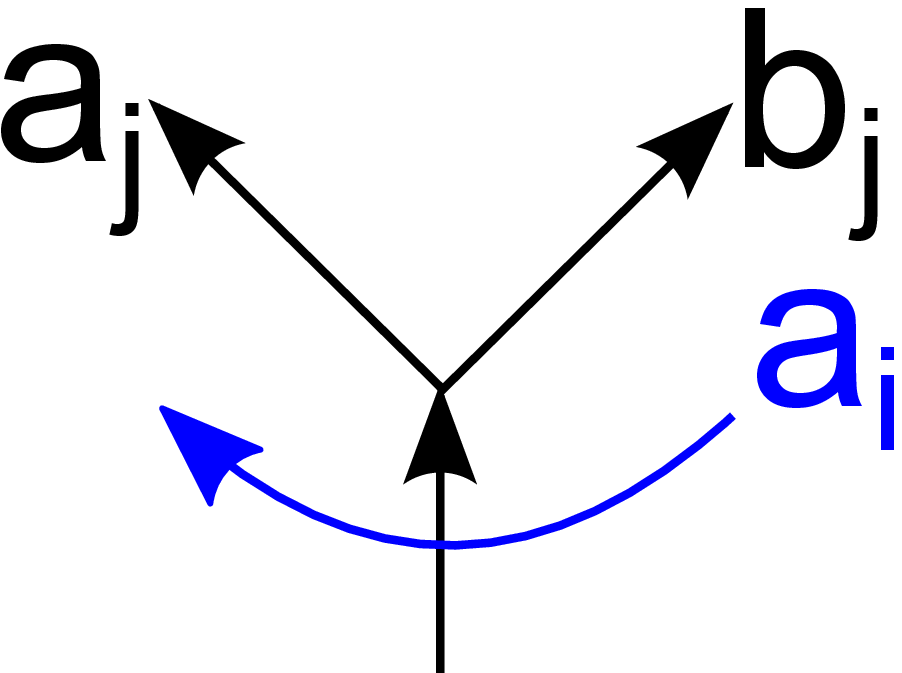}}%
\right) \text{ },  \label{rule2aa} \\
\Phi \left( \raisebox{-0.1in}{\includegraphics[height=0.3in]{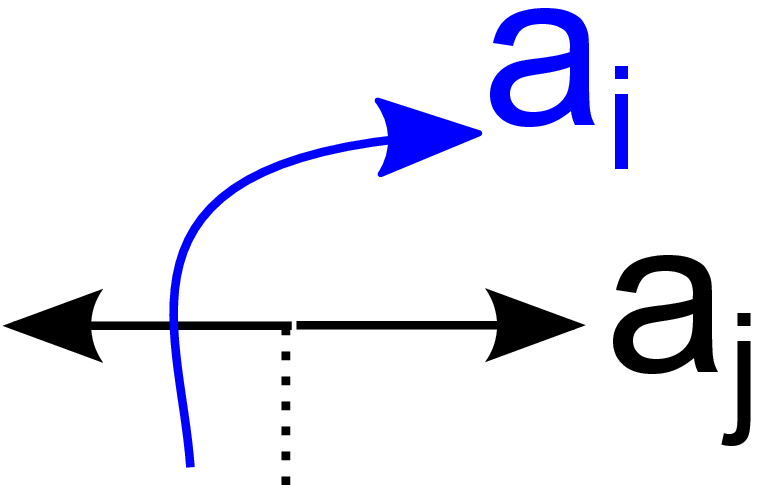}}%
\right) & =\kappa _{a_{i}}\left( a_{j}\right) \Phi \left( %
\raisebox{-0.1in}{\includegraphics[height=0.3in]{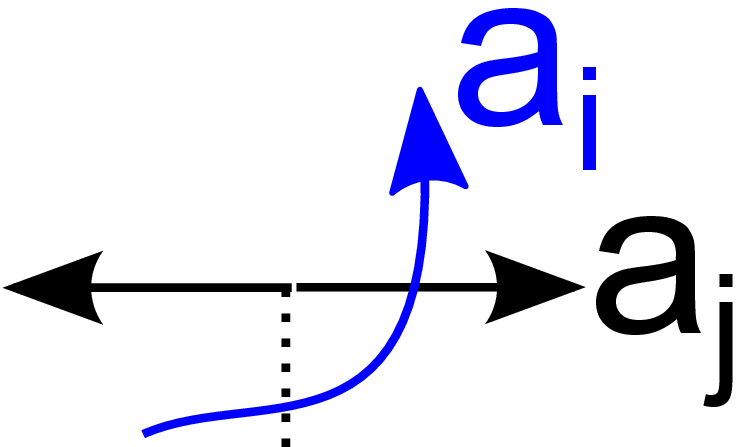}}\right) ,
\label{rule2b} \\
\Phi \left( \raisebox{-0.1in}{\includegraphics[height=0.27in]{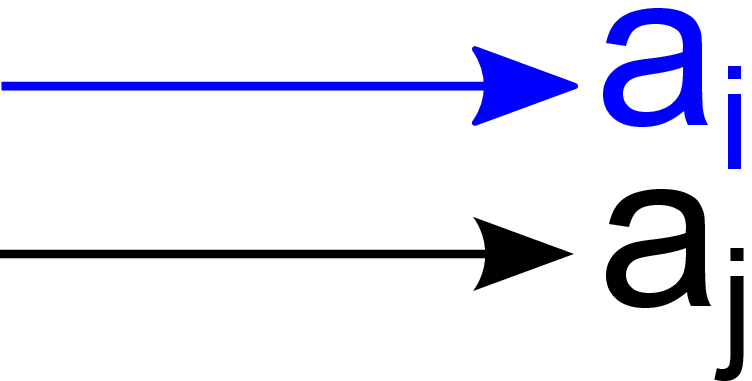}}%
\right) & =\eta _{a_{i}}\left( a_{j}\right) \Phi \left( \raisebox{-0.1in}{%
\includegraphics[height=0.27in]{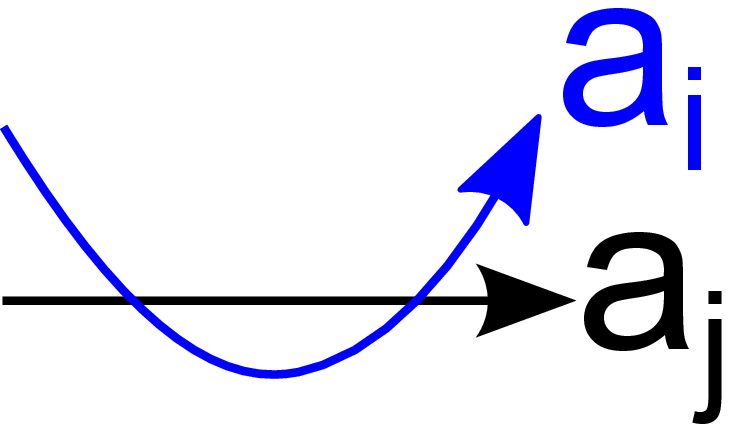}}\right) ,  \label{rule2c} \\
\Phi \left( \raisebox{-0.16in}{\includegraphics[height=0.4in]{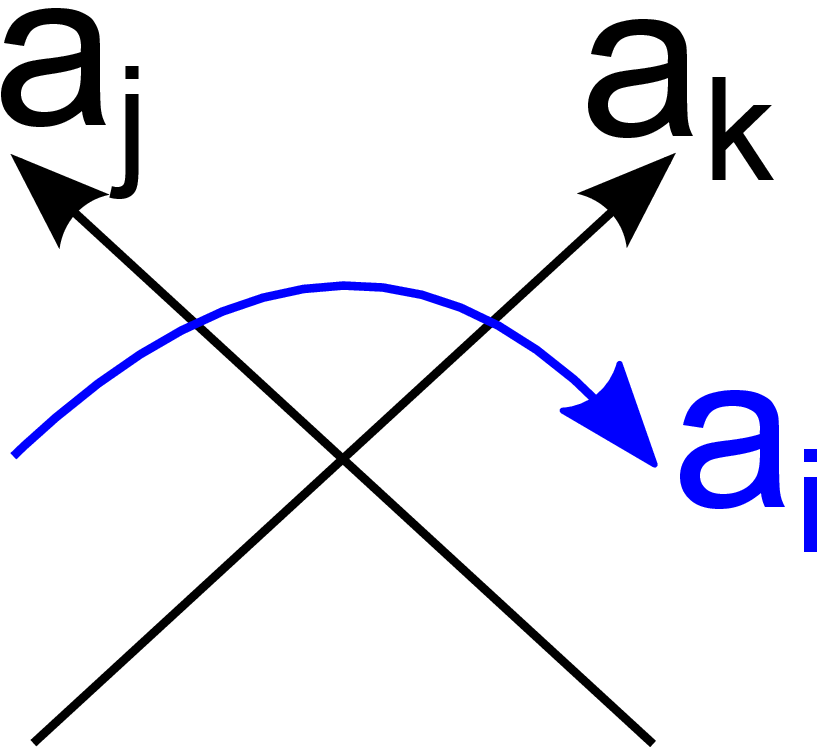}}%
\right) & =F_{a_{i}}^{\left( 3\right) }\left( a_{j},a_{k}\right) \Phi \left( %
\raisebox{-0.16in}{\includegraphics[height=0.4in]{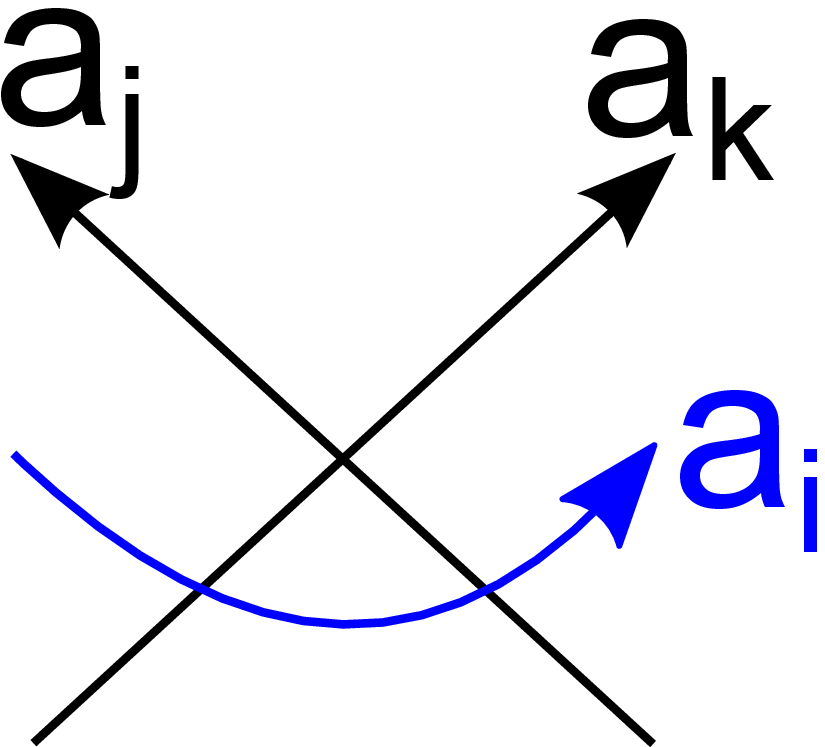}}\right) .
\label{rule2d}
\end{align}%
Here the subindices $i,j,k$ indicate the three different flavors of
string-nets. Note that we only draw the part of configuration which is
changed and neglect other part which is unchanged. Namely, the graphs in
these rules are understood as being with fixed end points which connect to
other unchanged part of the string-net configuration. The $F_{a_{i}}^{\left(
2\right) }\left( a_{j},b_{j}\right) ,$ $\bar{F}_{a_{i}}^{\left( 2\right)
}\left( a_{j},b_{j}\right) ,\kappa _{a_{i}}\left( a_{j}\right) $ and $\eta
_{a_{i}}\left( a_{j}\right) $ are complex numbers that depend on two flavors
of string-nets while the $F_{a_{i}}^{\left( 3\right) }\left(
a_{j},a_{k}\right) $ are complex numbers that depend on three flavors of
string-nets.

The first four rules (\ref{rule2a}--\ref{rule2c})involve two flavors of
string-nets while the last rule (\ref{rule2d}) involves three flavors of
string-nets. For these five rules, it is understood that the value of $\Phi$ 
depends only on the \emph{topology} of the intersecting string-net configurations.
That is, two configurations have the same value of $\Phi$ if one can be smoothly deformed into one another without changing the number of crossings between each two of strings 
This symmetry will put some constraints on the parameters $F^{\left( 2\right) },\bar{F}^{\left( 2\right)
},\kappa ,\eta ,F^{\left( 3\right) }$ which we will discuss in the next
section.

We now discuss the meanings of these new local rules. The first rule (\ref%
{rule2a}) says that one can glide a string $a_{i}$ across the vertex of the
string-nets of different flavor $\left\{ a_{j},b_{j},a_{j}+b_{j}\right\} $
with the amplitude of the final configuration related to the amplitude of
the original configuration by a phase $F_{a_{i}}^{\left( 2\right) }\left(
a_{j},b_{j}\right) .$ The second rule (\ref{rule2aa}) is similar to the
first rule (\ref{rule2a}) but with the orientation of $a_{i}$ reversed.
Similarly, the third rule (\ref{rule2b}) dictates that one can glide a
string $a_{i}$ across the vertex of the string-nets of different flavor with
one null string $\left\{ b_{j},b_{j}^{\ast },0\right\} $ with the relative
amplitude of two configurations being $\kappa _{a_{i}}\left( a_{j}\right) $.
The fourth rule (\ref{rule2c}) says that the amplitudes of two
configurations which are related by deforming one string $a_{i}$ across the
other string $a_{j}$ of different flavor differ by a phase $\eta
_{a_{i}}\left( a_{j}\right) $. The last rule (\ref{rule2d}) depicts that one
can glide a string $a_{i}$ across the intersection of the other two strings $%
a_{j},a_{k}$ where three strings are of different flavors. The amplitude of
resulting configuration is related to the amplitude of the original
configuration by a phase $F_{a_{i}}^{\left( 3\right) }\left(
a_{j},a_{k}\right) .$

By applying these new local rules (\ref{rule2a}--\ref{rule2d}) multiple
times, one can disentangle all different flavors of string-nets. Then we
apply the original local rules (\ref{rule1a}--\ref{rule1c}) for each flavor
of string-nets and relate the amplitude of any string-net configuration to
the amplitude of vacuum. Accordingly, the rules determine the wave functions
of intersecting string-nets once the parameters $\left\{ d,F,\gamma ,\alpha
,F^{\left( 2\right) },\bar{F}^{\left( 2\right) },\eta ,\kappa ,F^{\left(
3\right) }\right\} $ are given.

Before we discuss the constraints which these parameters have to satisfy, we
discuss some corollaries of (\ref{rule2a}--\ref{rule2c}).
First, the rules (\ref{gamma2}) and (\ref{rule2b}) imply that
\begin{equation}
\Phi \left( \raisebox{-0.1in}{\includegraphics[height=0.3in]{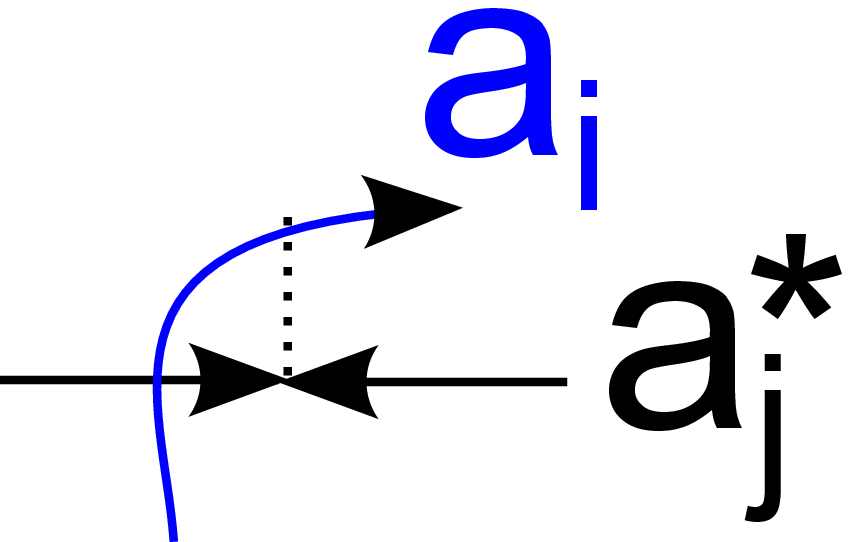}}%
\right) =\kappa _{a_{i}}\left( a_{j}\right) ^{-1}\Phi \left( %
\raisebox{-0.1in}{\includegraphics[height=0.3in]{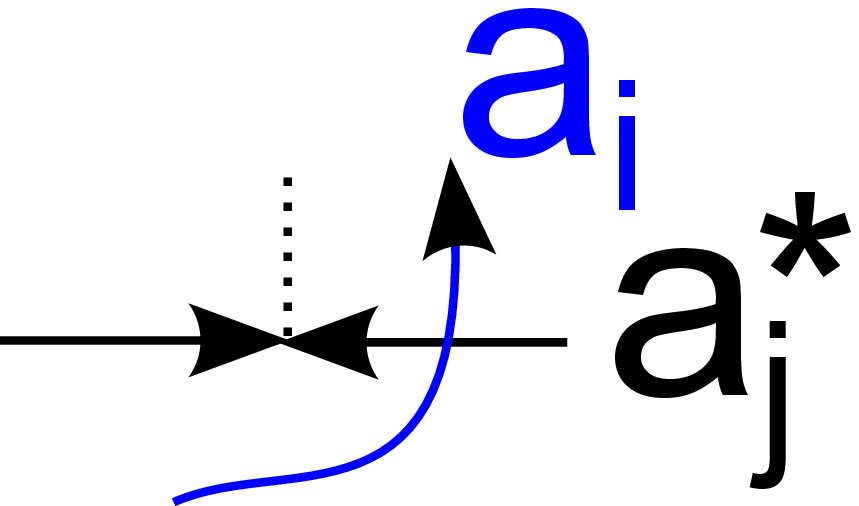}}\right) .
\label{rule2b1}
\end{equation}%
We can also flip the null string of (\ref{rule2b},\ref{rule2b1}) by applying
(\ref{gamma1}) to both sides of the two equations. 

Second, the rules (\ref{rule1b}) and (\ref{rule2c}) imply that
\begin{equation}
\Phi \left( \raisebox{-0.1in}{\includegraphics[height=0.27in]{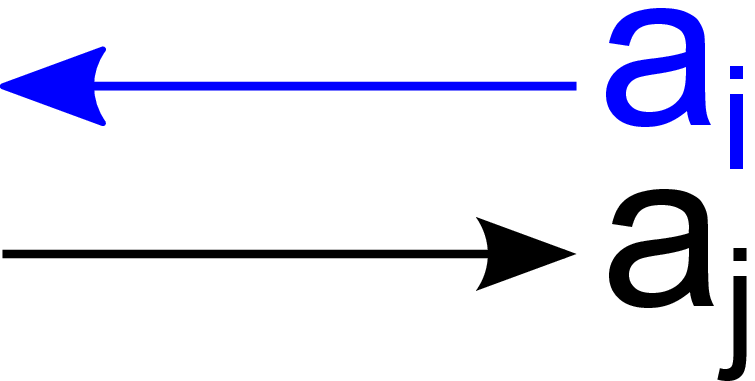}}%
\right) =\eta _{a_{i}}\left( a_{j}\right) \Phi \left( \raisebox{-0.1in}{%
	\includegraphics[height=0.27in]{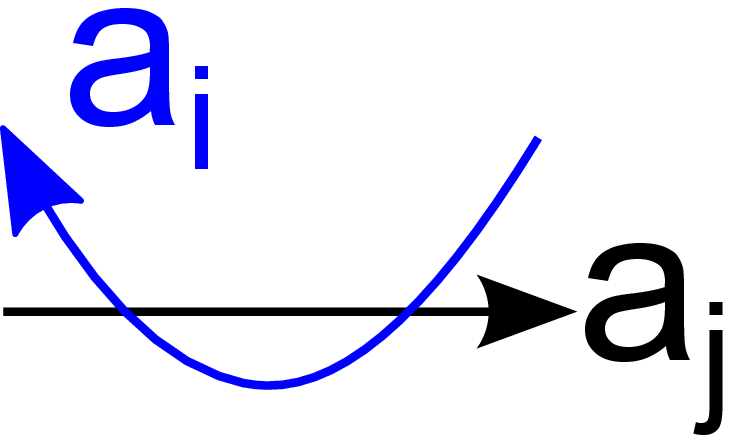}}\right) \label{rule2cc}.
\end{equation}
Together with the rules (\ref{rule2a},\ref{rule2aa},\ref{rule2c}), one can derive the following relations
\begin{gather}
\Phi \left( \raisebox{-0.16in}{\includegraphics[height=0.4in]{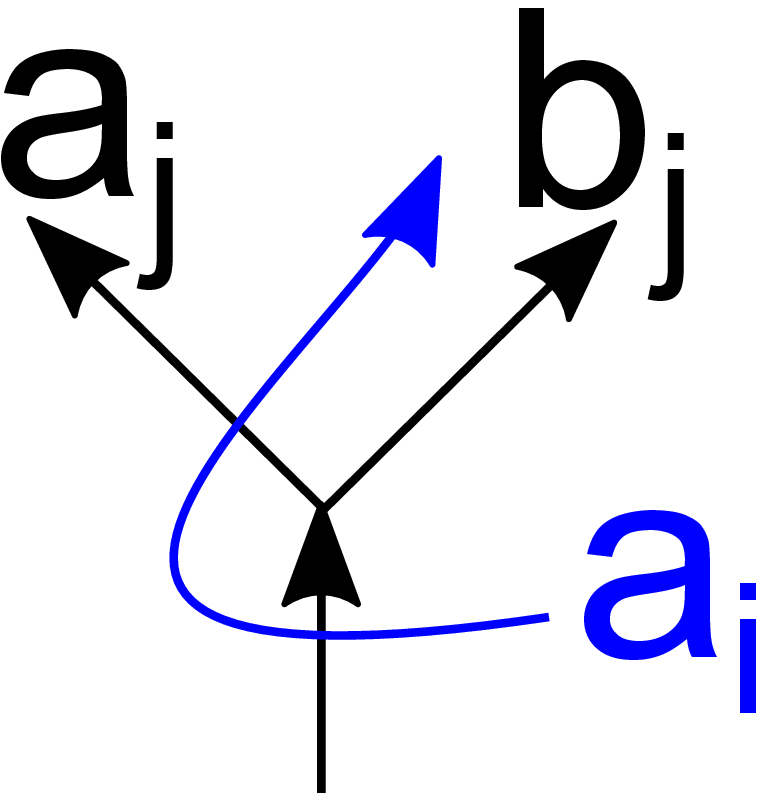}}%
\right) =\frac{F_{a_{i}}^{\left( 2\right) }\left( a_{j},b_{j}\right) \eta
_{a_{i}}\left( b_{j}\right) }{\eta _{a_{i}}\left( a_{j}+b_{j}\right) }\Phi
\left( \raisebox{-0.16in}{\includegraphics[height=0.4in]{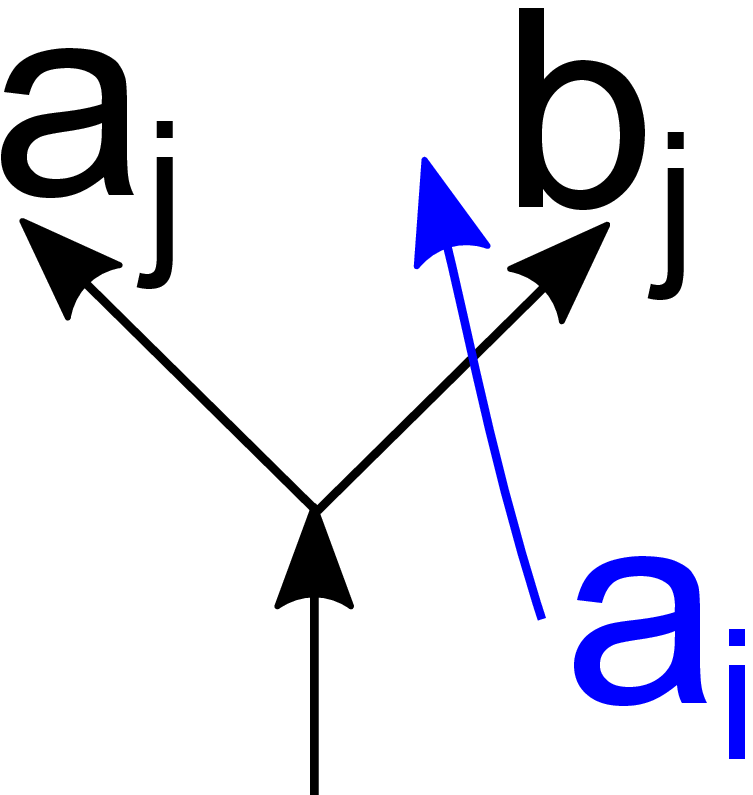}}%
\right) ,  \label{1c} \\
\Phi \left( \raisebox{-0.16in}{\includegraphics[height=0.4in]{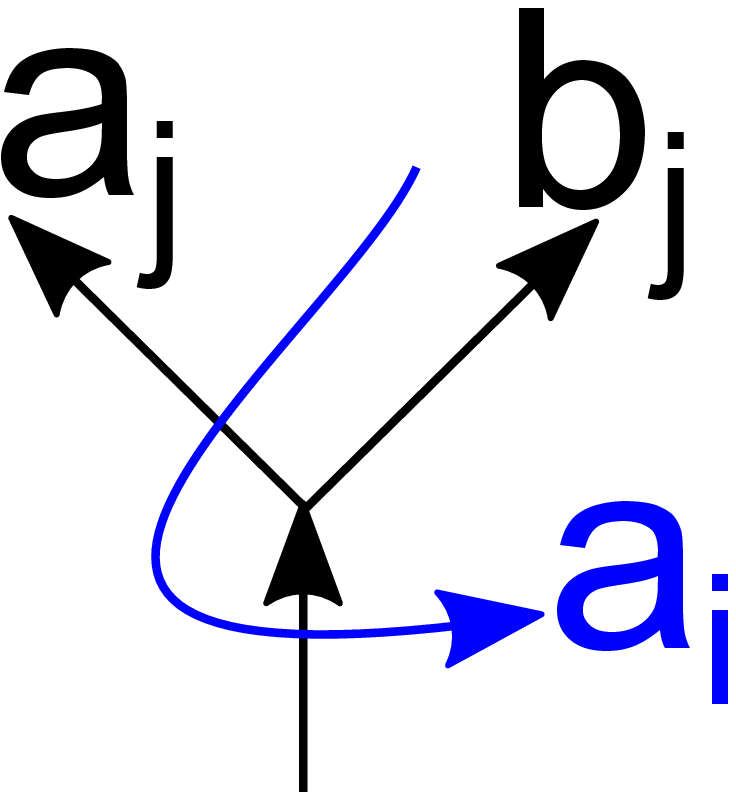}}%
\right) =\frac{\bar{F}_{a_{i}}^{\left( 2\right) }\left( a_{j},b_{j}\right)
\eta _{a_{i}}\left( b_{j}\right) }{\eta _{a_{i}}\left( a_{j}+b_{j}\right) }%
\Phi \left( \raisebox{-0.16in}{\includegraphics[height=0.4in]{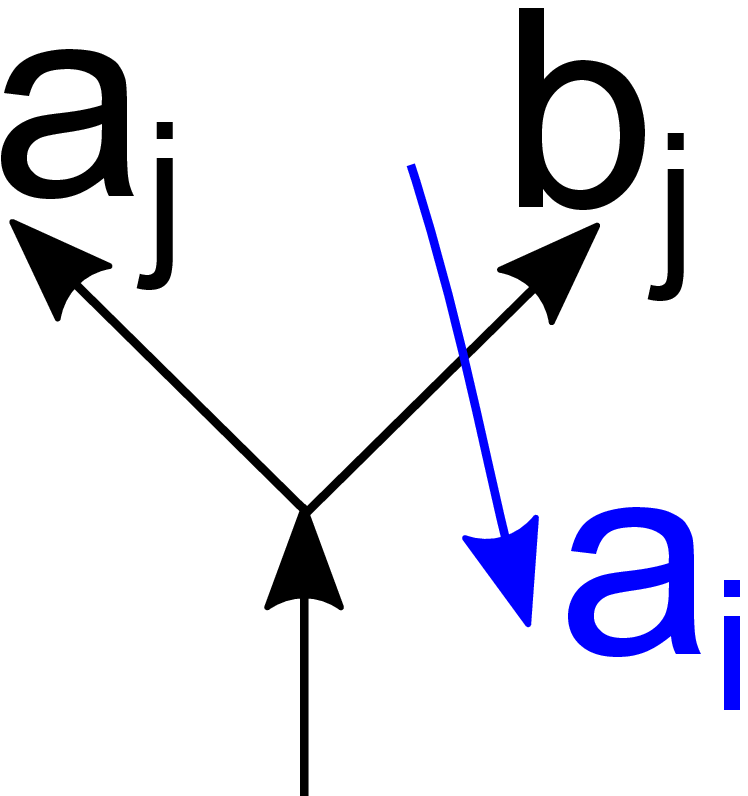}}%
\right) ,  \label{1d} \\
\Phi \left( \raisebox{-0.16in}{\includegraphics[height=0.4in]{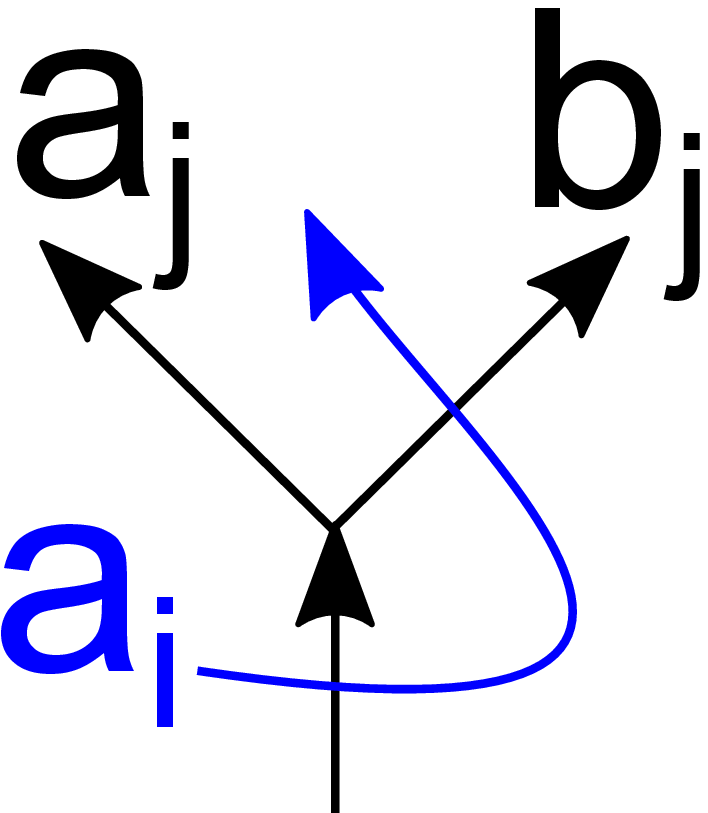}}%
\right) =\frac{\bar{F}_{a_{i}}^{\left( 2\right) }\left( a_{j},b_{j}\right)
\eta _{a_{i}}\left( a_{j}\right) }{\eta _{a_{i}}\left( a_{j}+b_{j}\right) }%
\Phi \left( \raisebox{-0.16in}{\includegraphics[height=0.4in]{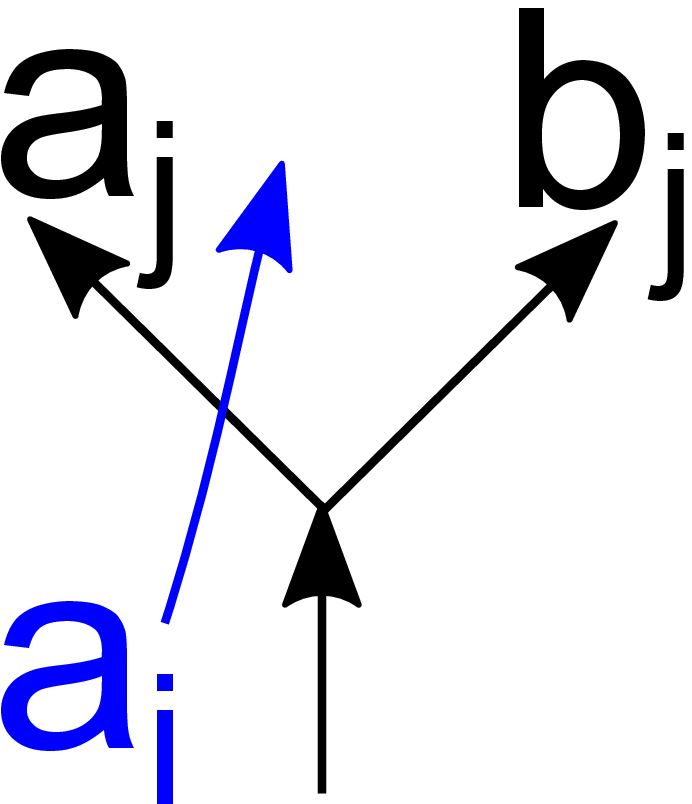}}%
\right) ,  \label{1e} \\
\Phi \left( \raisebox{-0.16in}{\includegraphics[height=0.4in]{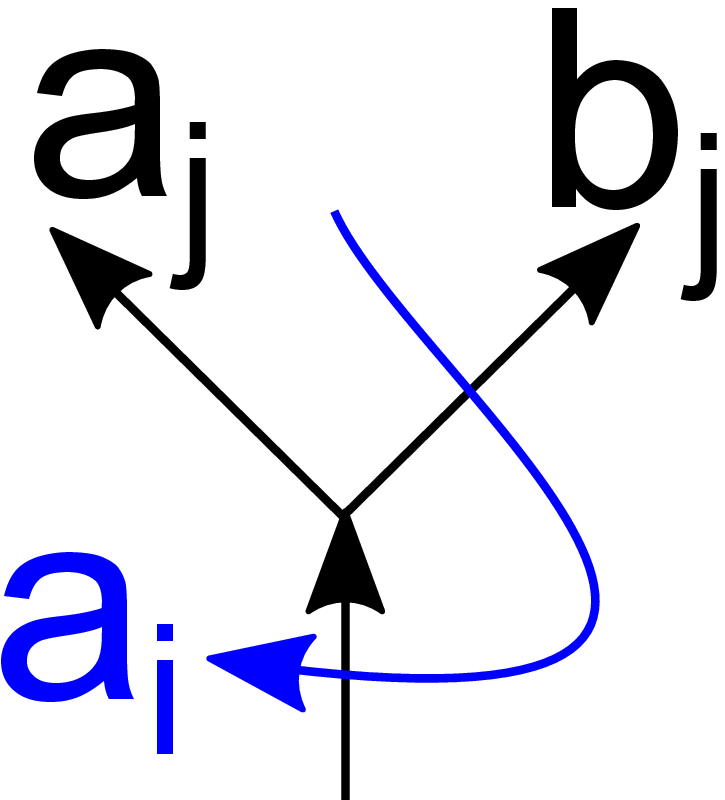}}%
\right) =\frac{F_{a_{i}}^{\left( 2\right) }\left( a_{j},b_{j}\right) \eta
_{a_{i}}\left( a_{j}\right) }{\eta _{a_{i}}\left( a_{j}+b_{j}\right) }\Phi
\left( \raisebox{-0.16in}{\includegraphics[height=0.4in]{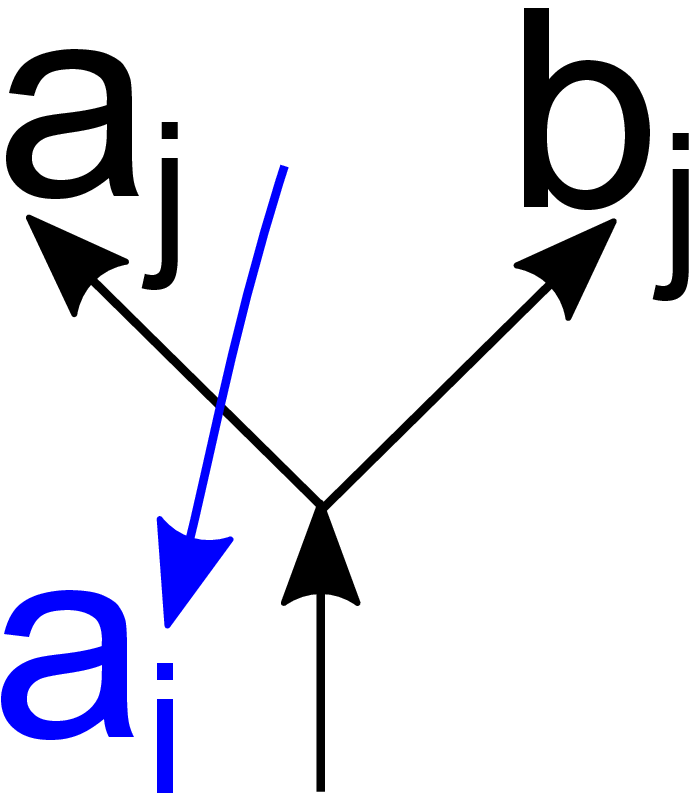}}%
\right) .  \label{1f}
\end{gather}%
These relations allow us to glide a string $a_{i}$ with various orientations
across the vertex of the string-nets of different flavor $\left\{
a_{j},b_{j},a_{j}+b_{j}\right\} $. 

The first equation (\ref{1c}) can be
shown by considering
\begin{gather*}
\Phi \left( \raisebox{-0.16in}{\includegraphics[height=0.4in]{rule11a.eps}}%
\right) =\eta _{a_{i}}\left( b_{j}\right) \Phi \left( \raisebox{-0.16in}{%
\includegraphics[height=0.4in]{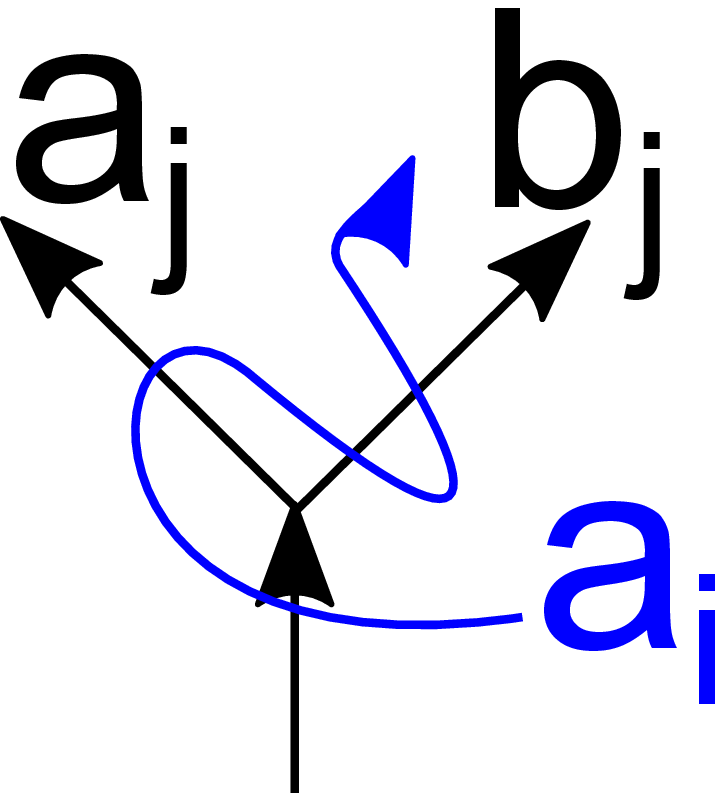}}\right) \\
=\eta _{a_{i}}\left( b_{j}\right) F_{a_{i}}\left( a_{j},b_{j}\right) \Phi
\left( \raisebox{-0.16in}{\includegraphics[height=0.4in]{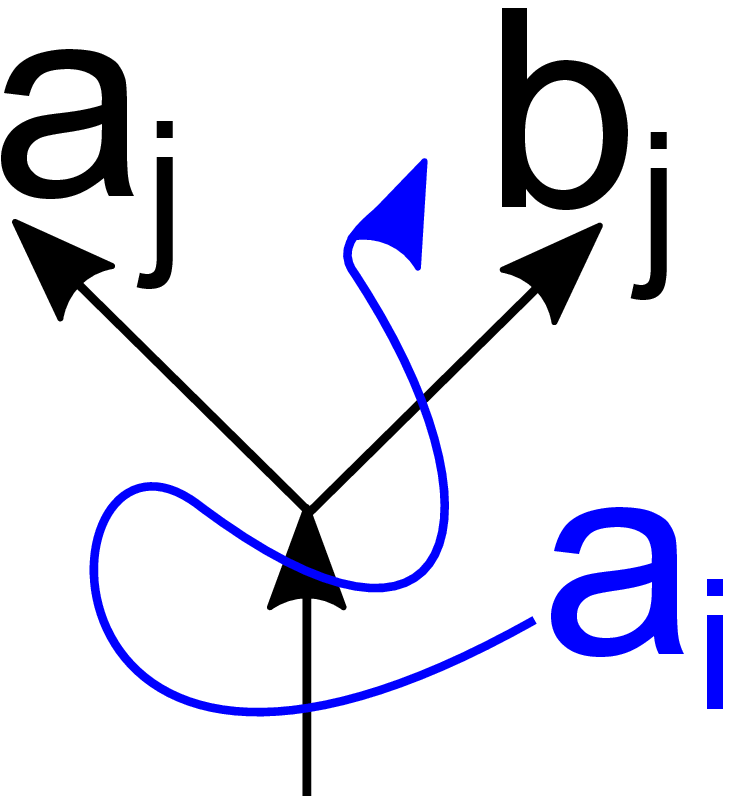}}\right)
\\
=F_{a_{i}}^{\left( 2\right) }\left( a_{j},b_{j}\right) \frac{\eta
_{a_{i}}\left( b_{j}\right) }{\eta _{a_{i}}\left( a_{j}+b_{j}\right) }\Phi
\left( \raisebox{-0.16in}{\includegraphics[height=0.4in]{rule11b.eps}}%
\right) .
\end{gather*}%
This shows (\ref{1c}). Similarly, (\ref{1d}-\ref{1f}) can be shown by the
same manner.

Finally, from the rules (\ref{rule2b},\ref{rule2d}) and (\ref{gamma2}), one can obtain
\begin{equation}
\Phi \left( \raisebox{-0.16in}{\includegraphics[height=0.4in]{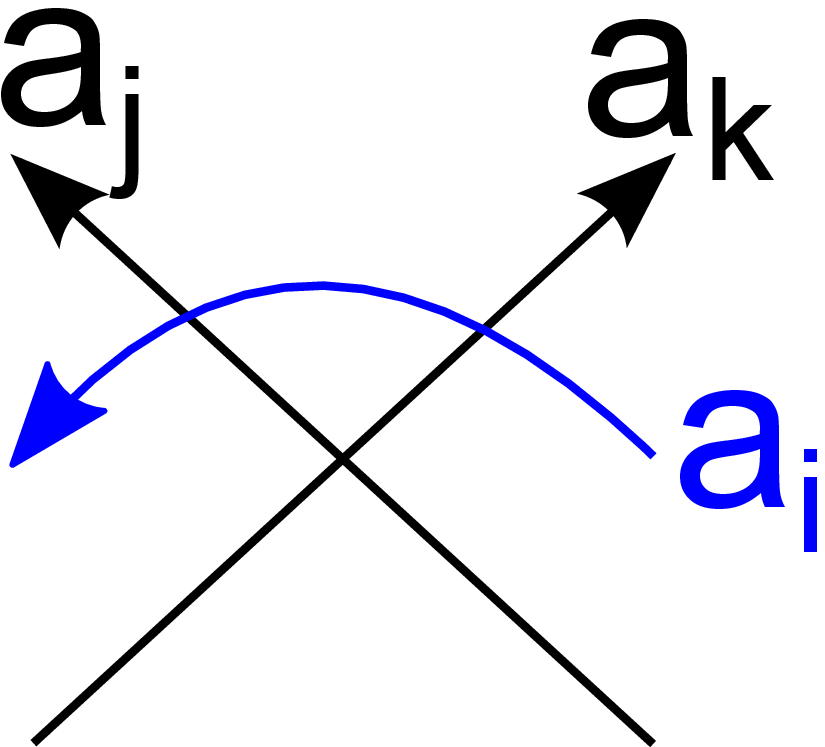}}%
\right) =F^{(3)}_{a_i^*}(a_j,a_k) \Phi \left( \raisebox{-0.16in}{%
	\includegraphics[height=0.4in]{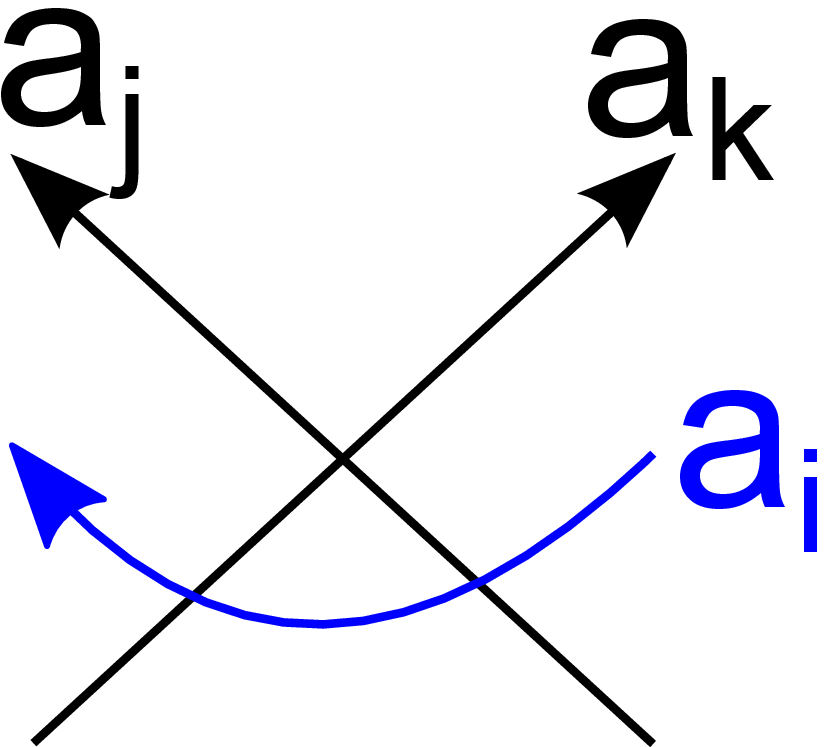}}\right) \label{rule2dd}.
\end{equation}

\subsection{Self-consistency conditions}

\begin{figure}[tbp]
\begin{center}
\includegraphics[width=0.9\columnwidth]{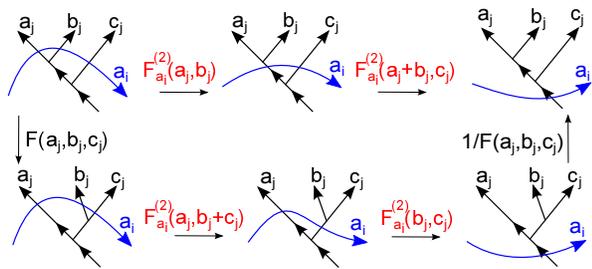}
\end{center}
\caption{The amplitude of the upper-left configuration can be related to the amplitude of the upper-right configuration in two different ways by the local rule (\ref{rule2a}). Self-consistency requires the two sequences of operation result in the same linear relations between the amplitudes of the two configurations.}
\label{figure:sfeq1}
\end{figure}

To have a well-defined intersecting string-net wave function $\Phi ,$ the
parameters $\left\{ d,F,\gamma ,\alpha \right\} $ need to satisfy (\ref%
{sfeqa}) for each flavor of string-nets and the parameters $\left\{
F^{\left( 2\right) },\bar{F}^{\left( 2\right) },\kappa ,\eta ,F^{\left(
3\right) }\right\} $ have to satisfy the following constraints:%
\begin{subequations}
\begin{gather}
F_{a_{i}}^{\left( 2\right) }\left( a_{j},b_{j}\right)  F_{a_{i}}^{\left(
2\right) }\left( a_{j}+b_{j},c_{j}\right)  \label{sfeq1} \\
\quad\ \ \ \ \ \ \ \ \ =F_{a_{i}}^{\left( 2\right) }\left( b_{j},c_{j}\right) F_{a_{i}}^{\left(
2\right) }\left( a_{j},b_{j}+c_{j}\right) , \notag \\
\bar{F}_{a_{i}}^{\left( 2\right) }\left( a_{j},b_{j}\right) =F_{a_{i}^{\ast
}}^{\left( 2\right) }\left( a_{j},b_{j}\right) \frac{\kappa _{a_{j}}\left(
a_{i}^{\ast }\right) \kappa _{b_{j}}\left( a_{i}^{\ast }\right) }{\kappa
_{a_{j}+b_{j}}\left( a_{i}^{\ast }\right) },  \label{fbar} \\
\frac{F_{a_{i}}^{\left( 2\right) }\left( a_{j},b_{j}\right) F_{b_{i}}^{\left(
2\right) }\left( a_{j},b_{j}\right)}{F_{a_{i}+b_{i}}^{\left(2\right) }\left( a_{j},b_{j}\right)}=\frac{\bar{F}_{a_{j}}^{\left( 2\right) }\left( a_{i},b_{i}\right) \bar{F}_{b_{j}}^{\left( 2\right) }\left( a_{i},b_{i}\right)}
{\bar{F}_{a_{j}+b_{j}}^{\left( 2\right)}\left( a_{i},b_{i}\right)}
 , \label{3f} \\
\kappa _{a_{i}}\left( a_{j}\right) \kappa _{a_{i}^{\ast }}\left( a_{j}^{\ast
}\right) \kappa _{a_{j}^{\ast }}\left( a_{i}\right) \kappa _{a_{j}}\left(
a_{i}^{\ast }\right) =1,  \label{sfeq2} \\
\kappa _{a_{i}}\left( a_{j}\right) =\eta _{a_{i}}\left( a_{j}\right)
F_{a_{i}}^{\left( 2\right) }\left( a_{j}^{\ast },a_{j}\right) ,
\label{sfeq4} \\
F_{a_{i}}^{\left( 3\right) }\left( a_{j},a_{k}\right) F_{a_{i}}^{\left(
3\right) }\left( b_{k}^{\ast },a_{j}\right) =F_{a_{i}}^{\left( 3\right)
}\left( a_{j},a_{k}+b_{k}\right) ,  \label{sfeq5} \\
F^{(3)}_{a_i}(a_j,a_k)F^{(3)}_{b_i}(a_j,a_k)=F^{(3)}_{a_i+b_i}(a_j,a_k),
\label{sfeq6} \\
\eta _{a_{i}}\left( a_{j}\right) =\eta _{a_{j}}\left( a_{i}\right) ,
\label{sfeq7} \\
F_{a_{i}}^{\left( 3\right) }\left( a_{j},a_{k}\right) =F_{a_{j}^{\ast
}}^{\left( 3\right) }\left( a_{k},a_{i}\right) ^{-1}=F_{a_{i}^{\ast
}}^{\left( 3\right) }\left( a_{j}^{\ast },a_{k}^{\ast }\right) ^{-1},
\label{sfeq61} \\
F_{a_{i}}^{\left( 2\right) }\left( a_{j},b_{j}\right) =F_{a_{i}}^{\left(
3\right) }\left( a_{j},a_{k}\right) =1\text{ if }a\text{ or }b=0.
\label{sfeq8}
\end{gather}%
\label{sfeqb}
\end{subequations}
The first equation (\ref{sfeq1}) can be understood by considering the
sequence of manipulations shown in Fig. \ref{figure:sfeq1}. The amplitude of
(c) can be obtained from (a) in two different ways. For the rule to be
consistent, $F^{\left( 2\right) }$ must satisfy (\ref{sfeq1}). The other
conditions (\ref{fbar}--\ref{sfeq6}) can be derived from similar consistency
requirement (see appendix \ref{app:sfeqs}). Eq. (\ref{sfeq7}) comes from the
symmetry of the roles of two strings of different flavors $a_{i},a_{j}$ in
the rule (\ref{rule2c}). Similarly, Eq. (\ref{sfeq61}) follows from the
symmetry of the roles of three strings of different flavors $%
a_{i},a_{j},a_{k}$ in the rule (\ref{rule2d}). The last condition (\ref%
{sfeq8}) simply says that gliding a null string around the other string
gives no phase factor to the amplitude of the final string-net configuration.

We see from (\ref{fbar}), it is sufficient to solve $\left\{ F^{\left(
2\right) },\kappa ,\eta ,F^{\left( 3\right) }\right\} .$ Once we have
solutions to $\left\{ F^{\left( 2\right) },\kappa ,\eta ,F^{\left( 3\right)
}\right\} ,$ the $\bar{F}^{\left( 2\right) }$ can determined by $\left( \ref%
{fbar}\right) .$ Specifically, Eqs. (\ref{sfeq1}--\ref{sfeq2}) determine $F^{(2)},\kappa$. With $F^{(2)},\kappa$, Eqs. (\ref{sfeq4},\ref{sfeq7}) then determine $\eta$. Finally, Eqs. (\ref{sfeq5},\ref{sfeq6},\ref{sfeq61}) determine $F^{(3)}$. The parameters $F^{(2)},F^{(3)}$ are subject to the normalization conditions (\ref{sfeq8}). Thus in the following discussion, we will mainly focus on
solving for $\left\{ F^{\left(2\right) },\kappa ,\eta ,F^{\left( 3\right) }\right\} .$

Similarly, to construct a consistent string-net model, we need one more
constraint%
\begin{equation}
\left\vert F^{\left( 2\right) }\right\vert =\left\vert F^{\left( 3\right)
}\right\vert =\left\vert \kappa \right\vert =1.  \label{norm1}
\end{equation}%
This constraint ensures the corresponding exactly soluble Hamiltonians to be
Hermitian.

\subsection{Gauge transformation}
Like the solution $\{F,d,\alpha,\gamma\}$ to (\ref{sfeqa}), given a solution $\left\{ F^{\left( 2\right) },\kappa ,\eta ,F^{\left(
3\right) }\right\} $ to the self-consistency conditions (\ref{sfeqb}), we can construct an infinite class of other solutions $\left\{ 
\tilde{F}^{\left( 2\right) },\tilde{\kappa},\tilde{\eta},\tilde{F}^{\left(
3\right) }\right\} $ by defining
\begin{align}
\tilde{F}_{a_{i}}^{\left( 2\right) }\left( a_{j},b_{j}\right) &
=F_{a_{i}}^{\left( 2\right) }\left( a_{j},b_{j}\right) \frac{f_{a_{i}}\left(
a_{j}\right) f_{a_{i}}\left( b_{j}\right) }{f_{a_{i}}\left(
a_{j}+b_{j}\right) },  \label{gauge3} \\
\tilde{\kappa}_{a_{i}}\left( a_{j}\right) & =\kappa _{a_{i}}\left(
a_{j}\right) \frac{f_{a_{i}}\left( a_{j}^{\ast }\right) }{f_{a_{i}}\left(
a_{j}\right) },  \notag \\
\tilde{\eta}_{a_{i}}\left( a_{j}\right) & =\eta _{a_{i}}\left( a_{j}\right) 
\frac{1}{f_{a_{i}}\left( a_{j}\right) ^{2}},  \notag \\
\tilde{F}_{a_{i}}^{\left( 3\right) }\left( a_{j},a_{k}\right) &
=F_{a_{i}}^{\left( 3\right) }\left( a_{j},a_{k}\right) .  \notag
\end{align}%
Here $f_{a}\left( b\right) $ is any complex function with%
\begin{equation*}
f_{a}\left( b\right) =f_{b}\left( a\right) ,\left\vert f_{a}\left( b\right)
\right\vert =1,f_{a}\left( b\right) =1\text{ if }a\text{ or }b=0.
\end{equation*}%
We refer to (\ref{gauge3}) as the gauge transformations and two sets of
solutions $\left\{ F^{\left( 2\right) },\kappa ,\eta ,F^{\left( 3\right)
}\right\} $ and $\left\{ \tilde{F}^{\left( 2\right) },\tilde{\kappa},\tilde{%
\eta},\tilde{F}^{\left( 3\right) }\right\} $ are called gauge equivalent if
they are related by such a transformation. One can show that the gauge
transformation can be implemented by a local unitary transformation which
can be generated by the time evolution of a local Hamiltonian over a finite
period of time. Thus, this implies that if two solutions to the
self-consistency conditions (\ref{sfeqb}) are related by a
gauge transformation, then the corresponding wave functions $\Phi ,\tilde{%
\Phi}$ describe the same quantum phase. Since we are primarily interested in
constructing different topological phases, then we only need to consider one
solution to (\ref{sfeqb}) within each gauge equivalence class.\ 

\section{Intersecting-string-net Hamiltonians \label{section:H}}

In this section, we will construct a large class of exactly soluble lattice
Hamiltonians that have the wave functions $\Phi $ as their ground states.
For a given solution $\left\{ d,F,\gamma ,\alpha \right\} $ and $\left\{
F^{\left( 2\right) },\kappa ,\eta ,F^{\left( 3\right) }\right\} $ to the
self-consistency conditions (\ref{sfeqa}) and (\ref{sfeqb}), we will construct an exactly soluble Hamiltonian whose ground
state $\left\vert \Phi _{latt}\right\rangle $ obeys the local rules (\ref%
{rule1a}--\ref{rule1c},\ref{rule2a}--\ref{rule2d}) on the lattice.

\subsection{Definition of the Hamiltonian \label{section:ham}}

\begin{figure}[tbp]
\begin{center}
\includegraphics[width=0.7\columnwidth]{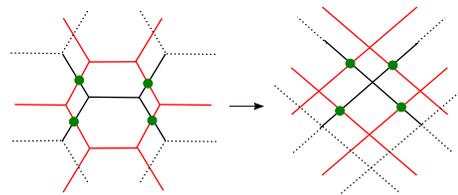}
\end{center}
\caption{We represent two intersecting honeycomb lattices (on the left) by
two intersecting square lattices (on the right) with proper lattice splitting at vertices of square lattices in mind. The green dots denote the intersections between two lattices.}
\label{figure:lattice1}
\end{figure}

Let us first specify the Hilbert space for our model. As the original
string-net model is a spin system with the spins located on the links of the
honeycomb lattice, the $L$-intersecting string-net model is defined on the $%
L $ copies of intersecting honeycomb lattices with each flavor of spins
living on the links of individual honeycomb lattice. The $L$ honeycomb
lattices are arranged in such a way that the $\left( i+1\right) $-th lattice
is obtained by shifting the $i$-th lattice by a small constant vector $v$ (see
Fig. \ref{figure:lattice1}). We assume that the overall shifting between the first and $L$-th honeycomb lattices is smaller than twice the lattice constant $2a$, namely$(L-1)v<2a$ for the sake of ordering the $L$ lattices. The spins of the $i$-th flavor can be $%
\left\vert G_{i}\right\vert $ different states which are labeled by elements
of the subgroup $G_{i}:$ $\left\{\left\vert a_{i}\right\rangle :a_{i}\in
G_{i}\right\} $ with $i=1,...,L.$ For the ease of graphical presentation, we
replace the honeycomb lattice by a square lattice with the proper lattice
splitting at vertices of square lattices in mind (see Fig. \ref{figure:lattice1}). 
As far as the intersections are concerned, intersecting square lattices capture all
the intersections on the original intersecting honeycomb lattices and it is
more transparent to see the intersections in the square lattices when we consider more intersecting lattices.
When a
spin of the $i$-th flavor is in state $\left\vert a_{i}\right\rangle ,$ we
regard the link as being occupied by a sting of type-$a_{i}$, oriented in a
certain direction. If the spin is in state $\left\vert 0\right\rangle ,$ the
link is occupied by the null string. In this way, each spin state can be
equivalently described as an intersecting string-net state.

\begin{figure}[tbp]
\begin{center}
\includegraphics[width=0.35\columnwidth]{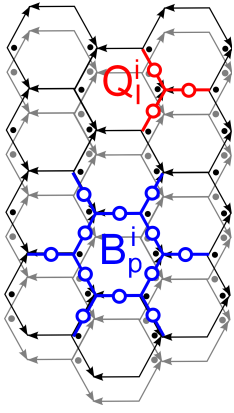}
\end{center}
\caption{Lattice spin model (\ref{ham}). The $Q_I^i$ operators act on three spins on the links which meet at vertex $I$ while the $B_p^i$ operators acts on $12$ spins adjacent to the plaquette $p$ and $4(L-1)$ more spins intersecting $p$ with $L$ being the number of lattices. Here $L=2$ since there are two intersecting honeycomb lattices.}
\label{figure:lattice2}
\end{figure}

The Hamiltonian is of the form%
\begin{equation}
H=-\sum_{i=1}^{L}\left( \sum_{I}Q_{I}^i+\sum_{p}B_{p}^i\right)
\label{ham}
\end{equation}%
where the first sum runs over all flavors of spins $i=1,...,L$ and the next
two sums run over the sites $I$ and the plaquettes $p$ of the $i$-th
honeycomb lattice. Here we label the sites and plaquettes of all other lattices according to the ones of the first lattice. 

The operator $Q_{I}^i$ acts on the 3 spins adjacent to
the site $I$ :%
\begin{equation*}
Q_{I}^i\left\vert \raisebox{-0.16in}{%
\includegraphics[height=0.4in]{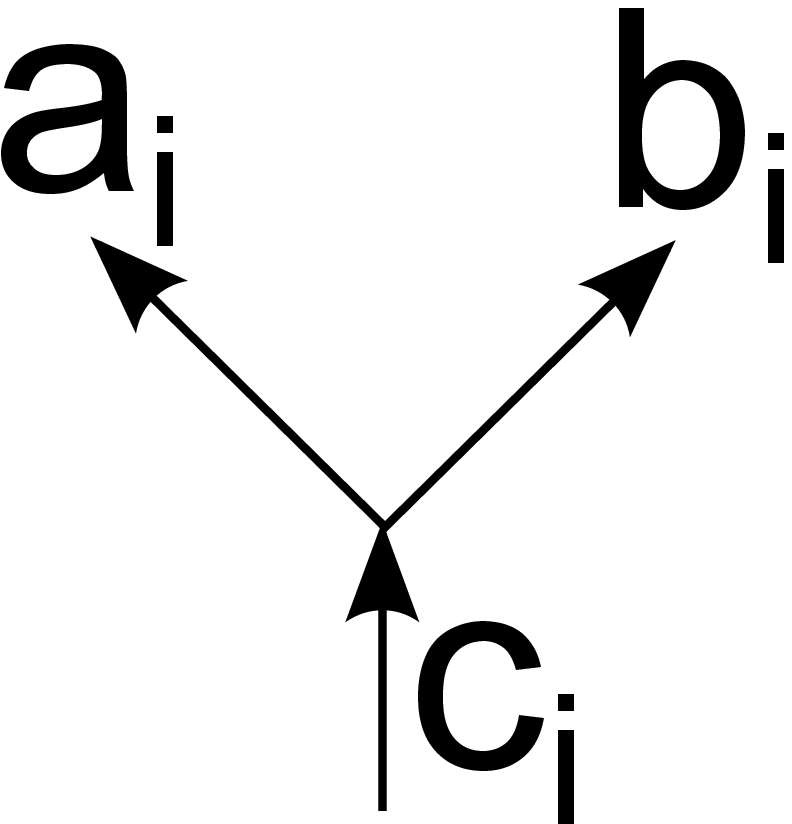}}\right\rangle =\delta
_{a_{i}b_{i}c_{i}}\left\vert \raisebox{-0.16in}{%
\includegraphics[height=0.4in]{qi.eps}}\right\rangle
\end{equation*}%
where%
\begin{equation*}
\delta _{a_{i}b_{i}c_{i}}=\left\{ 
\begin{array}{l}
1,\text{ if }a_{i}+b_{i}+c_{i}=0 \\ 
0,\text{ otherwise}%
\end{array}%
\right.
\end{equation*}%
(see Fig. \ref{figure:lattice2}). The $Q_{I}$ terms annihilate the states that do not
satisfy the branching rules.

The $B_{p}^i$ operator provides dynamics to the string-net configurations.
It can be written as a linear combination%
\begin{equation*}
B_{p}^i=\sum_{s_{i}\in G_{i}}a_{s_{i}}B_{p}^{s_{i}},
\end{equation*}
where $B_{p}^{s_{i}}$ describes a $12+4\left( L-1\right) $ spin
interaction involving the spins on the $12$ links surrounding the plaquette $%
p$ and $4$ links of each other $\left( L-1\right) $ honeycomb lattices
which intersect of the boundary of the plaquette $p.$ The $%
B_{p}^{s_{i}}$ operator describes three types of interactions among $L$
flavors of string-nets. First, it contains the 12-spin interaction for
each flavor of string-nets. For $L=1$ case, the above Hamiltonian reduces to
the original string-net model in Ref. \onlinecite{LinLevinstrnet}. For $%
L\geq2,$ it further contains 2-spin interaction at each crossing between 
any two different flavors of string-nets.
The $a_{s_{i}}$
are some complex coefficients satisfying $a_{s_{i}^{\ast}}=a_{s_{i}}^{\ast}.$

The operator $B_{p}^{s_{i}}$ has some special structures. First, it
annihilates any state that does not obey the branching rules at 6 vertices
surrounding the plaquette $p$. Second, the $B_{p}^{s_{i}}$ acts
non-trivially on the 6 inner spins along the boundary of $p$ and it does
not affect the outer $6$ spins and other intersecting spins of different
flavors. However, its matrix elements depend on the state of outer spins and
other intersecting spins of different flavors.

Specifically, the matrix elements are defined by 
\begin{align*}
\left\langle \raisebox{-0.25in}{\includegraphics[height=0.6in]{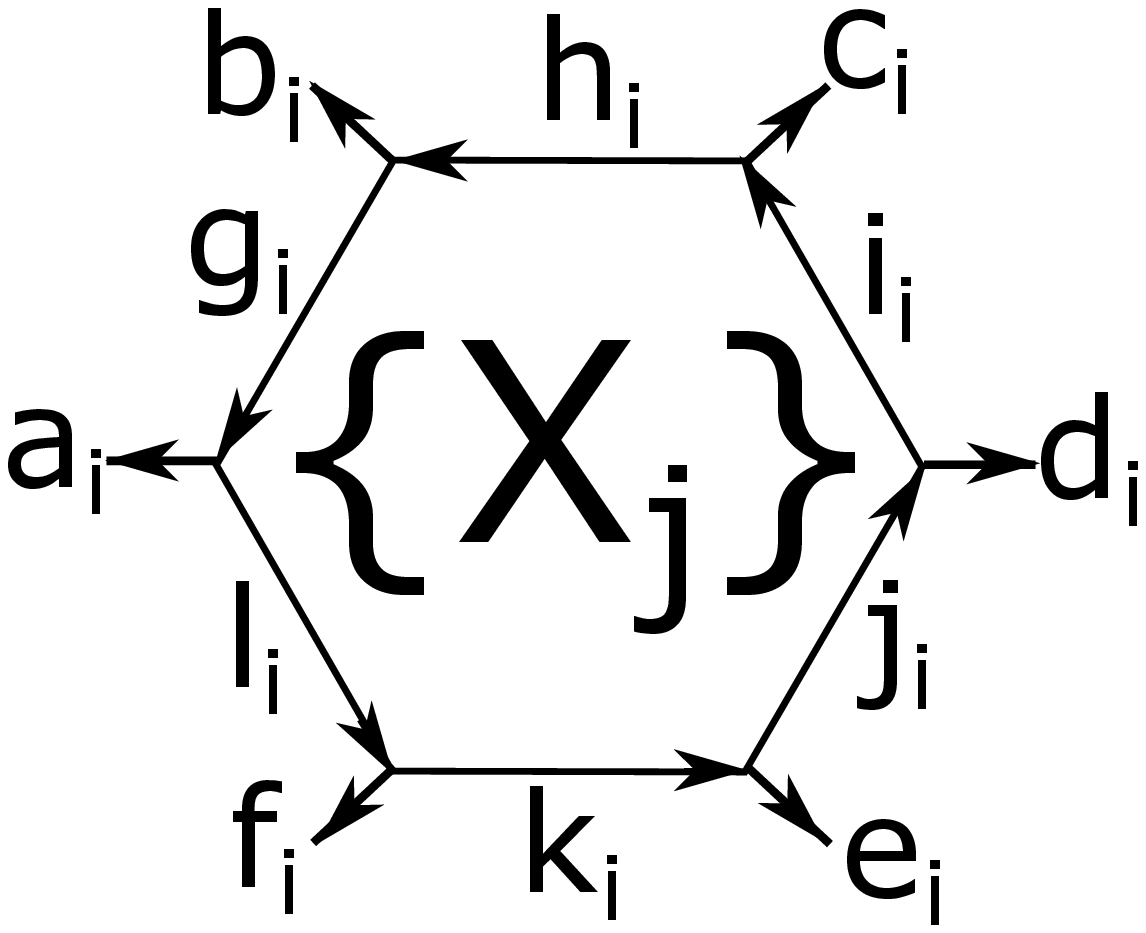}}
\right\vert B_{p}^{s_{i}} & 
\left\vert \raisebox{-0.25in}{\includegraphics[height=0.6in]{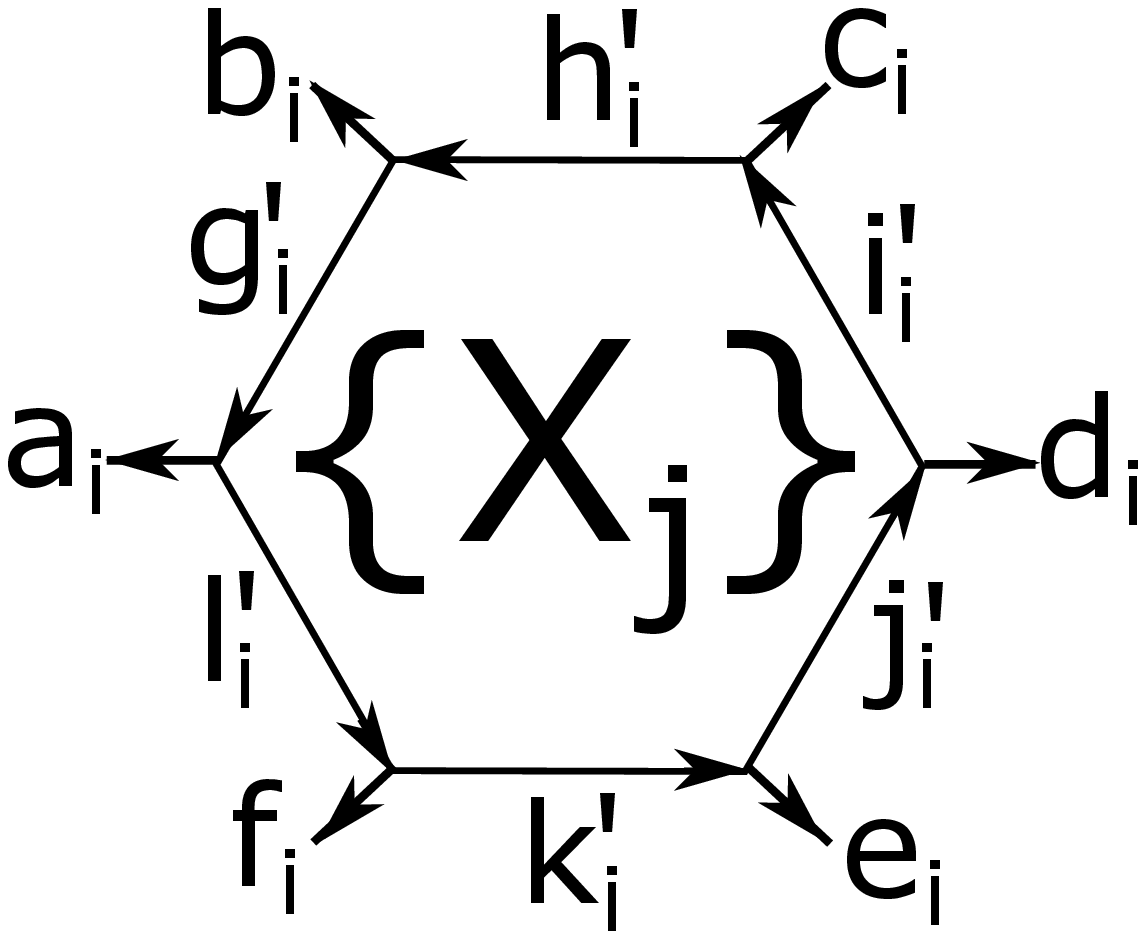}}
\right\rangle  \\
&=B_{p,g_i'...l_i'}^{s_i,g_i...l_i}\left( a_i...f_i;\sum_{i\neq j}{X_j} \right) 
\end{align*}
where%
\begin{equation}
B_{p,g_{i}^{^{\prime }}...l_{i}^{\prime }}^{s_{i},g_{i}...l_{i}}\left(
a_{i}...f_{i};\sum_{i\neq j}\left\{ X_{j}\right\} \right) =B_{p,x'x}^{s_i\left(
1\right) }B_{p,x'x}^{s_i\left( 2\right) }B_{p,x'x}^{s_i\left( 3\right) }  \label{bps}
\end{equation}%
with%
\begin{align}
B_{p,x'x}^{s_i\left( 1\right) }& =\delta _{g_{i}^{\prime }}^{g_{i}+s_{i}}\delta
_{h_{i}^{\prime }}^{h_{i}+s_{i}}\delta _{i_{i}^{\prime
}}^{i_{i}+s_{i}}\delta _{j_{i}^{\prime }}^{j_{i}+s_{i}}\delta
_{k_{i}^{\prime }}^{k_{i}+s_{i}}\delta _{l_{i}^{\prime
}}^{l_{i}+s_{i}}\times   \label{b1} \\
& F_{s_{i}^{\ast }g_{i}^{\prime }b_{i}}F_{s_{i}^{\ast }h_{i}^{\prime
}c_{i}}F_{s_{i}^{\ast }i_{i}^{\prime }d_{i}}F_{s_{i}^{\ast }j_{i}^{\prime
}e_{i}}F_{s_{i}^{\ast }k_{i}^{\prime }f_{i}}F_{s_{i}^{\ast }l_{i}^{\prime
}a_{i}},  \notag \\
B_{p,x'x}^{s_i\left( 2\right) }& =\prod_{j\neq i}
\eta_{s_i,i_j}\frac{F^{(2)}_{g_j,s_il_i}F^{(2)}_{b_j,s_ig_i}F^{(2)}_{c_j,s_ii_i}\bar{F}^{(2)}_{i_j,s_ij_i}}{\bar{F}^{(2)}_{s_i,g_jb_j}\bar{F}^{(2)}_{s_i,(g_j+b_j)c_j}}, \label{b2} \\
B_{p,x'x}^{s_i\left( 3\right) }& =\prod_{j\neq k\neq i}F_{s_{i}}^{\left( 3\right)
}\left( c_{k}^*,i_{j}\right) F_{s_{i^*}}^{\left( 3\right) -1}\left(
g_{j},b_{k}\right)   \label{b3}
\end{align}%
and $F_{abc}\equiv F\left( a,b,c\right) ,F_{a}^{\left( 2\right) }\left(
b,c\right) \equiv F_{a,bc}^{\left( 2\right) },\eta _{a,b}\equiv \eta
_{a}\left( b\right) .$ 
The $g_{i}...l_{i}$ and $a_{i}...f_{i}$ denote $g_{i}h_{i}i_{i}j_{i}k_{i}l_{i}$
and $a_{i}b_{i}c_{i}d_{i}e_{i}f_{i}$ respectively. The $x,x'$ denote the spin labels around the plaquette $p$ for the initial and final state configurations, i.e. $x=g_ih_i\dots$ and $x'=g_i'h_i'\dots$.
Here the action of $B_{p}^{s_{i}}$ is defined to
be on the bra state $\left\langle \raisebox{-0.25in}{%
\includegraphics[height=0.6in]{bp1.eps}}\right\vert $ and $\left\{
X_{j}\right\} $ in the bra state represents the configuration of other
flavors $j\neq i$ of string-nets which intersects the plaquette $p.$ Note that the above
expression is only valid if the initial and final states obey the branching
rules, i.e. $h_{i}=b_{i}+g_{i},h_{i}^{\prime }=b_{i}+g_{i}^{\prime },$ etc.
Otherwise, the matrix element of $B_{p}^{s_{i}}$ vanishes.

\begin{figure}[tbp]
\begin{center}
\includegraphics[width=0.7\columnwidth]{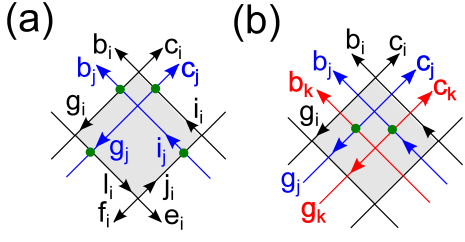}
\end{center}
\caption{(a) Intersections (green dots) between $i,j$ flavors of strings in the boundary of the plaquette $%
p$ (the grey region). (b) Intersections between $j,k$ flavors of strings  inside the plaquette $p$.}
\label{figure:23intersection}
\end{figure}

The matrix elements of $B_{p}^{s_{i}}$ are the product of three components.
The first component $B_{p,x'x}^{s_i\left( 1\right) }$ is the phase factor of creating a
closed loop $s_{i}$ on the plaquette $p$ in the absence of other flavors
of string-nets. 
The other two components $B_{p,x'x}^{s_i\left( 2\right) },B_{p,x'x}^{s_i\left( 3\right) }$ are the phase factors
associated with the intersections on and inside the plaquette $p$,
respectively. They account for the interactions between different flavors of
string-nets. The geometry of the intersections of string-nets on and inside
the plaquette $p$ is shown in Fig. \ref{figure:23intersection}.

Notice that the above matrix elements are computed for a particular orientation configuration in which the inner links of $i$-th lattice are oriented cyclically and the links of other lattices intersecting the plaquette $p$ are oriented as in Fig. \ref{figure:23intersection}. This choice of orientations make the matrix elements simple but however, this orientation configuration can not be extended to the whole lattices while preserving translational symmetry. If we instead choose a translationally invariant orientation configuration as in Fig. \ref{figure:lattice2}, the matrix elements are modified as
\begin{align*}
\left\langle \raisebox{-0.25in}{%
\includegraphics[height=0.6in]{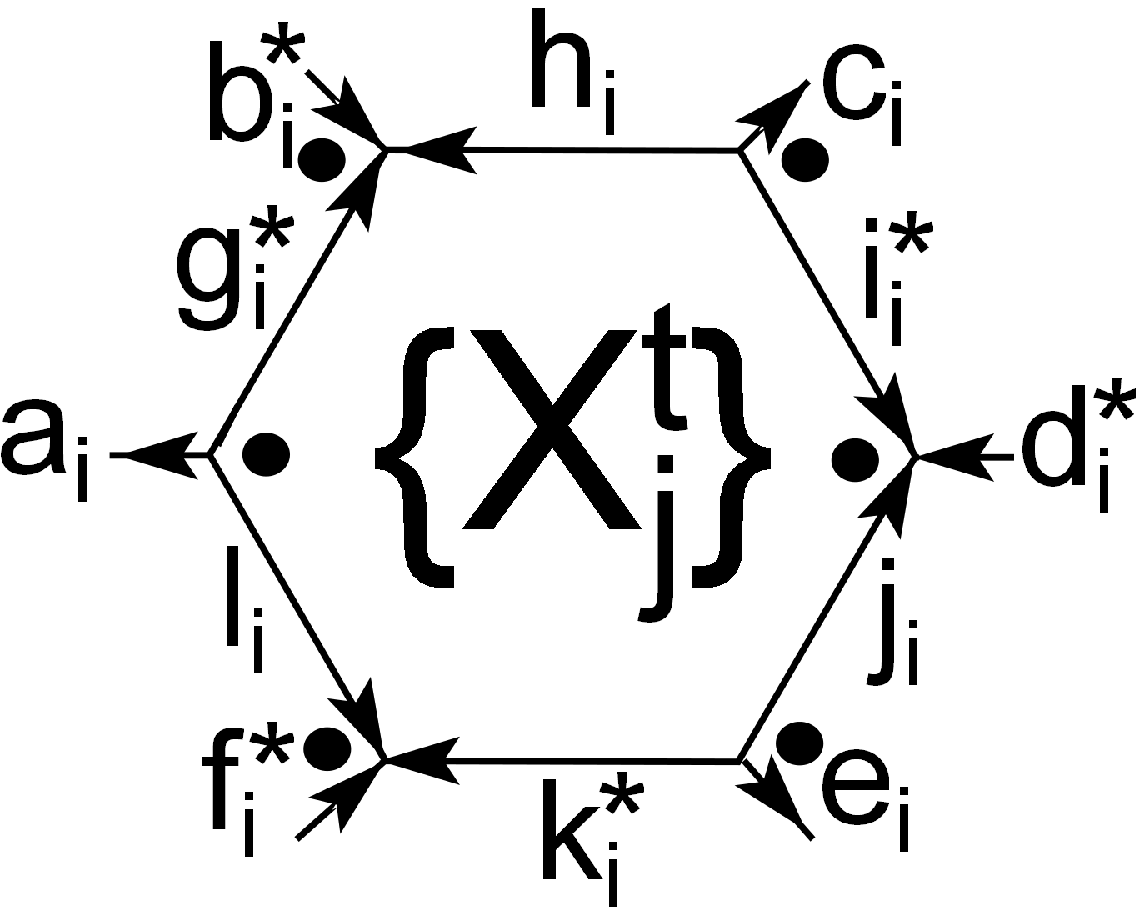}}\right\vert B_p^{s_i} &
\left\vert \raisebox{-0.25in}{%
\includegraphics[height=0.6in]{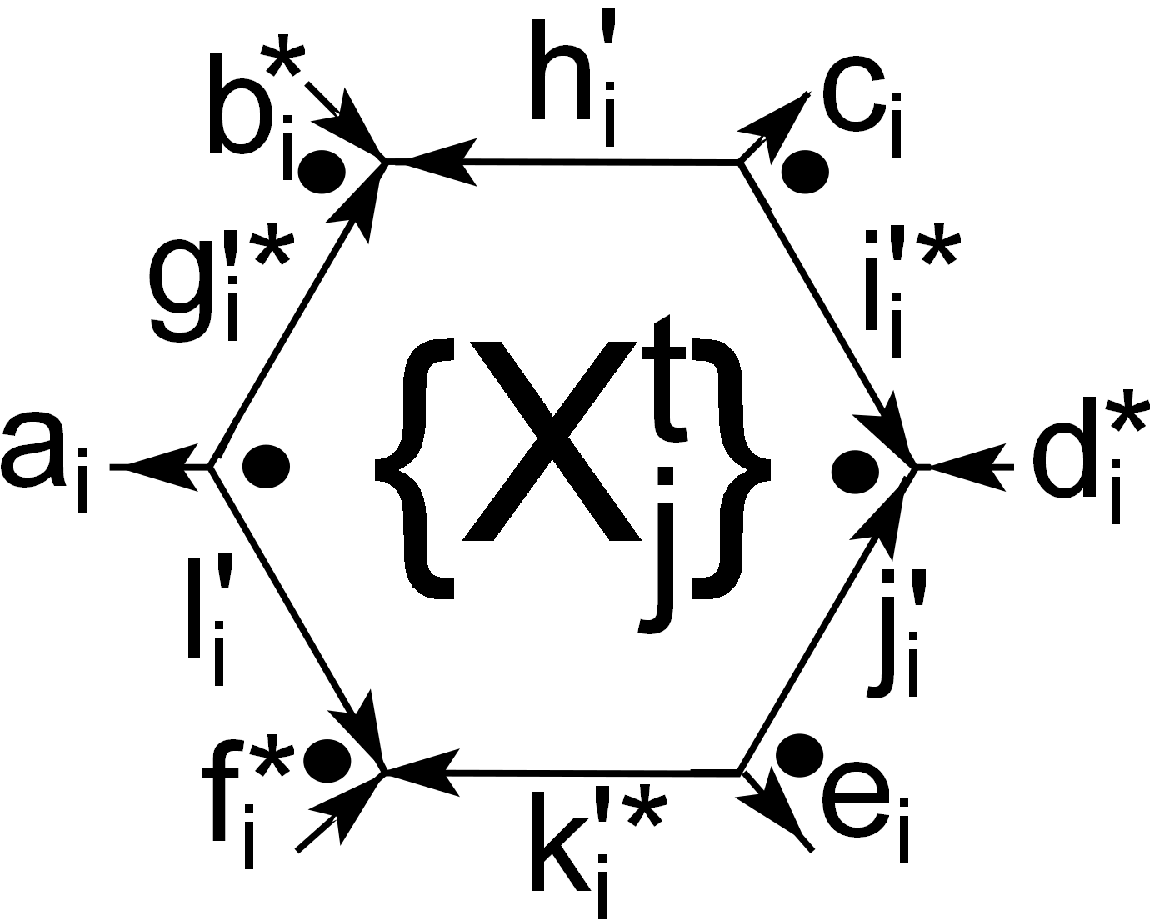}}\right\rangle \\
&=\mathbf{B}_{p,g_{i}^{^{\prime }}...l_{i}^{\prime
}}^{s_{i},g_{i}...l_{i}}\left( a_{i}...f_{i};\sum_{i\neq j}{
X_{j}^t} \right) 
	\label{}
\end{align*}
with 
\begin{align}
	\mathbf{B}_{p,g_{i}^{^{\prime }}...l_{i}^{\prime
}}^{s_{i},g_{i}...l_{i}}&( a_{i}...f_{i};\sum_{i\neq j}\left\{
X_{j}^t\right\}) =B_{p,g_{i}^{^{\prime }}...l_{i}^{\prime
}}^{s_{i},g_{i}...l_{i}}( a_{i}...f_{i};\sum_{i\neq j}\left\{
X_{j}^t\right\} )  \notag \\
&\times\frac{\al_{g_i^*l_i}\al_{h_i'c_i}\al_{j_i^*i_i}\al_{k_i'f_i}\ga_{g_i}\ga_{i_i}\ga_{k_i}}{\al_{g_i'^*l_i'}\al_{h_ic_i}\al_{j_i'^*i_i'}\al_{k_if_i}\ga_{g_i'}\ga_{i_i'}\ga_{k_i'}} \notag\\
&\times \frac{\ka_{l_i,g_j}\ka_{b_j,g_i'^*}\ka_{g_i'^*,b_j^*}\ka_{c_j,i_i'^*}\ka_{s_i,i_j^*}\ka_{j_i',i_j}}{\ka_{l_i',g_j}\ka_{b_j,g_i^*}\ka_{g_i^*,b_j^*}\ka_{c_j,i_i^*}\ka_{s_i,i_j}\ka_{j_i,i_j}}.
	\label{}
\end{align}
Here $\{X_j^t\}$ denotes the translationally invariant orientation configuration of other flavors of string-nets (see Fig. \ref{figure:lattice2}). 
The additional factors come from reversing the orientations on $g_i,i_i,k_i$ links and gliding $s_i$ through vertices involving one null string.

Although the algebraic definition of $B_{p}^{s_{i}}$ is complicated, we
provide with an alternative graphical representation for this operator which
is much simpler. In the graphical representation, the action of $%
B_{p}^{s_{i}}$ can be understood as adding a loop of the type-$s_{i}$
string inside the plaquette $p:$%
\begin{equation*}
\left\langle \raisebox{-0.25in}{\includegraphics[height=0.6in]{bp1.eps}}%
\right\vert B_{p}^{s_{i}}=\left\langle \raisebox{-0.25in}{%
\includegraphics[height=0.6in]{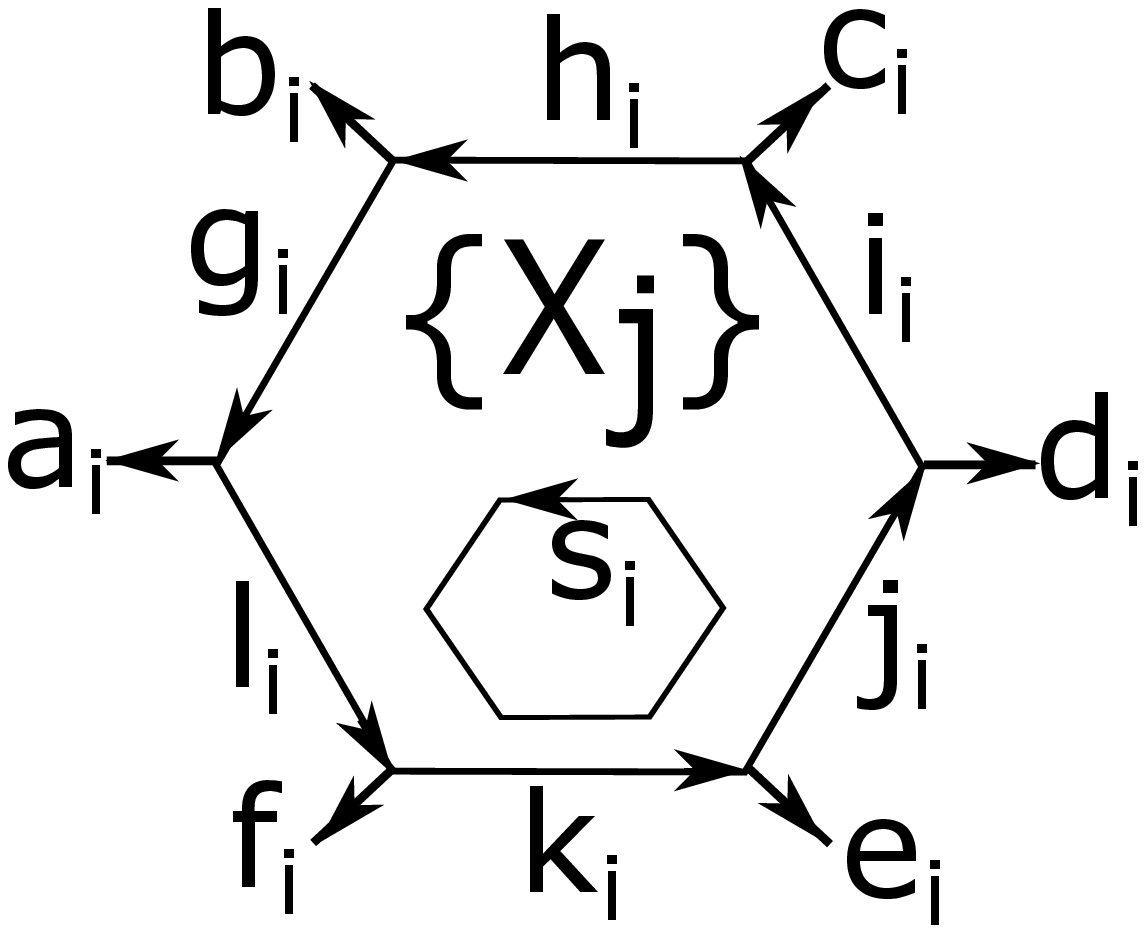}}\right\vert
\end{equation*}%
To obtain matrix elements of $B_{p}^{s_{i}},$ it requires two steps. The
first step is to use the local rules (\ref{rule2a}--\ref{rule2d}) to glide
the string $s_{i}$ to near the boundary of the plaquette. The second step is
then to use (\ref{rule1a}--\ref{rule1c}) to fuse the string $s_{i}$ onto the
links along the boundary of the plaquette. We obtain the phase factors $%
B_{p,x'x}^{s_i\left( 2\right) }$ and $B_{p,x'x}^{s_i\left( 3\right) }$ in the first step
and $B_{p,x'x}^{s_i\left( 1\right) }$ in the second step. In appendix \ref{app:h},
we show that this prescription reproduces the formula in equation (\ref{bps}%
).

\subsection{Properties of the Hamiltonian \label{section:prop}}

The Hamiltonian has many nice properties. The first property is that the
Hamiltonian is Hermitian if $a_{s_{i}^{\ast}}=a_{s_{i}}^{\ast}.$ This result
follow from the identity%
\begin{equation*}
\left( B_{p}^{s_{i}}\right) ^{\dagger}=B_{p}^{s_{i}^{\ast}}.
\end{equation*}
We derive this equality in appendix \ref{app:ham}.

The second property is that the $Q_{I}^i$ and $B_{p}^{s_{i}}$
operators commute with each other:%
\begin{equation*}
	\left[ Q_{I_1}^i,Q_{I_2}^{j}\right] =0,\quad \left[ Q_{I}^i,B_{p}^{s_{j}}%
\right] =0,\quad \left[ B_{p_1}^{s_{i}},B_{p_2}^{t_{j}}\right] =0.
\end{equation*}
The first two equalities can be shown easily from the definition of $%
Q_{I}^i,B_{p}^{s_{i}}.$ The proof of the third equality is given in appendix \ref{app:commute}.

Since every term in the Hamiltonian (\ref{ham}) commutes with one another,
the model is exactly soluble for any value of the coefficients $a_{s_i}.$ In
particular, we choose
\begin{equation*}
a_{s_{i}}=\frac{d_{s_{i}}}{\left\vert G_{i}\right\vert }.
\end{equation*}
With this choice of $a_{s_{i}},$ the $Q_{I}^i$ and $B_{p}^i$ are
projector operators which have eigenvalues $0,1$. It is easy to derive the
result for $Q_{I}^i;$ we show $B_{p}^i$ are projector operators in
appendix \ref{app:ham}.

As a result, we can derive the low energy properties of $H$. Let $\left\vert
\left\{ q_{I}^i,b_{p}^i\right\} \right\rangle $ denote the simultaneous
eigenvalues of $\left\{ Q_{I}^i,B_{p}^i\right\} $ with $i=1,..,L:$%
\begin{eqnarray*}
Q_{I}^i\left\vert \left\{ q_{I}^i,b_{p}^i\right\} \right\rangle
&=&q_{I}^i\left\vert \left\{ q_{I}^i,b_{p}^i\right\} \right\rangle , \\
B_{p}^i\left\vert \left\{ q_{I}^i,b_{p}^i\right\} \right\rangle
&=&b_{p}^i\left\vert \left\{ q_{I}^i,b_{p}^i\right\} \right\rangle .
\end{eqnarray*}%
The corresponding energies are%
\begin{equation*}
E=-\sum_{i=1}^{L}\left( \sum_{I}q_{I}^i+\sum_{p}b_{p}^i\right) .
\end{equation*}%
Since $q_{I}^i,b_{p}^i$ take values in $0$ or $1,$ thus the ground
state(s) have $q_{I}^i=b_{p}^i=1,$ while the excited states have $%
q_{I}^i=0$ or $b_{p}^i=0$ for at least one site $I$ or plaquette $%
p.$ We see that there is a finite energy gap separating the ground
states(s) from the excited states. All that remains is to determine the
ground state degeneracy. The degeneracy is simply the product of the
degeneracy associated with\ each flavor of string-nets. The degeneracy
depends on the global topology of our system. For a disk geometry with open
boundary conditions, there is a unique state with $q_{I}^i=b_{p}^i=1.$
On the other hand, for a periodic torus geometry, the number of degenerate
ground states is equal to the number of quasiparticles types (see Ref. \onlinecite{LinLevinstrnet} for the computation of the ground state degeneracy).

The final property of our model is that the ground state of lattice model in a disk geometry, $|\Phi_{latt}\>$, obeys the local rules (\ref{rule1a}--\ref{rule1c}) and (\ref{rule2a}--\ref{rule2d}). We establish this property in appendix (\ref{app:ham}). As a result, we conclude that $|\Phi_{latt}\>$ is identical to the continuum wave function $\Phi$ restricted to the string-net configurations on the lattice. From now on, we will use $|\Phi\>$ to denote both the lattice ground state and the continuum wave function.

\section{Quasiparticle excitations \label{section:particle}}

In this section, we discuss the topological properties of the quasiparticle
excitations of the string-net Hamiltonian (\ref{ham}). We generalize the
analysis in Ref. \onlinecite{LinLevinstrnet} to multiple flavors of
string-nets. We first find all the topologically distinct types of
quasiparticles by constructing the string operators which create the
quasiparticle excitations. Then we compute their braiding statistics by the
commutation algebra of the string operators.

\subsection{String operator picture}

For the topological phases, the excitations with nontrivial statistics are
generally created by string-like operators. For each topologically distinct
quasiparticle excitation $\alpha,$ there is a corresponding string operator $%
W_{\alpha}\left( P\right) $ where $P$ is the path along which the string
operator acts. If $P$ is an open path, then $W_{\alpha}\left( P\right) $ is
called an open string operator while if $P$ is a closed path, then $%
W_{\alpha }\left( P\right) $ is called an closed string operator.

The string operator $W_{\alpha}\left( P\right) $ has several properties.
First, an open string operator acting on the ground state $\left\vert
\Phi\right\rangle $ will create an excited state containing a pair of
quasiparticle $\alpha$ and its antiparticle at two ends of $P$%
\begin{equation*}
W_{\alpha}\left( P\right) \left\vert \Phi\right\rangle =\left\vert \Phi
_{ex}\right\rangle .
\end{equation*}
Furthermore, the excited state does not depend on the path of the string but
only on the end points of $P,$ that is%
\begin{equation*}
W_{\alpha}\left( P\right) \left\vert \Phi\right\rangle =W_{\alpha}\left(
P^{\prime}\right) \left\vert \Phi\right\rangle
\end{equation*}
for any two paths $P,P^{\prime}$ that have the same end points. Finally, a
closed string operator does not create any excitations: $W_{\alpha}\left(
P\right) \left\vert \Phi\right\rangle $ $\propto\left\vert \Phi\right\rangle 
$ .

Physically, one may think of an open string operator as describing a process
of creating a particle-antiparticle pair out of the ground state and then
bringing the two particles to the two ends of the string. Similarly, a
closed string operator describes a process of creating a pair of
quasiparticles and then moving one of them around the path of the string all
the way to its original position, where it annihilates its partner.
Throughout the discussion, we assume that the system is defined in a
topologically trivial geometry with a unique ground state.

\subsection{Constructing the string operators}

We follow the same strategy as Ref. \onlinecite{LinLevinstrnet} to construct
string operators that create each of the distinct quasiparticle excitations
of the Hamiltonian (\ref{ham}). We begin with our ansatz for constructing
string operators. The string operators are defined by specifying how $%
W_{\alpha}\left( P\right) $ acts on each string-net configuration. We
describe the action of $W_{\alpha }\left( P\right) $ using a graphical
representation and we use the convention that $W_{\alpha}\left( P\right) $
acts on a bra $\left\langle X\right\vert$. Specifically, when $W_{\alpha}\left( P\right) $
is applied to a string-net state $\left\langle X\right\vert ,$ it adds a
dashed string along the path $P$ under the preexisting string-nets:%
\begin{equation*}
\left\langle \raisebox{-0.16in}{\includegraphics[height=0.4in]{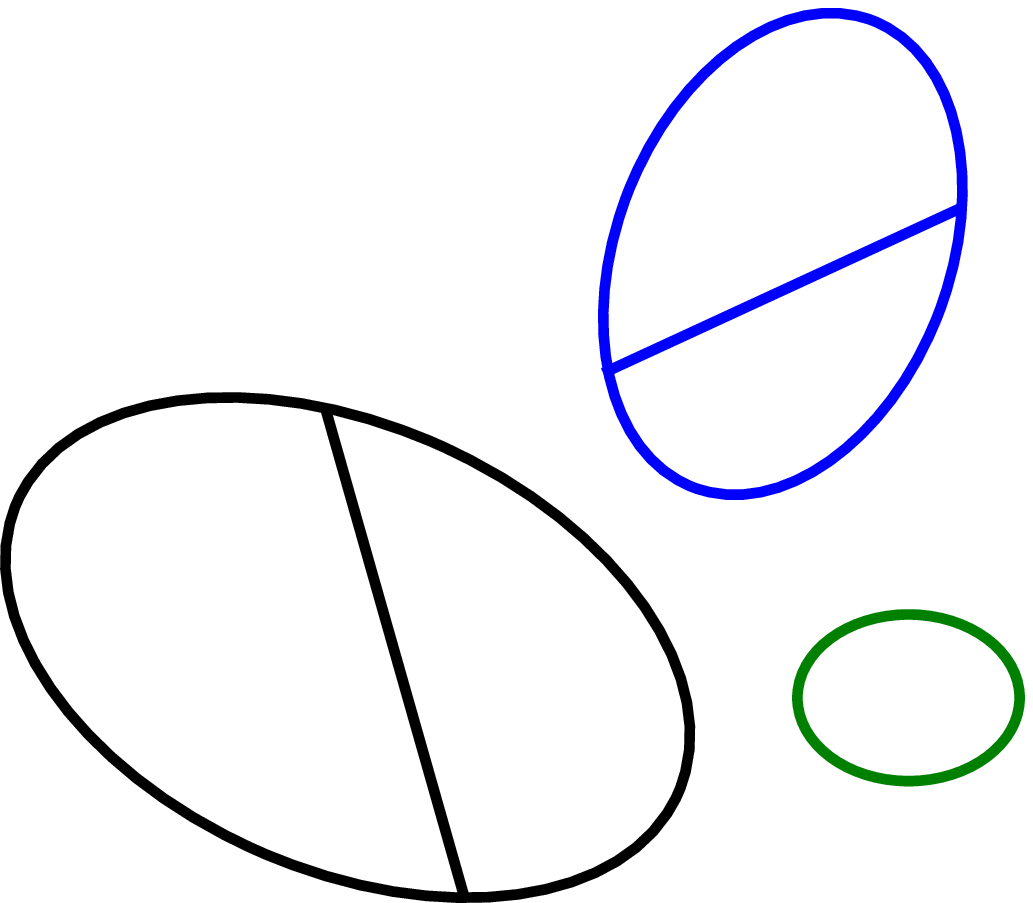}}\right\vert W\left( P\right) =\left\langle \raisebox{-0.16in}{\includegraphics[height=0.4in]{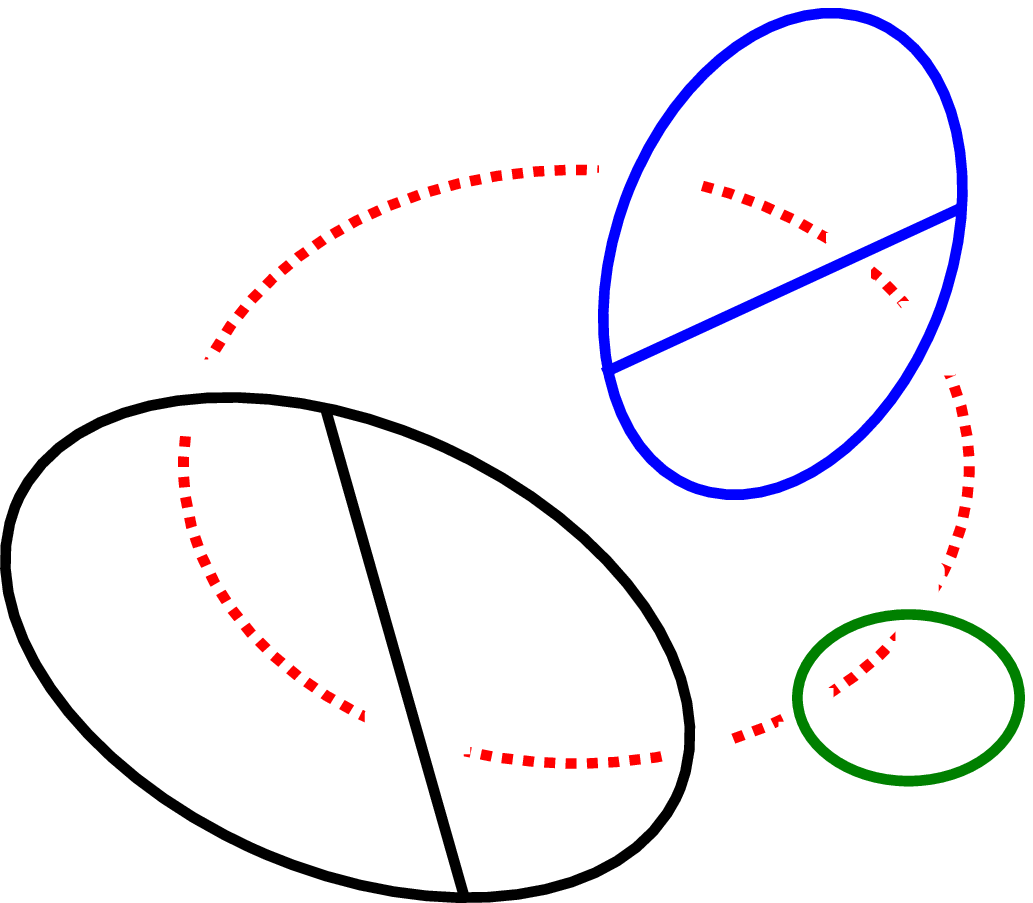}}\right\vert .
\end{equation*}
We then replace the dashed string with a type-$s_{i}$ and replace every
crossing using the rules:
\begin{align}
\left\langle \raisebox{-0.16in}{%
\includegraphics[height=0.4in]{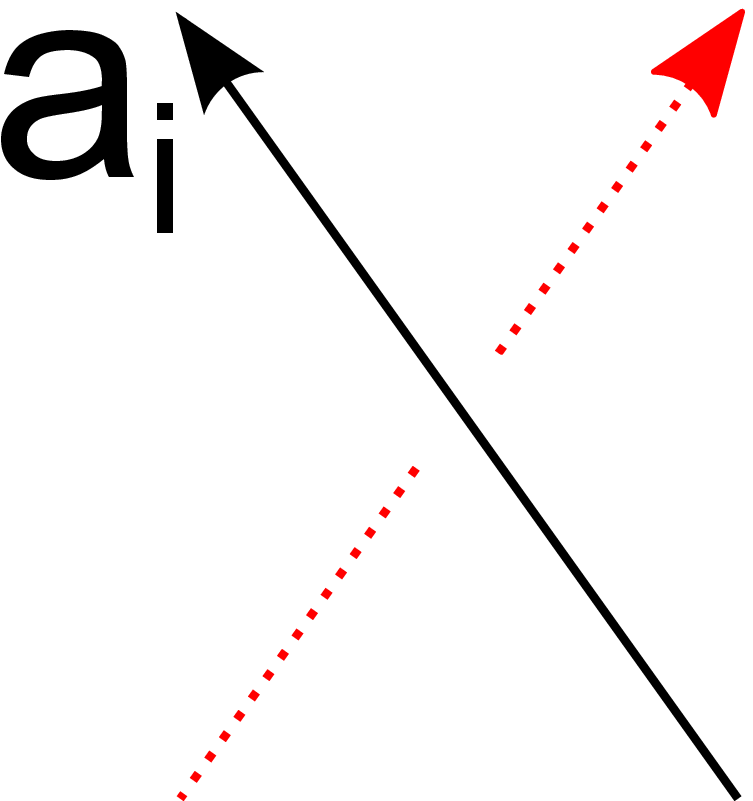}}\right\vert & =\omega
_{s_{i}}\left( a_{i}\right) \left\langle \raisebox{-0.16in}{%
\includegraphics[height=0.4in]{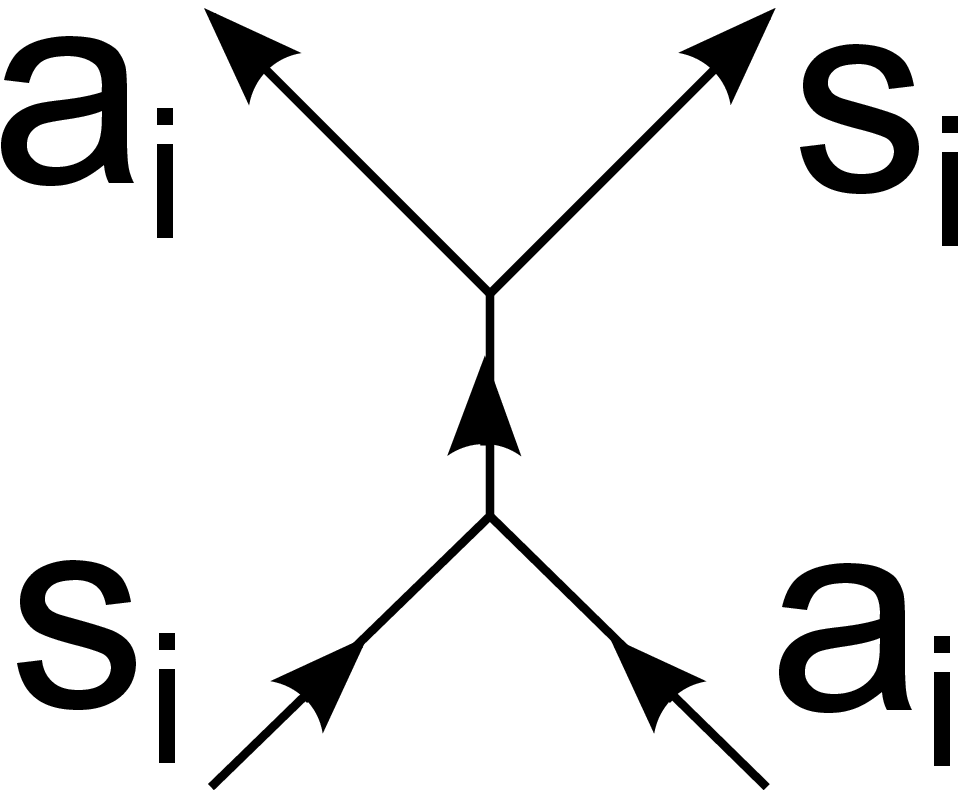}}\right\vert  \label{srule1} \\
\left\langle \raisebox{-0.16in}{%
\includegraphics[height=0.4in]{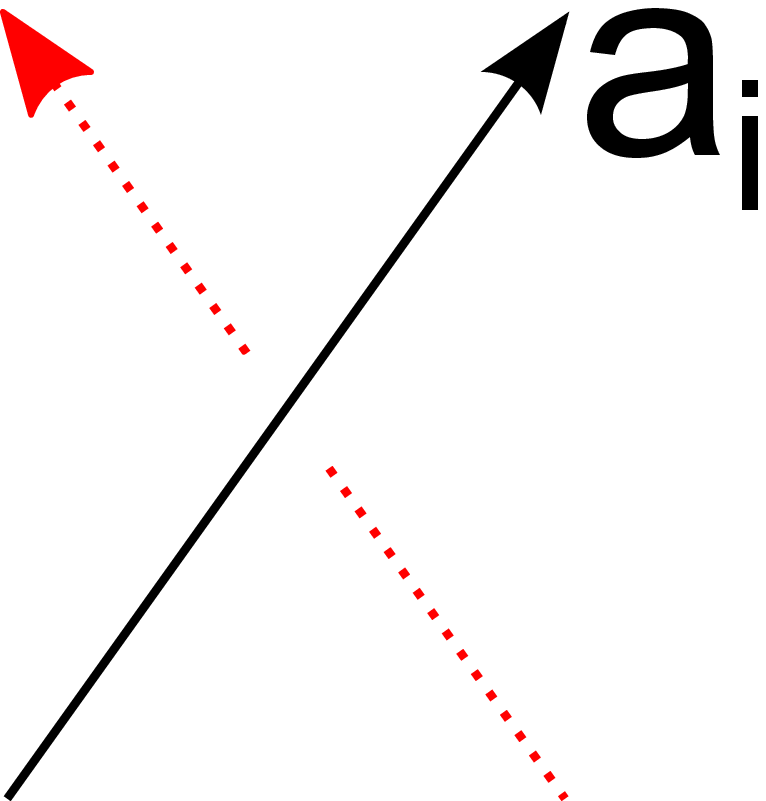}}\right\vert & =\bar{\omega}%
_{s_{i}}\left( a_{i}\right) \left\langle \raisebox{-0.16in}{%
\includegraphics[height=0.4in]{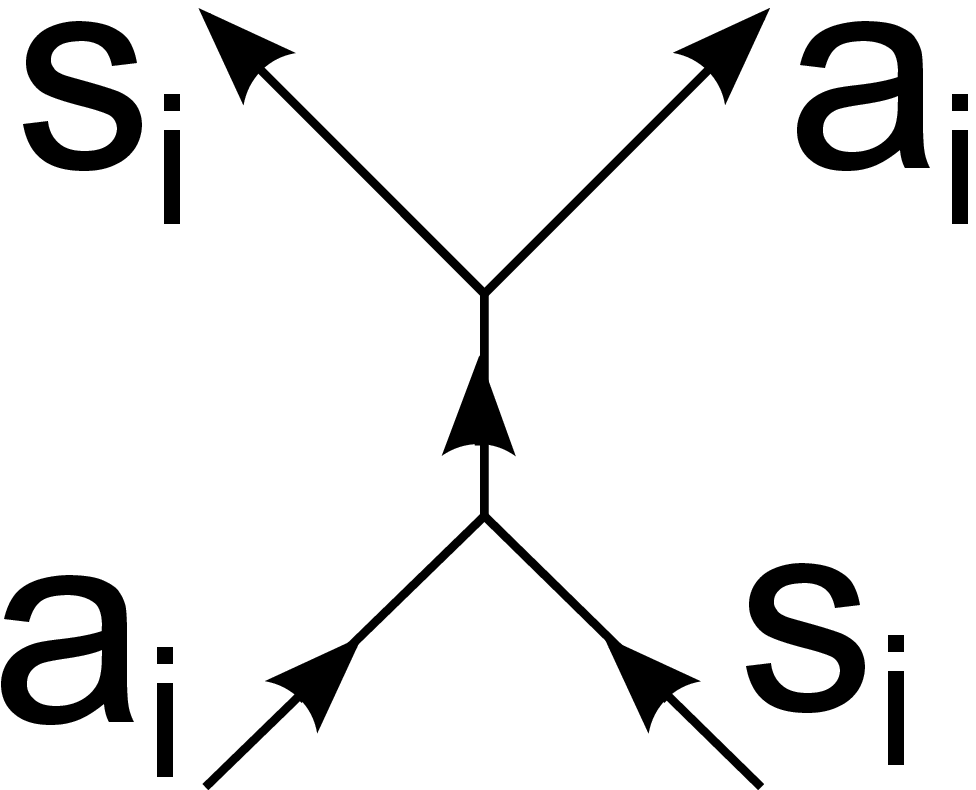}}\right\vert  \label{srule2} \\
\left\langle \raisebox{-0.16in}{%
\includegraphics[height=0.4in]{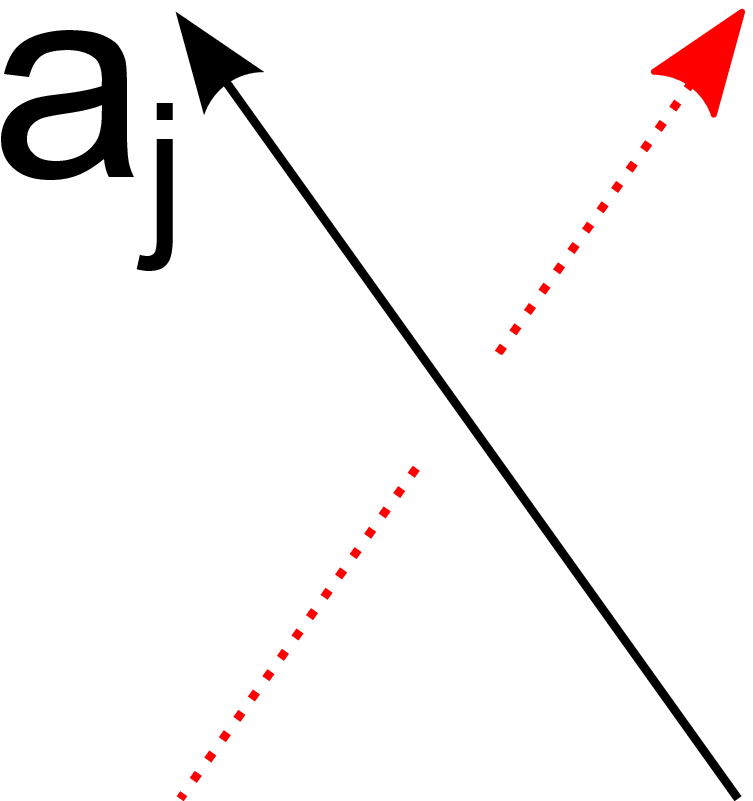}}\right\vert & =\Omega
_{s_{i}}\left( a_{j}\right) \left\langle \raisebox{-0.16in}{%
\includegraphics[height=0.4in]{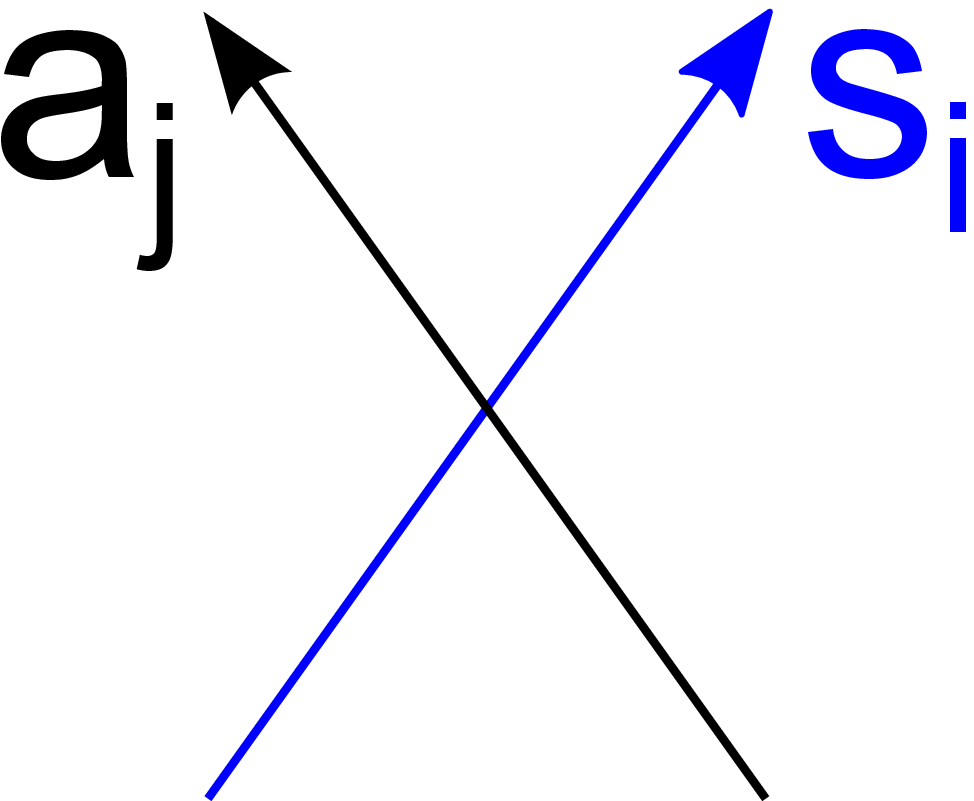}}\right\vert ,  \label{srule3}
\\
\left\langle \raisebox{-0.16in}{%
\includegraphics[height=0.4in]{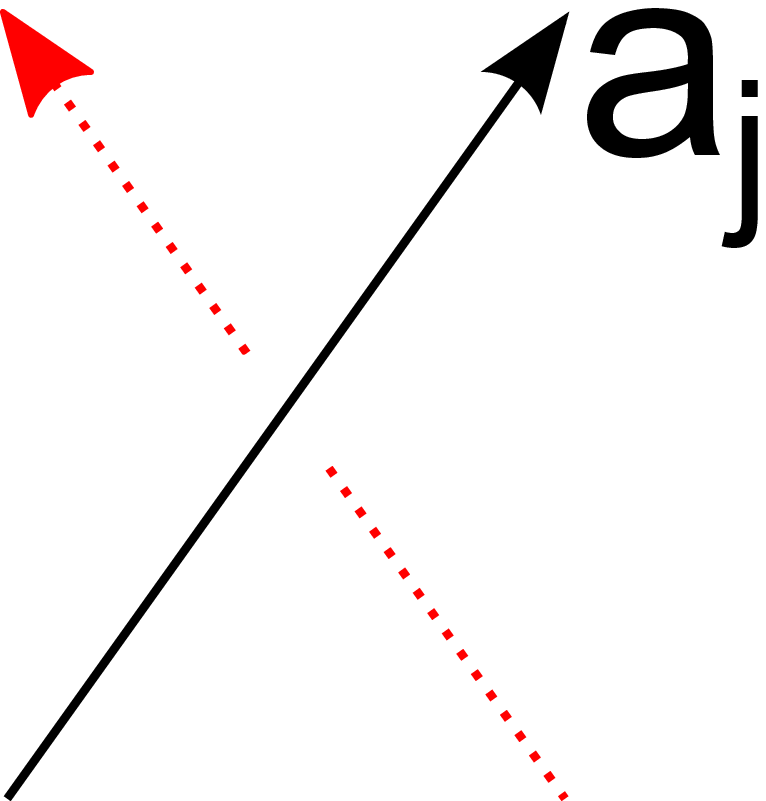}}\right\vert & =\bar{\Omega}%
_{s_{i}}\left( a_{j}\right) \left\langle \raisebox{-0.16in}{%
\includegraphics[height=0.4in]{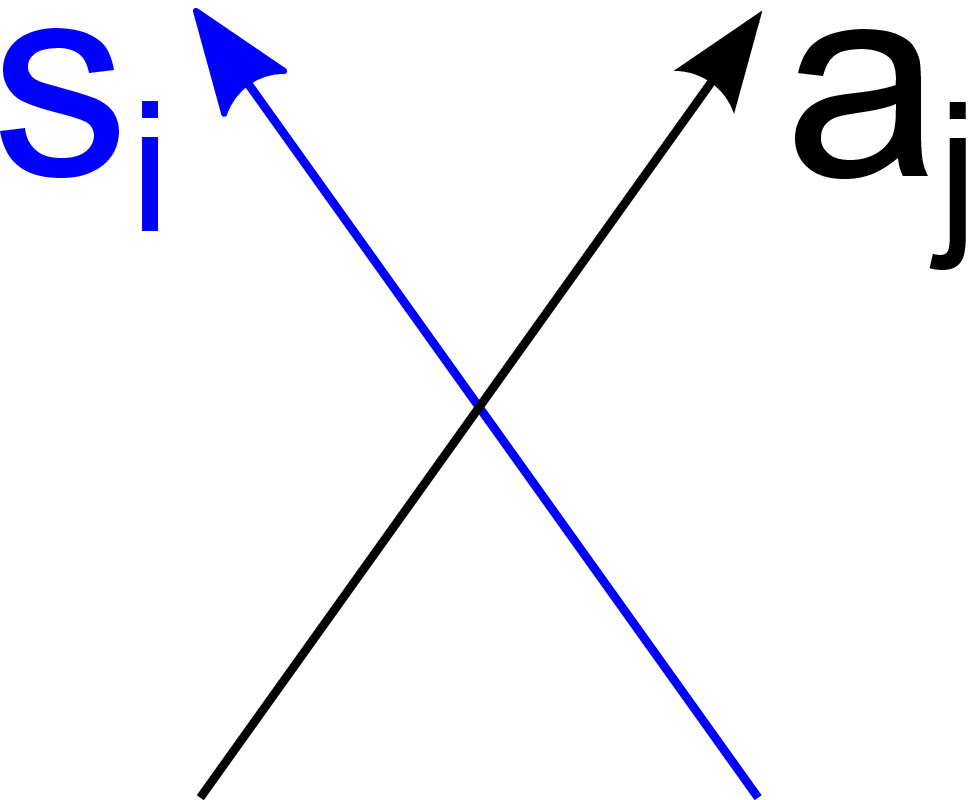}}\right\vert  \label{srule4}
\end{align}%
Here the first two equations specify the rules for crossings between
string-nets of the same flavor while the last two equations are rules for
crossings between string-nets of different flavors. The $\omega ,\bar{\omega}%
,\Omega ,\bar{\Omega}$ are four complex-valued functions defined on the
group $G$ with $\omega \left( 0\right) =\bar{\omega}\left( 0\right) =\Omega
\left( 0\right) =\bar{\Omega}\left( 0\right) =1.$ After making these
replacements, the resulting state $\left\langle X^{\prime }\right\vert $ is
multiplied by a product of $\omega ,\bar{\omega},\Omega ,\bar{\Omega}$ and
it is simply the matrix element of the string operator $\left\langle
X\left\vert W\left( P\right) \right\vert X^{\prime }\right\rangle .$ This
defines the string operator $W\left( P\right) .$

\subsection{Path independence constraints}

The string operators $W\left( P\right) $ must satisfy path independence so
that they can create deconfined quasiparticle excitations at two ends of the
path. Specifically, $W$ satisfies path independence if and only if%
\begin{align}
\left\langle \raisebox{-0.16in}{\includegraphics[height=0.4in]{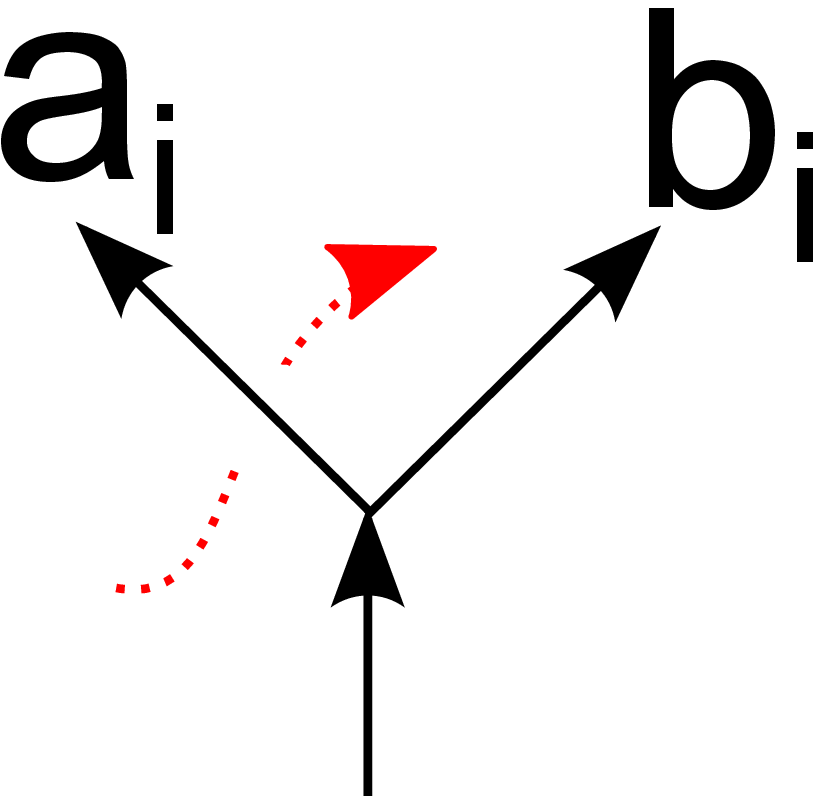}}%
\middle|\Phi \right\rangle & =\left\langle \raisebox{-0.16in}{%
\includegraphics[height=0.4in]{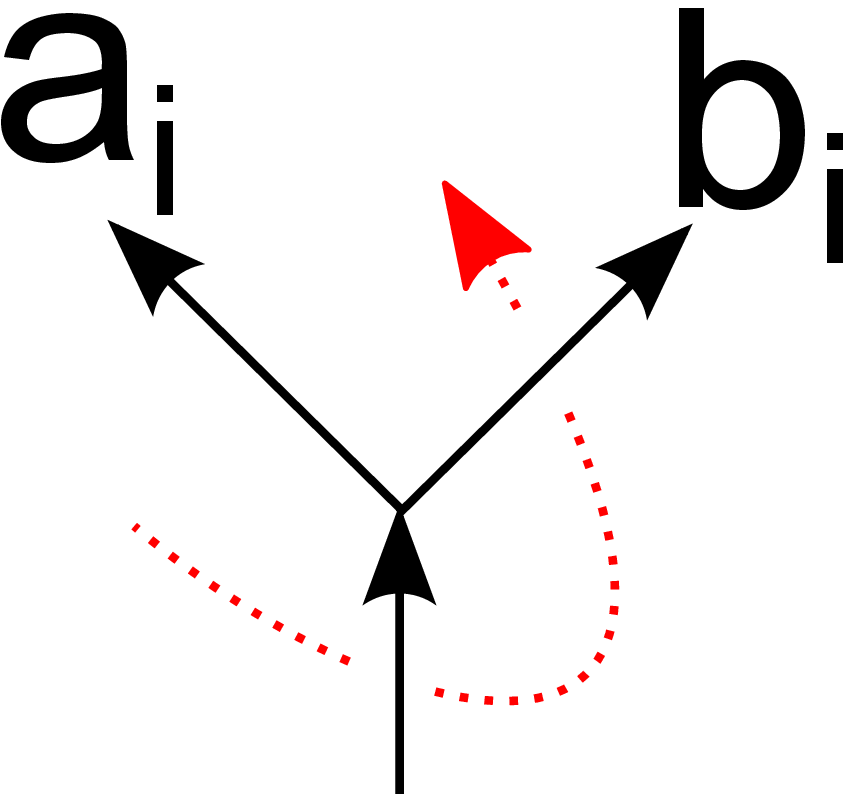}}\middle|\Phi \right\rangle \\
\left\langle \raisebox{-0.1in}{\includegraphics[height=0.23in]{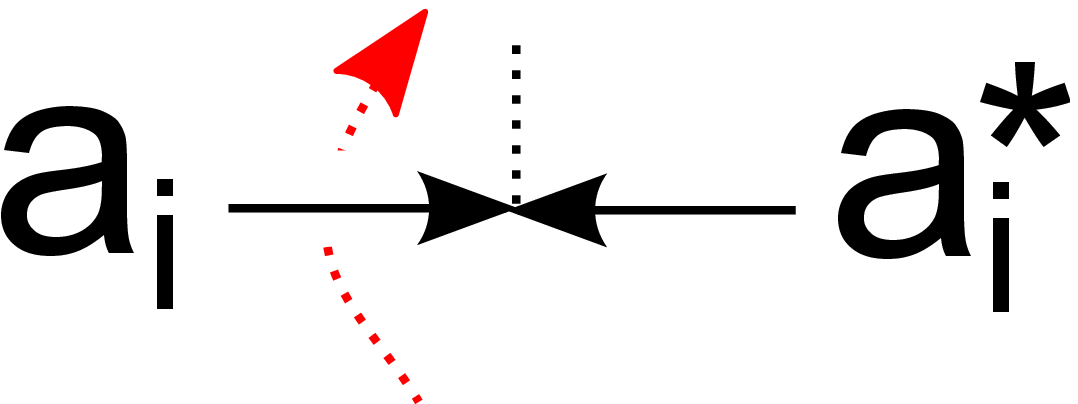}}%
\middle|\Phi \right\rangle & =\left\langle \raisebox{-0.1in}{%
\includegraphics[height=0.23in]{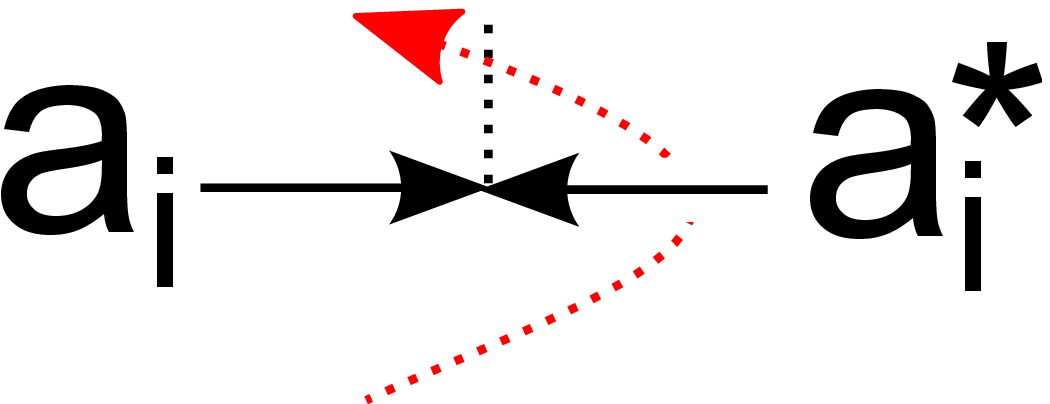}}\middle|\Phi \right\rangle \\
\left\langle \raisebox{-0.16in}{\includegraphics[height=0.4in]{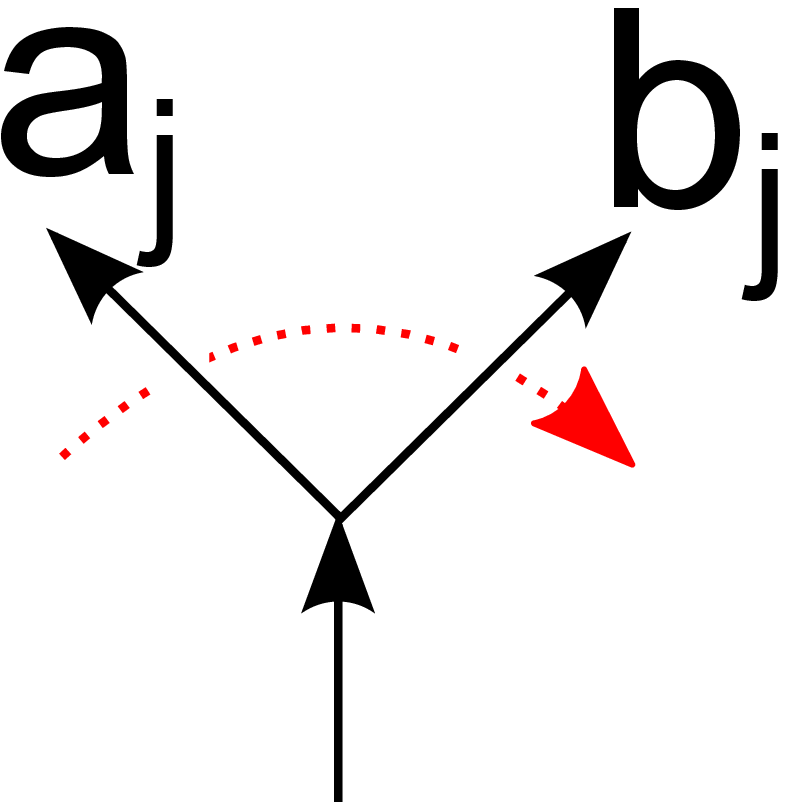}}%
\middle|\Phi \right\rangle & =\left\langle \raisebox{-0.16in}{%
\includegraphics[height=0.4in]{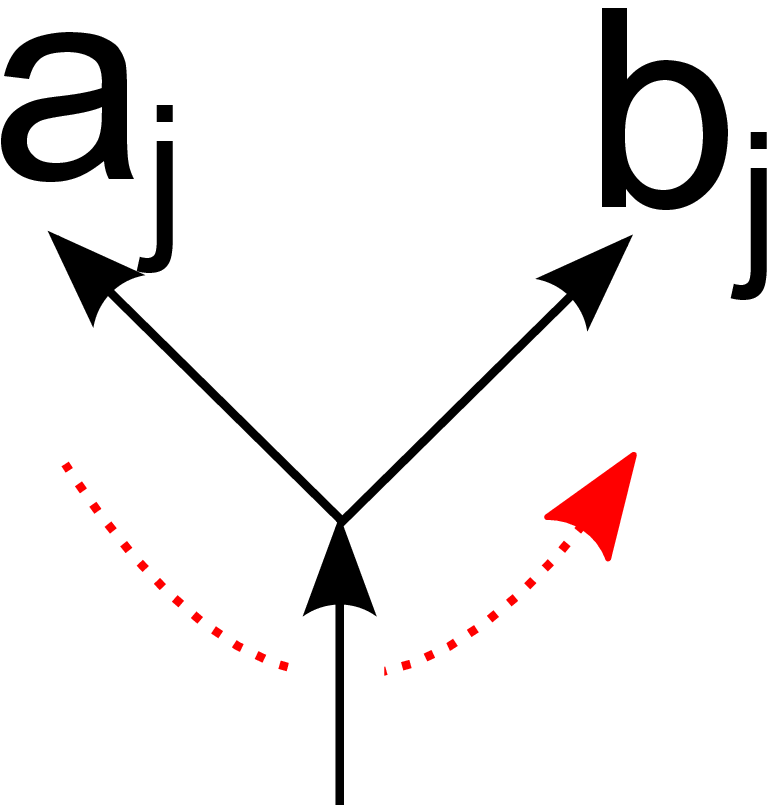}}\middle|\Phi \right\rangle
\label{path3} \\
\left\langle \raisebox{-0.06in}{\includegraphics[height=0.18in]{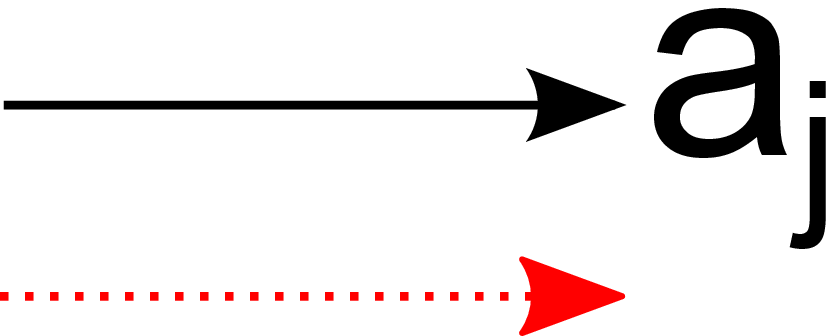}}%
\middle|\Phi \right\rangle & =\left\langle \raisebox{-0.06in}{%
\includegraphics[height=0.18in]{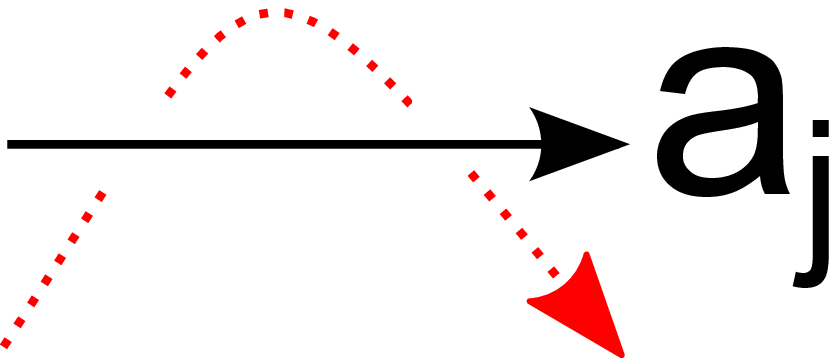}}\middle|\Phi \right\rangle \\
\left\langle \raisebox{-0.16in}{\includegraphics[height=0.4in]{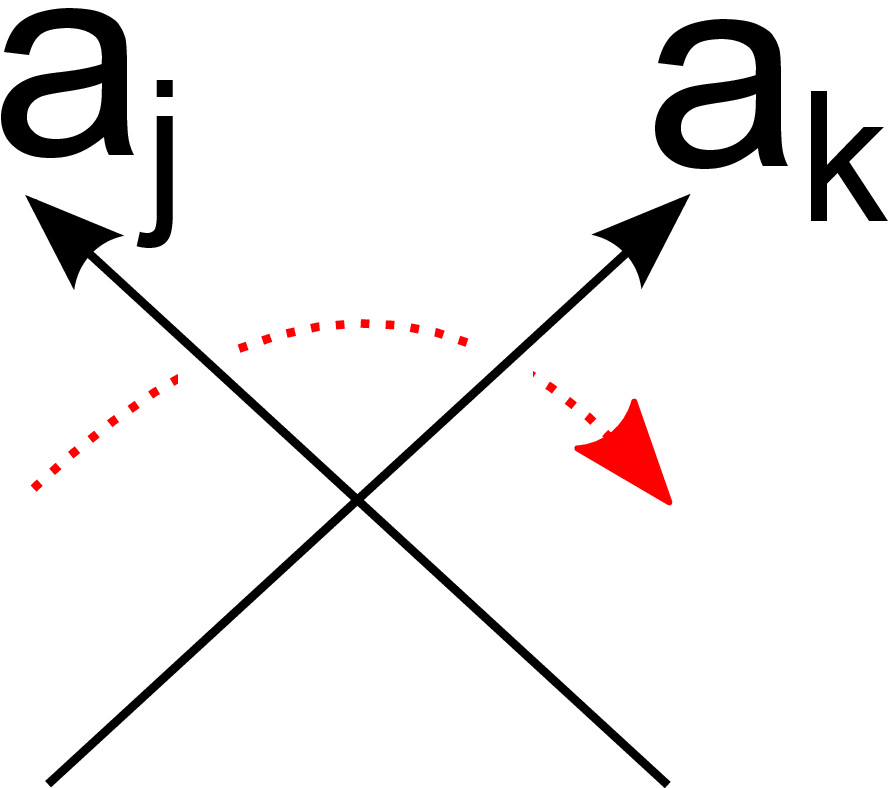}}%
\middle|\Phi \right\rangle & =\left\langle \raisebox{-0.16in}{
\includegraphics[height=0.4in]{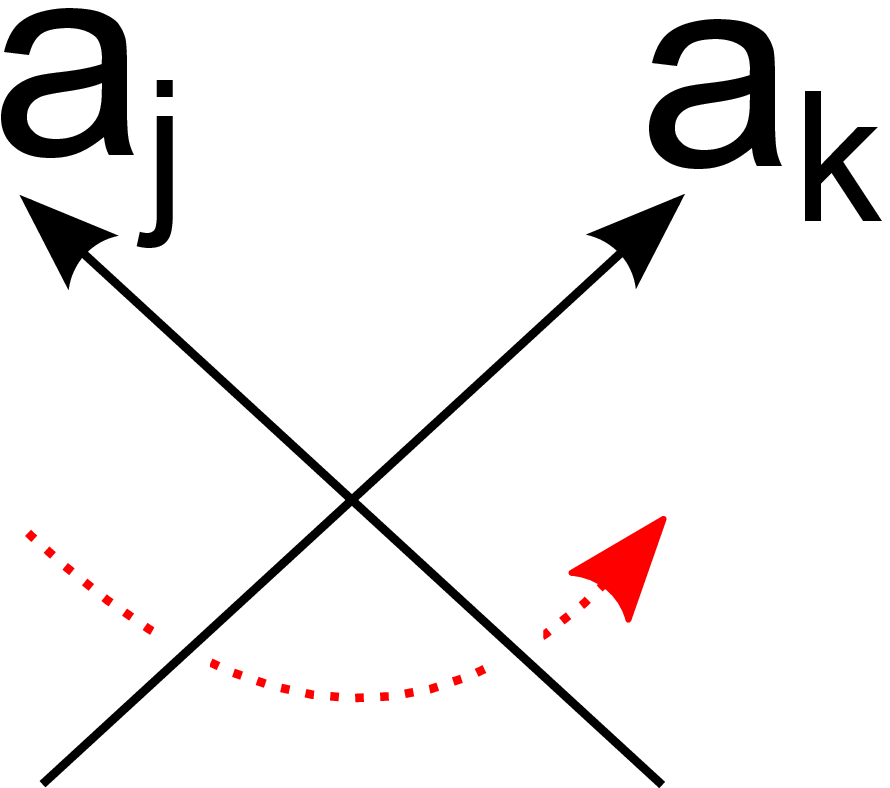}}\middle|\Phi \right\rangle
\label{path5}
\end{align}%
Other deformations of the path can be built out of these elementary
ones. These graphical relations can be translated into algebraic conditions
by using the local rules (\ref{rule1a}--\ref{rule1c},\ref{rule2a}--\ref{rule2d}). The result is 
\begin{subequations}
\begin{align}
\mathbf{w}_{s_{i}}\left( a_{i}\right) \mathbf{w}_{s_{i}}\left( b_{i}\right)
& =c_{s_{i}}\left( a_{i},b_{i}\right) \mathbf{w}_{s_{i}}\left(
a_{i}+b_{i}\right) ,  \label{string1} \\
\mathbf{\bar{w}}_{s_{i}}\left( a_{i}\right) & =\mathbf{w}_{s_{i}}\left(
a_{i}\right) ^{-1},  \label{string2} \\
\Omega _{s_{i}}\left( a_{j}\right) \Omega _{s_{i}}\left( b_{j}\right) &
=F_{s_{i}}^{\left( 2\right) }\left( a_{j},b_{j}\right) \Omega _{s_{i}}\left(
a_{j}+b_{j}\right) ,  \label{string3} \\
\bar{\Omega}_{s_{i}}\left( a_{j}\right) & =\Omega _{s_{i}}\left( a_{j}\right)^{-1} \eta_{s_{i}}\left( a_{j}\right) ,  \label{string4} \\
\Omega _{s_{i}}\left( a_{j}\right) \Omega _{s_{i}}\left( a_{k}\right) &
=F_{s_{i}}^{\left( 3\right) }\left( a_{j},a_{k}\right) \Omega _{s_{i}}\left(
a_{k}\right) \Omega _{s_{i}}\left( a_{j}\right) .  \label{string5}
\end{align}%
\label{string}
\end{subequations}
Here we define%
\begin{align}
\mathbf{w}_{s_{i}}\left( a_{i}\right) & =\omega _{s_{i}}\left( a_{i}\right)
F\left( s_{i},a_{i},a_{i}^{\ast }\right) ,  \notag \\
\mathbf{\bar{w}}_{s_{i}}\left( a_{i}\right) & =\bar{\omega}_{s_{i}}\left(
a_{i}\right) F\left( a_{i},s_{i},s_{i}^{\ast }\right) , \notag \\
c_{s_{i}}\left( a_{i},b_{i}\right) & =\frac{F\left( a_{i},b_{i},s_{i}\right)
F\left( s_{i},a_{i},b_{i}\right) }{F\left( a_{i},s_{i},b_{i}\right) }.
\label{cs}
\end{align}%
Notice that (\ref{string1},\ref{string2}), (\ref{string3},\ref{string4}), (%
\ref{string5}) involve one, two and three flavors of strings respectively.
The string operators constructed by the first two rules (\ref{string1},\ref%
{string2}) were studied in Ref. \onlinecite{LinLevinstrnet}.

To solve (\ref{string1}--\ref{string5}), we note that the self-consistency
condition (\ref{sfeqa}) implies that $c_{s_{i}}\left( a_{i},b_{i}\right) $
obey the identity%
\begin{equation}
c_{s_{i}}\left( a_{i},b_{i}\right) c_{s_{i}}\left( a_{i}+b_{i},c_{i}\right)
=c_{s_{i}}\left( b_{i},c_{i}\right) c_{s_{i}}\left( a_{i},b_{i}+c_{i}\right)
.  \label{2cocycle}
\end{equation}
This identity resembles the self-consistency condition (\ref{sfeq1}) for $%
F_{s_{i}}^{\left( 2\right) }\left( a_{j},b_{j}\right) $. 
The only difference is that 
$c_{s_{i}}\left( a_{i},b_{i}\right) $ is associated with one flavor of
string-nets while $F_{a_{i}}^{\left( 2\right) }\left( a_{j},b_{j}\right) $
is associated with two different flavors of string-nets. 
Eqs. (\ref{cs},\ref{sfeq1}) are also called the 2-cocycle condition. They are the factor systems of a projective representation\cite{ChenGuWenSPT}. Thus, solving (\ref{string1},\ref{string3}) is equivalent to finding a projective
representation corresponding to the factor systems $c_{s_{i}}\left(
a_{i},b_{i}\right) $ and $F_{a_{i}}^{\left( 2\right) }\left(
a_{j},b_{j}\right) $ respectively. Furthermore, one can see from (\ref%
{string5}) that if $F_{s_{i}}^{\left( 3\right) }\left( a_{j},a_{k}\right) $
is nontrivial, then $\Omega_{s_{i}}\left( a_{j}\right) $ require to be
higher dimensional objects, namely matrices.

\subsection{Unifying different flavors of strings \label{unify}}
So far in our construction, different flavors of strings cross but do not branch with one another. With this restriction, we obtain several self-consistency conditions (\ref{sfeqb}) and path independence conditions (\ref{string}). Some equations are similar except they are associated with different flavors of strings. 
For the ease of solving this set of equations, we like to first compactify them to a fewer equations. 

To this end, we combine different flavors of string labels into a vector $a=(a_1,\dots,a_i,\dots,a_L)$, namely we regard the flavor index $i$ as the $i$-th component of the vector $a$. In this way, we equivalently unify all different flavors of strings $a_i\in G_i$ into one flavor of strings with multiple components $a\in G=\prod_{i=1}^L G_i$. We relabel the strings by $L$-component vectors $a$. The strings can branch if each component satisfies the corresponding component-wise branch rules, namely $(a,b,c)$ is allowed if $a_i+b_i+c_i=0$ for all components $i$.

After unifying the flavors of strings, $F,F^{(2)}$ can now take different components of $a$ as inputs, e.g. $F^{(2)}_{a_i}(a_j,a_k)$. 
Moreover, all strings can branch. To implement this in our construction, we do the replacement at each crossing:
\begin{equation}
\left\langle \raisebox{-0.14in}{%
\includegraphics[height=0.35in]{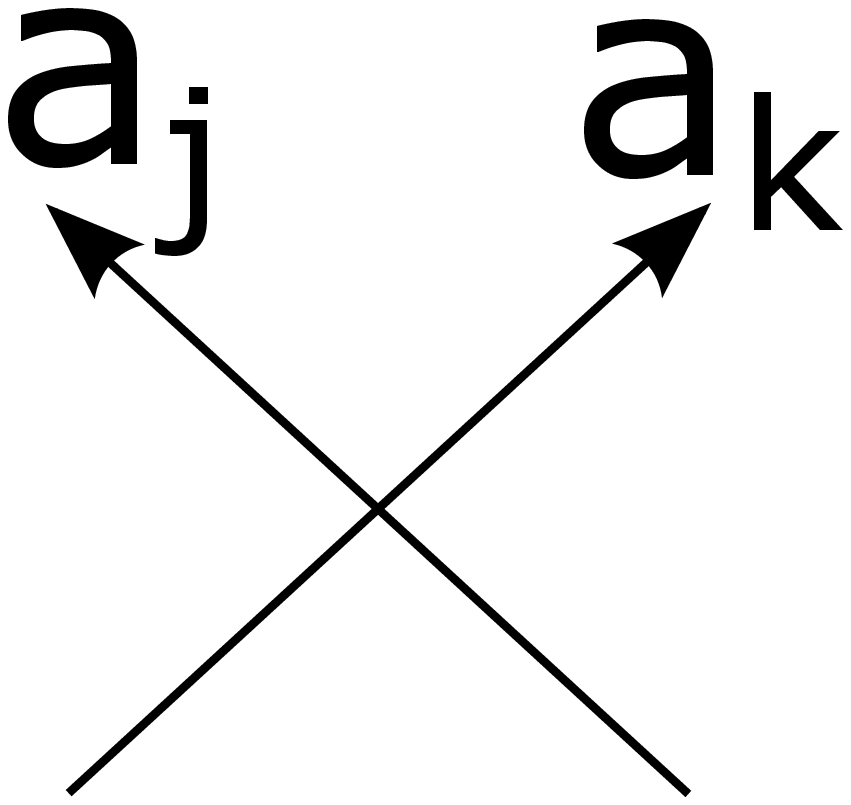}}|\Phi \right\rangle
=\left\langle \raisebox{-0.21in}{%
\includegraphics[height=0.45in]{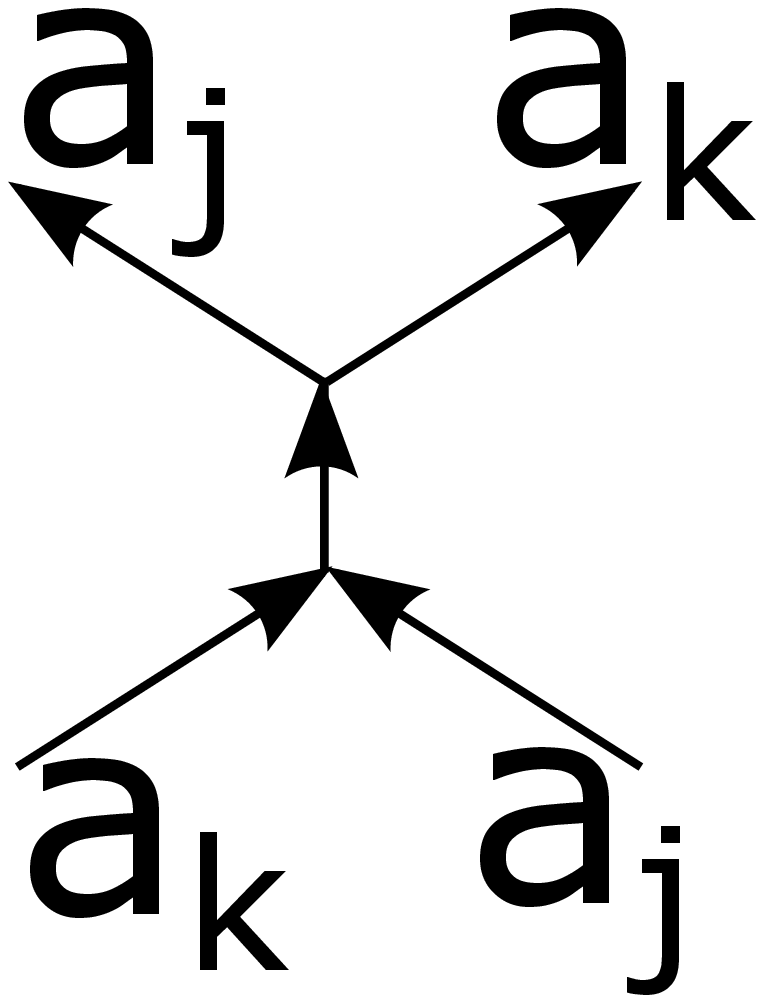}}|\Phi \right\rangle.
\label{fusion1}
\end{equation}%
Namely, each crossing is understood as fusing two strings and then splitting it into two with exchanged order.
Intuitively, one can think of this replacement as zooming in the crossing where $a_j,a_k$ branch and split.  
As a consequence of this replacement in the local rules (\ref{rule2a}--\ref{rule2d}), one can relate the parameters $F,F^{(2)},\kappa,\eta,F^{(3)}$ and reduce the number of independent parameters. 

Specifically, applying (\ref{fusion1}) to (\ref{rule2d}) gives the condition 
\begin{equation}
F_{s_{i}}^{\left( 3\right) }\left( a_{j},a_{k}\right) =\frac{F_{s_{i}}^{\left( 2\right) }\left( a_{j},a_{k}\right) }{F_{s_{i}}^{\left(2\right) }\left( a_{k},a_{j}\right) }.
\label{string5a}
\end{equation}
Similarly, by inserting (\ref{fusion1}) to the local rules (\ref{rule2a}--\ref{rule2c}), one can obtain
\begin{align}
	F^{(2)}_{a_i}(a_j,b_j)&=\frac{F(a_i,a_j,a_j^*)F(a_i,b_j,b_j^*)}
	{F(a_i,a_j+b_j,(a_j+b_j)^*)c_{a_i}(a_j,b_j)} \label{f2}\\
	\kappa_{a_i}(a_j)&=F(a_j^*,a_i,a_j)F(a_j^*,a_i+a_j,a_i^*)\\
	\eta_{a_i}(a_j)&=1/(F(a_i,a_j,a_j^*)F(a_j,a_i,a_i^*)). 
\end{align}
Since the parameters $F^{(2)},\kappa,\eta,F^{(3)}$ are functions of $F$, then solving (\ref{sfeqb}) reduces to solve (\ref{sfeqa}) with the strings labeled by the group $G=\prod_i G_i$. 

Furthermore, one can insert (\ref{string5a}--\ref{f2}) into (\ref{string}) and simplify the expressions by (\ref{sfeqa}).
We can then combine the conditions (\ref{string}) into
\begin{align}
\mathbf{w}_{s}\left( a\right) \mathbf{w}_{s}\left( b\right) & =c_{s}\left(
a,b\right) \mathbf{w}_{s}\left( a+b\right) ,  \label{string0} \\
\mathbf{\bar{w}}_{s}\left( a\right) & =\mathbf{w}_{s}\left( a\right) ^{-1},
\notag \\
c_{s}\left( a,b\right) & =\frac{F\left( a,b,s\right) F\left( s,a,b\right) }{%
F\left( a,s,b\right) },
\end{align}
with
\begin{align*}
	\mathbf{w}_{s_i}(a_i)=\mathbf{w}_s(a) \text{ with }s=s_i,a=a_i, \\
	\Omega_{s_i}(a_j)=\frac{F(s,a,a^*)}{\mathbf{w}_{s}(a)} \text{ with }s=s_i,a=a_j.
	\label{}
\end{align*}
Again, $s,a,b$ are elements of the group $G.$

Thus, by unifying the flavors of strings and applying (\ref{fusion1}) to each crossing of intersecting string-nets in our construction, we successfully 
reduce the equations (\ref{sfeqb}) and (\ref{string}) into the equations (\ref{sfeqa}) and (\ref{string0}) with strings labeled by $G=\prod_i G_i$. 
The last two are exactly the equations which determines ground state wave functions and string operators in the original construction in Ref. \onlinecite{LinLevinstrnet}. 
Therefore, we establish the equivalence between our new construction associated with multiple flavors of strings labeled by $\{G_i,i=1,\dots,L\}$ and the original construction associated with one flavor of strings labeled by $G=\prod_{i=1}^L G_i$.

In the rest of the discussion, we stick to the convention that string-nets of different flavors cross one another with the understanding that the crossings are resolved by (\ref{fusion1}). The convention, namely using $F^{(2)},F^{(3)}$, simplifies the description of the string-net states and models since $F^{(2)},F^{(3)}$ are complicated combinations of $F$ (see Eqs. (\ref{f2},\ref{string5a})). 
In the next section, we will analyze the string operators parametrized by $\mathbf{w}_s$ in Eq. (\ref{string0}).

\subsection{Solving (\ref{string0}) for the string parameters $\mathbf{w}$}
We want to find all complex valued functions $\mathbf{w}$ that satisfy
(\ref{string0}). It is sufficient to find $\mathbf{w}$ satisfying (\ref%
{string0}), $\mathbf{\bar{w}}$ can be obtained immediately from (\ref%
{string01}). There are two cases to consider: $c_{s}\left( a,b\right) $ is
symmetric in $a,b$ or it is non-symmetric. First, if $c_{s}\left( a,b\right) 
$ is symmetric, Eq. (\ref{string0}) has scalar solutions. It was solved in
Ref. \onlinecite{LinLevinstrnet} in great details so we will not repeat the
computation but only cite some relevant results for completeness. The
resulting particles in this case are abelian quasiparticles. If $c_{s}\left(
a,b\right) $ is non-symmetric, we can see that \ Eq. $\left( \ref{string0}%
\right) $ has no nonzero scalar solutions, since the left-hand side is
manifestly symmetric in $a,b$ while the right-hand side is non-symmetric. To
build a path independent string operator, we have to allow the parameters $%
\mathbf{w}$ to be matrices rather than scalars. \ Therefore, we need to look
for higher dimensional projective representations with factor system $%
c_{s}\left( a,b\right) .$ The resulting particles have non-abelian
statistics.

Let's first review the case with symmetric $c_{s}\left( a,b\right) .$ We
assume that the group is $G=\prod_{i=1}^{L}\mathbb{Z}_{N_{i}}.$ Let $e_{1},...,e_{L}$
be the generators of $G.$ Once we find the values of $\mathbf{w}_{s}\left(
e_{i}\right) $ for each generator $e_{i},$ then $\mathbf{w}_{s}$ is fully
determined by Eq. $\left( \ref{string0}\right) .$ To find the values of $%
\mathbf{w}_{s}\left( e_{i}\right) ,$ we set $a=e_{i}$ and $b=ye_{i}$ with
integer $y$ in (\ref{string0}) and take the product of the equation over $%
y=0,1,...,N_{i}-1.$ By using the fact $N_{i}e_{i}=0,$ we find%
\begin{equation}
\mathbf{w}_{s}\left( e_{i}\right) ^{N_{i}}=\prod_{y=0}^{N_{i}-1}c_{s}\left(
e_{i},ye_{i}\right) .  \label{ws}
\end{equation}
We can see that $\mathbf{w}_{s}\left( e_{i}\right) $ can take $N_{i}$
different values for each $i.$ Thus, there are $\prod_{k}N_{k}=\left\vert
G\right\vert $ solutions to Eq $\left( \ref{ws}\right) $ for a given $s.$
The parameter $s$ can also take $\left\vert G\right\vert $ different values,
so altogether we find $\left\vert G\right\vert ^{2}$ solutions, corresponding
to $\left\vert G\right\vert ^{2}$ path independent string operators. When we
apply them to the ground state, they will create quasiparticles at two ends
of the string. Thus, the above operators allow us to construct $\left\vert
G\right\vert ^{2}$ different quasiparticle excitations.

Now, we consider the case with non-symmetric $c_{s}\left( a,b\right) $.
Again, we fix $s=\left( s_{1},s_{2},...,s_{L}\right) $ and set $%
a=e_{j},b=e_{k}.$ (Let us remind that $s,a,b$ are $L$-component vectors and
the subindex $i$ denotes the $i$-th component.) Once we find the value of $%
\mathbf{w}_{s}\left( e_{i}\right) $, then $\mathbf{w}_{s}\left( a\right) $
is fully determined by equation (\ref{string0}). To proceed, it is useful to
define the ratio
\begin{equation}
C_{s}\left( e_{j},e_{k}\right) =\frac{c_{s}\left( e_{j},e_{k}\right) }{%
c_{s}\left( e_{k},e_{j}\right) }.
\label{ratio}
\end{equation}
When $C_{s}=1,$ namely $c_{s}\left( a,b\right) $ is symmetric, $\mathbf{w}%
_{s}$ has one dimensional representation as discussed above. If $C_{s}\neq1,$
$\mathbf{w}_{s}$ has higher dimensional representation as we now discuss.
Thus, we can rewrite $\left( \ref{string0}\right) $ as%
\begin{equation}
\mathbf{w}_{s}\left( e_{j}\right) \mathbf{w}_{s}\left( e_{k}\right)
=C_{s}\left( e_{j},e_{k}\right) \mathbf{w}_{s}\left( e_{k}\right) \mathbf{w}%
_{s}\left( e_{j}\right) .  \label{string01}
\end{equation}
By the fact that $C_{s}\left( e_{j},e_{k}\right) $ is antisymmetric in $%
e_{j},e_{k}$ (see Eq. (\ref{ratio})) and the linearity in $s$, we can further parametrize $C_{s}\left(e_{i},e_{j}\right) $ by%
\begin{equation*}
C_{s}\left( e_{j},e_{k}\right) =\exp\left( 2\pi i\sum_{i}\frac{s_{i}p_{ijk}}{%
N_{ijk}}\right)
\end{equation*}
where $p_{ijk}$ is an antisymmetric tensor taking values in $0,1,...,N_{ijk}$
with $N_{ijk}$ being the great common divisor of $N_{i},N_{j},N_{k}.$ 

Next, for a given $s$, we consider a fixed set of $\left\{ s_{i}p_{ijk}\text{ with }i,j,k=1,...,L\right\} $.  Let $N$ be the least common multiple
of the set $\left\{ N_{ijk}\text{ with }i,j,k=1,...,L\right\} $. We then rewrite%
\begin{equation}
C_{s}\left( e_{j},e_{k}\right) =\exp\left( 2\pi i\frac{t_{jk}}{N}%
\right) \label{cs1}
\end{equation}
with $t_{jk}=\sum_{i}\frac{N}{N_{ijk}}s_{i}p_{ijk}$ being the
elements of the $L\times L$ antisymmetric integer matrix $T=\left[
t_{jk}\right] .$ We can diagonalize $T$ into a skew-normal form by a
unimodular integer matrix $U,$ i.e. $\left\vert \det U\right\vert =1:$%
\begin{equation}
T=UDU^{T}  \label{tt}
\end{equation}
with%
\begin{equation}
D=\left( 
\begin{array}{cc}
0 & D_{1} \\ 
-D_{1} & 0%
\end{array}
\right) \oplus...\oplus\left( 
\begin{array}{cc}
0 & D_{s} \\ 
-D_{s} & 0%
\end{array}
\right) \oplus0_{L-2s}.  \label{D}
\end{equation}
Here $0_{L-2s}$ is an $\left( L-2s\right) \times\left( L-2s\right) $ null
matrix. 

One nice thing about the skew-normal form (\ref{tt}) is that it allows us to obtain the projective representations associated with the factor set (\ref{cs1}) from the ones with the diagonal factor set $exp(2\pi i D_{jk}/N)$ where $D_{jk}$ are the matrix elements of the block-diagonal matrix $D$ in (\ref{D})\cite{Jagannathan1985}.
More specifically, once we
find the representation of $\left\{ \mathbf{w}_{s}^{\prime}\left(
e_{i}\right) ,i=1,...,L\right\} $ which obeys 
\begin{equation}
\mathbf{w}_{s}^{\prime}\left( e_{j}\right) \mathbf{w}_{s}^{\prime}\left(
e_{k}\right) =\exp\left( 2\pi i\frac{D_{jk}}{N}\right) \mathbf{w}%
_{s}^{\prime}\left( e_{k}\right) \mathbf{w}_{s}^{\prime}\left( e_{j}\right)
\label{rep1}
\end{equation}
with $D_{jk}$ being the $jk$-th element of $D,$ the representation of $%
\left\{ \mathbf{w}_{s}\left( e_{i}\right) ,i=1,...,L\right\} $ which obey $%
\left( \ref{string01}\right) $%
\begin{equation*}
\mathbf{w}_{s}\left( e_{j}\right) \mathbf{w}_{s}\left( e_{k}\right)
=\exp\left( 2\pi i\frac{t_{jk}}{N}\right) \mathbf{w}_{s}\left(
e_{k}\right) \mathbf{w}_{s}\left( e_{j}\right)
\end{equation*}
can be obtained by taking the product of $\mathbf{w}'_s$:
\begin{equation}
\mathbf{w}_{s}\left( e_{i}\right) =c_{i}^{^{\prime}}\mathbf{w}_{s}^{\prime
}\left( e_{1}\right) ^{u_{i1}}\mathbf{w}_{s}^{\prime}\left( e_{2}\right)
^{u_{i2}}...\mathbf{w}_{s}^{\prime}\left( e_{L}\right) ^{u_{iL}},i=1,...,L.
\label{rep2}
\end{equation}
Here $c_{i}$ are proper complex scalars chosen such that $\mathbf{w}%
_{s}\left( e_{i}\right) ^{N_{i}}=I$ and $u_{ij}=\left[ U\right] _{ij}.$

All that remains is to determine the representation of $\left\{ \mathbf{w}%
_{s}^{\prime }\left( e_{i}\right) ,i=1,...,L\right\} $ obeying (\ref{rep1}).
To this end, let us first look at the pair $\left\{ \mathbf{w}_{s}^{\prime
}\left( e_{2l-1}\right) ,\mathbf{w}_{s}^{\prime }\left( e_{2l}\right)
\right\} $ with $l=1,...,s$ which obey%
\begin{equation*}
\mathbf{w}_{s}^{\prime }\left( e_{2l-1}\right) \mathbf{w}_{s}^{\prime
}\left( e_{2l}\right) =\exp \left( 2\pi i\frac{D_{l}}{N}\right) \mathbf{w}%
_{s}^{\prime }\left( e_{2l}\right) \mathbf{w}_{s}^{\prime }\left(
e_{2l-1}\right) .
\end{equation*}%
The pair $\left\{ \mathbf{w}_{s}^{\prime }\left( e_{2l-1}\right) ,\mathbf{w}%
_{s}^{\prime }\left( e_{2l}\right) \right\} $ can be represented by the
shift $S_{l}$ and clock $C_{l}$ matrices with dimension%
\begin{equation*}
d_{l}=\frac{N}{\gcd \left( D_{l},N\right) }.
\end{equation*}%
Here the shift $S_{l}$ and clock $C_{l}$ matrices are 
\begin{equation}
S_{l}=\left( 
\begin{array}{ccccc}
0 & 1 & 0 & ... & 0 \\ 
0 & 0 & 1 & ... & 0 \\ 
0 & 0 & ... & 1 & 0 \\ 
... & ... & ... & ... & ... \\ 
1 & 0 & 0 & ... & 0%
\end{array}%
\right) ,C_{l}=\left( 
\begin{array}{ccccc}
1 & 0 & 0 & ... & 0 \\ 
0 & \omega & 0 & ... & 0 \\ 
0 & 0 & \omega ^{2} & ... & 0 \\ 
... & ... & ... & ... & 0 \\ 
0 & 0 & 0 & ... & \omega ^{d_{l}-1}%
\end{array}%
\right)
\end{equation}%
with $\omega =\exp \left( \frac{2\pi i}{d_{l}}\right) .$ Since the matrix $D$
has the block diagonal form (\ref{rep1}), the full representation of $%
\left\{ \mathbf{w}_{s}^{\prime }\left( e_{i}\right) ,i=1,...,L\right\} $ can
be expressed as 
\begin{gather*}
\mathbf{w}_{s}^{\prime }\left( e_{2l-1}\right)  =I_{1}\otimes ...\otimes
I_{l-1}\otimes C_{l}\otimes I_{l+1}\otimes ...\otimes I_{s}, \\
\qquad  \mathbf{w}_{s}^{\prime }\left( e_{2l}\right)  =I_{1}\otimes ...\otimes
I_{l-1}\otimes S_{l}\otimes I_{l+1}\otimes ...\otimes I_{s}, \\
\qquad \qquad \qquad \qquad \qquad \qquad \text{ with }l =1,...,s, \\
\mathbf{w}_{s}^{\prime }\left( e_{m}\right)  =I_{1}\otimes ...\otimes I_{s},%
\text{ with }m>2s.
\end{gather*}%
Here $I_{l}$ is the $d_{l}\times d_{l}$ identity matrix. Thus, the dimension
of the representation $\left\{ \mathbf{w}_{s}^{\prime }\left( e_{i}\right)
,i=1,...,L\right\} $ is%
\begin{equation*}
d_{s}=\prod_{l=1}^{s}d_{l}=\frac{N^{s}}{\prod_{i=1}^{s}\gcd \left(
D_{l},N\right) }.
\end{equation*}%
Finally, Plugging $\left\{ \mathbf{w}_{s}^{\prime }\left( e_{i}\right)
,i=1,...,L\right\} $ to (\ref{rep2}), we then obtain the representation of $%
\left\{ \mathbf{w}_{s}\left( e_{i}\right) ,i=1,...,L\right\} .$ This
completes the solution to the equation (\ref{string0}).

After constructing all the string operators and thus all the quasiparticle
excitations, we now discuss the labeling scheme for quasiparticle
excitations. We label each type-$s$ solution $\mathbf{w}_{s}$ to Eqs. (\ref%
{string0}) by an ordered pair $\alpha=\left( s,m\right) $ where $s\in G$ and 
$m$ is the representation of $G.$ We denote this solution by $\mathbf{w}%
_{\alpha}$, the corresponding string operator by $W_{\alpha}$ and the
quasiparticle excitation created by $W_{\alpha}$ as $\alpha.$ We will also
call the $\left( s,0\right) $ excitations \textquotedblleft
fluxes\textquotedblright\ and the $\left( 0,m\right) $ excitations
\textquotedblleft charges\textquotedblright\ (The definition of the pure
fluxes $\left( s,0\right) $ is a matter of convention.). Likewise, we think
of a general excitation $\left( s,m\right) $ as a composite of a flux and a
charge.

\subsection{Braiding statistics of quasiparticles}

In the previous section, we constructed string operators $W_{\alpha }$ for
each quasiparticle excitation $\alpha =\left( s,m\right) $ where $s\in G$
and $m$ is a representation of $G$. In this section, we will compute the
braiding statistics of these quasiparticles. We first review the computation
of the exchange statistics $\theta _{\alpha }$ for every particle $\alpha $
and the mutual statistics $\theta _{\alpha \beta }$ for every pair of
quasiparticles $\alpha ,\beta $ in the abelian topological phases. We then
compute the braiding statistics of the non-abelian quasiparticle excitations. 
\begin{figure}[tbp]
\begin{center}
\includegraphics[width=0.5\columnwidth]{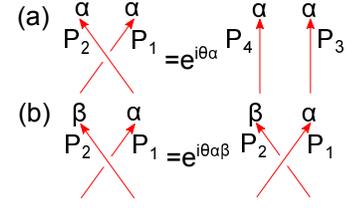}
\end{center}
\caption{(a) Computation of exchange statistics $e^{i\protect\theta _{%
\protect\alpha }}$ of quasiparticle $\protect\alpha $ created by $W_{\protect%
\alpha }.$ (b) Computation of mutual statistics $e^{i\protect\theta _{%
\protect\alpha \protect\beta }}$ between two quasiparticles created by $W_{%
\protect\alpha },W_{\protect\beta }.$}
\label{theta}
\end{figure}

To begin, we review the general relationship between braiding statistics and
the string operator algebra in the abelian topological phases. Let $%
\alpha,\beta$ be two abelian quasiparticles and let $W_{\alpha},W_{\beta}$
be the corresponding string operators. Then the exchange statistics $%
\theta_{\alpha}$ can be extracted from the algebra%
\begin{equation*}
W_{\alpha}\left( P_{2}\right) W_{\alpha}\left( P_{1}\right) \left\vert
\Phi\right\rangle =e^{i\theta_{\alpha}}W_{\alpha}\left( P_{4}\right)
W_{\alpha}\left( P_{3}\right) \left\vert \Phi\right\rangle
\end{equation*}
for any four paths $P_{1},P_{2},P_{3},P_{4}$ with the geometry of Fig. \ref%
{theta}(a) and $\left\vert \Phi\right\rangle $ denotes the ground state of
the system. By the path independence conditions and the local rules, one can
show that 
\begin{equation}
e^{i\theta_{\alpha}}=\mathbf{w}_{\alpha}\left( s\right) .  \label{exstat}
\end{equation}
On the other hand, the mutual statistics between two excitations $%
\alpha=\left( a,m\right) ,\beta=\left( t,n\right) $ are encoded in the
algebra (see Fig. \ref{theta}(b))%
\begin{equation*}
W_{\beta}\left( P_{2}\right) W_{\alpha}\left( P_{1}\right) \left\vert
\Phi\right\rangle =e^{i\theta_{\alpha\beta}}W_{\alpha}\left( P_{1}\right)
W_{\beta}\left( P_{2}\right) \left\vert \Phi\right\rangle .
\end{equation*}
In the same way, one can find that 
\begin{equation}
e^{i\theta_{\alpha\beta}}=\mathbf{w}_{\beta}\left( s\right) \mathbf{w}%
_{\alpha}\left( t\right) .  \label{mutualstat}
\end{equation}
Thus, in the abelian topological phases, $\left( \ref{exstat}\right) ,\left( %
\ref{mutualstat1}\right) $ allow us to compute the quasiparticle braiding
statistics after we solve $\mathbf{w}_{\alpha}$ in Eq. $\left( \ref{string0}%
\right) .$

Next, we compute the braiding statistics of particles in the non-abelian
topological phases. The computation of braiding statistics is similar as in
the abelian case. However, in this case, the string parameters $\mathbf{w}%
_{\alpha}$ are matrices. Thus, in order to extract the statistics phases, we
need to consider the whole braiding path (instead of segments of the
braiding path for the abelian phases) to contract the matrix indices.

The exchange statistics can be computed by comparing the two braiding
processes. In the first process, we create a pair of quasiparticles $\alpha,%
\bar{\alpha},$ exchange them and then annihilate the pair. In the second
process, we create them and annihilate the pair without exchange. The ratio
of the amplitudes for the two processes define the exchange statistics: 
\begin{equation}
e^{i\theta_{\alpha}}=\frac{Tr\left[ \mathbf{w}_{\alpha}\left( s\right) %
\right] }{\dim\left( \alpha\right) }  \label{exstat1}
\end{equation}
with $\dim\left( \alpha\right) $ being the dimension of the representation $%
\alpha.$

To describe the mutual statistics of two particles $\alpha =\left( s,\alpha
\right) ,\beta =\left( t,\beta \right) ,$ which is the matrix element $%
s_{\alpha \beta }$ of the S matrix, we consider the following process: We
create two pairs of quasiparticles $\alpha ,\bar{\alpha},\beta ,\bar{\beta},$
braid $\alpha $ around $\beta $ and then annihilate the two pairs. The
amplitude of this process divided by a factor $\left\vert G\right\vert $
defines the matrix element $s_{\alpha \beta }$ of the S matrix. More
specifically, we have%
\begin{equation}
s_{\alpha \beta }=\frac{\left\langle \Phi \left\vert \raisebox{-0.16in}{%
\includegraphics[height=0.4in]{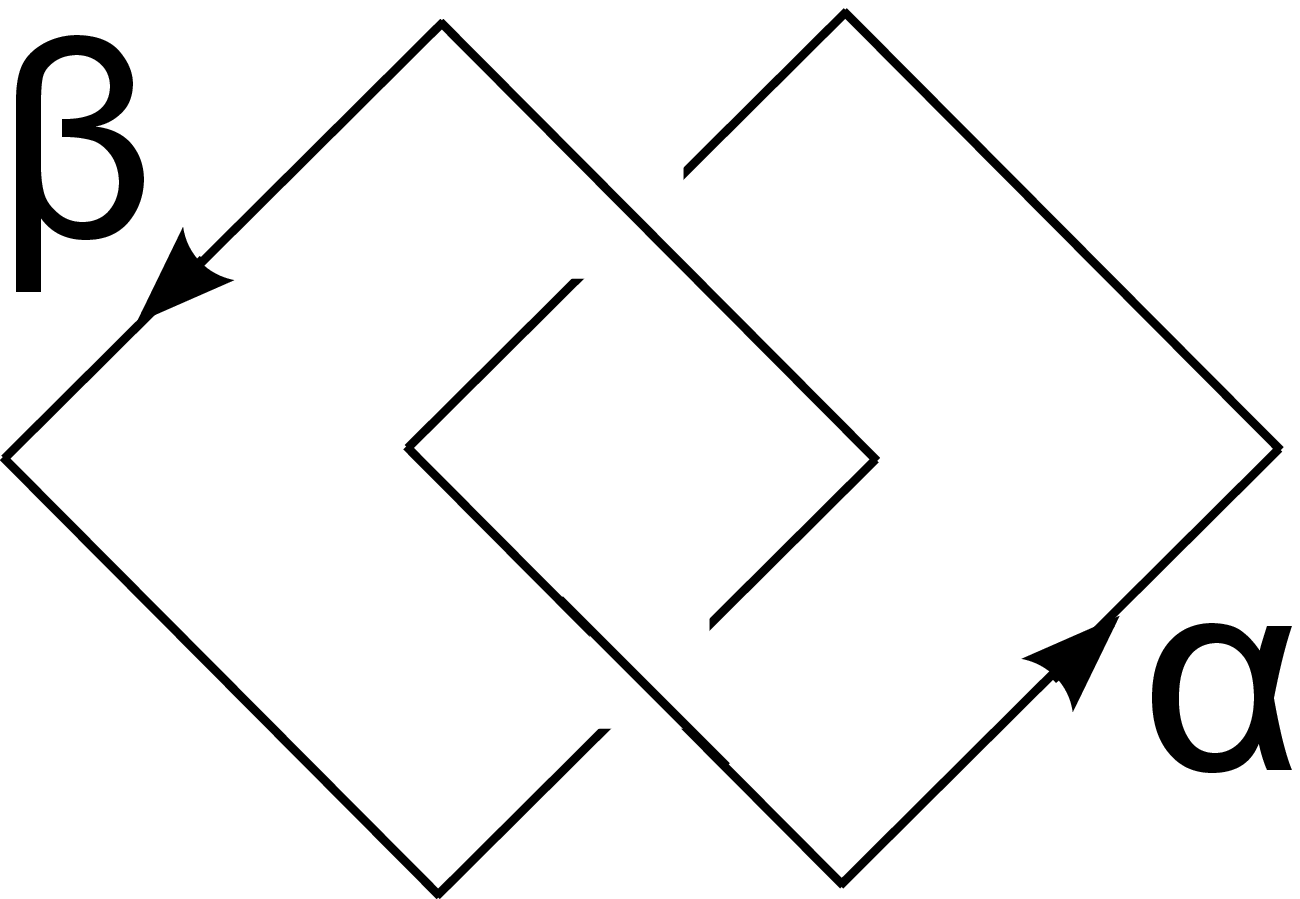}}\right\vert \Phi \right\rangle }{%
\left\vert G\right\vert }=\frac{Tr\left( \mathbf{w}_{\alpha }\left( t\right)
\right) Tr\left( \mathbf{w}_{\beta }\left( s\right) \right) }{\left\vert
G\right\vert }.  \label{mutualstat1}
\end{equation}
One can see that (\ref{exstat1},\ref{mutualstat1}) reduce to (\ref{exstat},\ref{mutualstat}) respectively for abelian quasiparticles $\alpha$ with $dim(\alpha)=1$ up to some normalization constants.

\section{QUASIPARTICLE STATISTICS OF GENERAL ABELIAN STRING-NET MODELS \label%
{section:statistics}}

After discussing the general framework for constructing multiple flavors of
abelian string-net models and computing the quasiparticle braiding
statistics, in this section we explicitly compute the braiding statistics of
these abelian string-net models. First, we review the string-net models with
one flavor of strings labeled by $a\in G=\mathbb{Z}_{N}.$ Then we analyze the general
case $G=\mathbb{Z}_{N_{1}}\times...\times\mathbb{Z}_{N_{L}}$ where
the models consist of $L$ flavors of strings, each of which is labeled by $a_i\in G_{i}=\mathbb{Z}_{N_i}.$ 

\subsection{$\mathbb{Z}_{N}$ string-net models}

We begin by reviewing the abelian string-net models with group $G=\mathbb{Z}%
_{N}.$ In the $\mathbb{Z}_{N}$ models, the string type $a$ has values in $%
\left\{ 0,1,...,N-1\right\} .$ The dual string type is defined by $a^{\ast
}=-a$ $\left( \text{mod }N\right) ,$ while the allowed branchings are the
triplets $\left( a,b,c\right) $ that satisfy $a+b+c=0$ $\left( \text{mod }%
N\right) .$

There are $N$ distinct solutions to the self-consistency conditions $\left( %
\ref{sfeqa}\right) $ parametrized by $p=0,...,N-1:$%
\begin{align}
F\left( a,b,c\right) & =e^{2\pi i\frac{pa}{N^{2}}\left( b+c-\left[ b+c\right]
\right) },  \label{sol1} \\
d_{a} & =1,\gamma_{a}=F\left( a^{\ast},a,a^{\ast}\right) ,  \notag \\
\alpha\left( a,b\right) & =F\left( a,b,\left( a+b\right) ^{\ast }\right)
\gamma_{a+b}.  \notag
\end{align}
The arguments $a,b,c$ can take values in the range $0,1,...,N-1$ and the
square bracket $\left[ b+c\right] $ denotes $b+c\left( \text{mod }N\right) $
with values also taken in the range $0,1,...,N-1.$ For each of the above $N$
solutions, we can construct a corresponding string-net model.

Next, we compute the braiding statistics of the quasiparticle excitations in
these models. By inserting the solution (\ref{sol1}) to $\left( \ref{string0}\right),$ we find 
\begin{align*}
c_{s}\left( a,b\right) & =e^{2\pi i\frac{ps}{N^{2}}\left( a+b-\left[ a+b%
\right] \right) }, \\
\mathbf{w}_{\left( s,m\right) }\left( a\right) & =e^{2\pi i\left( \frac{psa}{%
N^{2}}+\frac{ma}{N}\right) }.
\end{align*}
Then, the exchange statistics $\left( \ref{exstat}\right) $ of $%
\alpha=\left( s,m\right) $ is given by%
\begin{equation*}
\theta_{\left( s,m\right) }=2\pi\left( \frac{ps^{2}}{N^{2}}+\frac{ms}{N}%
\right)
\end{equation*}
and the mutual statistics $\left( \ref{mutualstat1}\right) $ of $%
\alpha=\left( s,m\right) $ and $\beta=\left( t,n\right) $ is 
\begin{equation*}
\theta_{\left( s,m\right) \left( t,n\right) }=2\pi\left( \frac {2pst}{N^{2}}+%
\frac{ns+mt}{N}\right) .
\end{equation*}

\section{$\mathbb{Z}_{N_{1}}\times...\times\mathbb{Z}_{N_{k}}$ string-net
models}

We now construct the string-net models with $G=\mathbb{Z}_{N_{1}}\times...%
\times\mathbb{Z}_{N_{L}}$. These models were constructed previously in Ref. %
\onlinecite{LinLevinstrnet}. Here we construct the models in a different
way. We think of the models consisting of $L$ flavors of string-nets. Each
flavor of string-nets form a $\mathbb{Z}_{N_{i}}$ string-net model as
reviewed in the previous section. The strings of each flavor are labeled by $%
a_{i}\in\mathbb{Z}_{N_{i}}$ and the dual string is defined by $%
a_{i}^{\ast}=-a_{i}$ $\left( \text{mod }N_{i}\right) .$ Each flavor of
string-nets satisfies the branching rules $a_{i}+b_{i}+c_{i}=0$ $\left( 
\text{mod }N_{i}\right) $ for $i=1,...,L$. 

As in the $\mathbb{Z}_{N}$ case discussed above, we have to find distinct
solutions to the self-consistency conditions (\ref{sfeqa},\ref{sfeqb}). This problem is closely related to the problem of computing the
cohomology group $H^{3}\left( \prod_{i}\mathbb{Z}_{N_{i}},U\left( 1\right)
\right) .$ This cohomology group has been calculated previously\cite%
{PropitiusThesis,ChenGuWenSPT,ChenGuWenSPT} and is given by%
\begin{equation}
H^{3}\left( \prod_{i}\mathbb{Z}_{N_{i}},U\left( 1\right) \right) =\prod_{i}%
\mathbb{Z}_{N_{i}}\prod_{i<j}\mathbb{Z}_{N_{ij}}\prod_{i<j<k}\mathbb{Z}%
_{N_{ijk}}  \label{h3}
\end{equation}
where $N_{ij}$ denotes the greatest common divisor of $N_{i}$ and $N_{j},$
and similarly for $\mathbb{Z}_{N_{ijk}}.$

In Ref. \onlinecite{LinLevinstrnet}, they focused on realizing the abelian topological phases and only discussed the $\prod_{i}\mathbb{Z}_{N_{i}}\prod_{i<j}\mathbb{Z}_{N_{ij}}$ distinct solutions and constructed the corresponding abelian
string-net models. In this paper, we will find all distinct solutions
including those realizing the non-abelian topological phases. Before going to
the details, it is worth mentioning that our new construction provides a
geometric interpretation of these distinct solutions. Specifically, Eq. $\left( \ref{h3}%
\right) $ can be understood by thinking of the multiple flavors of
string-nets, each of which is associated with $G_{i}=\mathbb{Z}_{N_{i}}.$
There are $N_{i}$ distinct string-net models corresponds to $N_{i}$ distinct
solutions $F_{i}\left( a_{i},b_{i},c_{i}\right) $ and altogether we have $%
\prod_{i}N_{i}$ distinct decoupled string-net models. Now we allow these
string-net models to intersect with one another. We need additional local
rules (\ref{rule2a}--\ref{rule2d}) to specify how they intersect. There are
two kinds of intersections to consider. One is intersections associated with
each two flavors of strings, which we will call 2-intersections and the
other is the intersections associated with three flavors of strings,
which we will call 3-intersections. As we will show later, there are $%
\prod_{i<j}N_{ij}$ distinct solutions $F_{a_{i}}^{\left( 2\right) }\left(
a_{j},b_{j}\right) $ for 2-intersections and $\prod_{i<j<k}N_{ijk}$ distinct
solutions $F_{a_{i}}^{3}\left( a_{j},b_{k}\right) $ for 3-intersections.
Putting everything together, we find all the solutions to the
self-consistency conditions and can be used to construct $\mathcal{N}%
=\prod_{i}N_{i}$ $\prod_{i<j}N_{ij}\prod_{i<j<k}N_{ijk}$ distinct string-net
models which realize $\mathcal{N}$ distinct topological phases.

More explicitly, the solutions to the self-consistency conditions (\ref{sfeqa},\ref{sfeqb}) are (for a particular
gauge choice)%
\begin{align}
F_{i}\left( a_{i},b_{i},c_{i}\right) & =e^{2\pi i\frac{p_i a_{i}}{N_{i}^{2}}%
\left( b_{i}+c_{i}-\left[ b_{i}+c_{i}\right] _{i}\right) },  \label{sol2} \\
d_{a_{i}} & =1,\gamma_{a_{i}}=F_{i}\left( a_{i}^{\ast},a_{i},a_{i}^{\ast
}\right) ,  \notag \\
\alpha_{i}\left( a_{i},b_{i}\right) & =F\left( a_{i},b_{i},\left(
a_{i}+b_{i}\right) ^{\ast}\right) \gamma_{a_{i}+b_{i}},  \notag \\
F_{a_{i}}^{\left( 2\right) }\left( b_{j},c_{j}\right) & =e^{2\pi ia_{i}\frac{%
p_{ij}}{N_{i}N_{j}}\left( b_{j}+c_{j}-\left[ b_{j}+c_{j}\right] _{j}\right)
},  \notag \\
\kappa_{a_{i}}\left( b_{j}\right) & =e^{2\pi ia_{i}\frac{p_{ij}}{N_{i}N_{j}}%
\left( b_{j}+\left[ b_{j}^*\right] _{j}\right) },  \notag \\
\eta_{a_{i}}\left( b_{j}\right) & =1,  \notag \\
F_{a_{i}}^{\left( 3\right) }\left( b_{j},c_{k}\right) & =e^{2\pi i\frac{%
p_{ijk}}{N_{ijk}}a_{i}b_{j}c_{k}}.  \notag
\end{align}
Here $p$ has values in $0,1,..,N_{i}$, $p_{ij}$ has values in $%
0,1,..,N_{ij}$ if $i<j$ and $p_{ij}=0$ if $i>j$ (Namely, we can gauge away the solutions with $p_{ij},i>j$.) and $p_{ijk}$ is an
antisymmetric tensor with the element of $p_{ijk}$ taking values in $%
0,1,...,N_{ijk}$. The square bracket $\left[ b_{i}+c_{i}\right] _{i}$
denotes $\left( b_{i}+c_{i}\right) $ $\left(\text{mod }N_{i}\right)
. $ As promised, there are $\prod_{i}N_{i}$ distinct $F_{i}\left(
a_{i},b_{i},c_{i}\right) ,$ $\prod_{i<j}N_{ij}$ distinct $F_{a_{i}}^{\left(
2\right) }\left( a_{j},b_{j}\right) $ and $\prod_{i<j<k}N_{ijk}$ distinct $%
F_{a_{i}}^{3}\left( a_{j},b_{k}\right) $ parametrized by $p,p_{ij}$ and $%
p_{ijk}.$

For comparison, in the original construction, the strings are labeled by
elements of $G$ parametrized by $L$-component integer vectors $a=\left(
a_{1},a_{2},...,a_{L}\right) $ where $0\leq a_{i}\leq N_{i}-1.$ The
solutions to the self-consistency conditions $\left( \ref{sfeqa}\right) $
are 
\begin{align}
F\left( a,b,c\right) & =e^{2\pi i\sum_{ij}\frac{P_{ij}a_{i}}{N_{i}N_{j}}%
\left( b_{j}+c_{j}-\left[ b_{j}+c_{j}\right] _{j}\right) +2\pi i\sum_{ijk}%
\frac{p_{ijk}}{N_{ijk}}a_{i}b_{j}c_{k}}  \label{sol3} \\
d_{a} & =1,\gamma_{a}=F\left( a^{\ast},a,a^{\ast}\right) ,  \notag \\
\alpha\left( a,b\right) & =F\left( a,b,\left( a+b\right) ^{\ast }\right)
\gamma_{a+b}.  \notag
\end{align}
Here $F$ is parametrized by the matrix $P_{ij}$ and the antisymmetric tensor 
$p_{ijk}.$ The matrix $P_{ij}$ is a $L\times L$ upper-triangular integer
matrix with diagonal elements $P_{ii}=p$ and off diagonal elements $%
P_{ij}=p_{ij}.$There are also $\mathcal{N}$ distinct solutions parametrized
by $P_{ij}$ and $p_{ijk}.$

It is now obvious that (\ref{sol2}) encodes the same information as (\ref{sol3}) as discussed in section \ref{unify}. In this regard, our new model can be viewed 
as an alternative construction of the original model while the new model provides a clear geometric picture of internal structure of the original construction.
In general, our construction provides various ways of building lattice models which realize topological order $G$ by different ways of encoding the information of $F(a,b,c)$ with $a,b,c\in G$, namely different decomposition of $G$.

The $\mathcal{N}$ solutions can be used to construct $\mathcal{N}$ different
lattice models (\ref{ham}). Our next task is to determine the braiding
statistics of the quasiparticle excitations in these models. To this end, we
can insert the solution $\left( \ref{sol3}\right) $ to (\ref{string0}) and
find $\mathbf{w}_{s}.$ 
After obtaining $\mathbf{w}%
_{s},$ the braiding statistics are given by (\ref{exstat1},\ref{mutualstat1}%
). Instead of giving the explicitly expression for the braiding statistics
of quasiparticles (we will compute the braiding statistics explicitly in the
example section), we want to make connection with the topological invariants
defined in Ref. \onlinecite{WangLevininv}. The topological invariants are a
set of three gauge invariant quantities $\left\{ \Theta_{i},\Theta
_{ij},\Theta_{ijk}\right\} $ associated with various braiding processes of
excitations with unit flux. Let $e_{i}$ denote the excitation which carries
a unit flux $\frac{2\pi}{N_{i}}$ associate $i$-th flavor of string-nets.
More explicitly, $\Theta_{i}$ is $N_{i}$ times of exchange statistics of $%
e_{i};$ $\Theta_{ij}$ is $N^{ij}$ times of mutual statistics of $e_{i},e_{j}$
with $N^{ij}$ being the least common multiple of $N_{i},N_{j}$ and $%
\Theta_{ijk}$ is the phase associated with braiding $e_{i}$ around $e_{j}$,
then around $e_{k},$ then around $e_{j} $ in an opposite direction and
finally around $e_{k}$ in an opposite direction. Then, the three topological
invariants can be expressed in terms of $c,F^{\left( 2\right) }$ and $%
F^{\left( 3\right) }:$ 
\begin{align*}
e^{i\Theta_{i}} & =\prod_{y=0}^{N_{i}-1}c_{e_{i}}\left( e_{i},ye_{i}\right) ,
\\
e^{i\Theta_{ij}} & =\prod_{y=0}^{N^{ij}-1}F_{e_{i}}^{\left( 2\right) }\left(
e_{j},ye_{j}\right) F_{e_{j}}^{\left( 2\right) }\left( e_{i},ye_{i}\right) ,
\\
e^{i\Theta_{ijk}} & =F_{e_{i}}^{\left( 3\right) }\left( e_{j},e_{k}\right) .
\end{align*}
One can check the quantities on the right hand side are indeed invariant
under the gauge transformations.
Thus, $\mathcal{N}$ topological phases are characterized by distinct topological invariants and therefore they belong to distinct quantum phases.

\section{Relation between different constructions \label{section:connection}}

We have constructed the multi-flavor string-net model by intersecting different flavors of string-net models, each of which is associated with $G_i.$ We show that the new model realizes the same topological order as the original model with group $G=\prod_i G_i$. This suggests that there are many ways of constructing models to realize the same topological phases with $G$ by intersecting string-net models with different choices of subgroups of $G$.
In this section, we will discuss the equivalence of these construction from a mathematical point of view.

We recall that in the original string-net models with group $G=\prod_{i}\mathbb{Z}_{N_{i}},$ there are $|H^{3}\left( G,U(1)\right)| $ distinct
solutions to the 3-cocycle conditions (\ref{sfeqa}) and thus there are $|H^{3}\left( G,U(1)\right)|$ models which realize distinct topological phases. On the
other hand, in the new construction, we construct multiple flavors of
intersecting string-net models, each of which belong to a group $%
G_{i}=\mathbb{Z}_{N_{i}}.$ In this way, we have to solve not only the
3-cocycle conditions $\left( \ref{sfeqa}\right) $ for each flavor of
strings but also the 2-cocycle conditions $\left( \ref{sfeq1}\right) 
$ and 1-cocycle conditions $\left( \ref{sfeq5}\right) .$ Correspondingly,
there are $\prod_{i}\mathbb{Z}_{N_{i}}$ distinct solutions to $\left( \ref%
{sfeqa}\right) ,$ $\prod_{i<j}\mathbb{Z}_{N_{ij}}$ distinct solutions to $%
\left( \ref{sfeq1}\right) $ and $\prod_{i<j<k}\mathbb{Z}_{N_{ijk}}$ distinct
solutions to $\left( \ref{sfeq5}\right) $ as parametrized by $%
p,p_{ij},p_{ijk}$ in $\left( \ref{sol2}\right) ,$ respectively.
Thus, the two constructions give rises to the same number of distinct topological phases.

It turns out that the relation between the two constructions of string-net
models are described by the mathematical relation called the K\text{\"u}nneth
formula. In $d$ spatial dimension, given a group $G=G_{1}\times G_{2},$ the
K\text{\"u}nneth formula is given by 
\begin{equation}
H^{d+1}\left( G,U(1)\right) =\sum_{k=0}^{d+1}H^{k}\left(
G_{1},H^{d+1-k}\left( G_{2},U\left( 1\right) \right) \right) .
\label{kformula}
\end{equation}
The formula expresses the cohomology group with $G$ in $d+1$ dimension in
terms of cohomology groups with its subgroups $G_{1},G_{2}$ in the lower
dimensions. Assuming the distinct topological phases with group $G$ in 
$d$ spatial dimension is classified by $H^{d+1}\left( G,U(1)\right) $ and
interpreting $H^{k}$ as dictating various rules for $\left( d-k\right) $%
-intersections among $\left( d-1\right) $-dimensional objects, then the
formula suggests how to construct models which realize $d$ dimensional
topological phases with $G$ by intersecting some simpler models which
realize topological phases with its subgroups $G_{1},G_{2}$. Notice that $G_1,G_2$ can be arbitrary two subgroups of $G$ such that $G=G_1 \times G_2$. For different choices of $G_1,G_2$, we have different decomposition of $H^{d+1}(G,U(1))$ which corresponds to intersecting models associated with different $G_1,G_2$. All these constructions give rise to the same topological phases classified by $H^{d+1}(G,U(1))$.
In this paper, we demonstrate the case with $d=2$ and $G=\prod_{i}G_i$ with $G_i=\mathbb{Z}_{N_i}$.

Let use consider $G=\mathbb{Z}_{N_{1}}\times \mathbb{Z}_{N_{2}}\times 
\mathbb{Z}_{N_{3}}$ and $d=2$ for example. We can choose $G=G_{1}\times
G_{2} $ with $G_{1}=\mathbb{Z}_{N_{1}}$ and $G_{2}=\mathbb{Z}_{N_{2}}\times 
\mathbb{Z}_{N_{3}}.$ Applying the K\text{\"u}nneth formula, we get%
\begin{eqnarray*}
H^{3}\left( G,U\left( 1\right) \right) &=&H^{3}\left( \mathbb{Z}%
_{N_{1}},U\left( 1\right) \right) +H^{3}\left( \mathbb{Z}_{N_{2}}\times 
\mathbb{Z}_{N_{3}},U\left( 1\right) \right) \\
&&+H^{2}\left( \mathbb{Z}_{N_{1}},\mathbb{Z}_{N_{2}}\times \mathbb{Z}%
_{N_{3}}\right) +H^{1}\left( \mathbb{Z}_{N_{1}},\mathbb{Z}_{N_{23}}\right) .
\end{eqnarray*}%
On the left hand side, it corresponds to the original string-net
construction with group $G$ and the model realizes $H^{3}\left(
G,U\left( 1\right) \right) $ distinct phases. On the right hand side, there
are four components. The first two terms counts the distinct phases realized
by two non-intersecting string-net models, which we call $G_{1}$-model and $%
G_{2}$-model. The last two terms counts the additional distinct phases which
arise from the intersections between the $G_{1}$-model and the $G_{2}$%
-model. Different rules for 2-intersections give $H^{2}\left( \mathbb{Z}_{N_{1}},\mathbb{Z}%
_{N_{2}}\times \mathbb{Z}_{N_{3}}\right) =\mathbb{Z}_{N_{12}}+\mathbb{Z}%
_{N_{13}}$ distinct phases while different rules for the 3-intersections give $H^{1}\left( 
\mathbb{Z}_{N_{1}},\mathbb{Z}_{N_{23}}\right) =\mathbb{Z}_{N_{123}}$
distinct phases. Totally, the intersecting models gives $|H^3(G,U(1))|$ distinct quantum phases. Alternatively, one can choose $G_1=\mathbb{Z}_{N_1}\times\mathbb{Z}_{N_2}$ and $G_2=\mathbb{Z}_{N_3}$ and get different models which realize these topological phases.

Furthermore, the K\text{\"u}nneth formula shows that the information of $F(a,b,c)\in H^3(G,U(1))$ with $G=\prod_i G_i$ can be encoded in some simpler objects $F(a_i,b_i,c_i),F^{(2)}_{a_i}(a_j,b_j),F^{(3)}_{a_i}(a_j,a_k)$ where $a_i\in G_i$ and $F^{(2)},F^{(3)}$ satisfy simpler algebraic conditions (\ref{sfeq1},\ref{sfeq5}). This is explicitly shown in Eqs. (\ref{string5a},\ref{f2}). Thus, our construction concretely demonstrates the K\text{\"u}nneth formula in building lattice models.

It is worth mentioning the other interesting application of the K\text{\"u}nneth formula is the decorated
domain wall construction in Ref. \onlinecite{decoratedDW} to realize
symmetry protected topological (SPT) phases. \ In that paper, they consider
the case that $G$ of the form $G=\mathbb{Z}_{2}\times G^{\prime}.$ They
proposed that SPT phases with symmetry $G$ in $d$ spatial dimensions can be
constructed by first decorating the $\mathbb{Z}_{2}$ domain walls with $d-1$
dimensional SPT phases with symmetry $G^{\prime}$ and then condensing these
decorated domain walls. This construction corresponds to $k=1$ term in the
K\text{\"u}nneth formula $\left( \ref{kformula}\right) $ which can realize $%
H^{1}\left( \mathbb{Z}_{2},H^{d}\left( G^{\prime},U\left( 1\right) \right)
\right) $ distinct SPT phases. (In general, $\left( \ref{kformula}\right) $
suggests even more ways to construct models for higher dimensional SPT
phases. One can construct $d$ dimensional SPT phases with symmetry $%
G=G_{1}\times G_{2}$ by decorating $d-k$ dimensional SPT phases with
symmetry $G_{2}$ onto $d-k$ dimensional $G_{1}$ symmetry defects and then
proliferating the $G_{1}$ symmetry defects.)

In addition, by the duality between SPT phases and topological phases\cite{LevinGu}, 
we can ungauge our intersecting string-net models with $G=G_{1}\times G_{2}$ and
obtain models which realize SPT phases with symmetry $G.$ However, our
construction is different from the decorated domain wall construction in
Ref. \onlinecite{decoratedDW}. The physical picture of our construction is
to assign $G_{1}$ charges to the vertices of $G_{2}$ domain walls and assign 
$G_{2}$ charges to the vertices of $G_{1}$ domain walls (see the rule \ref{rule2a}) and then proliferate
the $G_{1},G_{2} $ domain walls. A similar idea was proposed in Ref. %
\onlinecite{multikink} to realize SPT phases from the field theory
perspectives.

Finally, as suggested by the K\text{\"u}nneth formula, our construction can be generalized to non-abelian groups and higher dimensions. We leave the generalization for the future work.

\section{Examples \label{section:examples}}

In this section, we present some illustrative examples of intersecting
string-net models, namely $\mathbb{Z}_{2}\times\mathbb{Z}_{2}$ and $\mathbb{Z%
}_{2}\times\mathbb{Z}_{2}\times\mathbb{Z}_{2}$ string-net models. For each
example, we construct the wave functions and the corresponding Hamiltonian
and analyze braiding statistics of quasiparticles.

\subsection{$\mathbb{Z}_{2}\times\mathbb{Z}_{2}$ string-net model}

We consider $\mathbb{Z}_{2}\times\mathbb{Z}_{2}$ string-nets as two flavors
of $\mathbb{Z}_{2}$ string-nets intersecting with one another. In this way,
the Hilbert space of the $\mathbb{Z}_{2}\times\mathbb{Z}_{2}$ models
involves two string types $\left\{ 0=\left( 0,0\right) ,b=\left( 1,0\right)
\right\} $ for the first flavor and the other two string types $\left\{
0=\left( 0,0\right) ,r=\left( 0,1\right) \right\} $ for the second flavor.
Each $\mathbb{Z}_{2}$ string-nets obeys $\mathbb{Z}_{2}$ branching rules.

To construct the wave function and Hamiltonian, we solve the
self-consistency conditions for $\{F,d,\gamma,\alpha,F^{(2)},\kappa,\eta\} .$ We use the general solution (\ref{sol2}):%
\begin{gather*}
F\left( b,b,b\right) =\left( -1\right) ^{p_{1}},d_{b}=\left( -1\right)
^{p_{1}}, \\
F\left( r,r,r\right) =\left( -1\right) ^{p_{2}},d_{r}=\left( -1\right)
^{p_{2}}, \\
F_{b}^{\left( 2\right) }\left( r,r\right) =\kappa_{b}\left( r\right) =\left(
-1\right) ^{p_{12}}, \\
\text{other }F,F^{\left( 2\right) },d,\gamma,\alpha,\eta,\kappa=1.
\end{gather*}
Here $p_{1},p_{2},p_{12}=0,1.$ The $p_i=0$ solution corresponds to
the toric code model while the $p_i=1$ solution corresponds to the double
semion model for the $i$-th flavor of string-nets with $i=1,2$.
The $p_{12}=0,1$ solutions indicate if two flavors of string-nets are decoupled or coupled.
Altogether
there are $8$ distinct phases for $\mathbb{Z}_{2}\times\mathbb{Z}_{2}$
string-net models: each flavor of string-nets realizes two distinct phases
and two different ways of intersections between them realize two more phases.

The toric code and double semion models are well studied in Ref. %
\onlinecite{LevinWenstrnet}. Here we will focus on the new phases from the
intersections. For simplicity, we consider the model with $p_{1}=p_{2}=0,$
namely two intersecting toric code models. Then the $p_{12}=0$ solution corresponds to two decoupled Toric code models while the $p_{12}=1$ solution gives the other interesting model as we will construct.

The nontrivial local rules for the $p_{12}=1$ case are given by

\begin{gather*}
\Phi \left( \raisebox{-0.16in}{\includegraphics[height=0.4in]{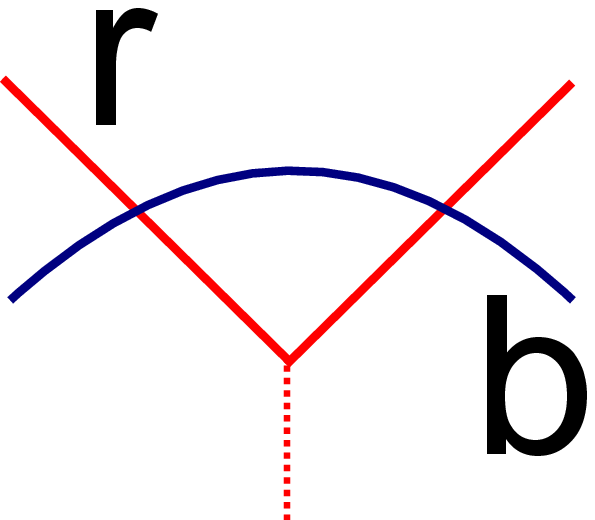}}%
\right) =-\Phi \left( \raisebox{-0.16in}{%
\includegraphics[height=0.4in]{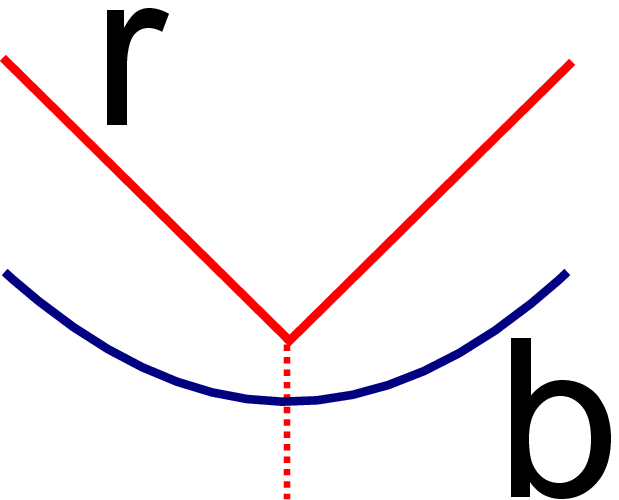}}\right) , \\
\Phi \left( \raisebox{-0.1in}{\includegraphics[height=0.32in]{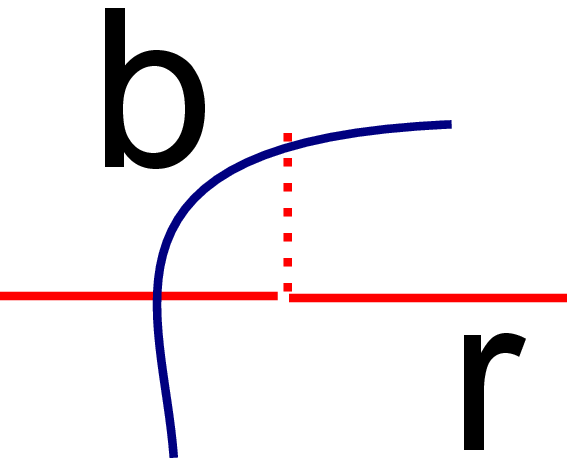}}%
\right) =-\Phi \left( \raisebox{-0.1in}{%
\includegraphics[height=0.32in]{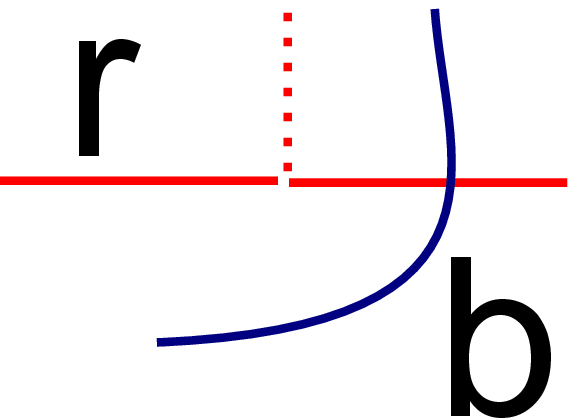}}\right) .
\end{gather*}%
The ground state consists of closed blue and red loops and their
intersections. The wave function, namely the solution for the local rules, is%
\begin{equation*}
\Phi \left( X\right) =\left( -1\right) ^{N_{ij}\left( X\right) }.
\end{equation*}%
Here $N_{ij}$ is the total number of red null strings inside the blue loops.

\begin{figure}[tbp]
\begin{center}
\includegraphics[width=0.3\columnwidth]{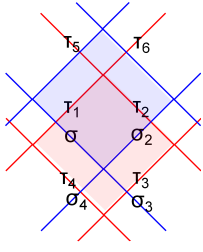}
\end{center}
\caption{The schematic lattice for the model (\protect\ref{ex1h}). The
lattice contains two intersecting square lattices with blue $\sigma$ and red $\tau$ spins
sitting on the links of each lattice.}
\label{fig:ex2lattice}
\end{figure}

The model is made up of two species of spins: blue spins $\sigma _{i}$ and
red spins $\tau_{i}.$ The blue spins $\sigma _{i}$ live on the
links $i$ of square lattice while the red spins $\tau _{i}$ live on the link 
$i$ of the shifted square lattice. We regard a link with $\sigma _{i}^{x}=-1$
as being occupied by a blue string and the state $\sigma _{i}^{x}=1$ as
being unoccupied. Similarly, if $\tau _{i}^{x}=-1$ then $i$ is occupied by a
red string while if $\tau _{i}^{x}=1$, then $i$ is empty. The Hamiltonian is
of the form 
\begin{equation}
H=-\sum_{I}Q_{I}^{b}-\sum_{I}Q_{I}^{r}-\sum_{p}B_{p}^{b}-\sum_{p}B_{p}^{r}
\label{ex1h}
\end{equation}%
with%
\begin{align*}
Q_{I}^{b}& =\frac{1}{2}\left( 1+\prod_{i\in I}\sigma _{i}^{x}\right)
,Q_{I}^{r}=\frac{1}{2}\left( 1+\prod_{i\in I}\tau _{i}^{x}\right) , \\
B_{p}^{b}& =\frac{1}{2}\left( 1+\prod_{i\in \{1,2,5,6\}}\sigma _{i}^{z}\left(
-1\right) ^{\frac{1-\tau _{1}^{x}}{2}\frac{1-\tau _{5}^{x}}{2}+\frac{1-\tau
_{2}^{x}}{2}\frac{1-\tau _{6}^{x}}{2}}\right) , \\
B_{p}^{r}& =\frac{1}{2}\left( 1+\prod_{i\in \{1,2,3,4\}}\tau _{i}^{z}\left(
-1\right) ^{\sum_{i\in \left\{ 1,2,3,4\right\} }\frac{1-\sigma _{i}^{x}}{2}%
\frac{1-\tau _{i}^{x}}{2}}\right) .
\end{align*}%
The link indices are labeled as in Fig. \ref{fig:ex2lattice}. 

Next we compute the quasiparticle excitations for the above Hamiltonian.
Each flavor of spins realize a toric code model which has four distinct
quasiparticles: $\left\{ 1_{i},e_{i},m_{i},e_{i}m_{i}\right\} $ with $i=b,r$
where $1_{i}$ is trivial quasiparticle, $e_{i}$ is the charge excitation, $%
m_{i}$ is the magnetic flux excitation and $e_{i}m_{i}$ is the bound state
of $e_{i}$ and $m_{i}$. The $e_{i}m_{i}$ particle is a fermion and any two
distinct nontrivial quasiparticles have semionic mutual braiding statistics.
What is left is to determine the mutual statistics between quasiparticles of
different flavors. By (\ref{mutualstat}), we find that 
\begin{align*}
\theta_{e_{1}e_{2}}&=0,\quad \theta_{e_{1}m_{2}}=0,\quad \theta_{m_{1}e_{2}}=0,\\
\theta_{e_1m_1}&=\pi,\quad \theta_{e_2m_2}=\pi,\quad \theta_{m_{1}m_{2}}=\frac{\pi}{2}.
\end{align*}
Other mutual statistics can be determined by linearity of the statistical
phases.

The 8 distinct $\mathbb{Z}_{2}\times\mathbb{Z}_{2}$ string-net models are
parametrized by $p_{1},p_{2},p_{12}=0,1.$ Each of the models has $%
4\times4=16 $ quasiparticles, which can be labeled by ordered pairs $\left(
s_{1},s_{2},m_{1},m_{2}\right) $ with $s_{i},m_{i}=0,1.$ These quasiparticle
statistics can be described by four component $U\left( 1\right) $
Chern-Simons theory with $K$ matrix%
\begin{equation*}
K=\left( 
\begin{array}{cccc}
0 & 0 & 2 & 0 \\ 
0 & 0 & 0 & 2 \\ 
2 & 0 & -2p_{1} & -p_{12} \\ 
0 & 2 & -p_{12} & -2p_{2}%
\end{array}
\right) .
\end{equation*}

\subsection{$\mathbb{Z}_{2}\times\mathbb{Z}_{2}\times\mathbb{Z}_{2}$
string-net model}

Our next example is the $\mathbb{Z}_{2}\times\mathbb{Z}_{2}\times \mathbb{Z}%
_{2}$ string-net model. The corresponding Chern-Simons theory with $G=%
\mathbb{Z}_{2}\times\mathbb{Z}_{2}\times\mathbb{Z}_{2}$ was studied in Ref. %
\onlinecite{PropitiusThesis} and \onlinecite{Z2Z2Z2}. In this section, we will construct the ground
state wave function and the lattice model. Then we analyze the quasiparticle
braiding statistics. The model is interesting because they can realize the
topological phase with non-abelian quasiparticles. 

The Hilbert space for the $\mathbb{Z}_{2}\times\mathbb{Z}_{2}\times \mathbb{Z%
}_{2}$ string-net models involves three flavors of strings, each of which
has two string types labeled by $\left\{ 0_{i},1_{i}\right\} $ with flavor
index $i=1,2,3$. These strings are self-dual and obey $\mathbb{Z}_{2}$
branching rules.

To construct the Hamiltonians and wave functions for the $\mathbb{Z}%
_{2}\times\mathbb{Z}_{2}\times\mathbb{Z}_{2}$ string-net models, we solve
the self-consistency conditions (\ref{sfeqa},\ref{sfeqb}) for $\left\{ F,d,\gamma,\alpha,F^{\left( 2\right)
},F^{\left( 3\right) },...\right\} .$ The general solution (\ref{sol2})
tells us that there are $2^{3}\times2^{3}\times2=128$ solutions labeled by $%
p_i,p_{ij},p_{ijk}=0,1.$ The first factor $2^{3}$ means that each flavor
of string-nets can be in toric code or double semion models. The second
factor $2^{3}$ means that there can be nontrivial intersections among any
two flavors of string-nets which give new topological phases as discussed in
the previous example. The last factor $2$ indicates that there is one
additional new phase which results from a nontrivial intersections among
three flavors of string-nets.

In this example, we will mainly focus on the last new phase, namely the
phase corresponding to $p_i=p_{ij}=0$ and $p_{ijk}=1.$ In this case, the
only nontrivial local rule is given by
\begin{equation*}
\Phi \left( \raisebox{-0.16in}{\includegraphics[height=0.4in]{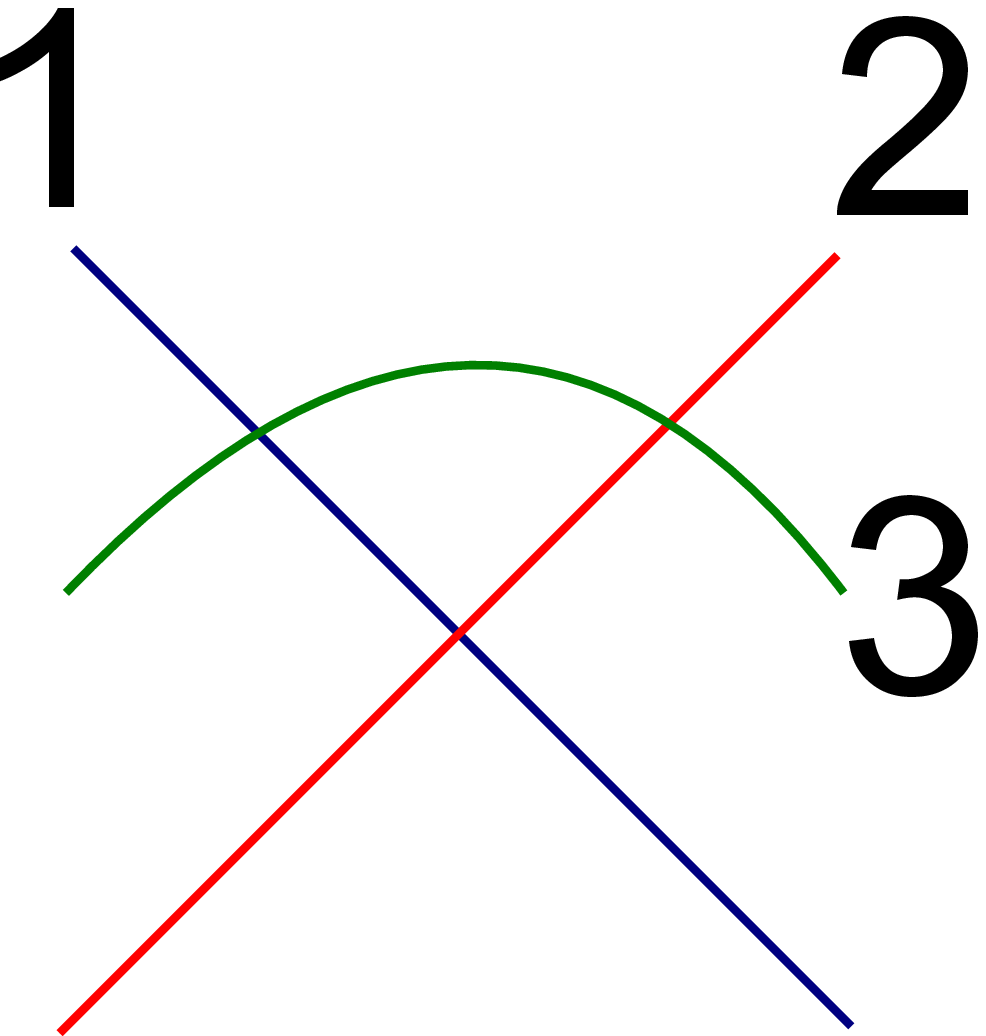}}%
\right) =-\Phi \left( \raisebox{-0.16in}{%
\includegraphics[height=0.4in]{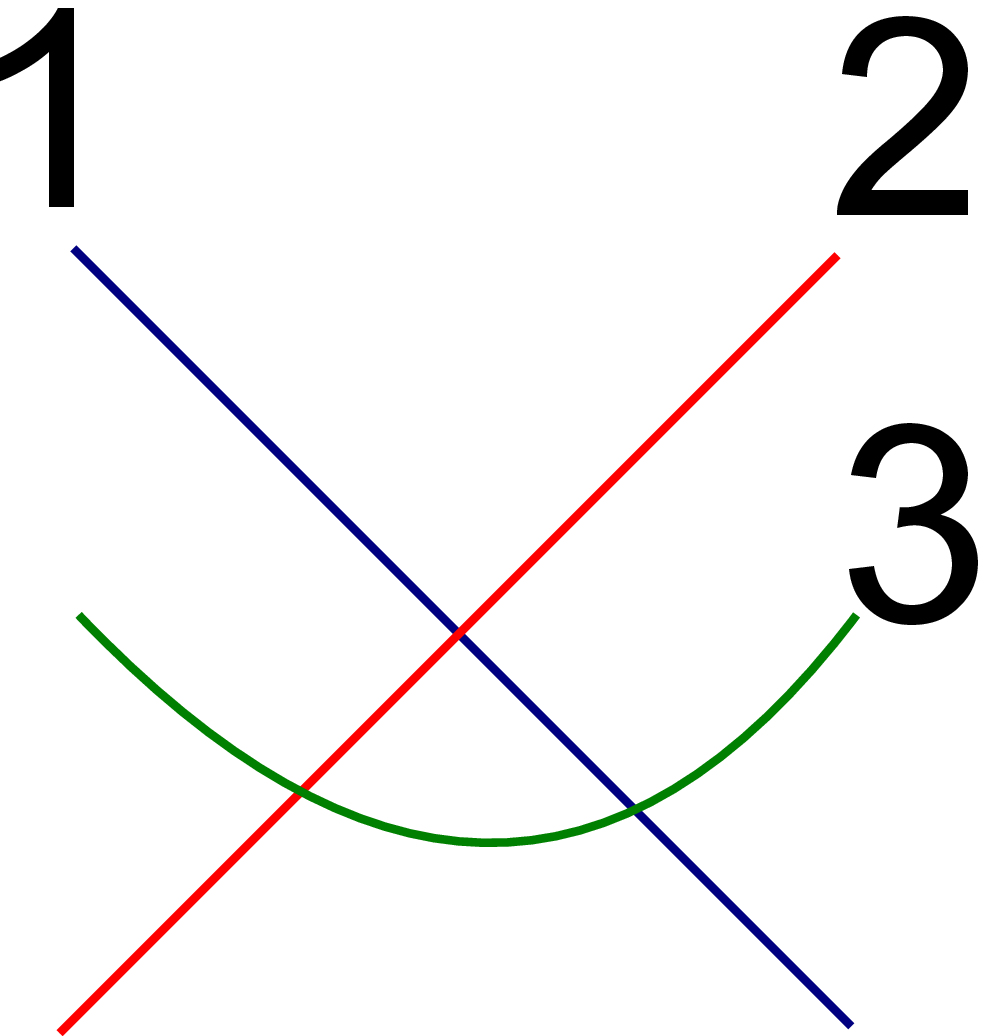}}\right) .
\end{equation*}%
The ground state wave function is%
\begin{equation*}
\Phi \left( X\right) =\left( -1\right) ^{N_{ijk}\left( X\right) }
\end{equation*}%
for any closed string-net configuration $X$. Here $N_{ijk}\left( X\right) $
is the number of crossings between two of the three flavors $i,j$ of strings
inside the loops of the third flavor $k$. Notice that $N_{ijk},N_{jki},N_{kij}$ have the same parity. Thus, to count $N_{ijk}(X)$, we can choose any one flavor of loops and count the number of crossings of the other two flavors of strings inside the loops.  
For example, let $X=\left(\raisebox{-0.16in}{\includegraphics[height=0.4in]{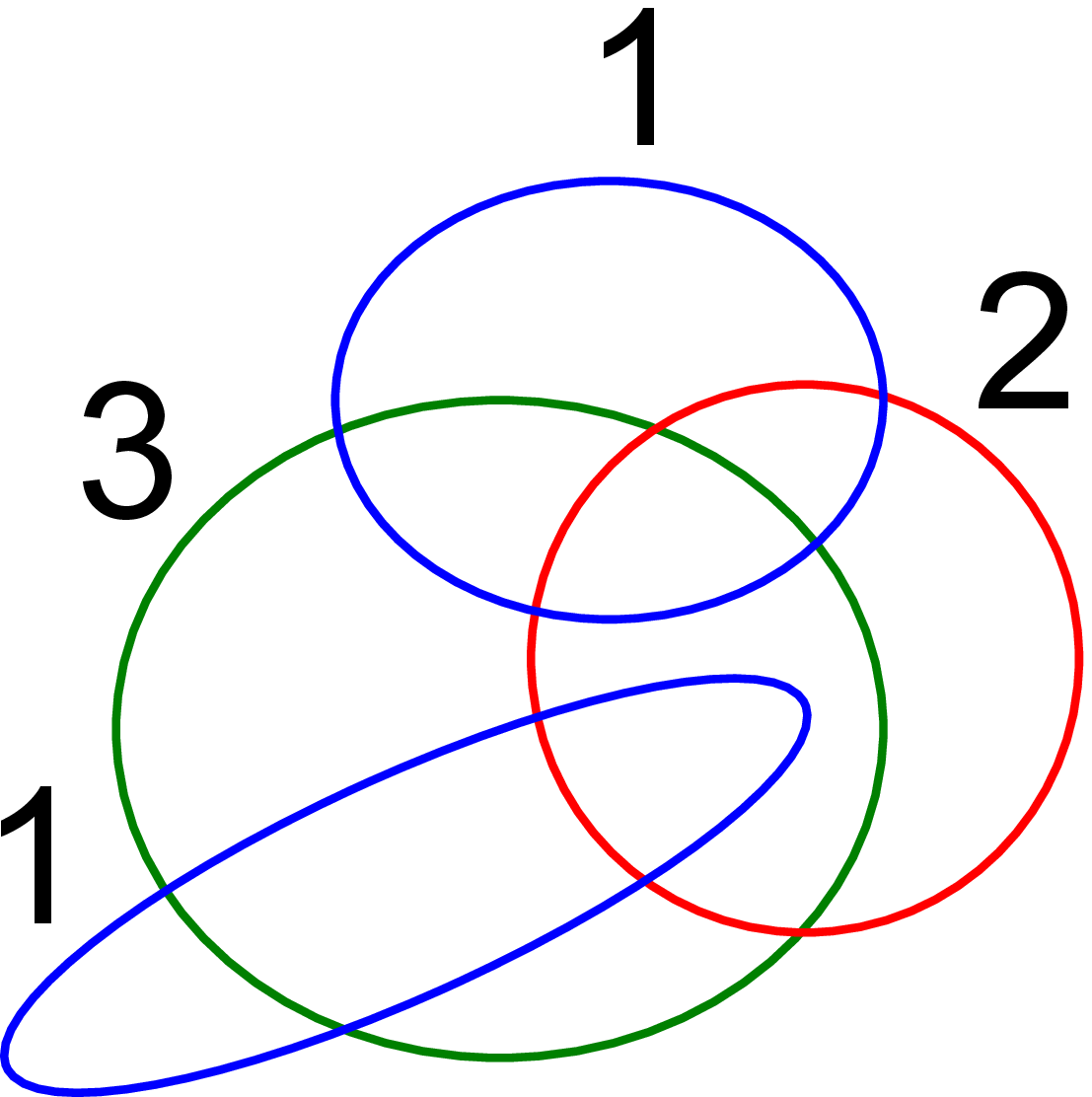}}\right)$. Then $N_{123}(X)=3$ because there are 3 1-2-crossings in the 3-loop. Similarly, one can see $N_{231}(X)=N_{312}(X)=1$. Thus they have the same parity.

\begin{figure}[tbp]
\begin{center}
\includegraphics[width=0.35\columnwidth]{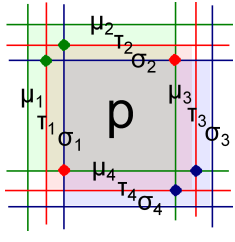}
\end{center}
\caption{The schematic lattice for the model (\protect\ref{ex2h}). The
lattice contains three intersecting square lattices with three types of
spins sitting on the links of each lattice.}
\label{fig:bpz2z2z2}
\end{figure}

Next, we want to construct the Hamiltonian. The model is made up of three
flavors of spins: 1-spins $\sigma _{i}$, 2-spins $\tau _{i}$ and 3-spins $%
\mu _{i}.$ The three kinds of spins live on the links $i$ of three square
lattices as shown in Fig. \ref{fig:bpz2z2z2}. Again, if $\sigma _{i}^{x}=-1,$
then $i$ is occupied by a 1-string while if $\sigma _{i}^{x}=1$, then $i$ is
empty. Similarly for the 2-spins $\tau _{i}$ and 3-spin $\mu _{i}.$ The
Hamiltonian is of the form 
\begin{equation}
H=\sum_{i=1,2,3}\left( -\sum_{I}Q_{I}^{i}-\sum_{p}B_{p}^{i}\right)
\label{ex2h}
\end{equation}%
with%
\begin{align*}
Q_{I}^{1}& =\frac{1}{2}\left( 1+\prod_{i\in I}\sigma _{i}^{x}\right)
,Q_{I}^{2}=\frac{1}{2}\left( 1+\prod_{i\in I}\tau _{i}^{x}\right) , \\
Q_{I}^{3}& =\frac{1}{2}\left( 1+\prod_{i\in I}\mu _{i}^{x}\right) , \\
B_{p}^{1}& =\frac{1}{2}\left( 1+\prod_{i\in p}\sigma _{i}^{z}\left(
-1\right) ^{f\left( \tau _{3}^{x},\mu _{4}^{x}\right) +f\left( \tau
_{4}^{x},\mu _{3}^{x}\right) }\right) , \\
B_{p}^{2}& =\frac{1}{2}\left( 1+\prod_{i\in p}\tau _{i}^{z}\left( -1\right)
^{f\left( \sigma _{2}^{x},\mu _{3}^{x}\right) +f\left( \sigma _{1}^{x},\mu
_{4}^{x}\right) }\right) , \\
B_{p}^{3}& =\frac{1}{2}\left( 1+\prod_{i\in p}\mu _{i}^{z}\left( -1\right)
^{f\left( \sigma _{1}^{x},\tau _{2}^{x}\right) +f\left( \sigma _{2}^{x},\tau
_{1}^{x}\right) }\right) .
\end{align*}%
Here we define%
\begin{equation*}
f\left( a,b\right) =\frac{1-a}{2}\frac{1-b}{2}.
\end{equation*}%
The phase factors associated with $B_{p}^{i}$ terms count the number of
crossings of the two $j,k$ flavors of strings inside the plaquette $p$ of
the $i$-lattice with $i\neq j\neq k$ (see Fig. \ref{fig:bpz2z2z2}). 

Now, we compute the braiding statistics of quasiparticles. 
Recall that we label the quasiparticles
by $\alpha=\left( s,m\right) $ where $s$ is the group element of $G=\mathbb{Z}_2\times\mathbb{Z}_2\times\mathbb{Z}_2$ and $m$
is the representation of $\mathbf{w}.$ We will label $s=\left(
s_{1},s_{2},s_{3}\right) \equiv s_{1}s_{2}s_{3}$ with $%
s_{1},s_{2},s_{3}=0,1. $

To proceed, we use the general solution

\begin{align*}
F\left( a,b,c\right) & =\exp \left( i\pi \left( 
\begin{array}{c}
a_{1}b_{2}c_{3}+b_{1}c_{2}a_{3}+c_{1}a_{2}b_{3} \\ 
-a_{1}c_{2}b_{3}-b_{1}a_{2}c_{3}-c_{1}b_{2}a_{3}%
\end{array}%
\right) \right) , \\
c_{s}\left( a,b\right) & =\exp \left( i\pi \left( 
\begin{array}{c}
s_{1}a_{2}b_{3}+a_{1}b_{2}s_{3}+b_{1}s_{2}a_{3} \\ 
-s_{1}b_{2}a_{3}-a_{1}s_{2}b_{3}-b_{1}a_{2}s_{3}%
\end{array}%
\right) \right) .\text{ }
\end{align*}%
Inserting the solution to (\ref{string0}), we can then solve the string
parameters $\mathbf{w}_{s}\left( a\right) .$ Let $%
e_{1}=100,e_{2}=010,e_{3}=001$ be the generators of $G.$ To find $\mathbf{w}%
_{s}\left( a\right) ,$ it is sufficient to solve $\mathbf{w}_{s}\left(
e_{1}\right) ,\mathbf{w}_{s}\left( e_{2}\right) ,\mathbf{w}_{s}\left(
e_{3}\right) $ and $\mathbf{w}_{s}\left( a\right) $ can be obtained from $%
\left( \ref{string0}\right) .$

There are four cases to consider. First, when $s=000,$ $c_{0}\left(
a,b\right) =1$ and thus the string parameter is one dimensional
representation of $G$ parametrized by integer $m_{1},m_{2},m_{3}=0,1:\mathbf{%
w}_{\left( 0,m_{1}m_{2}m_{3}\right) }\left( a\right) =\exp \left( \pi
i\left( m_{1}a_{1}+m_{2}a_{2}+m_{3}a_{3}\right) \right) $. The corresponding
particles are charges which are bosons. 

The second case is when $s\in
\left\{ 100,010,001\right\} .$ In particular, $\mathbf{w}_{100}$ satisfies 
\begin{gather*}
\mathbf{w}_{100}\left( e_{1}\right) ^{2}=\mathbf{w}_{100}\left( e_{2}\right)
^{2}=\mathbf{w}_{100}\left( e_{3}\right) ^{2}=I, \\
\left[ \mathbf{w}_{100}\left( e_{2}\right) ,\mathbf{w}_{100}\left(
e_{3}\right) \right] _{+}=0, \\
\left[ \mathbf{w}_{100}\left( e_{2}\right) ,\mathbf{w}_{100}\left(
e_{1}\right) \right] _{-}=0, \\
\left[ \mathbf{w}_{100}\left( e_{3}\right) ,\mathbf{w}_{100}\left(
e_{1}\right) \right] _{-}=0,
\end{gather*}%
Here $I$ is a $2\times 2$ identity matrix and $\left[ \mathbf{w}_{s}\left(
a\right) ,\mathbf{w}_{s}\left( b\right) \right] _{\pm }=\mathbf{w}_{s}\left(
a\right) \mathbf{w}_{s}\left( b\right) \pm \mathbf{w}_{s}\left( b\right) 
\mathbf{w}_{s}\left( a\right) $. One can find two inequivalent two
dimensional representations for $\mathbf{w}_{\left( 100,\pm \right) }$
labeled by $\pm $:%
\begin{equation*}
\mathbf{w}_{\left( 100,\pm \right) }\left( e_{1}\right) =\pm I,\mathbf{w}%
_{\left( 100,\pm \right) }\left( e_{2}\right) =\sigma _{x},\mathbf{w}%
_{\left( 100,\pm \right) }\left( e_{3}\right) =\sigma _{z}.
\end{equation*}%
Similarly, for $s=010,001,$ one can find 
\begin{align*}
\mathbf{w}_{\left( 010,\pm \right) }\left( e_{1}\right) & =\sigma _{z},%
\mathbf{w}_{\left( 010,\pm \right) }\left( e_{2}\right) =\pm I,\mathbf{w}%
_{\left( 010,\pm \right) }\left( e_{3}\right) =\sigma _{x}, \\
\mathbf{w}_{\left( 001,\pm \right) }\left( e_{1}\right) & =\sigma _{x},%
\mathbf{w}_{\left( 001,\pm \right) }\left( e_{2}\right) =\sigma _{z},\mathbf{%
w}_{\left( 001,\pm \right) }\left( e_{3}\right) =\pm I.
\end{align*}%
The third case is when $s\in \left\{ 110,101,011\right\} .$ Let's first
consider\ $\mathbf{w}_{110}$ which satisfies%
\begin{gather*}
\mathbf{w}_{110}\left( e_{1}\right) ^{2}=\mathbf{w}_{110}\left( e_{2}\right)
^{2}=\mathbf{w}_{110}\left( e_{3}\right) ^{2}=I, \\
\left[ \mathbf{w}_{110}\left( e_{2}\right) ,\mathbf{w}_{110}\left(
e_{3}\right) \right] _{+}=0, \\
\left[ \mathbf{w}_{110}\left( e_{1}\right) \mathbf{w}_{110}\left(
e_{3}\right) \right] _{+}=0, \\
\left[ \mathbf{w}_{110}\left( e_{1}\right) ,\mathbf{w}_{110}\left(
e_{2}\right) \right] _{-}=0.
\end{gather*}%
Then $\mathbf{w}_{\left( 110,\pm \right) }$ can be represented by%
\begin{equation*}
\mathbf{w}_{\left( 110,\pm \right) }\left( e_{1}\right) =\sigma _{x},\mathbf{%
w}_{\left( 110,\pm \right) }\left( e_{2}\right) =\pm \sigma _{x},\mathbf{w}%
_{\left( 110,\pm \right) }\left( e_{3}\right) =\sigma _{z}.
\end{equation*}%
In the same manner, we can find 
\begin{align*}
\mathbf{w}_{\left( 101,\pm \right) }\left( e_{1}\right) & =\pm \sigma _{x},%
\mathbf{w}_{\left( 101,\pm \right) }\left( e_{2}\right) =\sigma _{z},\mathbf{%
w}_{\left( 101,\pm \right) }\left( e_{3}\right) =\sigma _{x}, \\
\mathbf{w}_{\left( 011,\pm \right) }\left( e_{1}\right) & =\sigma _{z},%
\mathbf{w}_{\left( 011,\pm \right) }\left( e_{2}\right) =\sigma _{x},\mathbf{%
w}_{\left( 011,\pm \right) }\left( e_{3}\right) =\pm \sigma _{x}.
\end{align*}

Finally, when $s=111,\mathbf{w}_{111}$ satisfies%
\begin{gather*}
\mathbf{w}_{111}\left( e_{1}\right) ^{2}=I,\mathbf{w}_{111}\left(
e_{2}\right) ^{2}=I,\mathbf{w}_{111}\left( e_{3}\right) ^{2}=I, \\
\left[ \mathbf{w}_{111}\left( e_{1}\right) ,\mathbf{w}_{111}\left(
e_{2}\right) \right] _{+}=0, \\
\left[ \mathbf{w}_{111}\left( e_{1}\right) ,\mathbf{w}_{111}\left(
e_{3}\right) \right] _{+}=0, \\
\left[ \mathbf{w}_{111}\left( e_{2}\right) ,\mathbf{w}_{111}\left(
e_{3}\right) \right] _{+}=0.
\end{gather*}%
Then $\mathbf{w}_{\left( 111,\pm \right) }$ can be represented by%
\begin{equation*}
\mathbf{w}_{\left( 111,\pm \right) }\left( e_{1}\right) =\pm \sigma _{x},%
\mathbf{w}_{\left( 111,\pm \right) }\left( e_{2}\right) =\pm \sigma _{y},%
\mathbf{w}_{\left( 111,\pm \right) }\left( e_{3}\right) =\pm \sigma _{z}.
\end{equation*}%
Thus, we find 8 one dimensional particles labeled by $\left(
0,m_{1}m_{2}m_{3}\right) $ and 14 two dimensional particles labeled by $%
\left( s,\pm \right) $ with $s\in G.$

We are ready to compute the braiding statistics of quasiparticles. We begin
with the exchange statistics. By use of $\left( \ref{exstat1}\right) ,$ we
find that 
\begin{equation*}
e^{i\theta _{\alpha =\left( s,\pm \right) }}=\left\{ 
\begin{array}{l}
1,\text{ for }s=000 \\ 
\pm 1,\text{ for }s\in \left\{ 100,010,001\right\} \\ 
\pm 1,\text{ for }s\in \left\{ 110,101,011\right\} \\ 
\mp i,\text{ for }s=111%
\end{array}%
\right. .
\end{equation*}%
On the other hand, the S matrix can be computed by $\left( \ref{mutualstat1}%
\right) $ which is a $22\times 22$ matrix.

\section{Conclusion \label{section:conclusion}}

In this paper, we generalize the string-net construction to multiple flavors of strings, each of which is labeled by the elements of an abelian group $G_i$. Strings of the same flavor can branch while strings of different flavors can cross one another and thus they form intersecting string-nets. We systematically construct the exactly soluble lattice models and the ground state wave functions for the intersecting string-net condensed phases. We analyze the braiding statistics of the low energy quasiparticle excitations. We find that our model realizes the all the topological phases as the string-net model with group $G=\prod_i G_i$. 

In this regard, our intersecting string-net model with $\{G_i\}$ can be viewed as an alternative construction of the original string-net model with $G$. Furthermore, our construction also provides different ways of constructing lattice models with topological order $G$, which correspond to different decomposition of $G=\prod_i G_i$ or equivalently different choices of flavors of string-nets. 
In fact, the equivalence between these different models concretely demonstrates the K\text{\"u}nneth formula.

One important feature of our construction is that we encode the information of $F(a,b,c)$ with $a,b,c\in G$ in the original construction into a set of objects $\{F^{(2)}_{a_i}(a_j,b_j),F^{(3)}_{a_i}(a_j,b_j)\}$ with $a_i,b_i,c_i\in G_i$ which satisfy simpler conditions (\ref{sfeq1},\ref{sfeq5}). This property allows us to construct the models associated with a complicated group $G$ in two steps. First, we stack/intersect a set of simpler models associated with the subgroups of $G$. Then we turn on the interactions $F^{(2)},F^{(3)}$ between these simpler models and obtain models which can realize the topological phases with group $G=\prod_i G_i$.

As an application, our models can be easily modified to realize symmetry
enriched topological phases\cite{MesarosRan13,SETLevin,SETCheng,SymmetryDefect}. 
For example, if we want to have a model
which has symmetry $G_{1}$ and topological order $G_{2}$, we can first
construct the intersecting string-net model with $G=G_{1}\times G_{2}$.
Next, we ungauge the subgroup $G_{1}.$ The resulting model will have symmetry 
$G_{1}$ and maintain the $G_{2}$ topological order. In addition, we can further ungauge $G_2$ and obtain models which realize symmetry protected topological phases by $G_1\times G_2$.
The protection comes from the fact that we assign $G_1$ charges to the vertices of $G_2$ domain walls and vice versa which is implemented in the local rule (\ref{rule2a}).

In this work, we only focus on the multi-flavor string-net models associated with abelian groups. It would be interesting to extend our construction to the non-abelian case. 

\begin{acknowledgments}
C-H Lin is indebted to M. Levin for inspiring discussions. This research was supported by the Canada Research Chair Program (CRC).
\end{acknowledgments}

\appendix

\section{Derivation of self-consistency conditions \label{app:sfeqs}}

In this section, we show that the local rules (\ref{sfeqb}) are
self-consistent if and only if the parameters $\left\{ F^{\left( 2\right) },%
\bar{F}^{\left( 2\right) },\kappa ,\eta ,F^{\left( 3\right) }\right\} $
satisfy the following conditions
\begin{gather}
F_{a_{i}}^{\left( 2\right) }\left( a_{j},b_{j}\right) F_{a_{i}}^{\left(
2\right) }\left( a_{j}+b_{j},c_{j}\right)  
 =F_{a_{i}}^{\left( 2\right) }\left( b_{j},c_{j}\right) F_{a_{i}}^{\left(
2\right) }\left( a_{j},b_{j}+c_{j}\right) , \label{sfeq1a} \\
\bar{F}_{a_{i}}^{\left( 2\right) }\left( a_{j},b_{j}\right) =F_{a_{i}^{\ast
}}^{\left( 2\right) }\left( a_{j},b_{j}\right) \frac{\kappa _{a_{j}}\left(
a_{i}^{\ast }\right) \kappa _{b_{j}}\left( a_{i}^{\ast }\right) }{\kappa
_{a_{j}+b_{j}}\left( a_{i}^{\ast }\right) },  \label{fbara} \\
\frac{F_{a_{i}}^{\left( 2\right) }\left( a_{j},b_{j}\right) F_{b_{i}}^{\left(
2\right) }\left( a_{j},b_{j}\right)}{F_{a_{i}+b_{i}}^{\left(2\right) }\left( a_{j},b_{j}\right)}=\frac{\bar{F}_{a_{j}}^{\left( 2\right) }\left( a_{i},b_{i}\right) \bar{F}_{b_{j}}^{\left( 2\right) }\left( a_{i},b_{i}\right)}
{\bar{F}_{a_{j}+b_{j}}^{\left( 2\right)}\left( a_{i},b_{i}\right)}
 , \label{3fa} \\
\kappa _{a_{i}}\left( a_{j}\right) \kappa _{a_{i}^{\ast }}\left( a_{j}^{\ast
}\right) \kappa _{a_{j}^{\ast }}\left( a_{i}\right) \kappa _{a_{j}}\left(
a_{i}^{\ast }\right) =1,  \label{sfeq2a} \\
\kappa _{a_{i}}\left( a_{j}\right) =\eta _{a_{i}}\left( a_{j}\right)
F_{a_{i}}^{\left( 2\right) }\left( a_{j}^{\ast },a_{j}\right) ,
\label{sfeq4a} \\
F_{a_{i}}^{\left( 3\right) }\left( a_{j},a_{k}\right) F_{a_{i}}^{\left(
3\right) }\left( b_{k}^{\ast },a_{j}\right) =F_{a_{i}}^{\left( 3\right)
}\left( a_{j},a_{k}+b_{k}\right) ,  \label{sfeq5a} \\
F^{(3)}_{a_i}(a_j,a_k)F^{(3)}_{b_i}(a_j,a_k)=F^{(3)}_{a_i+b_i}(a_j,a_k),
\label{sfeq6a} \\
\eta _{a_{i}}\left( a_{j}\right) =\eta _{a_{j}}\left( a_{i}\right) ,
\label{sfeq7a} \\
F_{a_{i}}^{\left( 3\right) }\left( a_{j},a_{k}\right) =F_{a_{j}^{\ast
}}^{\left( 3\right) }\left( a_{k},a_{i}\right) ^{-1}=F_{a_{i}^{\ast
}}^{\left( 3\right) }\left( a_{j}^{\ast },a_{k}^{\ast }\right) ^{-1},
\label{sfeq61a} \\
F_{a_{i}}^{\left( 2\right) }\left( a_{j},b_{j}\right) =F_{a_{i}}^{\left(
3\right) }\left( a_{j},a_{k}\right) =1\text{ if }a\text{ or }b=0.
\label{sfeq8a}
\end{gather}%

\begin{figure}[tbp]
\begin{center}
\includegraphics[width=0.9\columnwidth]{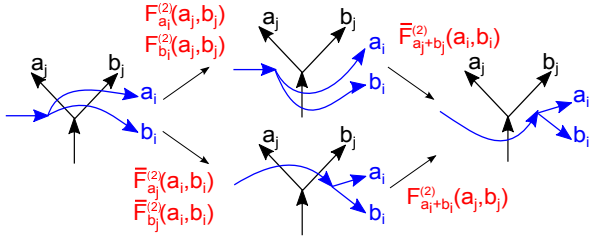}
\end{center}
\caption{Self-consistency in sequences requires the condition (\protect\ref%
{3fa}).}
\label{fig:3fa}
\end{figure}

\begin{figure}[tbp]
\begin{center}
\includegraphics[width=0.7\columnwidth]{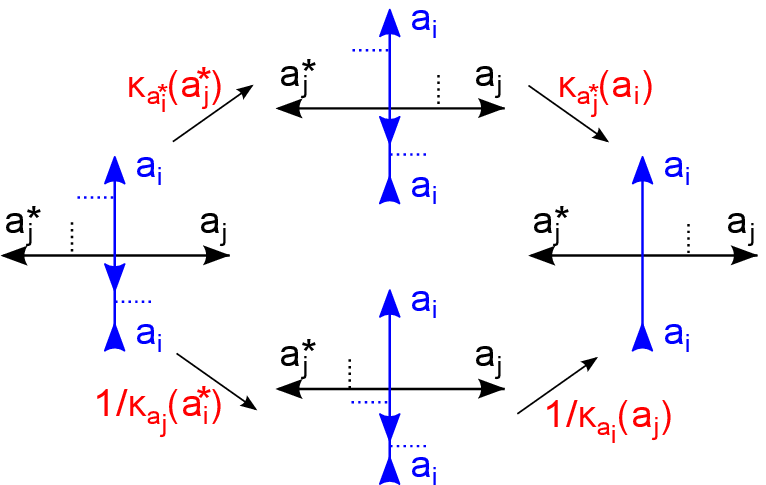}
\end{center}
\caption{Self-consistency in sequences requires the condition (\protect\ref%
{sfeq2a}).}
\label{figure:sfeq2}
\end{figure}

\begin{figure}[tbp]
\begin{center}
\includegraphics[width=0.45\columnwidth]{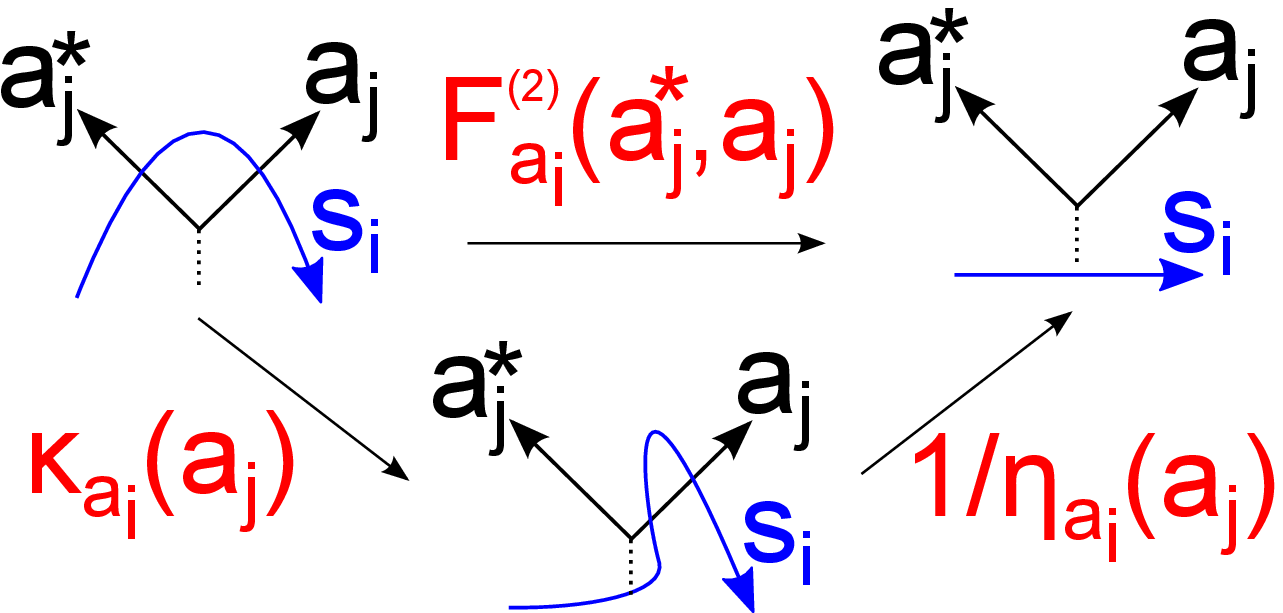}
\end{center}
\caption{Self-consistency in sequences requires the condition (\protect\ref%
{sfeq4a}).}
\label{figure:sfeq4}
\end{figure}

\begin{figure}[tbp]
\begin{center}
\includegraphics[width=1\columnwidth]{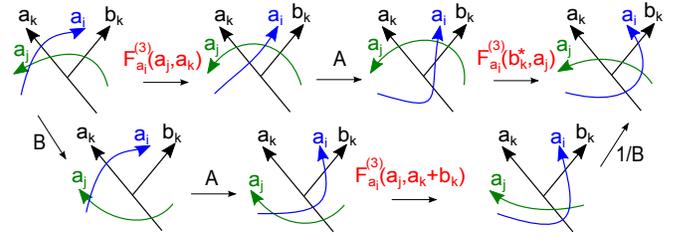}
\end{center}
\caption{Self-consistency in sequences requires the condition (\protect\ref%
{sfeq5a}). Here $A=\frac{\protect\eta _{a_{i}}\left( a_{k}+b_{k}\right) }{%
\protect\eta _{a_{i}}\left( b_{k}\right) \bar{F}_{a_{i}}^{\left( 2\right)
}\left( a_{k},b_{k}\right) }$ and $B=\bar{F}_{a_{j}}^{\left( 2\right)
}\left( a_{k},b_{k}\right) .$}
\label{figure:sfeq5}
\end{figure}

First, we show the above conditions are necessary for self-consistency. The
first condition (\ref{sfeq1a}) was derived in the text (see Sec \ref%
{section:mwavefunction}). The second equation (\ref{fbara}) can can shown by 
\begin{gather*}
\Phi \left( \raisebox{-0.16in}{\includegraphics[height=0.4in]{rule10a.eps}}%
\right) =\Phi \left( \raisebox{-0.16in}{%
\includegraphics[height=0.4in]{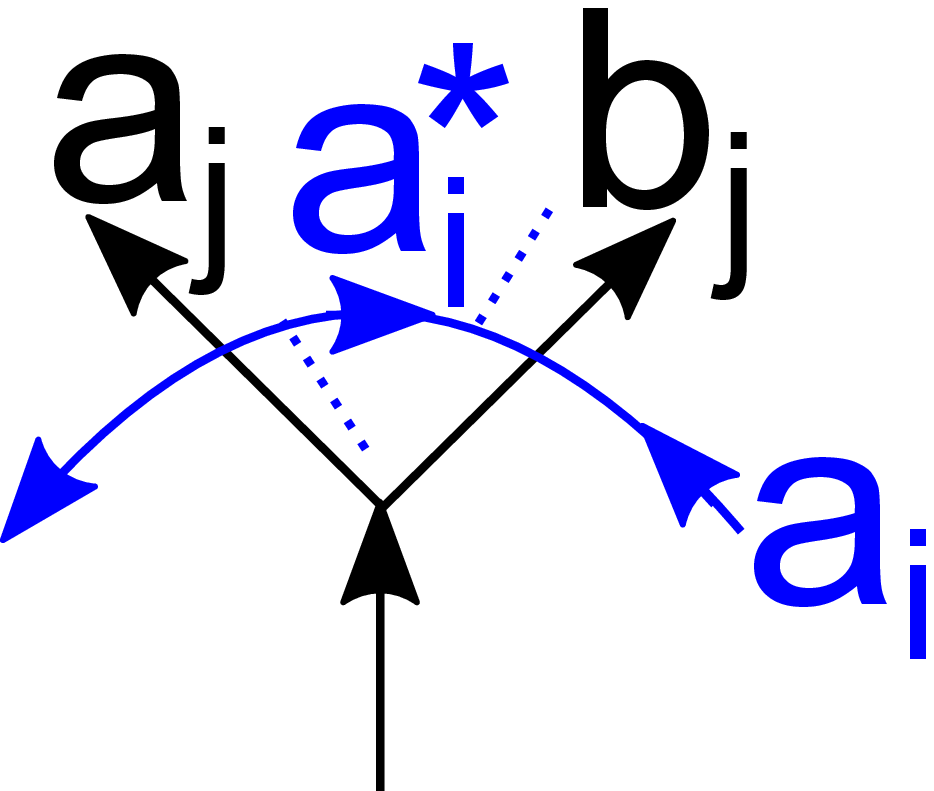}}\right) \\
=\kappa _{a_{j}}\left( a_{i}^{\ast }\right) \kappa _{b_{j}}\left(
a_{i}^{\ast }\right) \Phi \left( \raisebox{-0.16in}{%
\includegraphics[height=0.4in]{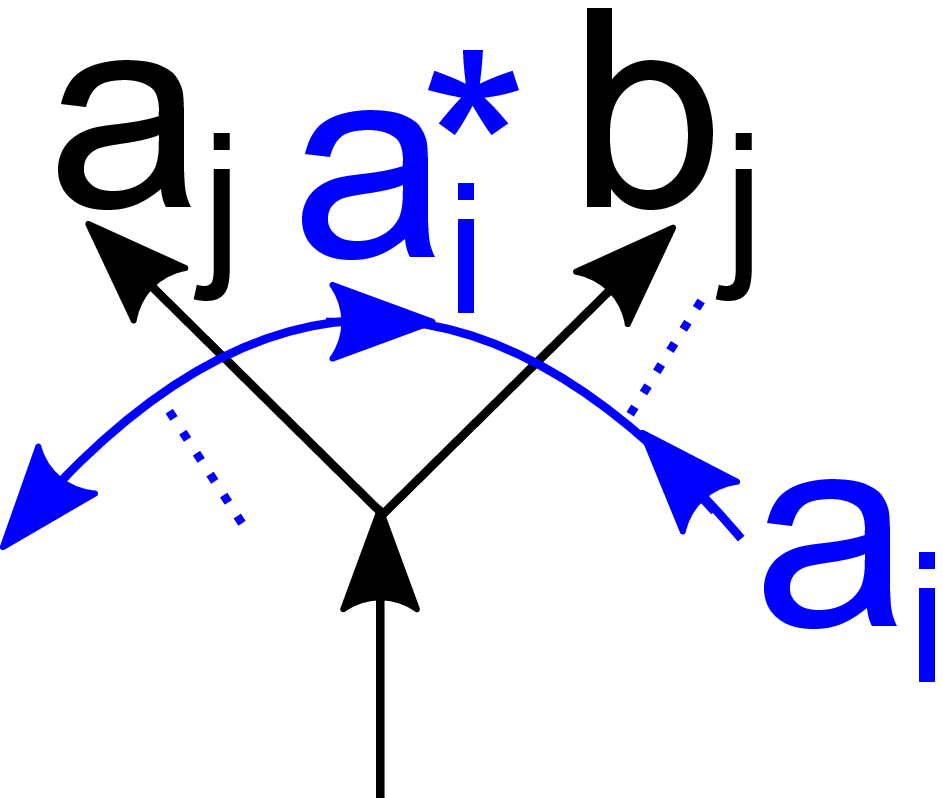}}\right) \\
=\kappa _{a_{j}}\left( a_{i}^{\ast }\right) \kappa _{b_{j}}\left(
a_{i}^{\ast }\right) F_{a_{i}^{\ast }}^{\left( 2\right) }\left(
a_{j},b_{j}\right) \Phi \left( \raisebox{-0.16in}{%
\includegraphics[height=0.4in]{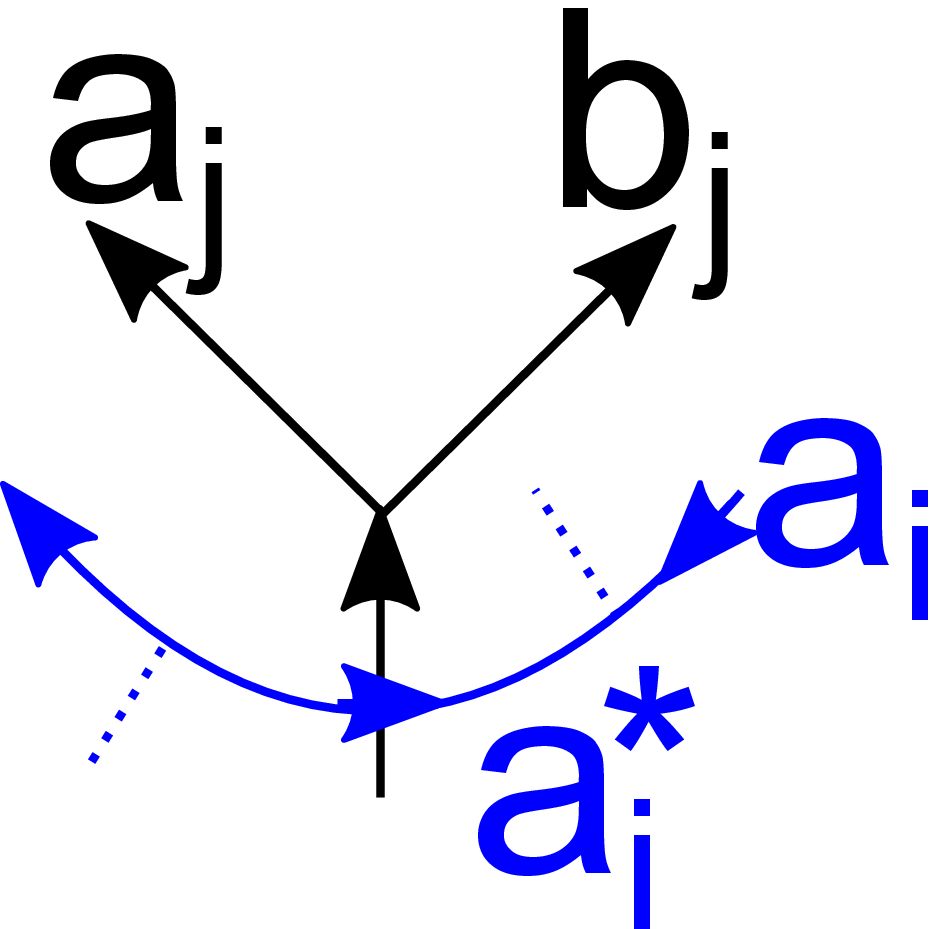}}\right) \\
=F_{a_{i}^{\ast }}^{\left( 2\right) }\left( a_{j},b_{j}\right) \frac{\kappa
_{a_{j}}\left( a_{i}^{\ast }\right) \kappa _{b_{j}}\left( a_{i}^{\ast
}\right) }{\kappa _{a_{j}+b_{j}}\left( a_{i}^{\ast }\right) }\Phi \left( %
\raisebox{-0.16in}{\includegraphics[height=0.4in]{rule10b.eps}}\right) .
\end{gather*}%
The third condition (\ref{3fa}) can be understood by considering the two
sequences in Fig. \ref{fig:3fa}. Eq. (\ref{3fa}) must be satisfied in order
for the rules (\ref{rule2a},\ref{rule2aa}) to be consistent. Similarly, the
conditions (\ref{sfeq2a}--\ref{sfeq6a}) follow from considering the
string-net configurations shown in Figs. \ref{figure:sfeq2}--\ref{figure:sfeq5} and demanding consistency between the two sequences of moves
shown there. The conditions (\ref{sfeq7a},\ref{sfeq61a}) come from the symmetry in the roles of strings of different flavors in the rules (\ref{rule2c}) and (\ref{rule2d}). Finally, (\ref{sfeq8a}) is the normalization condition for $F^{(2)},F^{(3)}$.

\begin{figure}[tbp]
\begin{center}
\includegraphics[width=0.8\columnwidth]{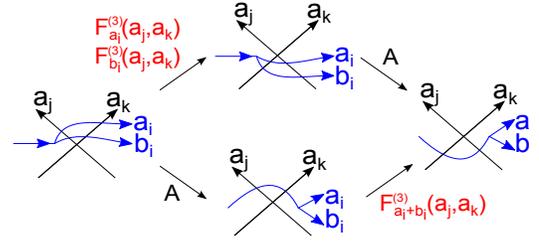}
\end{center}
\caption{Self-consistency in sequences requires the condition (\protect\ref%
	{sfeq6a}). Here $A=\bar{F}^{(2)}_{a_j}(a_i,b_i)\bar{F}^{(2)}_{a_k}(a_i,b_i) .$}
\label{figure:sfeq6}
\end{figure}

We have shown that Eqs. (\ref{sfeq1a}--\ref{sfeq8a}) are necessary
conditions for the rules to be self-consistent. To show that they are
sufficient, one can follow the same argument in Ref. %
\onlinecite{LinLevinstrnet}. Suppose that $\{F^{(2)},\kappa,\eta,F^{(3)}\}$
satisfy Eqs. (\ref{sfeq1a}--\ref{sfeq8a}) and
use them to construct an exactly soluble lattice Hamiltonian $H$ $\left( \ref%
{ham}\right) .$ Then the ground state of $H$ satisfies the local rules $%
\left( \ref{rule2a}\text{--\ref{rule2d}}\right) $ on the lattice. By using
this fact, we can prove that the rules are self-consistent. Suppose, on the
contrary, that the rules $\left( \ref{rule2a}\text{--\ref{rule2d}}\right) $
are not self-consistent. Then there exists two sequences of moves which
relate the same initial and final continuum string-net states $X_{1},X_{2}$
but with different proportionality constants. These two sequences of moves
can be adapted from the continuum to the lattice if we make the lattice
sufficiently fine. However, this leads to a contradiction since the ground
state wave function $\Phi_{latt}$ of $H$ obeys the rules on the lattice (see section \ref{section:prop}). Thus, our assumption is false. We conclude that the rules must be self-consistent.

\section{Graphical representation of the Hamiltonian \label{app:h}}
In this section, we demonstrate the graphical representation of $B_{p}^{s_{i}}$ leads to the matrix elements in equation (\ref{bps}).

By the graphical representation, the action of the operator $B_{p}^{s_{i}}$ on a string-net state is to add a loop of type-$s_{i}$
string. We can fuse the string $s_{i}$ onto the links along the boundary of
the plaquette by using the local rules (\ref{rule1a}--\ref{rule1c},\ref{rule2a}--\ref{rule2d}). This allows us to obtain the matrix elements of $B_{p}^{s_{i}}.$ Specifically, there are two steps in this fusion
process. The first step is to glide the string $s_{i}$ over all other
string-nets with different flavors inside the plaquette $p$. This gives
the factors $B_{p,x'x}^{s_i\left( 2\right) }$ and $B_{p,x'x}^{s_i\left( 3\right) }$. Then,
the second step is to fuse $s_{i}$ to the boundary links of $p$ as in
the original construction which gives the factor $B_{p,x'x}^{s_i\left( 1\right) }$.
The computation of $B_{p,x'x}^{s_i(1)}$ is exactly the same as in the
Appendix D of Ref. \onlinecite{LinLevinstrnet} by neglecting string-nets of
all other different flavors and thus we will not repeat here. We will
present the graphical representation of computing $B_{p,x'x}^{s_i(2)}$
and $B_{p,x'x}^{s_i(3)}$.

Let us start with computing $B_{p,x'x}^{s_i(2)}$. Since we
focus on the phase factors from gliding the string $s_{i}$, we represent the
honeycomb lattices by square lattices with proper splitting at vertices in
mind (see Fig. \ref{figure:lattice1}). It is sufficient to consider gliding
the string $s_{i}$ over the $j$-favor of string-nets, namely the
2-intersection between the $i$- and $j$-flavor of string-nets. Let $B_{ij}^{\left( 2\right) }$ be the phase associated with gliding $s_{i}$ to the boundary of $p$. We first compute%
\begin{widetext}
\begin{align*}
\left\langle \raisebox{-0.23in}{\includegraphics[height=0.7in]{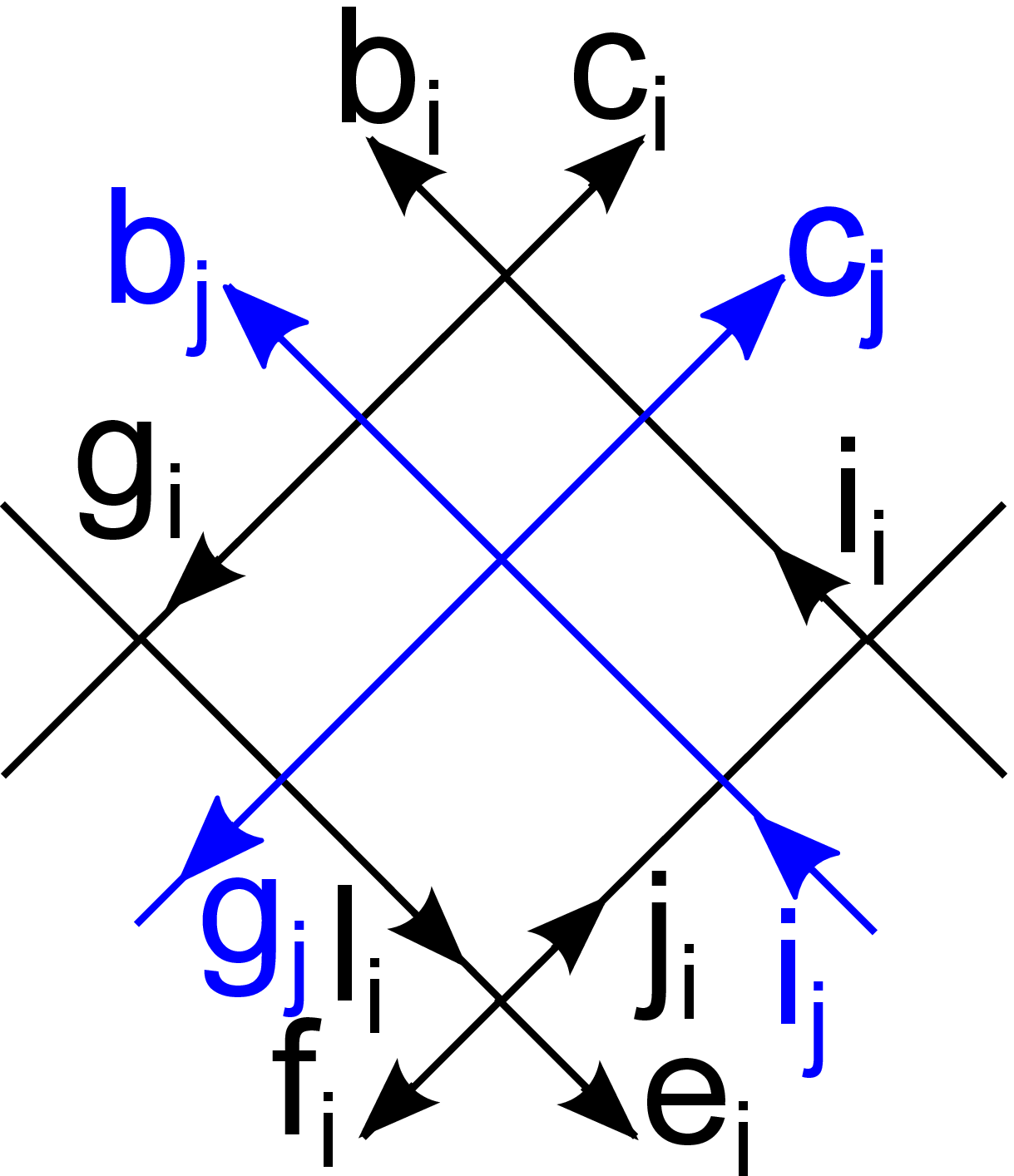}}%
\right\vert B_{p}^{s_{i}}& =\left\langle \raisebox{-0.23in}{%
\includegraphics[height=0.7in]{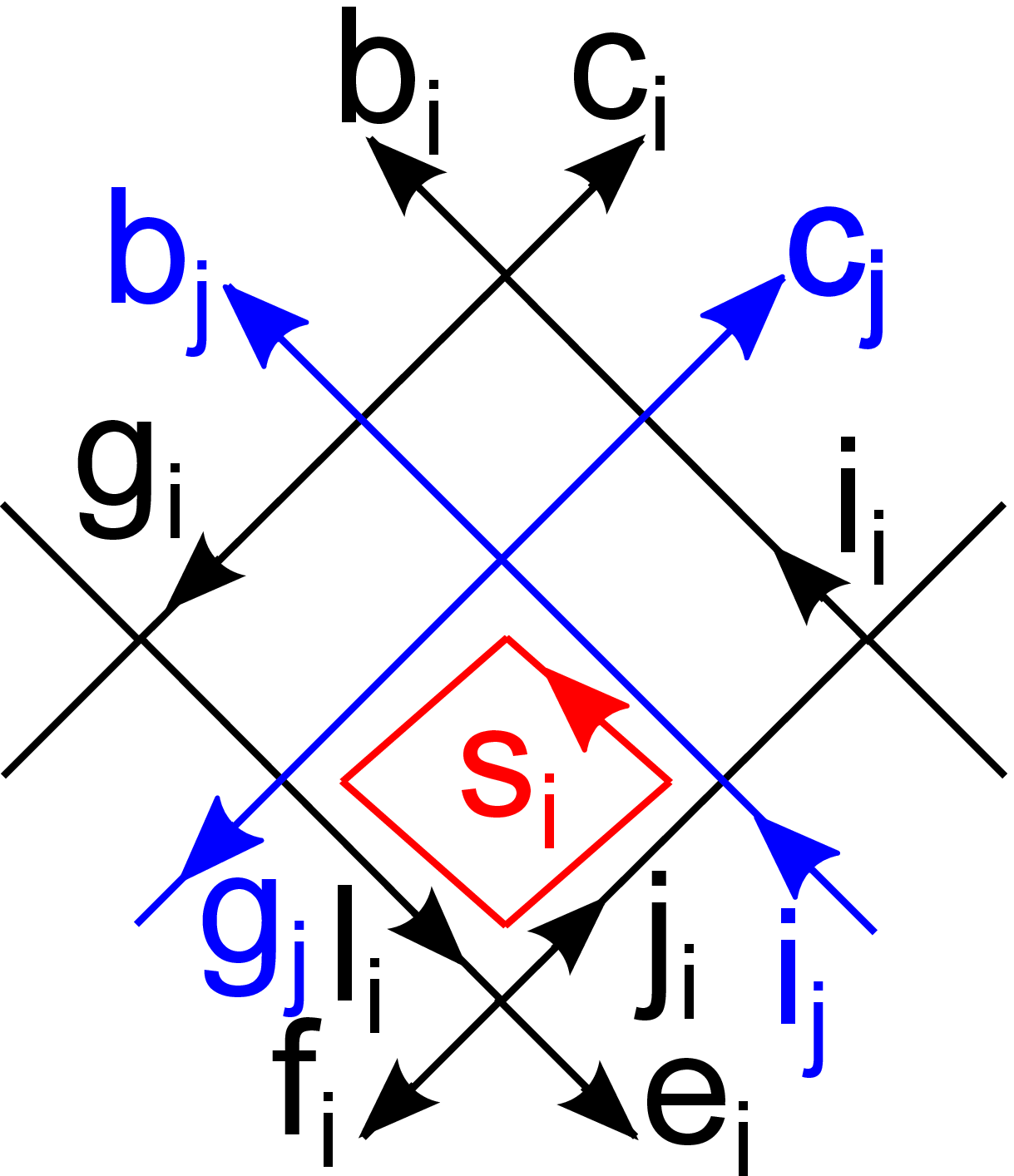}}\right\vert =\eta _{s_{i},g_{j}}\eta
_{s_{i},b_{j}}\eta _{s_{i},c_{j}}F_{s_{i},g_{j}b_{j}}^{\left( 2\right)
}F_{s_{i},\left( g_{j}+b_{j}\right) c_{j}}^{\left( 2\right) }\left\langle %
\raisebox{-0.23in}{\includegraphics[height=0.7in]{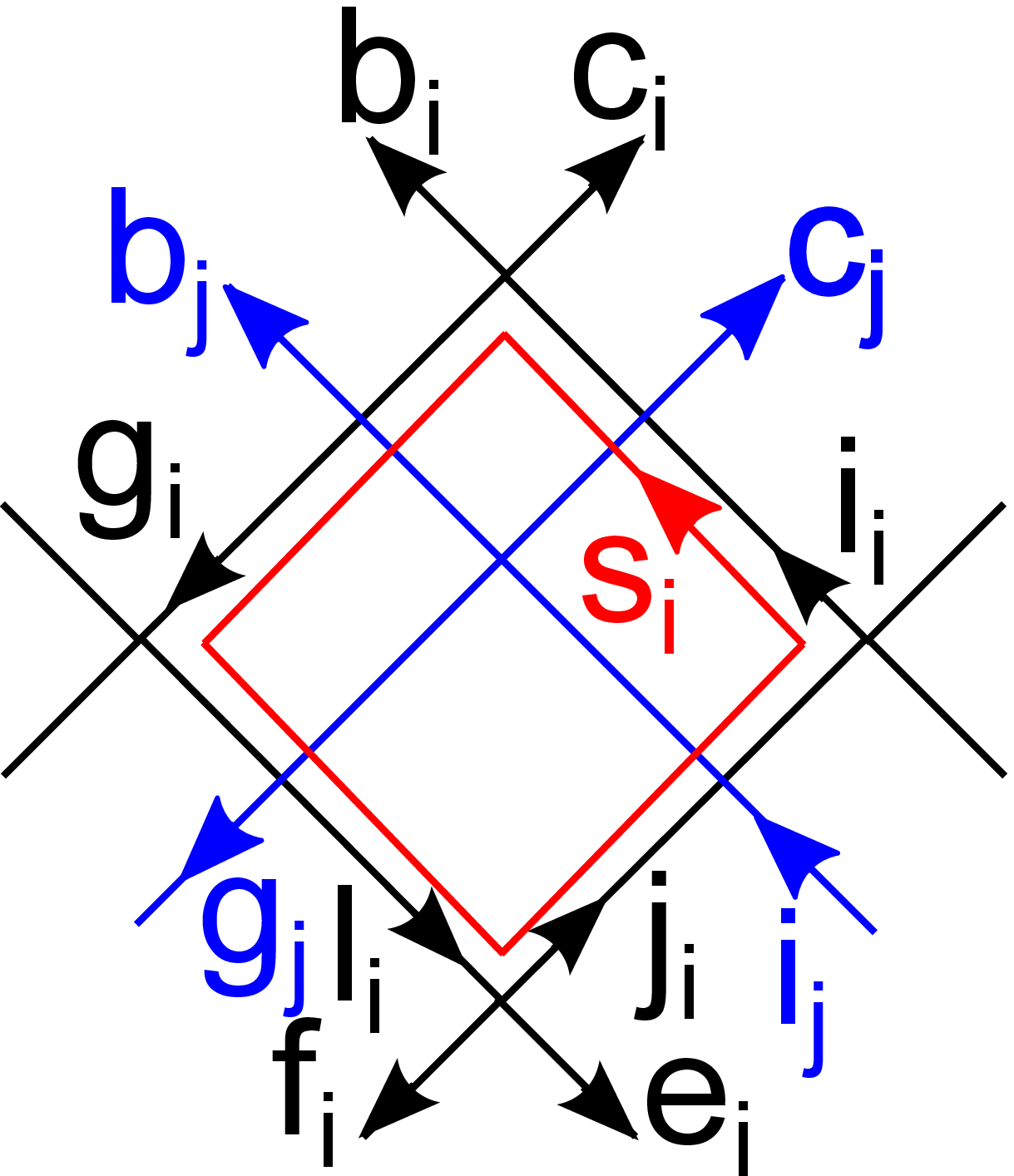}}\right\vert  \\
& \sim \eta _{s_{i},g_{j}}\eta _{s_{i},b_{j}}\eta
_{s_{i},c_{j}}F_{s_{i},g_{j}b_{j}}^{\left( 2\right) }F_{s_{i},\left(
g_{j}+b_{j}\right) c_{j}}^{\left( 2\right) }\left\langle \raisebox{-0.23in}{%
\includegraphics[height=0.7in]{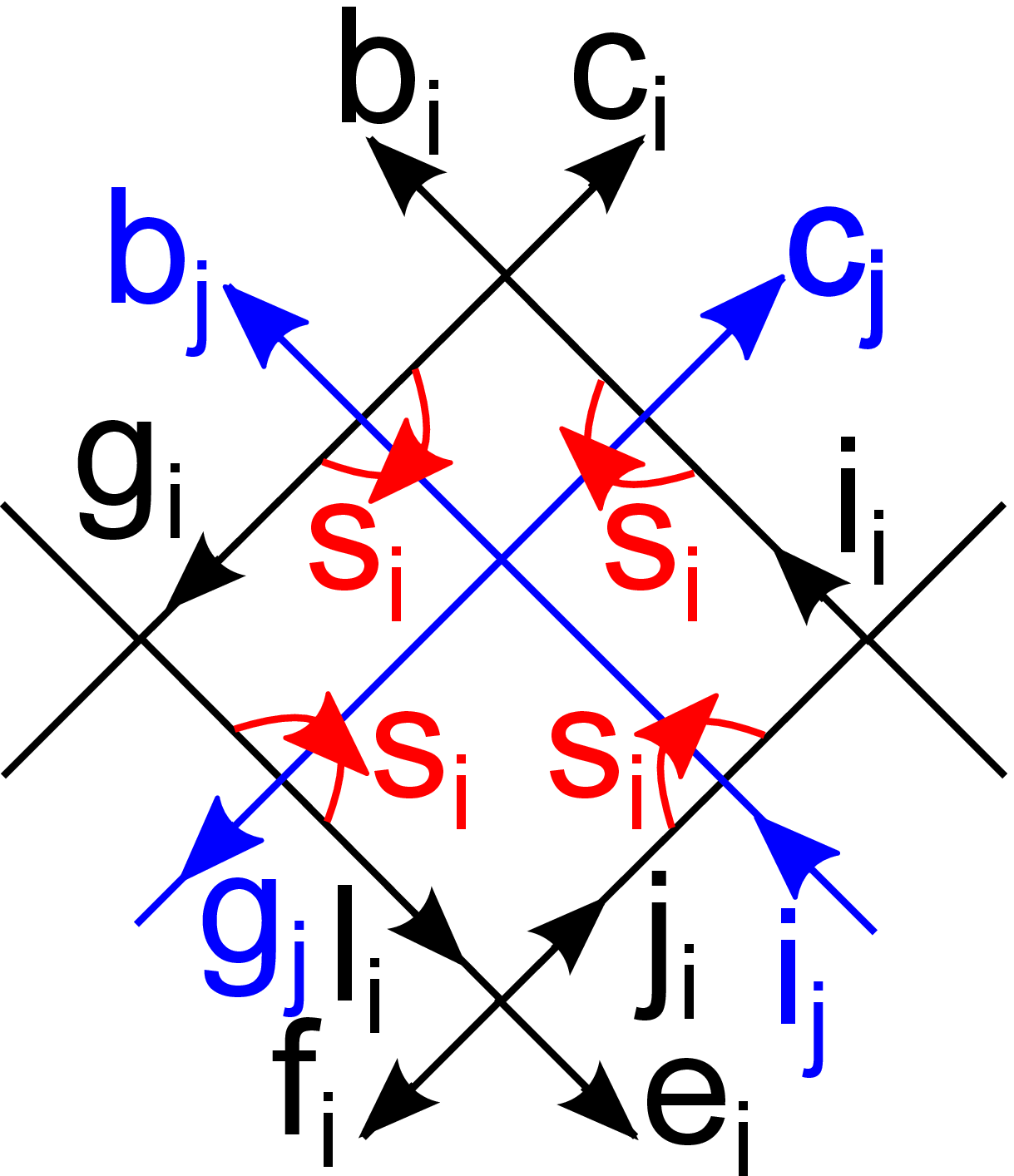}}\right\vert  \\
& \sim \eta _{s_{i},g_{j}}\eta _{s_{i},b_{j}}\eta
_{s_{i},c_{j}}F_{s_{i},g_{j}b_{j}}^{\left( 2\right) }F_{s_{i},\left(
g_{j}+b_{j}\right) c_{j}}^{\left( 2\right) }F_{g_{j},s_{i}l_{i}}^{\left(
2\right) }F_{b_{j},s_{i}g_{i}}^{\left( 2\right)
}F_{c_{j},s_{i}i_{i}}^{\left( 2\right) }\tilde{F}_{i_{j},s_{i}j_{i}}^{\left(
2\right) }\left\langle \raisebox{-0.23in}{%
\includegraphics[height=0.7in]{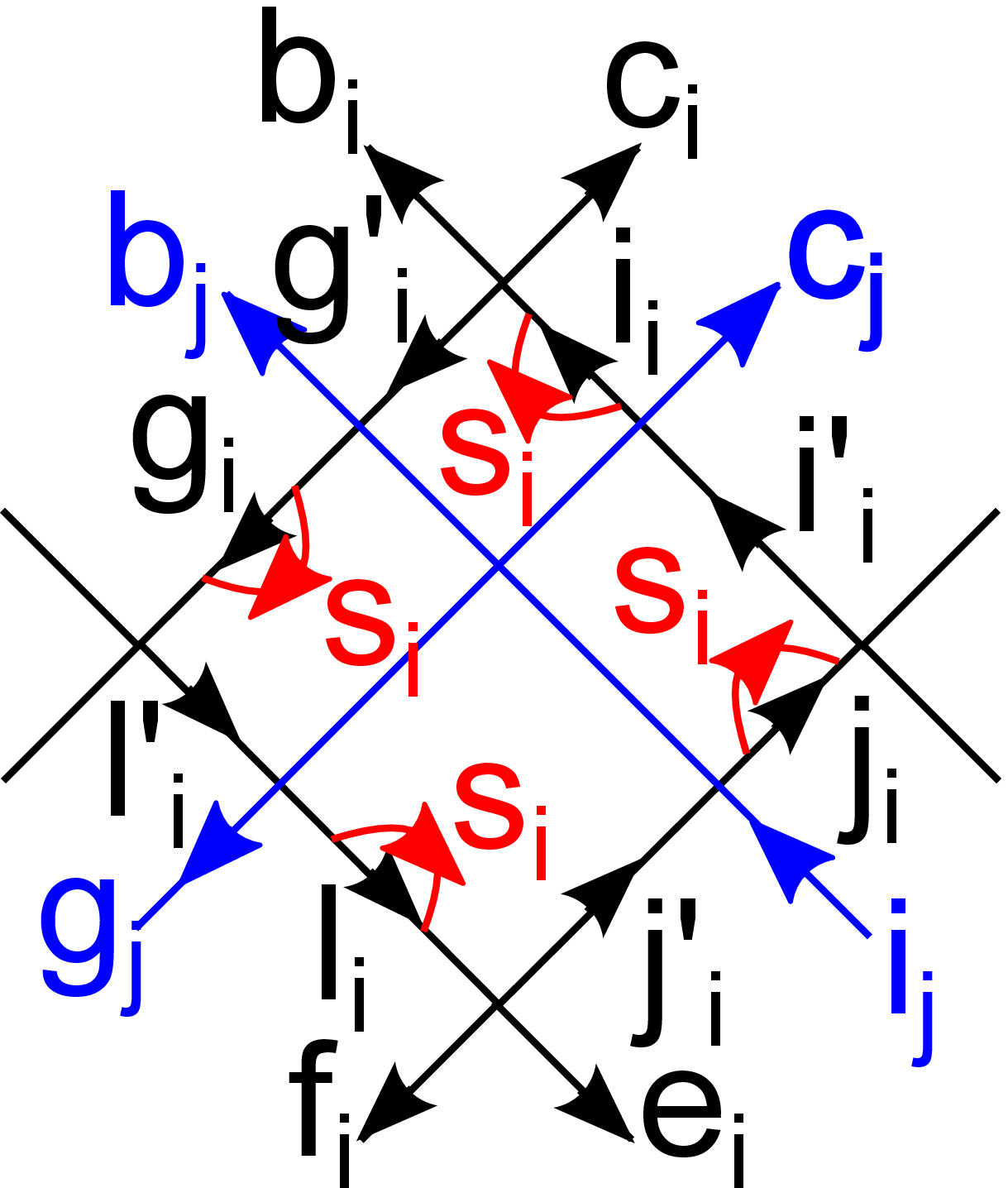}}\right\vert  \sim B_{ij}^{\left( 2\right) }\left\langle \raisebox{-0.23in}{%
\includegraphics[height=0.7in]{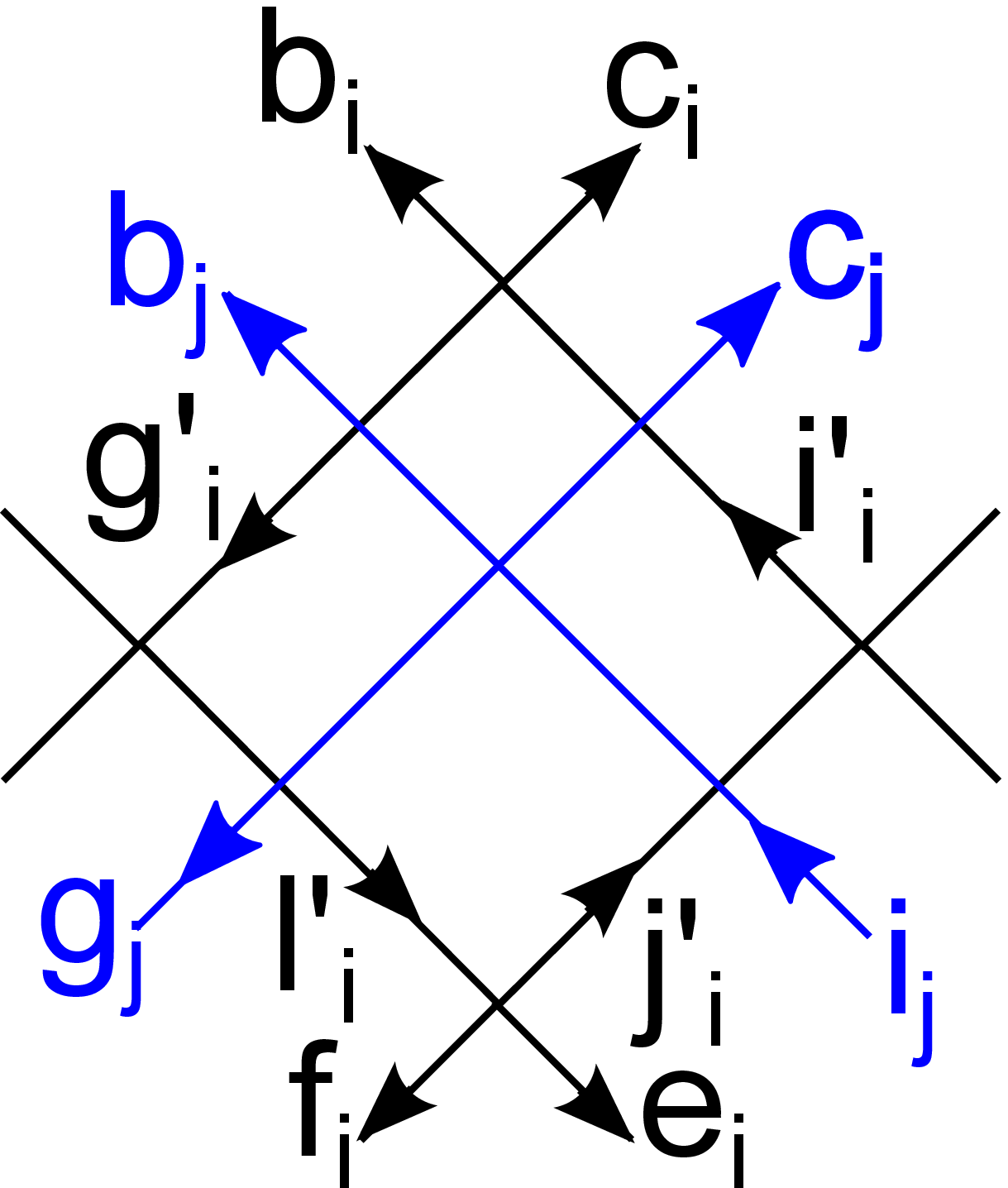}}\right\vert 
\end{align*}%
with%
\begin{align*}
B_{ij}^{\left( 2\right) }=\eta _{s_{i},g_{j}}\eta _{s_{i},b_{j}}\eta
_{s_{i},c_{j}}F_{s_{i},g_{j}b_{j}}^{\left( 2\right) }F_{s_{i},\left(
g_{j}+b_{j}\right) c_{j}}^{\left( 2\right) }F_{g_{j},s_{i}l_{i}}^{\left(
2\right) }F_{b_{j},s_{i}g_{i}}^{\left( 2\right)
}F_{c_{j},s_{i}i_{i}}^{\left( 2\right) }\tilde{F}_{i_{j},s_{i}j_{i}}^{\left(
2\right) }
=\eta_{s_i,i_j}\frac{F^{(2)}_{g_j,s_il_i}F^{(2)}_{b_j,s_ig_i}F^{(2)}_{c_j,s_ii_i}\bar{F}^{(2)}_{i_j,s_ij_i}}{\bar{F}^{(2)}_{s_i,g_jb_j}\bar{F}^{(2)}_{s_i,(g_j+b_j)c_j}}.
\end{align*}%
\end{widetext}
Here the symbol $\sim$ means equality up to phases in $B_{p,x'x}^{s_i(1)}$ associated with fusing $s_{i}$ to the boundary
links of $p$ and $x_i'=x_i+s_i$ in the last two states. 
The last equality follows from the identity (\ref{eq4}).
The $B_{p,x'x}^{s_i(2)}$ is then obtained by taking into
account all types of 2-intersections inside the plaquette $p$, namely, $B_{p,x'x}^{s_i(2)}=\prod_{j\neq i}B_{ij}^{\left( 2\right) }$ as in Eq. (\ref{b2}).

Similarly, to obtain $B_{p,x'x}^{s_i(3)}$, it suffices to consider
gliding the string $s_{i}$ over the $j$- and $k$-flavor of string-nets,
namely the 3-intersection among these three flavors of string-nets. Let $B_{ijk}^{\left( 3\right) }$ be the phase factor associated with gliding $s_{i}$ over
the intersections between $j$- and $k$-flavors of string-nets:%
\begin{gather*}
\left\langle \raisebox{-0.29in}{\includegraphics[height=0.7in]{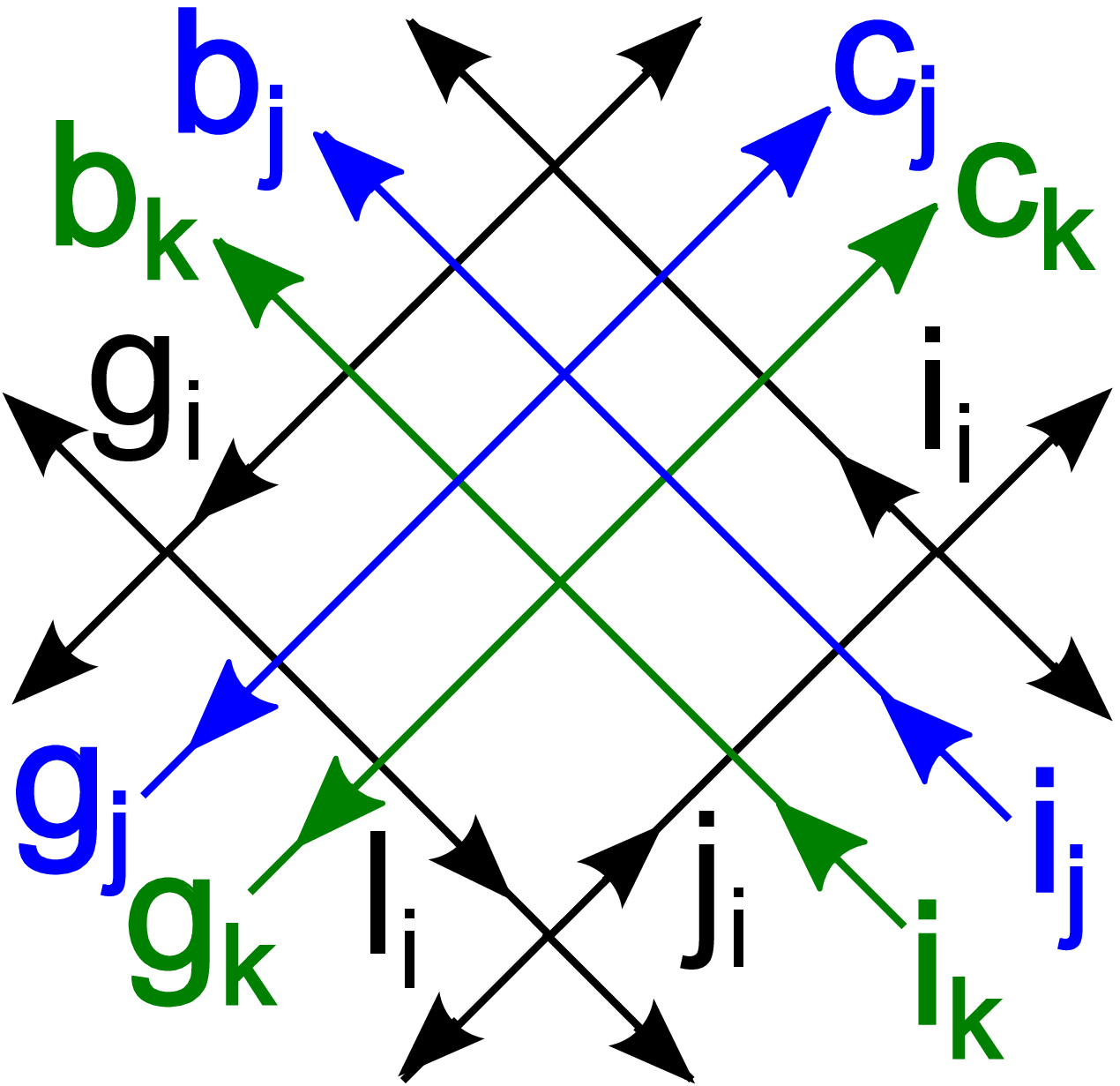}}%
\right\vert B_{p}^{s_{i}}=\left\langle \raisebox{-0.29in}{%
\includegraphics[height=0.7in]{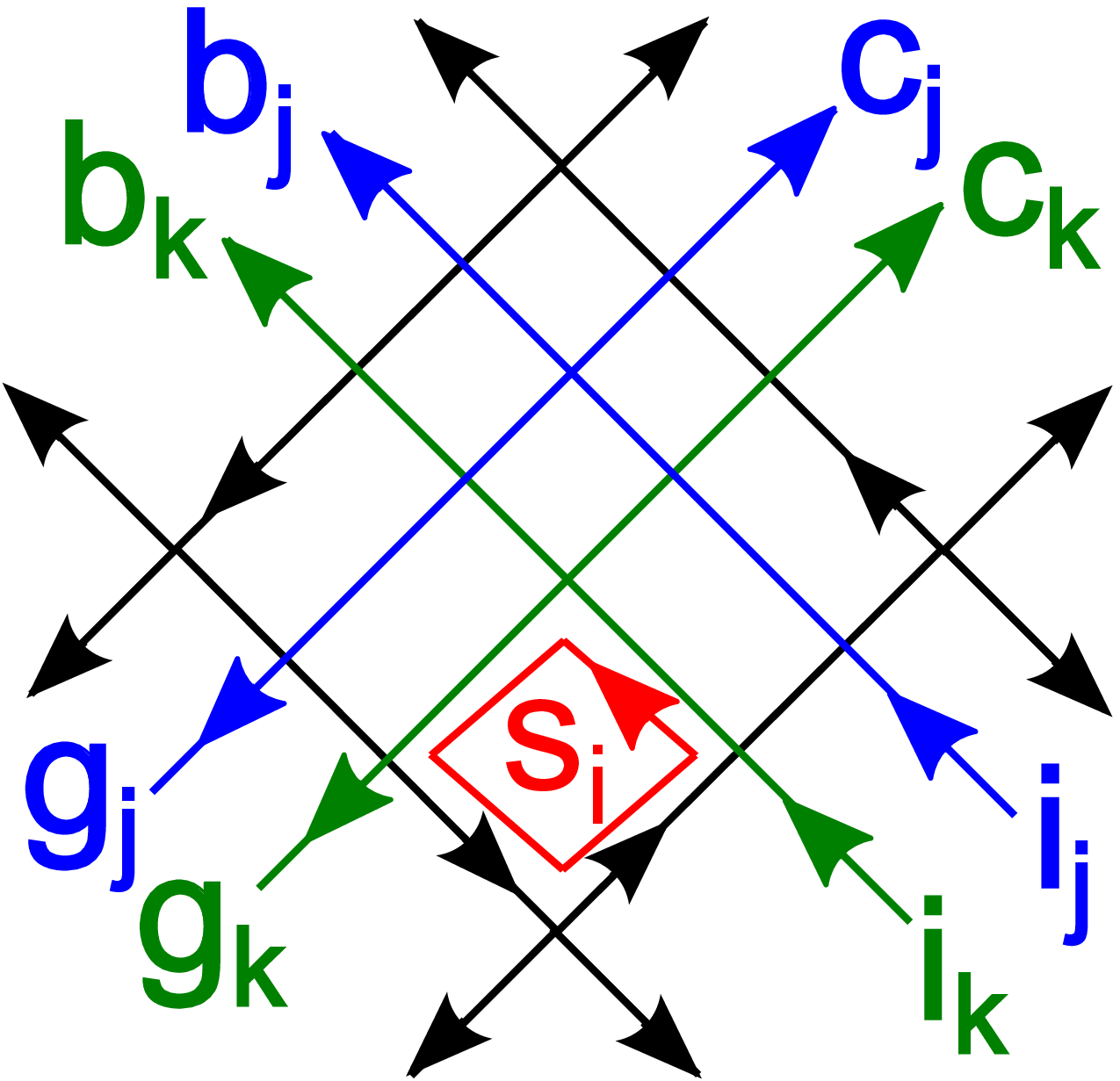}}\right\vert \sim \left\langle %
\raisebox{-0.29in}{\includegraphics[height=0.7in]{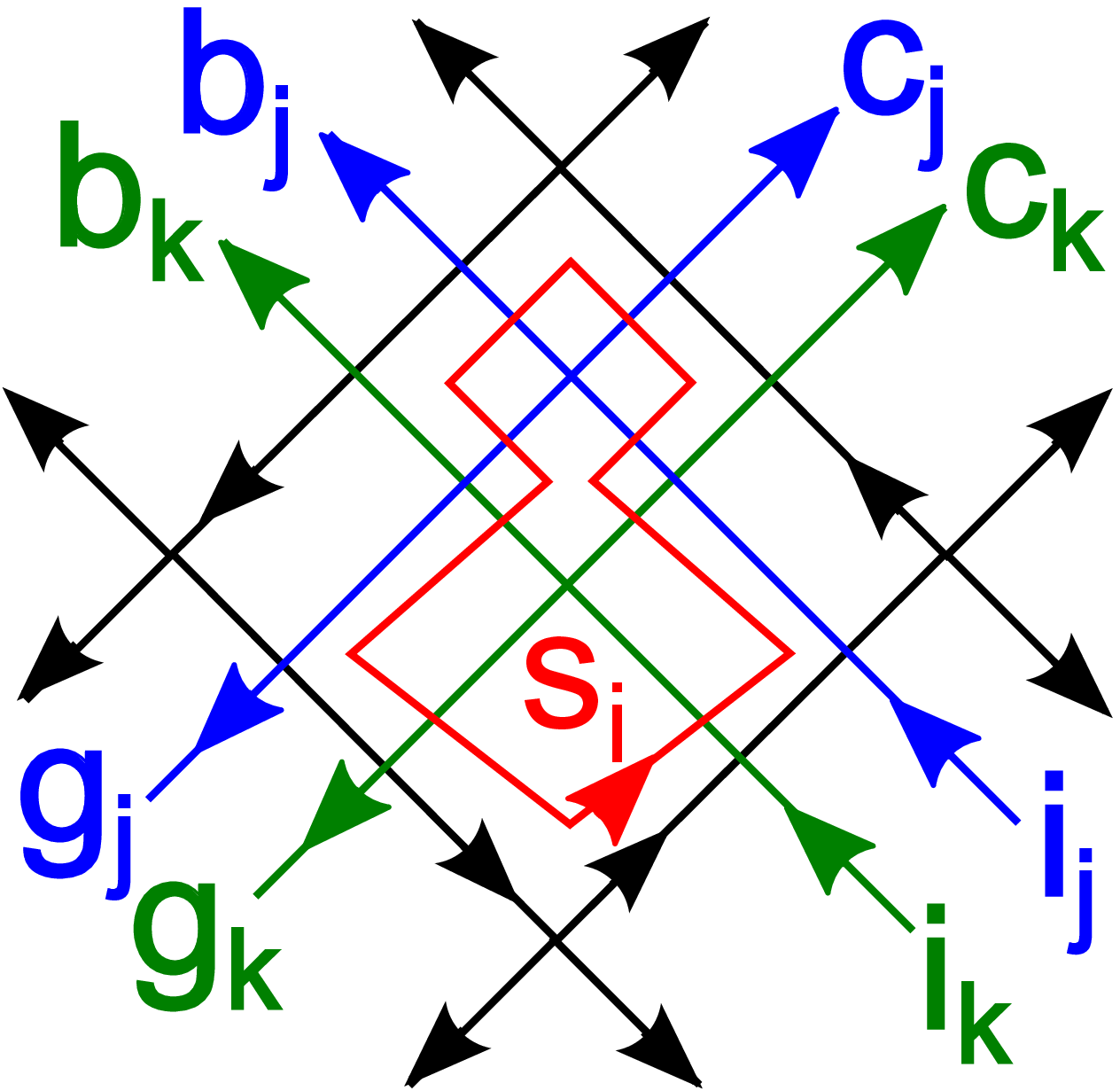}}\right\vert \\
\sim B_{ijk}^{(3)}
\left\langle \raisebox{-0.29in}{%
\includegraphics[height=0.7in]{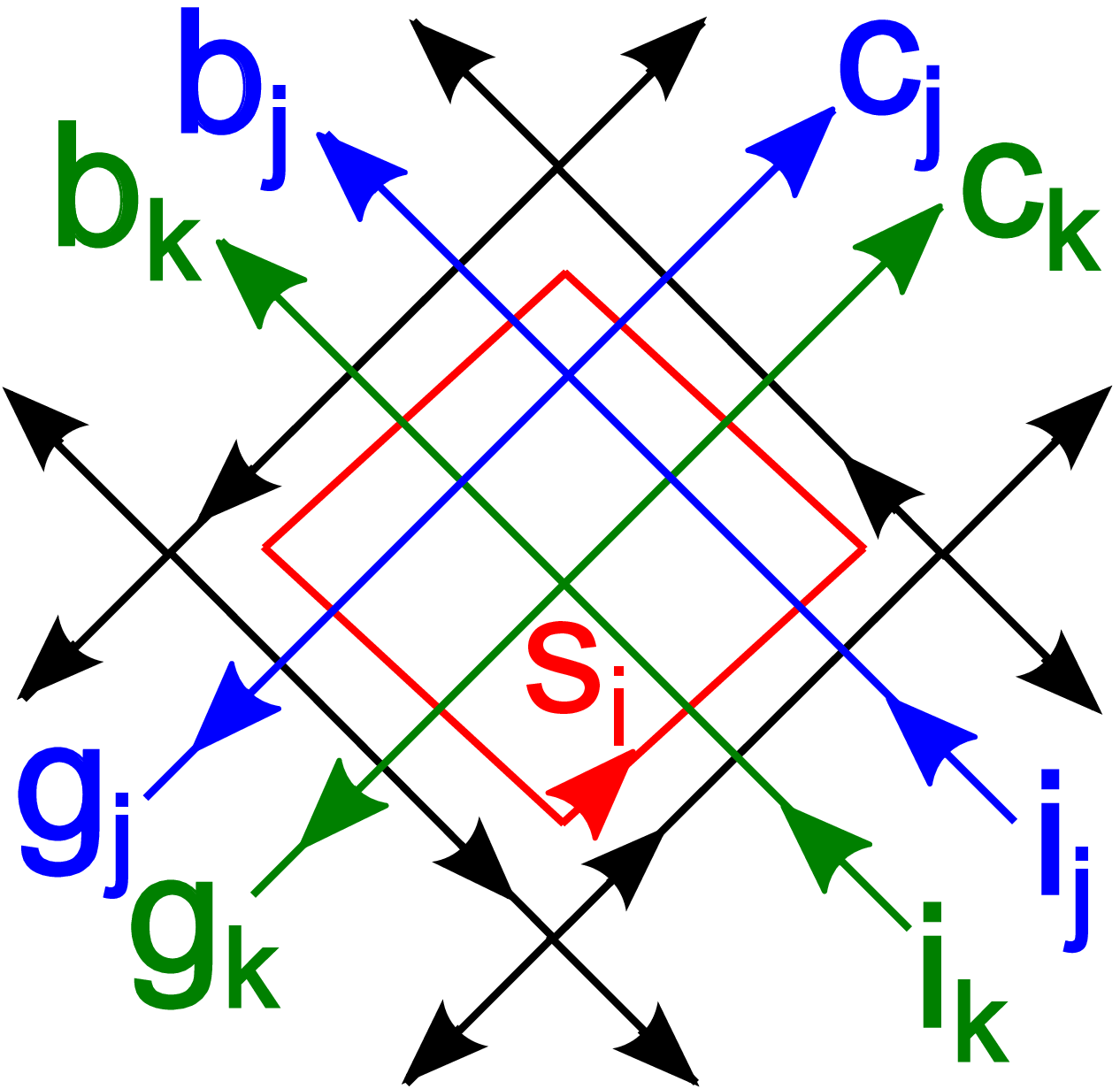}}\right\vert 
\sim B_{ijk}^{(3)}
\left\langle \raisebox{-0.29in}{%
\includegraphics[height=0.7in]{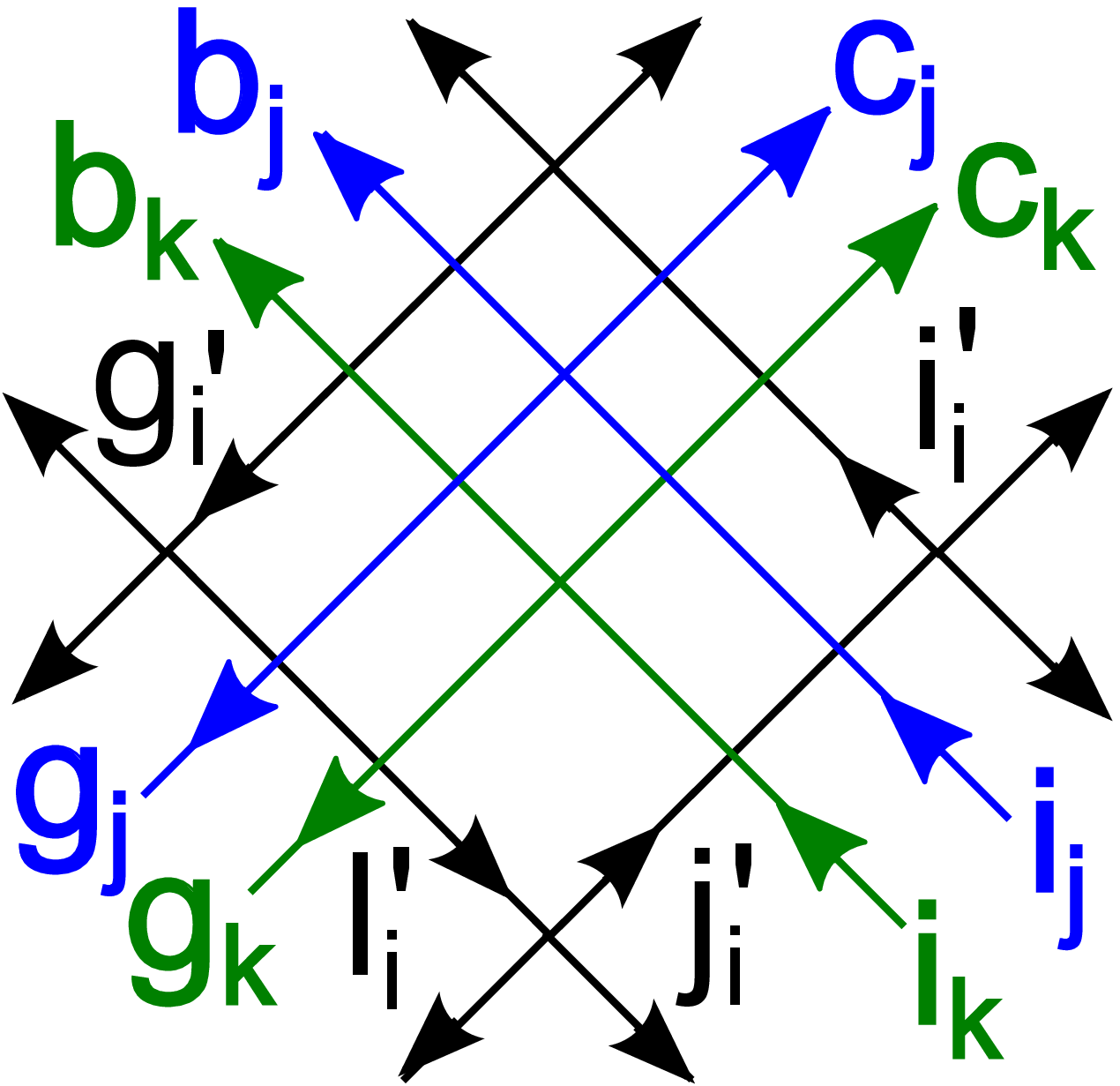}}\right\vert 
\end{gather*}%
with 
\begin{equation}
	B_{ijk}^{(3)}=F^{(3)}_{s_i}(c_k^*,i_j)F^{(3)-1}_{s_i^*}(g_j,b_k).
	\label{}
\end{equation}
Here the symbol $\sim$ indicates equality up to phases in $
B_{p,x'x}^{s_i(2)}$ and $B_{p,x'x}^{s_i(1)}$ associated with gliding $s_{i}$ over 2-intersections in $p$ and fusing $s_i$ to the boundary links of $p$. Therefore, $B_{p,x'x}^{s_i(3)}$ is obtained
by taking into account all kinds of 3-intersections inside $p$, namely, $B_{p,x'x}^{s_i(3)}=\prod_{j\neq k\neq i}B_{ijk}^{\left( 3\right) }$ as in Eq. (\ref%
{b3}). This completes our derivation of the Hamiltonian.

\section{Showing $B_{p_{1}}^{s_{i}}$ and $B_{p_{2}}^{s_{j}}$
commute \label{app:commute}}

\begin{figure}[tbp]
\begin{center}
\includegraphics[width=0.25\columnwidth]{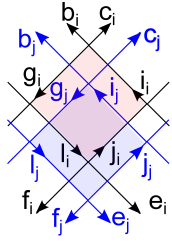}
\end{center}
\caption{Two plaquette operators $B_{p}^{s_{i}}$ and $B_{p}^{s_{j}}$
act on two overlapped (red and blue shaded) plaquettes.}
\label{figure:bpcommute}
\end{figure}

In this section, we will show that the operators $B_{p_{1}}^{s_{i}}$ and $%
B_{p_{2}}^{s_{j}}$ commute with one another. We only have to consider three cases. The first two cases are when $i,j$ are of the same flavor and $p_1=p_2=p$ or $p_1,p_2$ are adjacent. The third case is when $i,j$ are of different flavors and $p_1=p_2=p$. In last case, the two plaquettes at $p$ on different lattices are overlapped as shown in Fig. \ref{figure:bpcommute}. When $p_1,p_2$ are further apart, the two operators will commute.

The first case is when $i=j$ and $p_1=p_2=p$. To prove $[B_p^{s_{i}},B_p^{t_{i}}]=0$, it is sufficient to show that 
\begin{equation}
B_{p,x'y}^{s_{i}(m)}B_{p,yx}^{t_{i}(m)}
	=B_{p,x'z}^{t_{i}(m)}B_{p,zx}^{s_{i}(m)}
\label{bbm}
\end{equation}
holds for each component $m=1,2,3$. Here $x,x'$ denote a set of spin labels for initial and final state configurations and $y,z$ denote a set of spin labels for the intermediate state configurations after acting $B_p^{t_i}$ and $B_p^{s_i}$ to the initial state, respectively. 
For the $m=1$ component, the proof is identical to the one given in Ref. \onlinecite{LinLevinstrnet}. They showed that
\begin{equation}
	B_{p,x'y}^{s_{i}(1)}B_{p,yx}^{t_{i}(1)}
	=B_{p,x'z}^{t_{i}(1)}B_{p,zx}^{s_{i}(1)}=B_{p,x'x}^{s_{i}+t_{i}(1)}
	\label{}
\end{equation}

Similarly, for the $m=2$ component, one can show that 
\begin{equation}
		B_{p,x'y}^{s_{i}(2)}B_{p,yx}^{t_{i}(2)}
	=B_{p,x'z}^{t_{i}(2)}B_{p,zx}^{s_{i}(2)}=B_{p,x'x}^{s_{i}+t_{i}(2)}.
	\label{bb2}
\end{equation}
To see this, one can write down the expression of $B_{p,x'y}^{s_{i}(2)}B_{p,yx}^{t_{i}(2)}$:
\begin{align*}
	B_{p,x'y}^{s_{i}(2)}B_{p,yx}^{t_{i}(2)}=&\eta_{t_i,i_j}\frac{F^{(2)}_{g_j,t_il_i}F^{(2)}_{b_j,t_ig_i}F^{(2)}_{c_j,t_ii_i}\bar{F}^{(2)}_{i_j,t_ij_i}}{\bar{F}^{(2)}_{t_i,g_jb_j}\bar{F}^{(2)}_{t_i,(g_j+b_j)c_j}} \times \\
	& \eta_{s_i,i_j}\frac{F^{(2)}_{g_j,t_il_i'}F^{(2)}_{b_j,t_ig_i'}F^{(2)}_{c_j,t_ii_i'}\bar{F}^{(2)}_{i_j,t_ij_i'}}{\bar{F}^{(2)}_{s_i,g_jb_j}\bar{F}^{(2)}_{s_i,(g_j+b_j)c_j}}
	\label{}
\end{align*}
where $a_i'=a_i+t_i$ with $a_i\in\{l_i,g_i,i_i,j_i\}$. By using (\ref{sfeq1}) $F^{(2)}_{a_j,t_ia_i}F^{(2)}_{a_j,s_ia_i'}=F^{(2)}_{a_j,(s_i+t_i)a_i}F^{(2)}_{a_j,s_it_i}$ and (\ref{fbar}) $\bar{F}^{(2)}_{t_i,a_jb_j}\bar{F}^{(2)}_{s_i,a_jb_j}=\bar{F}^{(2)}_{(s_i+t_i),a_jb_j}F^{(2)}_{a_j,s_it_i}F^{(2)}_{b_j,s_it_i}/F^{(2)}_{(a_j+b_j),s_it_i}$ and (\ref{eq4}) $\bar{F}^{(2)}_{a_j,s_it_i}F^{(2)}_{a_j,s_it_i}=\eta_{i_j,{s_i+t_i}}/\eta_{i_j,s_i}\eta_{i_j,t_i}$, the above expression can be rewritten as
\begin{align*}
	B_{p,x'y}^{s_{i}(2)}B_{p,yx}^{t_{i}(2)}&=\eta_{u_i,i_j}\frac{F^{(2)}_{g_j,u_il_i}F^{(2)}_{b_j,u_ig_i}F^{(2)}_{c_j,u_ii_i}\bar{F}^{(2)}_{i_j,u_ij_i}}{\bar{F}^{(2)}_{u_i,g_jb_j}\bar{F}^{(2)}_{u_i,(g_j+b_j)c_j}}=B_{p,x'x}^{u_{i}(2)}
	\label{}
\end{align*}
with $u_i=s_i+t_i$. This shows (\ref{bb2}).

Finally, it is easy to see from (\ref{b3}) and (\ref{sfeq6}) that (\ref{bbm}) also holds for $m=3$. Thus, putting everything together, we conclude that
\begin{equation}
	B_p^{s_i}B_p^{t_i}=B_p^{t_i}B_p^{s_i}=B_p^{s_i+t_i}
	\label{bbbb}
\end{equation}
showing the commutativity when $i=j$ and $p_1=p_2=p$.

The second case is when $i=j$ and $p_1,p_2$ are adjacent plaquettes. We want to show $[B_{p_1}^{s_{i}},B_{p_2}^{t_{i}}]=0$. To prove this, it is sufficient to show that 
\begin{equation}
B_{p_1,x'y}^{s_{i}(m)}B_{p_2,yx}^{t_{i}(m)}
	=B_{p_2,x'z}^{t_{i}(m)}B_{p_1,zx}^{s_{i}(m)}
\label{bbm2}
\end{equation}
holds for $m=1,2,3$. The proof for $m=1$ is identical to the one in the appendix E of Ref. \onlinecite{LinLevinstrnet}. It is also easy to see that (\ref{bbm2}) holds for $m=2,3$ since the intersections between different flavors of strings are away from the link shared by two plaquettes $p_1,p_2$. This establishes $[B_{p_1}^{s_{i}},B_{p_2}^{t_{i}}]=0$.

Finally, we consider the case where $i\neq j$ and $p_1=p_2=p$. We want to show $[B_p^{s_{i}},B_p^{s_{j}}]=0$ with $B_{p}^{s_{i}},B_{p}^{s_{j}}$ acting on two overlapping plaquettes shown in Fig. \ref{figure:bpcommute}.
To prove $[B_p^{s_{i}},B_p^{s_{j}}]=0$, it is sufficient to show that 
\begin{equation}
B_{p,x'y}^{s_{i}(m)}B_{p,yx}^{s_{j}(m)}
	=B_{p,x'z}^{s_{j}(m)}B_{p,zx}^{s_{i}(m)}
\label{bbm3}
\end{equation}
for $m=1,2,3$. Again, $x,x'$ denote a set of spin labels for the initial and final state configurations while $y,z$ denote a set of spin labels for the intermediate state configurations after acting $B_p^{s_j}$ and $B_p^{s_i}$ to the initial state. In this case, it is easy to see that (\ref{bbm3}) holds for $m=1,3$ by writing down the phase factors on each side using (\ref{b1},\ref{b3}). Thus all that remains is to show that (\ref{bbm3}) holds for $m=2$. 


We can write down each side of (\ref{bbm3}) for $m=2$ by (\ref{b2}):
\begin{align}
LHS&=\bar{F}_{i_{j},s_{i}j_{i}}^{\left( 2\right) }F_{g_{j},s_{i}l_{i}}^{\left(
2\right) }F_{s_{i},g_{j}b_{j}}^{\left( 2\right) }F_{s_{i},\left(
g_{j}+b_{j}\right) c_{j}}^{\left( 2\right) }\eta_{s_{i},g_{j}}\times  \notag
\\
 &F_{j_{i}^{\prime},s_{j}i_{j}}^{\left( 2\right) }\bar{F}_{l_{i}^{\prime
},s_{j}g_{j}}^{\left( 2\right) }F_{s_{j},j_{i}^{\prime}e_{i}}^{\left(
2\right) }F_{s_{j},\left( j_{i}^{\prime}+e_{i}\right) f_{i}}^{\left(
2\right) }\eta_{s_{j},j_{i}^{\prime}}  \label{e1} \\
RHS&= F_{j_{i},s_{j}i_{j}}^{\left( 2\right) }\bar{F}_{l_{i},s_{j}g_{j}}^{\left(
2\right) }F_{s_{j},j_{i}e_{i}}^{\left( 2\right) }F_{s_{j},\left(
j_{i}+e_{i}\right) f_{i}}^{\left( 2\right) }\eta_{s_{j},j_{i}}\times  \notag
\\
& \bar{F}_{i_{j}^{\prime},s_{i}j_{i}}^{\left( 2\right) }F_{g_{j}^{\prime
},s_{i}l_{i}}^{\left( 2\right) }F_{s_{i},g_{j}^{\prime}b_{j}}^{\left(
2\right) }F_{s_{i},\left( g_{j}^{\prime}+b_{j}\right) c_{j}}^{\left(
2\right) }\eta_{s_{i},g_{j}^{\prime}}.  \label{e2}
\end{align}
where $a_i'=a_i+s_i$ with $a_i\in \{j_i,l_i\}$ and $a_j'=a_j+s_j$ with $a_j\in\{i_j,g_j\}$. 
Here we only keep the relevant factors which involve $i_{j},g_{j},l_{i},j_{i}$ strings. 
To show $\left( \ref{e1}\right) =\left( \ref{e2}\right) ,$ we first use $%
\left( \ref{3f}\right) $ to simplify both sides so that all phase factors
have subindices $s_{i}$ or $s_{j}.$ Then to show the equality is equivalent
to show $C_{L}=C_{R}$ with%
\begin{equation*}
C_{L}=\frac{F_{s_{i},g_{j}b_{j}}^{\left( 2\right) }F_{s_{i},\left(
g_{j}+b_{j}\right) c_{j}}^{\left( 2\right) }F_{s_{i},s_{j}i_{j}}^{\left(
2\right) }\bar{F}_{s_{i},s_{j}g_{j}}^{\left( 2\right) }\eta_{s_{i},g_{j}}}{%
F_{s_{i},g_{j}^{\prime}b_{j}}^{\left( 2\right) }F_{s_{i},\left(
g_{j}^{\prime}+b_{j}\right) c_{j}}^{\left( 2\right)
}\eta_{s_{i},g_{j}^{\prime}}}
\end{equation*}
and%
\begin{equation*}
C_{R}=\frac{F_{s_{j},j_{i}e_{i}}^{\left( 2\right) }F_{s_{j},\left(
j_{i}+e_{i}\right) f_{i}}^{\left( 2\right) }F_{s_{j},s_{i}l_{i}}^{\left(
2\right) }\bar{F}_{s_{j,}s_{i}j_{i}}^{\left( 2\right) }\eta_{s_{j},j_{i}}}{%
F_{s_{j},j_{i}^{\prime}e_{i}}^{\left( 2\right) }F_{s_{j},\left(
j_{i}^{\prime}+e_{i}\right) f_{i}}^{\left( 2\right)
}\eta_{s_{j},j_{i}^{\prime}}}.
\end{equation*}
To proceed, we use $\left( \ref{sfeq1}\right) $ and the facts $%
j_{i}+e_{i}+f_{i}=l_{i}$ and $g_{j}+b_{j}+c_{j}=i_{j}$ to obtain%
\begin{align*}
C_{L} & =F_{s_{i},s_{j}g_{j}}^{\left( 2\right) }\bar{F}_{s_{i},s_{j}g_{j}}^{%
\left( 2\right) }\frac{\eta_{s_{i},g_{j}}}{\eta_{s_{i},g_{j}^{\prime}}}, \\
C_{R} & =F_{s_{j},s_{i}j_{i}}^{\left( 2\right) }\bar{F}_{s_{j},s_{i}j_{i}}^{%
\left( 2\right) }\frac{\eta_{s_{j},j_{i}}}{\eta_{s_{j},j_{i}^{\prime}}}.
\end{align*}
Furthermore, we use (\ref{eq4}) to rewrite%
\begin{equation}
	C_L=\frac{1}{\eta_{s_i,s_j}}, \quad C_R=\frac{1}{\eta_{s_j,s_i}}.
	\label{}
\end{equation}
Finally, by (\ref{sfeq7}), we show $C_{L}=C_{R}$. This
completes the proof that $B_{p}^{s_i},B_{p}^{s_j}$ commute with one another.

\section{Properties of the Hamiltonian (\ref{ham}) \label{app:ham}}
In this section, we establish the following properties of the Hamiltonian (\ref{ham}):
\begin{enumerate}
	\item $(B_{p}^{s_i})^{\dagger}=B_{p}^{s_i}$.
	\item $B_{p}^i$ is a projection operator, namely, $B_{p}^{i2}=B_{p}^i$.
	\item The ground state wave function on the honeycomb lattice satisfies the local rules (\ref{rule1a}--\ref{rule1b},\ref{rule2a}--\ref{rule2d}).
\end{enumerate}
Let us show them in order. To show the first equality, we need to show the matrix elements on both sides are equal, namely $(B_{p,xx'}^{s_i})^*=B_{p,x'x}^{s^*_i}$ with $x,x'$ being the spin labels for the initial and final states after the action of $B_{p}^{s_i}$ on the initial state. 
By (\ref{norm1}), we can rewrite the equality as $B_{p,xx'}^{s^*_i-1}=B_{p,x'x}^{s_i}$. Since $B_{p,xx'}^{s_i}$ can be written as a product of three components $B_{p,xx'}^{s_i(1)}B_{p,xx'}^{s_i(2)}B_{p,xx'}^{s_i(3)}$ (see Eq. (\ref{bps}), thus it is sufficient to show the equality to hold for each component. 

Ref. \onlinecite{LinLevinstrnet} showed that $B_{p,xx'}^{s_i^*(1)-1}=B_{p,x'x}^{s_i(1)}$ (see Appendix F therein). It is also easy to see $B_{p,xx'}^{s_i^*(3)-1}=B_{p,x'x}^{s_i(3)}$ by using $F^{(3)}_{a_i,a_ja_k}=F^{(3)-1}_{a^*_i,a_ja_k}$ (Eq. (\ref{sfeq6}) with $b_i=a_i^*$) to simplify $B_{p,xx'}^{s_i^*(3)-1}$. All that remains is to show that $B_{p,xx'}^{s_i^*(2)-1}=B_{p,x'x}^{s_i(2)}$. To this end, we write down $B_{p,xx'}^{s_i^*(2)-1}$ explicitly
\begin{equation}
	B_{p,xx'}^{s_i^*(2)-1}=\eta_{s^*_i,i_j}\frac{F^{(2)}_{g_j,s_i^*l_i'}F^{(2)}_{b_j,s_i^*g_i'}F^{(2)}_{c_j,s_i^*i_i'}\bar{F}^{(2)}_{i_j,s_i^*j_i'}}{\bar{F}^{(2)}_{s_i^*,g_jb_j}\bar{F}^{(2)}_{s_i^*,(g_j+b_j)c_j}}
	\label{1/b}
\end{equation}
where $a_i'=a_i+s_i$ with $a_i\in\{g_i,l_i,j_i\}$.
To proceed, we use the identity $F^{(2)}_{a_j,s_i^*a_i'}=F^{(2)}_{a_j,s_i^*s_i}/F^{(2)}_{a_j,s_ia_i}$ (which can be obtained from (\ref{sfeq1})) and a similar identity for $\bar{F}^{(2)}$ to reexpress $F^{(2)},\bar{F}^{(2)}$ in the numerator of (\ref{1/b}) while we use (\ref{fbar}) to rewrite $\bar{F}^{(2)}$ in the denominator. Next, we use (\ref{sfeq4},\ref{eq3}) to express $F^{(2)}_{a_j,s_i^*s_i},\bar{F}^{(2)}_{a_j,s_i^*s_i}$ in terms of $\eta,\kappa$. Finally, we use (\ref{eq2}) and (\ref{eq4}) to simplify the expression to $B_{p,x'x}^{s_i(2)}$. This establishes the first property. 

To prove the second result, we use the identity (\ref{bbbb}) to derive
\begin{equation}
	B_p^{i2}=\sum_{s_i,t_i}\frac{d_{s_i}d_{t_i}}{|G|^2}B_p^{s_i}B_p^{t_i}
	=\sum_{s_i,t_i}\frac{d_{s_i}d_{t_i}}{|G|^2}B_p^{s_i+t_i}.
	\label{}
\end{equation}
We then use (\ref{sfeqa}) to write $d_{s_i}d_{t_i}=d_{s_i+t_i}$. After changing variables to $u_i=s_i+t_i$, we derive
\begin{equation}
	B_p^{i2}=\sum_{u_i}\frac{d_{u_i}}{|G|}B_p^{u_i}=B_p^i.
	\label{}
\end{equation}

Finally, we show that the ground state $\Phi_{latt}$ of $H$ obeys the local rules (\ref{rule1a}--\ref{rule1c}) and (\ref{rule2a}--\ref{rule2d}). The proof that $\Phi_{latt}$ obeys (\ref{rule1a}--\ref{rule1c}) is identical to the one in appendix F of Ref. \onlinecite{LinLevinstrnet}. Here we only show that $\Phi_{latt}$ obeys (\ref{rule2a}--\ref{rule2d}). To see this, we use the fact that $B_p^i|\Psi_{latt}\>=|\Phi_{latt}\>$ together with the following relations:
\begin{align}
	\left\langle \raisebox{-0.25in}{\includegraphics[height=0.65in]{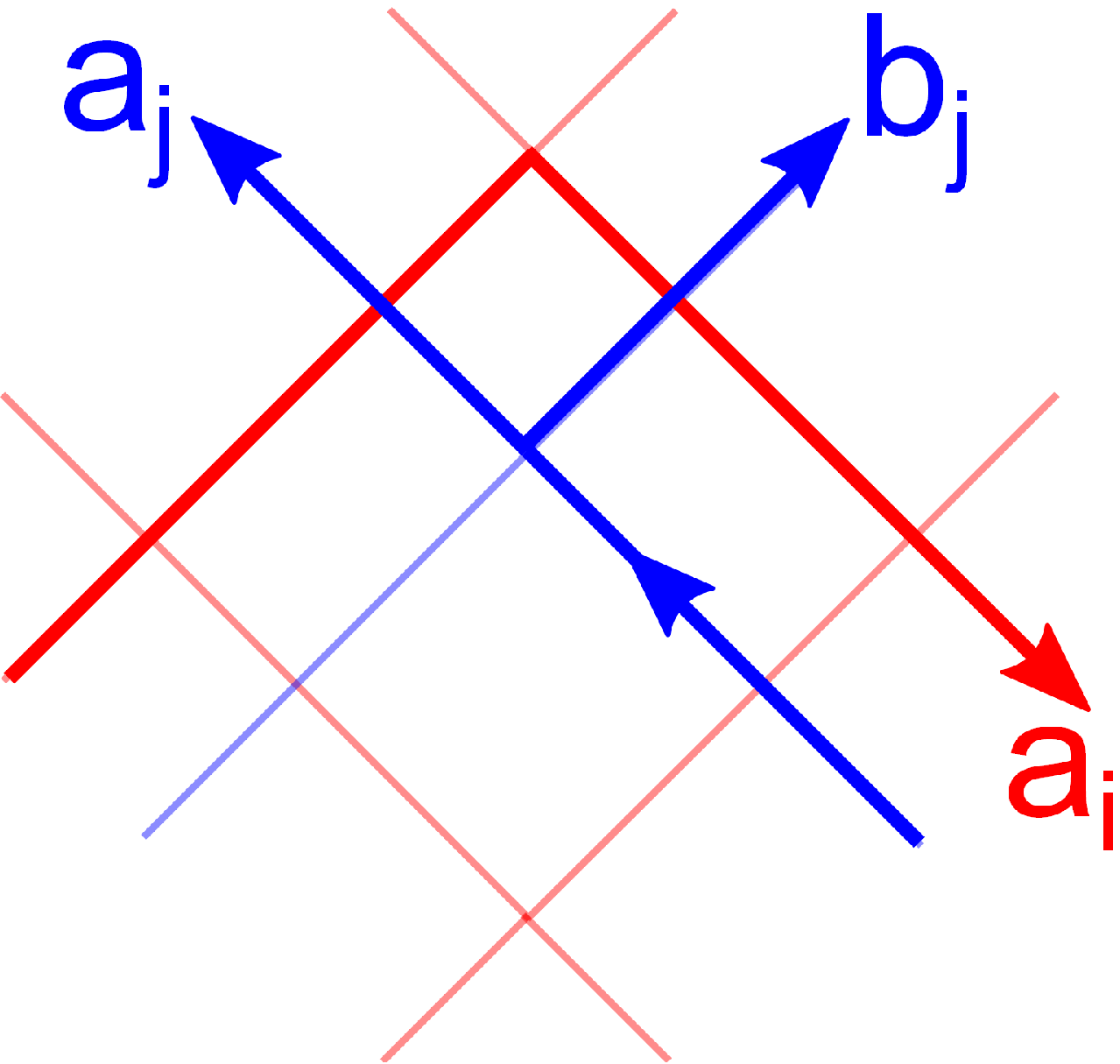}}\right\vert B_p^i
	&=F^{(2)}_{a_i}(a_j,b_j) \left\langle \raisebox{-0.25in}{\includegraphics[height=0.65in]{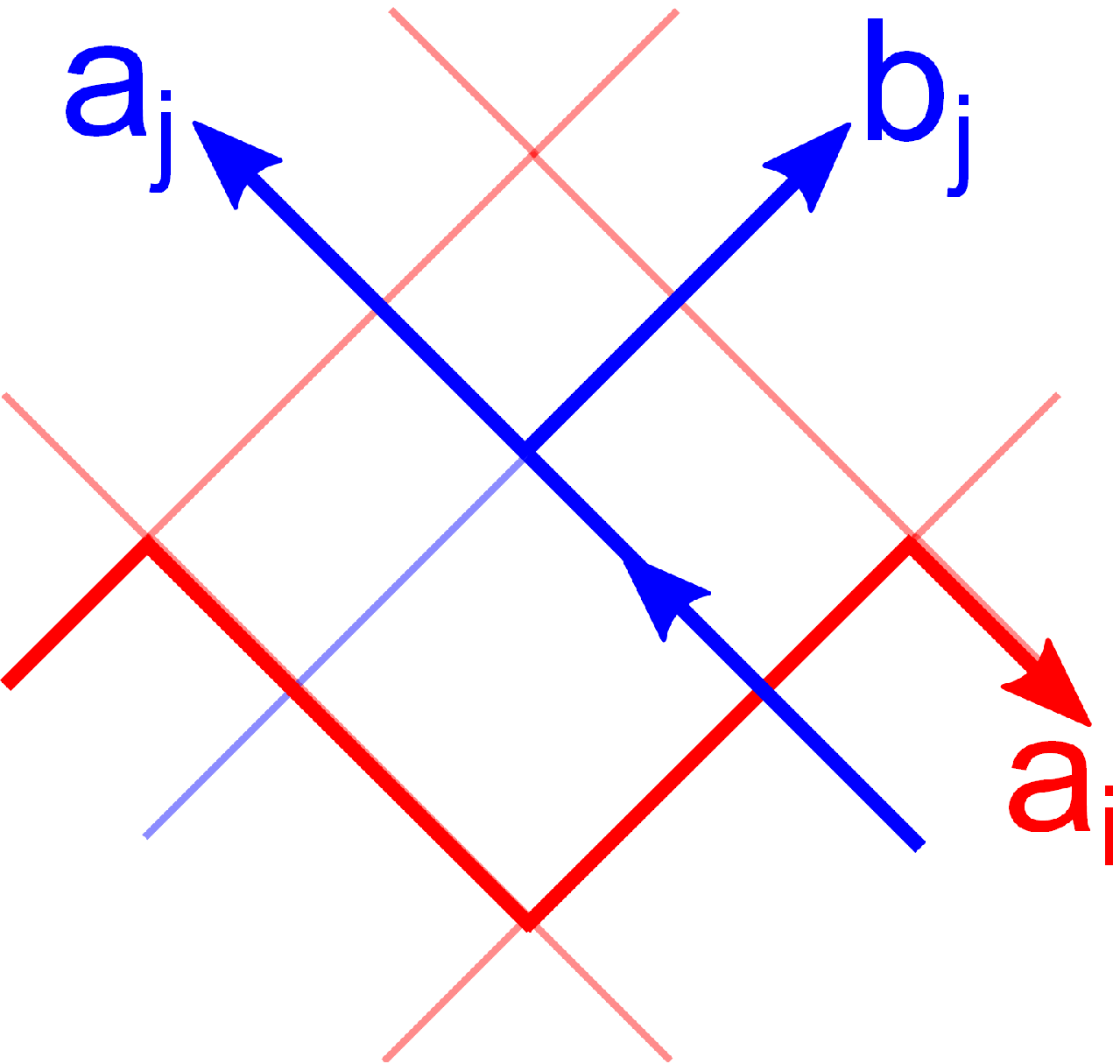}}\right\vert B_p^i, \label{5eqs} \\
	\left\langle \raisebox{-0.25in}{\includegraphics[height=0.65in]{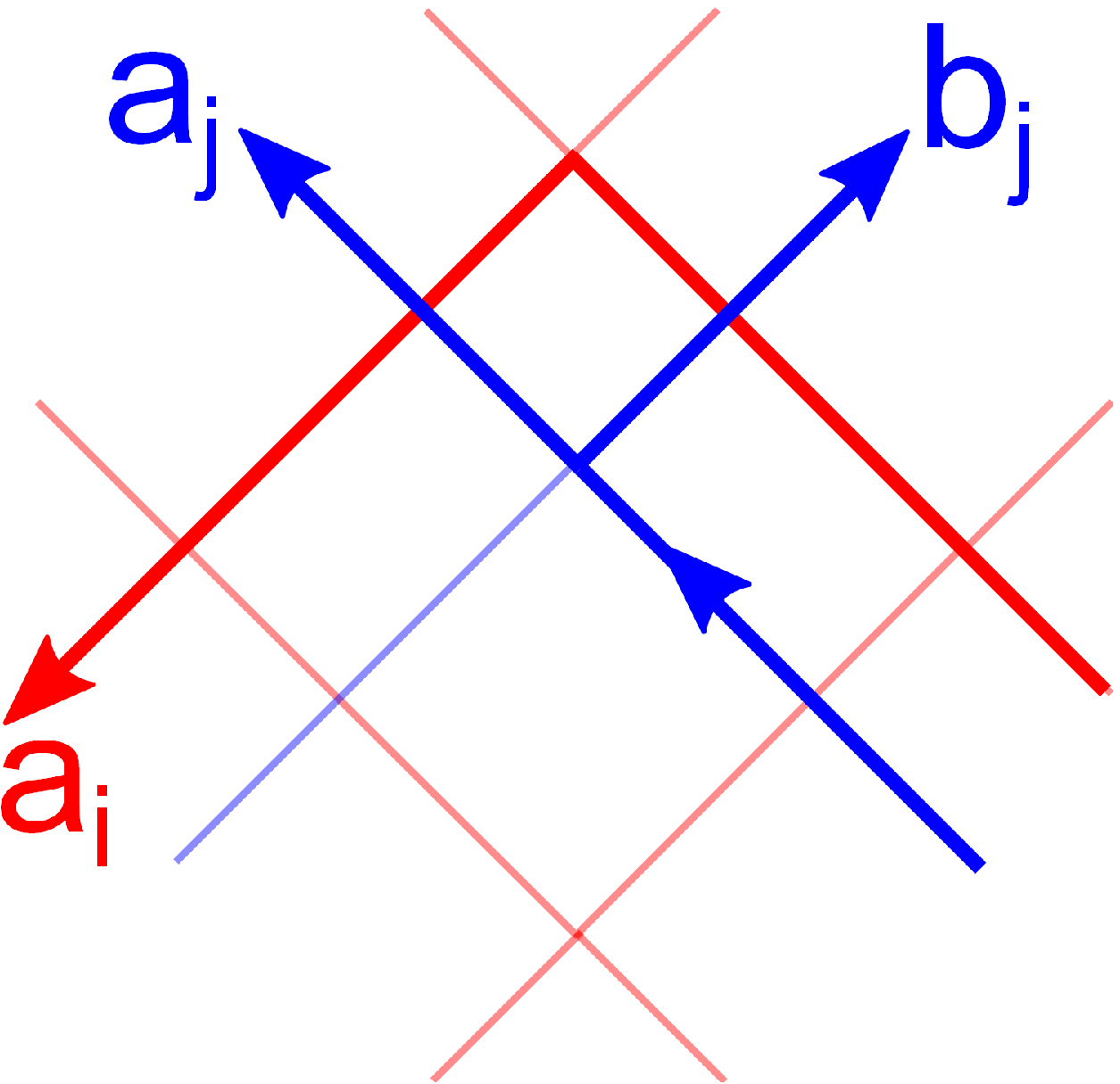}}\right\vert B_p^i
	&=\bar{F}^{(2)}_{a_i}(a_j,b_j)\left\langle \raisebox{-0.25in}{\includegraphics[height=0.65in]{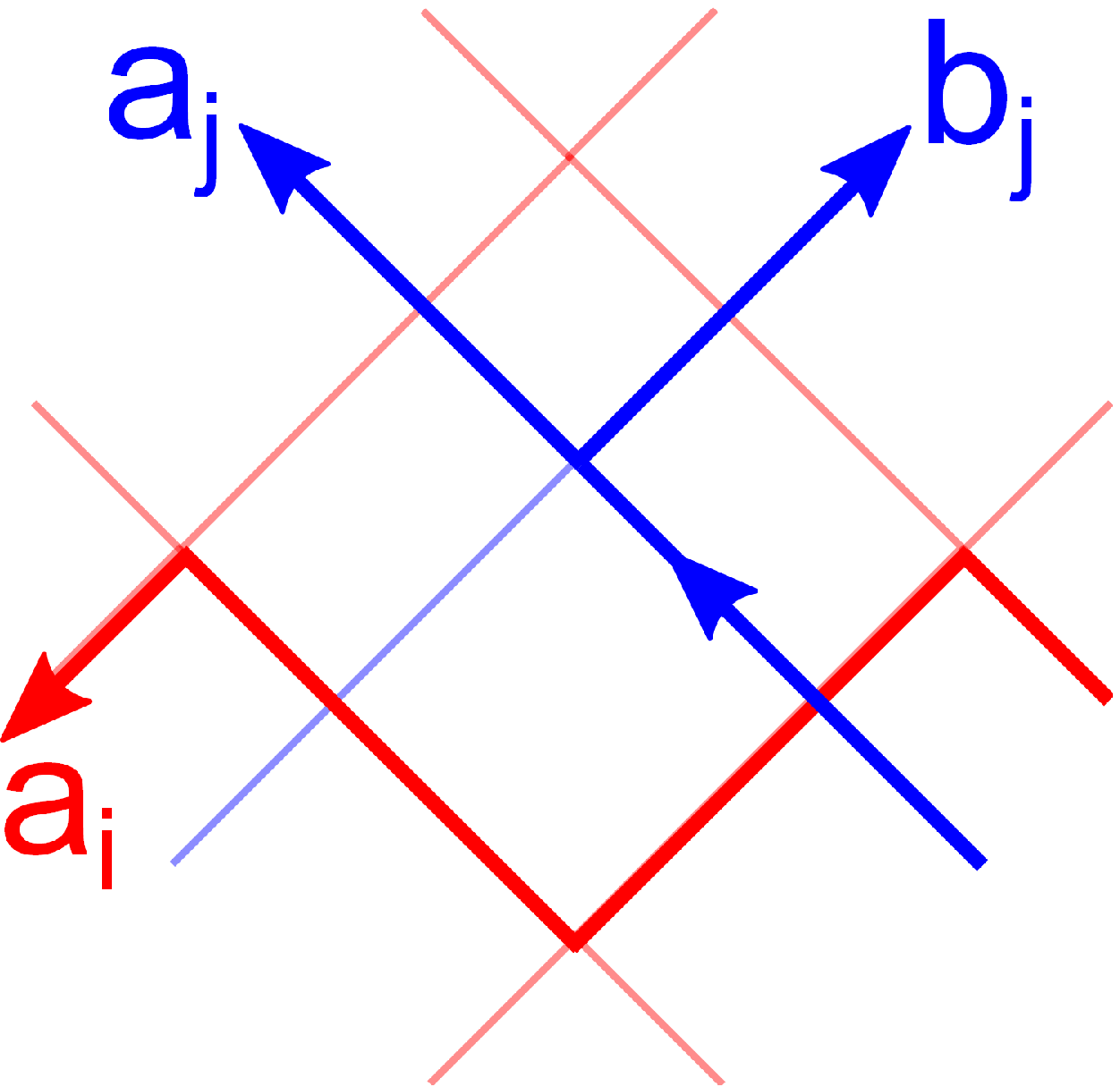}}\right\vert B_p^i, \notag \\
	\left\langle \raisebox{-0.25in}{\includegraphics[height=0.65in]{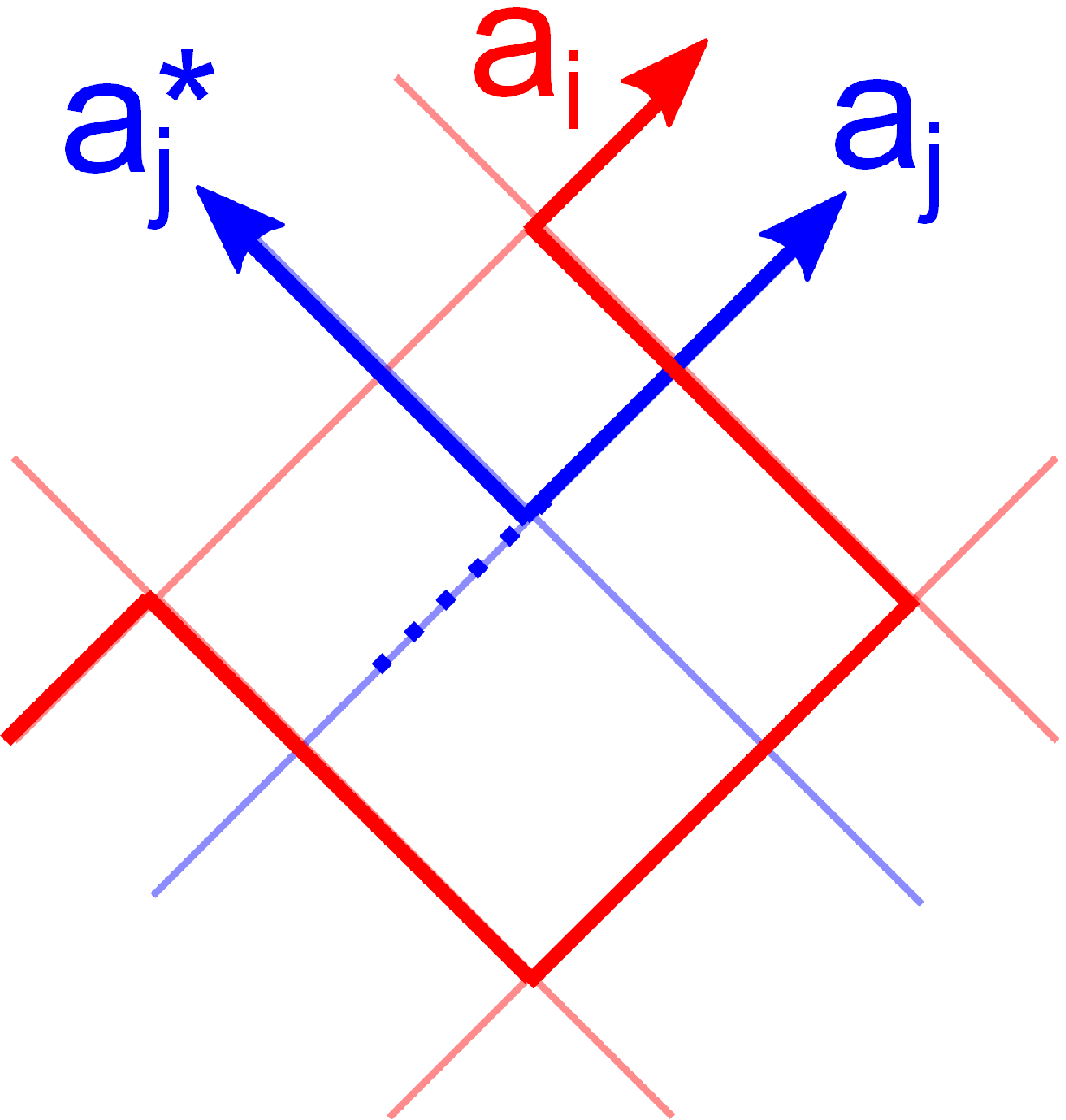}}\right\vert B_p^i
	&=\kappa_{a_i}(a_j)\left\langle \raisebox{-0.25in}{\includegraphics[height=0.65in]{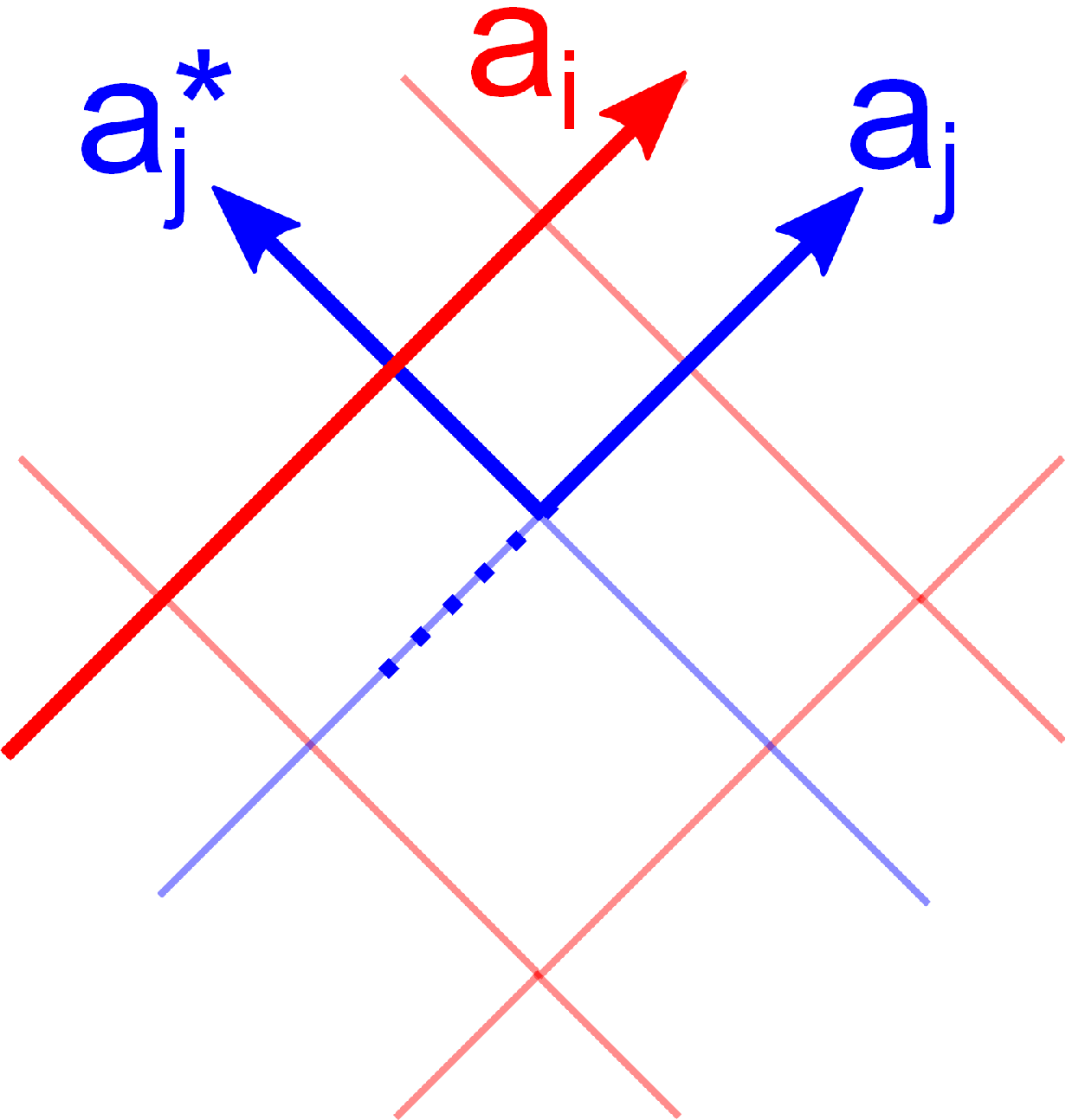}}\right\vert B_p^i, \notag \\
	\left\langle \raisebox{-0.25in}{\includegraphics[height=0.65in]{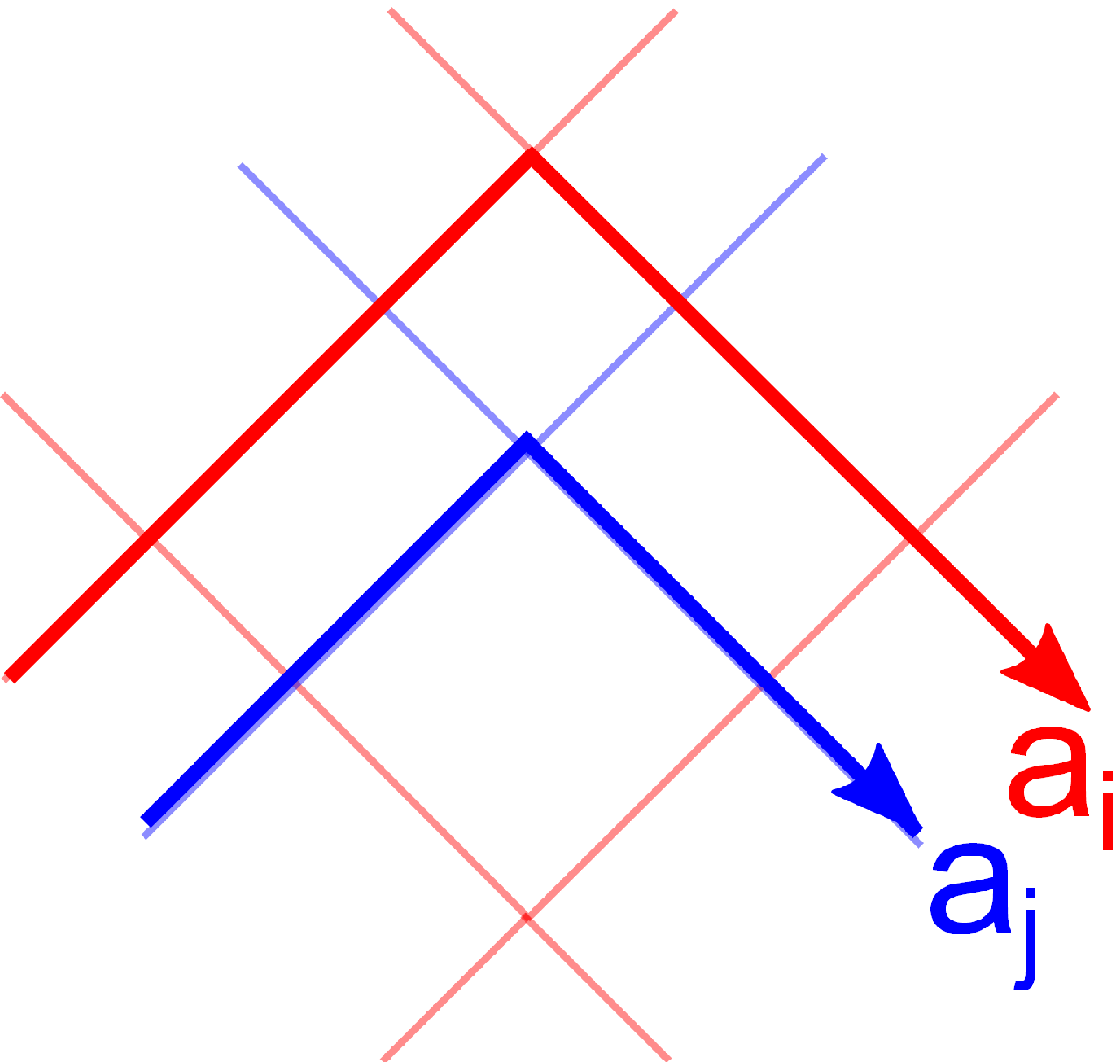}}\right\vert B_p^i
	&=\eta_{a_i}(a_j)\left\langle \raisebox{-0.25in}{\includegraphics[height=0.65in]{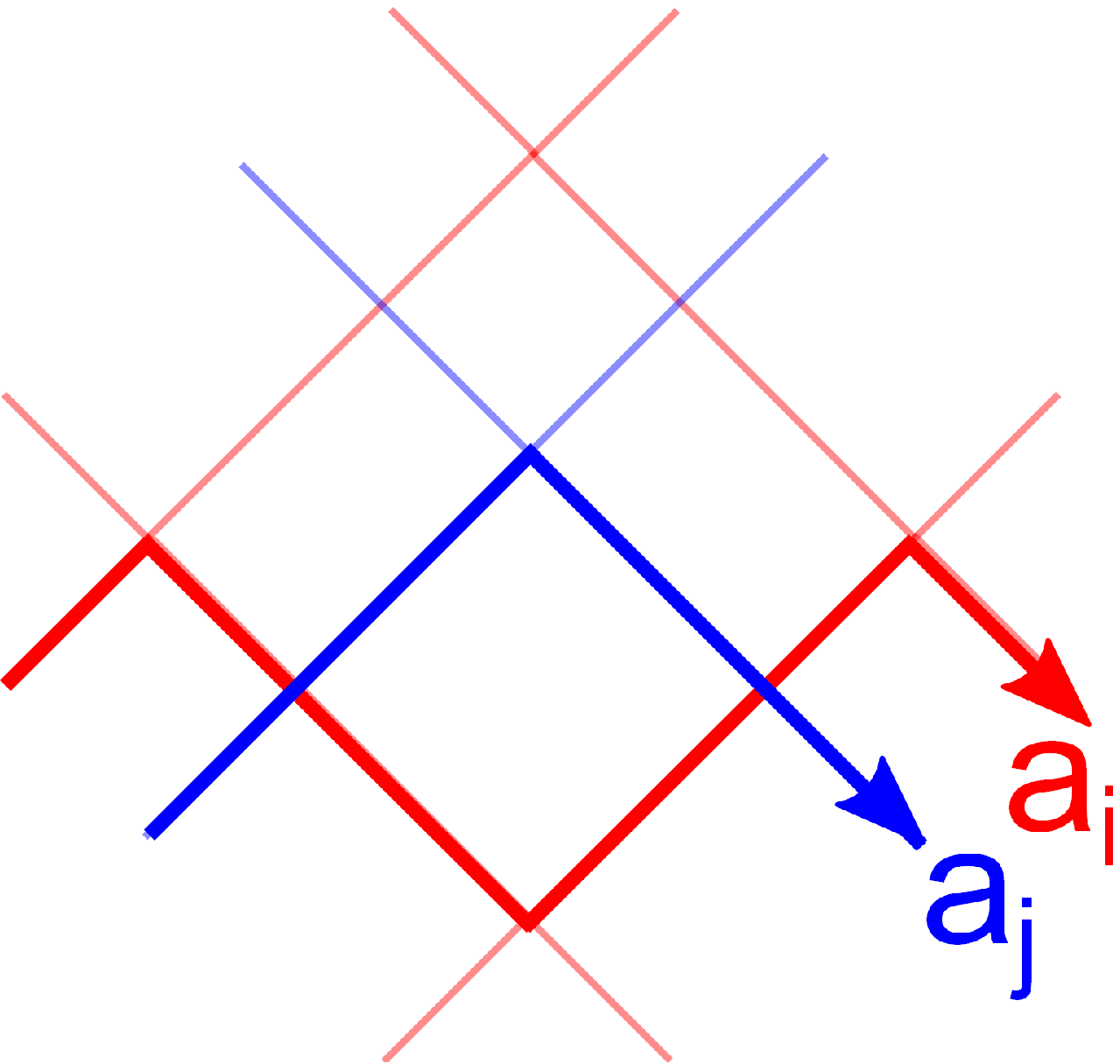}}\right\vert B_p^i, \notag \\
	\left\langle \raisebox{-0.25in}{\includegraphics[height=0.65in]{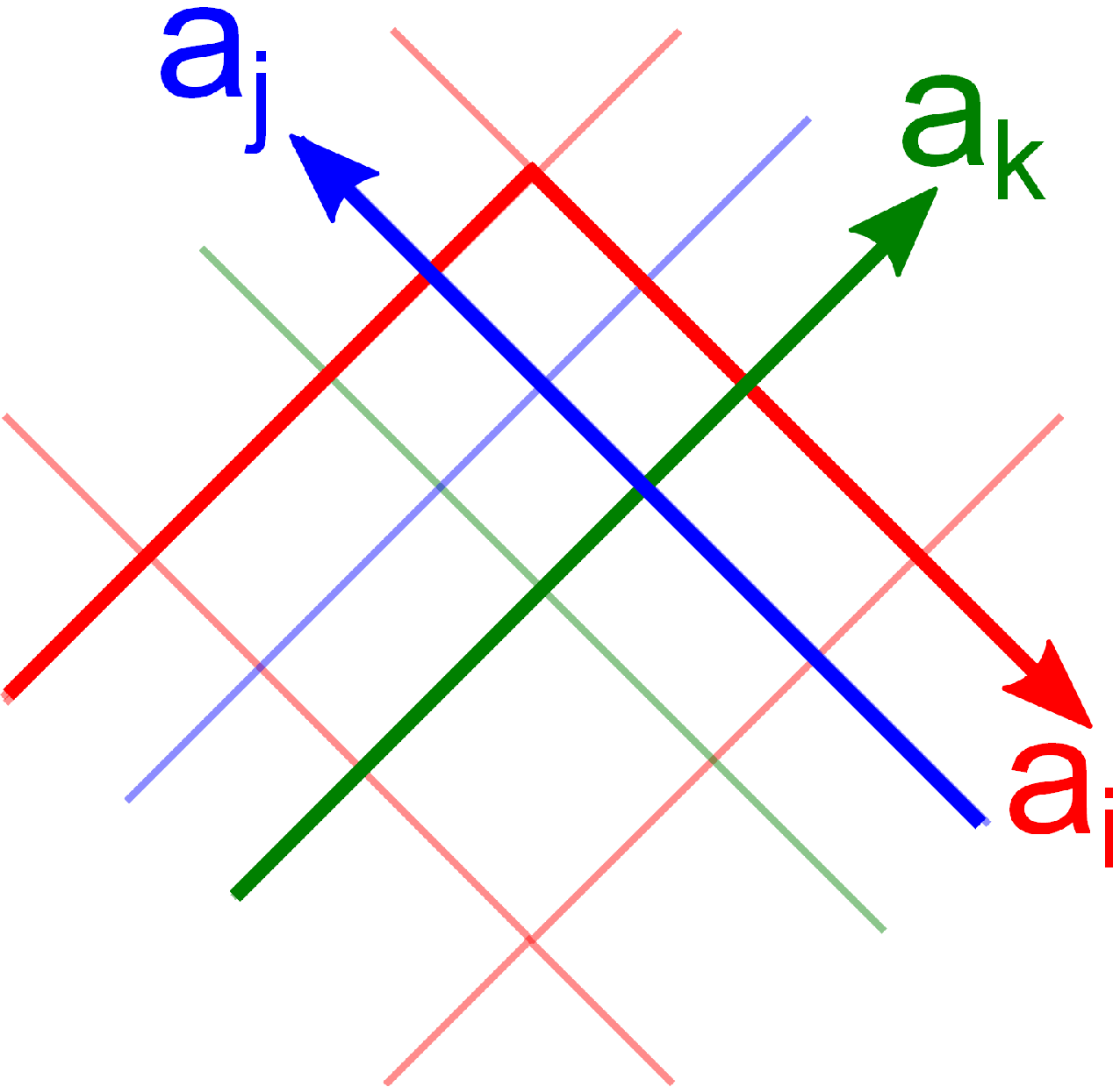}}\right\vert B_p^i
	&=F^{(3)}_{a_i}(a_j,a_k)\left\langle \raisebox{-0.25in}{\includegraphics[height=0.65in]{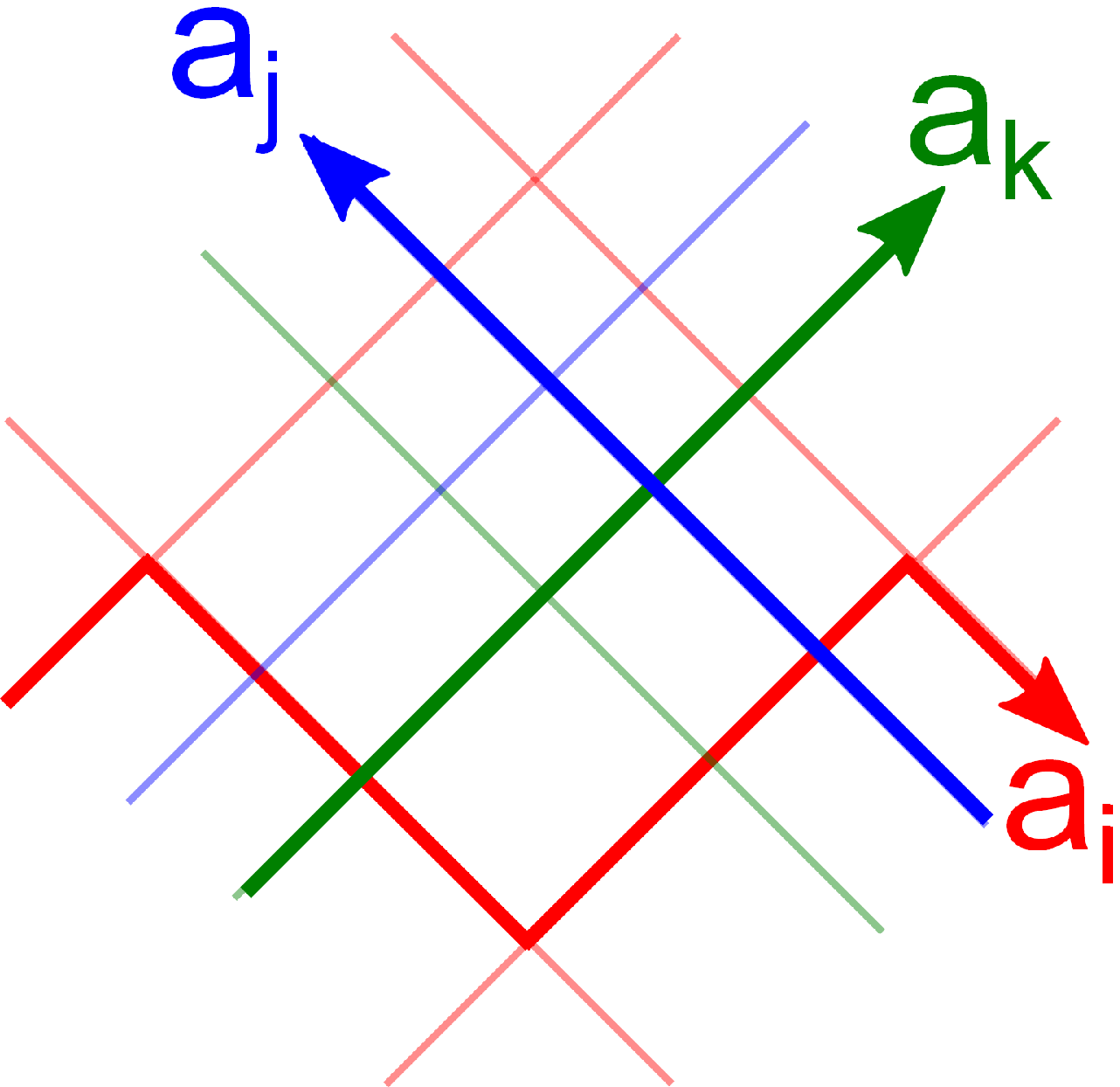}}\right\vert B_p^i \notag.
\end{align}
Multiplying the above equations by the ground state $|\Phi_{latt}\>$, one can immediately see that the ground state wave function $\Phi_{latt}(X)=\<X|\Phi_{latt}\>$ satisfies the local rules (\ref{rule2a}--\ref{rule2d}).

The relations (\ref{5eqs}) can be shown by using the expression for the matrix elements of $B_p^{s_i}$ in (\ref{bps}) together with (\ref{sfeq1}--\ref{sfeq8}). Here we prove the first equation for example. First, we expand the left hand side as
\begin{align*}
	&\left\langle \raisebox{-0.25in}{\includegraphics[height=0.65in]{state1a.eps}}\right\vert B_p^i=\sum_{s_i}\frac{d_{s_i}}{|G|} 
	\left\langle \raisebox{-0.25in}{\includegraphics[height=0.65in]{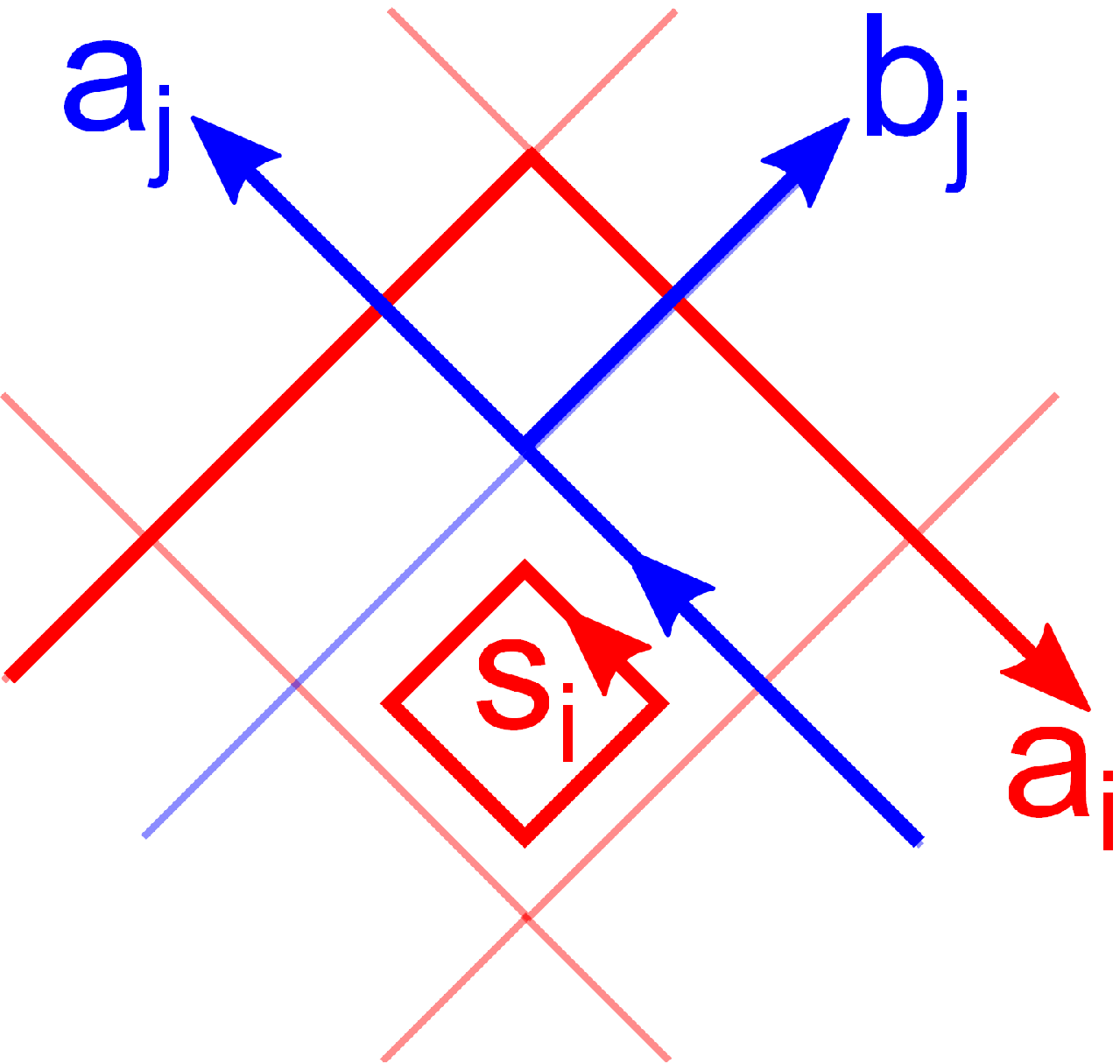}}\right\vert\\
	&=\sum_{s_i}\frac{d_{s_i}}{|G|}B^{(1),a_i^*a_i^*a_i^*000}_{a_i^{*\prime}a_i^{*\prime}a_i^{*\prime}s_is_is_i}(a_i^*00a_i00)\times \\
	&\eta_{s_i,a_j+b_j}\frac{F^{(2)_{a_j,s_ia_i*}}F^{(2)}_{b_j,s_ia_i^*}}{\bar{F}^{(2)}_{s_i,a_jb_j}}\kappa_{a_j,a_i}\kappa_{b_j,a_i} \left\langle \raisebox{-0.25in}{\includegraphics[height=0.65in]{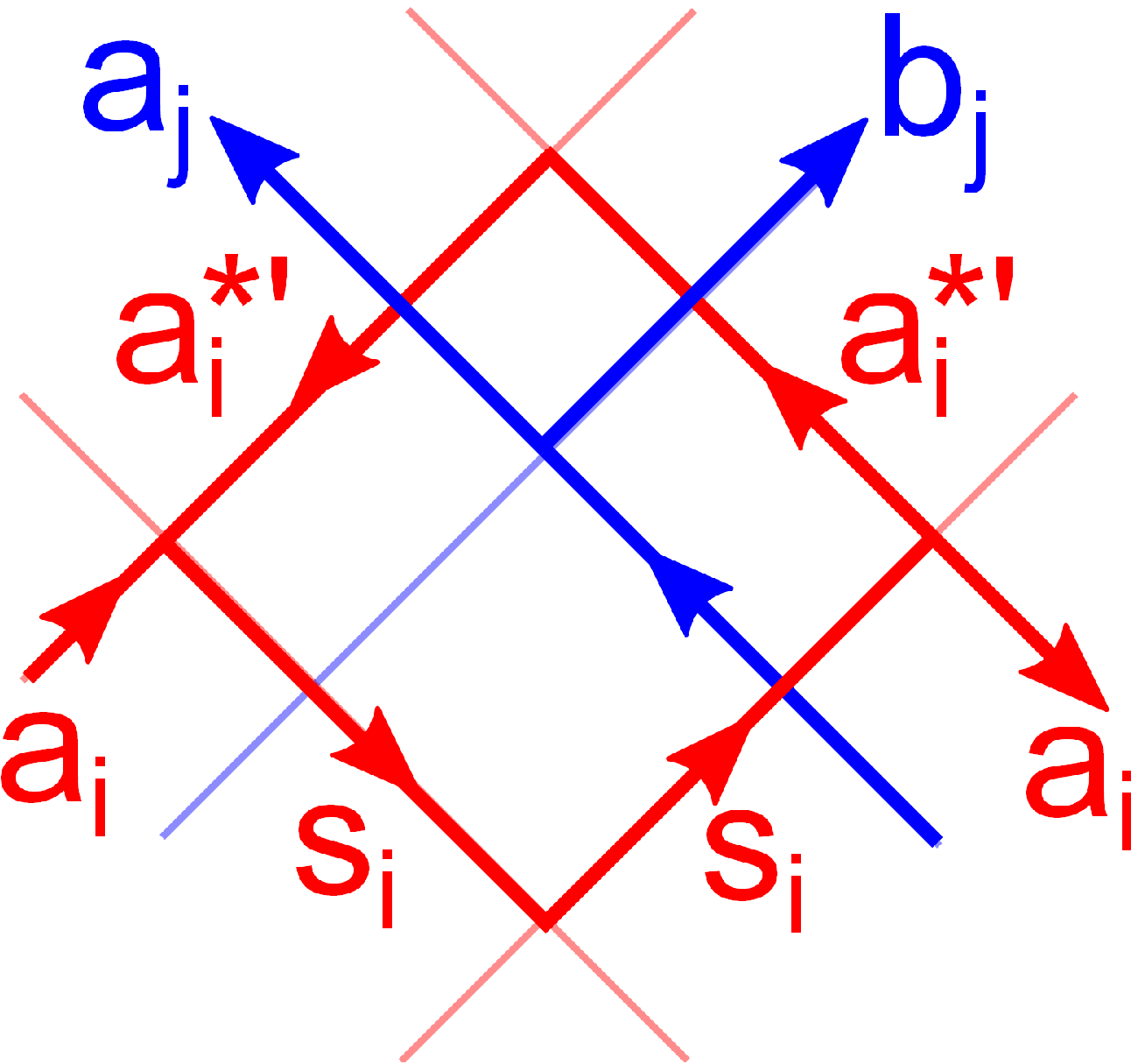}}\right\vert.
	\label{}
\end{align*}
Here $B_p^{(1)}$ is defined in (\ref{b1}) and $a'_i=a_i+s_i$. Similarly, the right hand side is given by
\begin{align*}
	&\left\langle \raisebox{-0.25in}{\includegraphics[height=0.65in]{state1b.eps}}\right\vert B_p^i=\sum_{s_i}\frac{d_{s_i}}{|G|} 
	\left\langle \raisebox{-0.25in}{\includegraphics[height=0.65in]{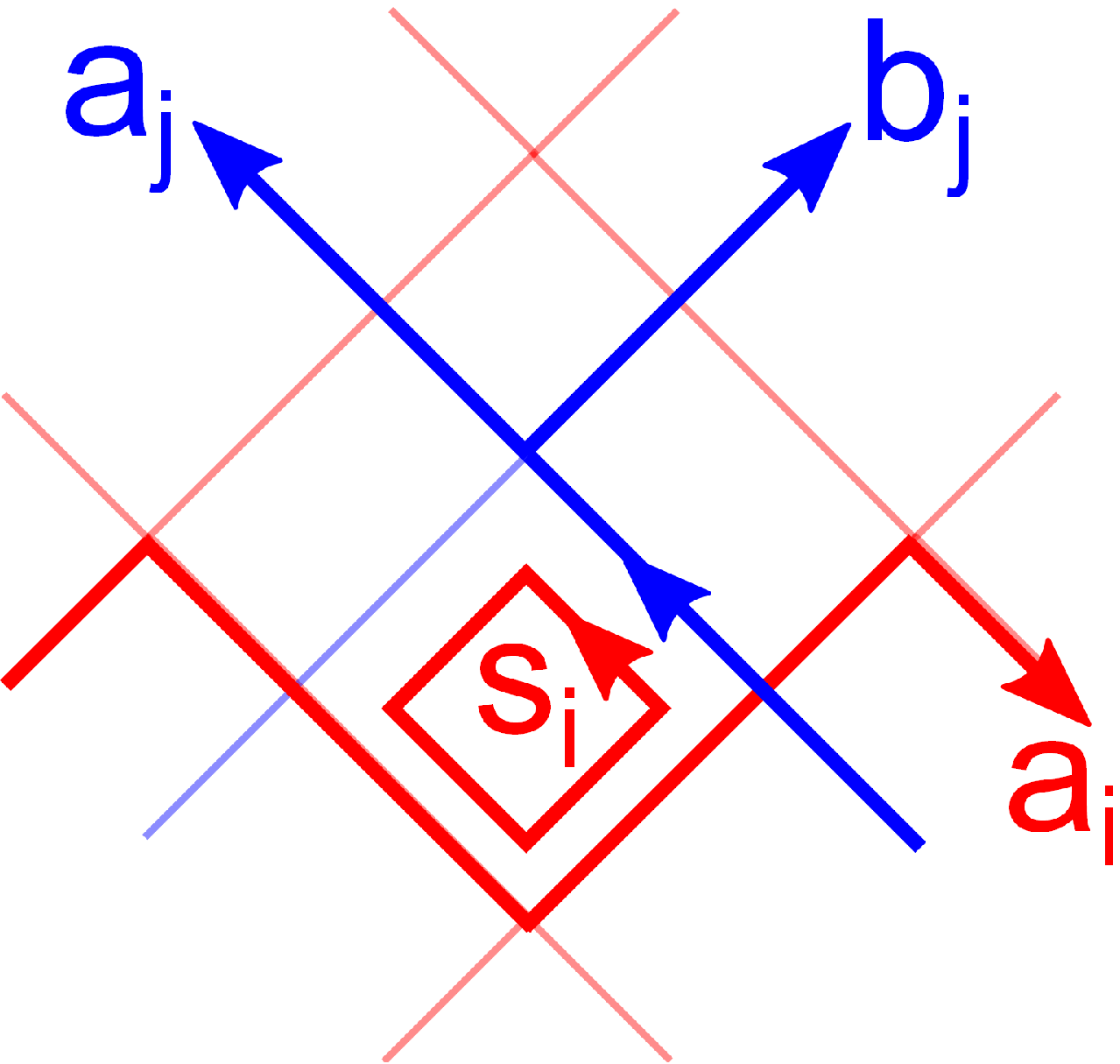}}\right\vert \\
	&=\sum_{s_i}\frac{d_{s_i}}{|G|}B^{(1),000a_ia_ia_i}_{s_is_is_ia_i'a_i'a_i'}(a_i^*00a_i00)\times \\
	&\eta_{s_i,a_j+b_j}\frac{\bar{F}^{(2)_{a_j+b_j,s_ia_i}}}{\bar{F}^{(2)}_{s_i,a_jb_j}} \left\langle \raisebox{-0.25in}{\includegraphics[height=0.65in]{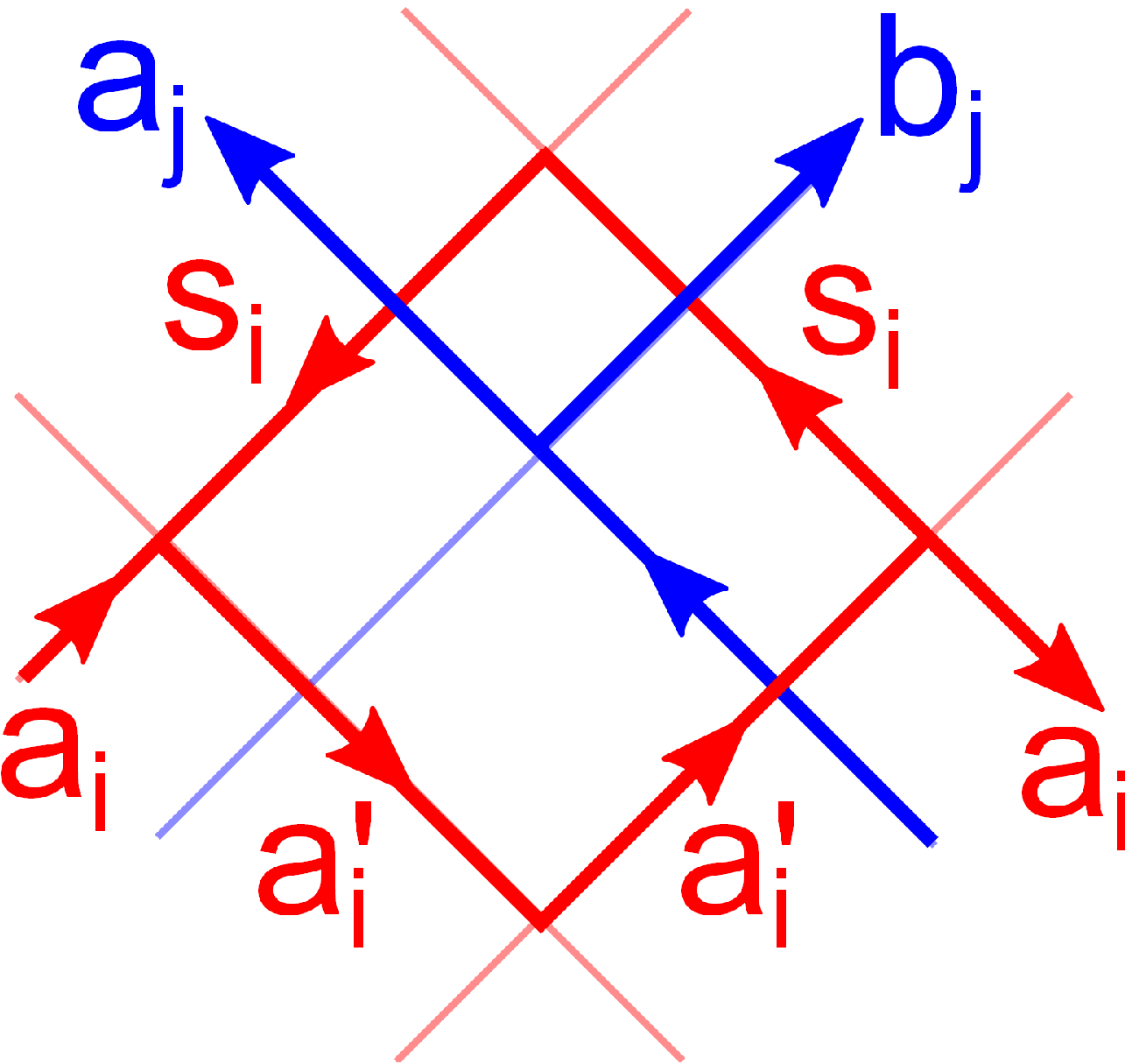}}\right\vert.
	\label{}
\end{align*}
By changing the dummy variables $s_i$ to $s_i+a_i$ in the first expression, we can see the final states for the two expressions are the same. We then compute the ratio of the two corresponding amplitudes. By using (\ref{sfeq1}--\ref{3f}), this ratio can be simplified to $F^{(2)}_{a_i}(a_j,b_j)$. This justifies the first equation in (\ref{5eqs}). The other equations can be shown in the same manner.

\section{Some useful identities}
In this section, we collect some useful identities which are used in previous appendices.

First, let $b_{j}=a_{j}^{\ast }=c_{j}^{\ast }$ in (\ref{sfeq1}) and get
\begin{equation}
F_{a_{i}}^{\left( 2\right) }\left( a_{j},a_{j}^{\ast }\right)
=F_{a_{i}}^{\left( 2\right) }\left( a_{j}^{\ast },a_{j}\right) .
\label{eq1}
\end{equation}%

Second, from (\ref{sfeq4}) and (\ref{eq1}), we get%
\begin{equation}
\frac{\eta _{x}\left( a^{\ast }\right) }{\eta _{x}\left( a\right) }=\frac{%
\kappa _{x}\left( a^{\ast }\right) }{\kappa _{x}\left( a\right) }.
\label{eq2}
\end{equation}%

Third, let $b_j=a_j^*$ in (\ref{fbar}), use (\ref{sfeq4}) to express $F^{(2)}$ in terms of $\eta,\kappa$ and finally use (\ref{sfeq2}) to simplify the expression. We get the analog of (\ref{sfeq4}) for $\bar{F}^{(2)}$:
\begin{align}
\kappa _{a_{i}}^{-1}\left( a_{j}^{\ast }\right) & =\bar{F}_{a_{i}}^{\left(
2\right) }\left( a_{j}^{\ast },a_{j}\right) \eta _{a_{i}}\left( a_{j}\right).
\label{eq3}
\end{align}%

Finally, let $b_i=a_i^*$ in (\ref{3f}) and use (\ref{eq3}) to express $\bar{F}^{(2)}$ in terms of $\eta,\kappa$ and finally use (\ref{sfeq2}) to simplify the expression. We get an alternative expression for $\bar{F}^{(2)}$ (c.f. Eq. (\ref{fbar})):
\begin{align}
\bar{F}_{a_{i}}^{\left( 2\right) }\left( a_{j},b_{j}\right) 
=F_{a_{i}}^{\left( 2\right) -1}\left( a_{j},b_{j}\right) \frac{\eta
_{a_{i}}\left( a_{j}+b_{j}\right) }{\eta _{a_{i}}\left( a_{j}\right) \eta
_{a_{i}}\left( b_{j}\right) }.
\label{eq4}
\end{align}
\bibliography{stringnet2}
\end{document}